\documentstyle[12pt,epsfig]{report}

\title{Microscopic Models for Ultrarelativistic Heavy Ion Collisions}

\author{S.~A.~Bass, M.~Belkacem, M.~Bleicher, M.~Brandstetter, L.~Bravina, 
C.~Ernst, \\ 
L.~Gerland, M.~Hofmann, S.~Hofmann, J.~Konopka, G.~Mao, L.~Neise, S.~Soff, \\
\smallskip
C.~Spieles, H.~Weber, L.~A.~Winckelmann, H.~St\"ocker  and W.~Greiner \\ 
\small{Institut f\"ur Theoretische Physik}\\
\small{Robert-Mayer-Strasse 10}\\
\small{Johann Wolfgang Goethe - Universit\"at}\\
\small{D-60054 Frankfurt am Main}\\
\\
\smallskip
Ch.~Hartnack and J.~Aichelin\\
\small{SUBATECH, Ecole des Mines     }\\
\small{4, rue Alfred Kastler           }\\
\small{La Chantrerie                     }\\
\small{F-44070 Nantes, Cedex 03, France  }\\
\\
\smallskip
N.~Amelin\\
\small{Joint Institute for Nuclear Research (JINR) Dubna}\\
\small{141980 Moscow region} \\
\small{Russia }
}

\begin{document}
\bibliographystyle{h-physrev}

\maketitle

\tableofcontents

\pagebreak

\chapter{Introduction}

The investigation of nuclear matter under extreme conditions
is one of the major research topics of nuclear and high
energy physics (a short and necessarily incomplete overview may be obtained
in refs.
\cite{scheid68a,chapline73a,shuryak80a,bjorken83a,csernai86a,stock86a,
stoecker86a,clare86a,schuermann87a,cassing90a,koCM96b,harris96a}).
The motivation for this is the unique opportunity to investigate the
equation of state of nuclear matter and the search for
phase transitions  (such as  the liquid gas or
the quark gluon plasma (QGP) phase transition) and the possible restoration
of chiral symmetry.
Also the general understanding of the dynamics of heavy
ion collisions over a vast energy range
from the Coulomb barrier (several MeV per nucleon) to the
highest energies currently available or planned for the future
is interesting in itself.
Here one can check the current understanding of the theory of
strong interactions (QCD) and different effective theories based
on hadronic/quark degrees of freedom. The intriguing role
of color coherence phenomena (transparency and opacity) for fluctuations,
stopping and charmonium production can be studied best at these energies.

The measurements above several hundreds of AMeV are done at 
experimental heavy-ion
facilities in three energy regimes: {\it i}) at about 1 AGeV at  
BEVALAC in Berkeley or SIS (SchwerIonenSynchrotron) at GSI-Darmstadt;
{\it ii}) the AGS (Alternating Gradient Synchrotron, in Brookhaven) 
energy regime at about 2-15 AGeV; {\it iii})
and the SPS (Super Proton Synchrotron, at CERN) energy regime of 40-200 AGeV. 
Much higher energies will be available in the
future with the Relativistic Heavy
Ion Collider (RHIC) in Brookhaven ($\sqrt{s} \approx$ 200 AGeV) and 
the Large Hadron Collider (LHC) at CERN ($\sqrt{s} \approx$ 6 ATeV).

Unfortunately, there exists presently no theoretical model that provides
a {\bf{consistent}} understanding of the reaction dynamics of heavy
ion collisions over the whole energy range.
One has to deal with quite different
reaction mechanisms from compound nucleus formation and deep
inelastic scattering at the Coulomb barrier over particle and
resonance production at intermediate energies up to string-excitation
and -fragmentation or parton scattering at (ultra-)relativistic energies.
While at low and intermediate energies, descriptions in terms of
hadrons (resonances) are appropriate, at high energies the
quark and gluon degrees of freedom enter the game.

One aim of this paper is the description of a theoretical
model that incorporates these different reaction
mechanisms and that is able to yield
observables. The model is dubbed UrQMD. Such a microscopic model is based on a
phase space description of the reaction. It contains a lot of
unknown parameters, which will have to be checked and fixed by
experimental data or by further model assumptions. This 
theoretical approach allows to pin down physical
ingredients that determine the values of certain observables like particle
abundances, collective flow of hadrons, rapidity distributions, 
cluster formation, etc. Thus certain
experimental observables can either be uniquely traced to a physical
parameter or if they are
described equally well by different physical assumptions.
This is of utmost importance if one wants to find evidence for new physical
phenomena like the phase transition to the quark gluon plasma. It must be
clarified whether the measured data could also be understood in terms of a
purely hadronic scenario or by assuming only a partial dynamical
deconfinement, which is not accompanied by complete equilibration.
One example is the vividly discussed suppression of $J/\Psi$ or 
$\Psi'$
production in Pb+Pb collisions as compared to p+p or p+A collisions, which
can be explained in different physical scenarios. 

It is one of the main tasks of theoretical heavy ion physics to
link experimental observables to the different phases
and manifestations of nuclear or -- in more general terms --
hadronic matter. For this, a detailed understanding of the dynamics
of heavy ion reactions is essential. 
Transport theory has played an important role in the interpretation
of experimental results and in predicting new interesting effects in
relativistic heavy ion reactions. 
It is particularly well suited for
the non-equilibrium situation, rapid time-dependence of the system
(even the use of the term ``state'' seems questionable), finite
size effects, non-homogeneity, $N$-body phase space, particle/resonance
production and freeze-out as well as collective dynamics.
Microscopic and macroscopic (hydrodynamical) transport 
models attempt to describe 
the full time-evolution
from the initial state of the heavy ion reaction (i.e. the two
colliding nuclei) up to the freeze-out of all initial and produced
particles after the reaction.
Simplified thermal equilibrium models neglect most of these dynamical effects, 
but make physical assumptions on the initial part
of the reaction, e.g. thermalization or plasma creation.

\begin{figure}[htb]
\centerline{\epsfig{figure=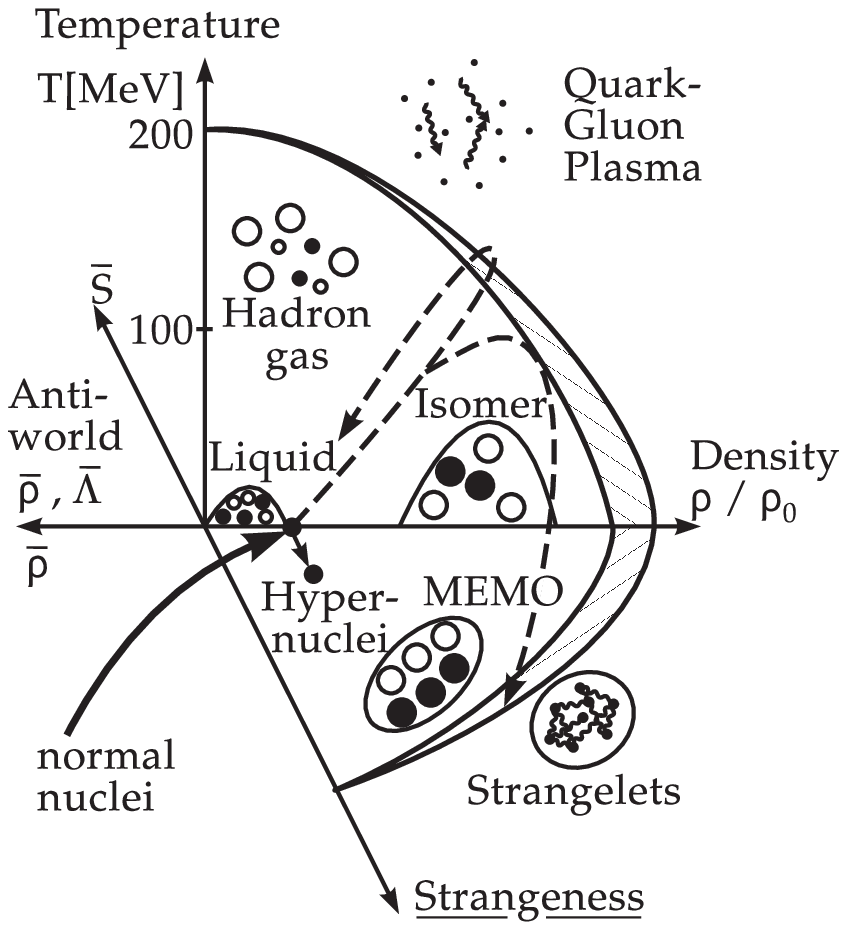,width=13cm}}
\caption{\label{phasedia}
Phase diagram of hadronic matter: Only the point at ground state density
is well known; the dashed lines show areas probed in the course of
heavy ion reactions.}
\end{figure}

It should be kept in mind that notions like `phase transition` are in
principle restricted to equilibrated systems in the thermodynamic limit
($N,V \to \infty, N/V=$const.). However, in heavy ion collisions 
we have to deal with
finite, dynamical systems, which makes the applicability of thermodynamic
models doubtful, even though they might successfully describe some global
observables like particle abundances. Nevertheless, thermodynamic
concepts are quite useful for a global characterization of the hot and
compressed hadronic matter. In figure~\ref{phasedia} 
we therefore present a conjectural
phase diagram of nuclear matter where only the point at 
T=0, $\rho_0=0.16$ fm$^{-3}$, realized
in nuclei, is well known. For low excitation energies 
(low densities $\rho<\rho_0$
and low temperatures $T<T_C \sim$ 15 MeV), one expects a phase 
transition from a
hadronic gas to liquefied nuclei.
This possibility results from the
resemblance of the nuclear interaction to that of atoms in gases. 
Both exhibit a short range repulsion and a
long range attraction. There are indications that this phase transition may
have been recently observed \cite{pochodzalla95a}. 

At higher densities (2-3 $\rho_0$) the possible
existence of density isomers was proposed \cite{stoecker78a,chapline73a}. 
At even higher densities and/or temperatures, 
one expects a phase transition or even a smooth cross over to the QGP.
However, neither the exact critical density and temperature are known,
nor the order of the transition. The latter is important, 
since a coexistence
region of plasma (strongly interacting quarks and gluons not confined to
well separated hadrons) and hadron gas might lead to observable effects in
the flow excitation function 
\cite{hofmann76a,stoecker86a,amelin92a,bravina94a,rischke95b} 
and other observables. 

Figure~\ref{phasedia} shows the generalization of the 
phase diagram to net-strangeness and net-antibaryon. Hypermatter and antimatter 
can only be produced in energetic heavy ion collisions. 
The production of (multi-)strange
nuclei containing one or more $\Lambda$-particles
\cite{juric73a,wilkinson59a,mondal79a,aokiS93a},
as well as the production of anti-nuclei (anti-matter cluster) have been 
measured \cite{rotondo96a,klingenberg96a}. 
Furthermore, the possibility of metastable exotic multi strange
 objects (MEMOS) \cite{schaffner92a,baltz94a} and strangelets 
\cite{ivanenko65a,bodmer71a,chin79a,farhi84a,witten84a,michel88a,greinerC87a}
might serve as observables for deconfinement during heavy ion
collisions. Ultra-relativistic heavy ion collisions offer the unique
opportunity to probe these different states of highly excited hot and
dense nuclear matter under controlled laboratory conditions.
The dashed lines in figure~\ref{phasedia} visualize how different 
regions in the phase
diagram may be probed simultaneously in the course of one heavy ion collision
event, due to distillation and local fluctuations \cite{spieles96b}.

Note that in a hot and compressed hadron gas
at high excitation energies we deal not only with nucleons
but with many different baryon ($>$ 50) and meson ($>$ 30) 
species.
With increasing mass, these particles are often unstable -- 
the width of the resonances increases.
Many of the presently available microscopic models treat
resonances as 
quasi-particles which
propagate and scatter during their lifetime and then decay into other
resonances or stable hadrons (with respect to strong interactions).
This concept becomes doubtful when the width of the resonances
is so large that no well defined quasi particle state
(i.e. width/mass $\approx 1$) exists.
However, the resonances turn
out to be important for the microscopic description of heavy
ion collisions at bombarding energies above 1 GeV per nucleon.
On the other hand, the concept of resonances cannot be used for internal
excitation energies above 2 GeV.
Here the quark degrees of freedom must be explicitly taken into account.
Therefore some of the models presently available, including the
UrQMD model presented in detail in this review, replace the resonances
by continuous string excitations for higher energies.

A microscopic dynamics description of heavy ion collisions is usually 
based on transport theory. Here a sequence of propagations of particles
is simulated numerically (representing baryons (resonances) and mesons with or without interaction
with a potential and subsequent scattering processes or decays).
The main ingredients in this description of heavy-ion collisions 
are the cross sections, the two-body potentials and decay widths.
It is obvious that -- at least in principal -- these models should be
obtained consistently from an underlying theory.
Since the particles propagate in a hot and dense and even not equilibrated
medium of highly excited hadrons, the properties of the particles might
change significantly.
Consequently, properties like effective masses, effective momenta,
in-medium cross sections and decay widths should be calculated for
the actual local situation in which the particle propagates.
Unfortunately this is very complicated and from the numerical point of
view also very time consuming, so that in most of the models drastic
approximations are made.

Furthermore, the basic requirements for the applicability of transport
theoretical models might not be fulfilled in certain realistic situations.
First of all the quasi particle limit, i.e. on shell propagation 
between subsequent scatterings, can be a bad approximation in strongly
interacting, dense matter.
Between two subsequent scatterings of a particle with time difference
$\Delta t$, the energy of the state of the particle can only be fixed up to
$\Delta E = \hbar / \Delta t$.
 For short collision times this uncertainty can become rather large
($\Delta t = 1 $ fm/c $\to \Delta E$ = 200 MeV).
Even without collisional broadening the quasi particles acquire a finite
life time, which requires the introduction of nontrivial spectral functions. 
Consequently one needs to introduce the energy of the 
quasi particles as an additional independent variable, i.e. the full 
8-dimensional phase space description of transport theory is required.
Secondly, transport theory relies on the description of the dynamics of
the system as a Markovian process.
 This means the particles completely 'forget' about their critical state
before each scattering.
At least in the energy regime of several GeV per nucleon the so called
memory effect does not seem to be negligible \cite{greinerC94a}.
Moreover, pure transport models describe the system by the one body
phase space distribution function, which does not contain any
information about correlations of the particles.
 These are, however, the most 
important for the description of fragment or cluster formation
(hypernuclei, antinuclei, etc.).
Such clusters can be obtained from the one body phase space distributions
only with the help of a statistical coalescence model.
In the UrQMD model described below, this shortcoming is avoided, since the
particles can interact by individual two body forces at least at low
energies.

Unfortunately, the generalization of such two body forces to the relativistic
region is not a simple task and is not incorporated in the present
UrQMD model.
In principle, the interaction must be mediated by fields, which are
propagated according to wave equations or one must make use of the
so called constraint Hamiltonian dynamics \cite{currie63a,sorge89a}.
Many other models avoid the propagation of fields using local
density approximation.

Finally, we would like to mention an additional problem arising from the
practical realization of transport models:
Even though the underlying transport equations are relativistically
invariant, the actual realization by a sequence of propagations, scatterings
and decays is not (see section~\ref{collcrit}).
This leads to a frame dependence of the observables which contributes
to the systematic errors of the respective model predictions.

This review is organized as follows:
Chapter~\ref{manybody} contains a brief introduction into 
the many-body theory of nuclear collisions. 
In chapter~\ref{urqmdkap} we discuss in great detail one specific
microscopic transport model, the Ultra-relativistic Quantum Molecular Dynamics
(UrQMD) model. Chapter~\ref{hotdense} focuses on the exploration
of hot and dense nuclear matter with microscopic transport approaches.
Different effects and observables are discussed and dependencies to
model parameters are pointed out. The main focus lies on baryon stopping, 
collective flow and particle production.
A brief summary and conclusions are found in chapter~\ref{conclusion}.

\newcommand{\e}{{\rm e}}
\renewcommand{\j}{{\rm j}}
\newcommand{\D}{{\rm d}}
\newcommand{\abs}{\hbox{abs}}
\renewcommand{\vec}[1]{\mbox{\boldmath$#1$}}
\newcommand{\Dp}[2]{\frac{\partial #1}{\partial #2}}
\newcommand{\nn}{\nonumber}
\newcommand{\be}{\begin{equation}}
\newcommand{\ee}{\end{equation}}
\newcommand{\ba}{\begin{array}}
\newcommand{\ea}{\end{array}}
\newcommand{\bt}{\begin{tabular}}
\newcommand{\et}{\end{tabular}}
\newcommand{\bea}{\begin{eqnarray}}
\newcommand{\eea}{\end{eqnarray}}
\newcommand{\gk}{\stackrel{>}{<}}
\newcommand{\dpar}[1]{\frac{\partial}{\partial #1}}
\newcommand{\grad}{\vec{\nabla}}
\chapter{Transport theory of nuclear collisions}
\label{manybody}

\section{Non-relativistic transport theory}
The ultimate goal of a theoretical description of heavy ion
collisions is a derivation and solution of 
equations of motion for the elementary degrees of freedom,
the quarks and gluons.
Since non perturbative dynamical solutions of
QCD are not possible yet, one is forced to separate the 
description into two parts:
 The low energy part ($\sqrt{s} <$ several GeV), 
where a description in terms of QCD-quasi particles 
(hadrons) is possible and a high energy part where the quark and
gluon degrees of freedom are treated explicitly.
For the low energy part one can derive solvable equations of
motion from phenomenological effective Lagrangians, if strong but 
controllable approximations are made.
This means that at least in principle one knows what has been neglected in the
approximation scheme as long as one accepts an effective Lagrangian with
baryons and mesons as degrees of freedom as a starting point.
One example is the $\sigma - \omega -$ model \cite{duerr56a,walecka74a}, 
where the interaction of
nucleons is described by the exchange of a scalar and a vector meson.
While the $\omega$-meson has a clear physical significance, it is  not quite
clear if the $\sigma$-meson can be interpreted as a real physical particle or
rather must be taken as an effective description of the interaction.
This model must be extended to include other baryonic and mesonic degrees of
freedom (e.g. $\Delta$, $N^*$, $\pi$, $K$, $\rho$ etc.) since already at
moderate energies of several hundred MeV per nucleon a significant
production of new particles sets in (mainly pions but also other
(pseu\-do-)scalar and (pseu\-do-)vec\-tor mesons as well as a lot of baryon resonances
and even anti-baryons).
The general theoretical framework for the derivation of practically solvable
equations of motion based on effective Lagrangians is transport
theory.
The basic approximation which is made here is the description of the
dynamics in terms of a semi-classical single particle phase space distribution
instead of $N$-body non-equilibrium Green's functions.

 In order to sketch the corresponding approximation scheme, we start with a
simple non-relativistic potential model for nucleons and then later on simply
write down the proper relativistic generalizations.
 The Hamilton operator shall be given by
\begin{eqnarray}
H& =& \int {\rm d}^3 \vec{x} \; \psi_H^\dagger (\vec{x}_1\;,t)
\left( - \frac{\hbar^2}{2m} \Delta_1 \right)
\psi_H( \vec{x}_1\;,t)
   +\frac{1}{2} \int {\rm d}^3\vec{x}_1\; {\rm d}^3\vec{x}_2 \;
                  {\rm d}^3\vec{x}_1'\; {\rm d}^3\vec{x}_2'
\nonumber \\ & &
\langle \vec{x}_1\,,\,\vec{x}_2 \mid V \mid \vec{x}_1'\,,\,\vec{x}_2' \rangle \,
\psi_H^\dagger( \vec{x}_1\,,t)
\psi_H^\dagger( \vec{x}_2\,,t)
\psi_H ( \vec{x}_2'\,,t)
\psi_H ( \vec{x}_1'\,,t)\, ,
\label{eq1}
\end{eqnarray}
where the nucleon field operator $\psi_H^\dagger (\vec{x}_1,t)$ and
$\psi_H(\vec{x}_1\;,t)$ are given in the Heisenberg representation.
 For simplicity we neglect here the complications introduced by a spin or
isospin dependent interaction as well as three body forces, since we only want
to review the principle steps of the derivation.
For details we will refer to the corresponding literature as well as to the
derivations in the relativistic case.
The non-relativistic  non-equilibrium Green's functions ($g^c,g^a$) and
correlation functions ($g^>$,$g^<$) are defined by:
\begin{eqnarray}
g^c (\vec{1},\vec{1'})&=&\frac{1}{{\rm i} \hbar} \langle T^c \psi_H(\vec{1})\,\psi_H^\dagger(\vec{1'}) \rangle\,,
\nonumber \\
g^a (\vec{1},\vec{1'})&=&\frac{1}{{\rm i} \hbar} \langle T^a \psi_H(\vec{1})\,\psi_H^\dagger(\vec{1'}) \rangle\,,
\nonumber \\
g^> (\vec{1},\vec{1'})&=&\frac{1}{{\rm i} \hbar} \langle  \psi_H(\vec{1})\,\psi_H^\dagger(\vec{1'})\rangle\,,
\nonumber \\
g^< (\vec{1},\vec{1'})&=&-\frac{1}{{\rm i} \hbar} \langle \psi_H^\dagger(\vec{1'})\,\psi_H(\vec{1})\rangle\,,
\label{eq2}
\end{eqnarray}
with {\bf{1}} a short hand notation for (${\vec{x_1},t_1}$) and/or spin
or isospin indices and $T^c,T^a$ are the chronological and
anti-chronological time ordering
operators.
 In the relativistic case $\psi^\dagger$ must be replaced by $\bar{\psi}$.
 The four two point functions can be written in a more compact form as
a path ordered Green's function ( $2 \times 2$ matrix denoted by the underline)
 \begin{eqnarray}
   \underline{G}(\vec{1},\vec{1'}) = \frac{1}{{\rm i} \hbar} \langle P \, \psi(\vec{1}) \, 
\psi^\dagger(\vec{1'}) \rangle.
 \label{eq3}
 \end{eqnarray}
Here, additionally to the times $t_1,t_{1'}$, one has to specify if these times are
on the upper or lower branch of the integration contour, which yields the four
combinations given in Eq. (\ref{eq2}). Originally, the theory was formulated by
Schwinger \cite{schwinger61a} and the first application to transport problems
is due to Kadanoff and Baym \cite{kadanoff62a}.
For an introduction to the formalism of path ordered non equilibrium Green's 
functions we refer the reader to the works of Danielewicz 
\cite{danielewicz84b} and 
Botermans and Malfliet \cite{botermans90a}.
To some extent we will follow their work and notations.
From the equations of motion for the Heisenberg field operators one can derive
equations of motion for the Green's functions.
However these equations for the one body Green's functions couple to the two
body Green's functions and so on.
This set of coupled equations of motion is usually called the Martin Schwinger
hierarchy \cite{martin59a}.
In order to solve the problem completely one has to solve the whole set of 
coupled equations for all n-body Green's functions.
However, one can replace  in the equation for the one body Green's function
the unknown two body Green's function by the also unknown self-energy.
This leads to the Dyson equation in differential form:
\be
s(\vec{1}) \, \underline{G}(\vec{1},\vec{1'}) \,=\, \underline{\delta}(\vec{1}-\vec{1'})
+ \, \int\limits_C 
{\rm d}\vec{1''} \,\underline{\Sigma}(\vec{1},\vec{1''}) \,\underline{G}(\vec{1''},\vec{1'})
\label{eq4}
\ee
here $s(\vec{1})$ is the differential operator
\be
s(\vec{1}) \,=\, {\rm i} \hbar \partial_{t_1} + \frac{\hbar^2}{2m} \Delta_{x_1} \quad.
\ee
The $\delta$-function is generalized  to the integration contour
\be
\int\limits_C {\rm d}\vec{1} \,=\, \int\limits_{-\infty}^{\infty} {\rm d}\vec{1} \,
(\mbox{upper}) - \int\limits_{-\infty}^{\infty} {\rm d}\vec{1} \, (\mbox{lower})
\ee
such that 
$ \int\limits_C {\rm d}\vec{1'} \,\underline{\delta}(\vec{1}-\vec{1'}) \,\underline{F}(\vec{1'}) = 
\underline{F}(\vec{1})$, namely:
\be
\underline{\delta}(t_1 - t_2) = \left\{ 
\ba{cl}
\delta(t_1 -t_2) & \mbox{both $t_1$ and $t_2$ on the upper branch} \\
-\delta(t_1 -t_2) & \mbox{both $t_1$ and $t_2$ on the lower branch}\\
0 & \mbox{otherwise}
\ea
\right. \quad .
\ee
Of course the proper self-energy $\underline{\Sigma}(\vec{1},\vec{1'})$ is also
a two by two matrix defined on the contour. In more detail, equation~(\ref{eq4})
has four components:
\begin{eqnarray}
s(\vec{1})g^c(\vec{1},\vec{1'}) &=& \delta (\vec{1}-\vec{1'}) + \int d\vec{1''}\left[
\Sigma^c (\vec{1}\;,\;\vec{1''})\;g^c(\vec{1''},\vec{1'})\, - \,\Sigma^< (\vec{1},\vec{1''})\;g^>(\vec{1''},\vec{1'})
\right]   \,,
\nonumber \\
s(\vec{1})g^a(\vec{1},\vec{1'}) &=&-\delta (\vec{1}-\vec{1'}) + \int d\vec{1''}\left[
\Sigma^> (\vec{1}\;,\;\vec{1''})\;g^<(\vec{1''},\vec{1'})\, -\, \Sigma^a (\vec{1},\vec{1''})\;g^a(\vec{1''},\vec{1'})
\right]   \,,
\nonumber \\
s(\vec{1})g^>(\vec{1},\vec{1'}) &=& \int d\vec{1''}\left[
\Sigma^> (\vec{1}\;,\;\vec{1''})\;g^c(\vec{1''},\vec{1'})\, -\, \Sigma^a (\vec{1},\vec{1''})\;g^>(\vec{1''},\vec{1'})
\right]   \,,
\nonumber \\
s(\vec{1})g^<(\vec{1},\vec{1'}) &=& \int d\vec{1''}\left[
\Sigma^c (\vec{1}\;,\;\vec{1''})\;g^<(\vec{1''},\vec{1'})\, -\, \Sigma^< (\vec{1},\vec{1''})\;g^a(\vec{1''},\vec{1'})
\right]   \,,
\label{eq8}
\end{eqnarray}
which are known as Kadanoff-Baym equations \cite{kadanoff62a}.
In integral form equation~(\ref{eq4}) is the Dyson equation on the contour
\be
\underline{G}(\vec{1},\vec{1'}) \,=\, \underline{G_0}(\vec{1},\vec{1'}) + 
\int\limits_C {\rm d}\vec{1''} \, \int\limits_C {\rm d} \vec{1'''} \, 
\underline{G_0}(\vec{1},\vec{1'''}) \, \underline{\Sigma}(\vec{1'''},\vec{1''}) \underline{G}(\vec{1''},\vec{1})
\label{eq9}
\ee
where the zeroth order Green's function inverts $s(\vec{1})$
\be
s(\vec{1}) \, \underline{G_0}(\vec{1},\vec{1'}) \,=\, \underline{\delta}(\vec{1}- \vec{1'}) \quad.
\ee
The equations of motion~(\ref{eq4}) or~(\ref{eq8}), respectively, contain
in principle the unknown self energies which have to be calculated
by an appropriate approximation scheme, e.g. in perturbation theory
or in T- or G-matrix approximation. However, the corresponding expressions
will in turn contain the unknown Green's functions, such that one obtains
a coupled set of equations for the $\underline{G}$'s and 
$\underline{\Sigma}$'s.
E.g. in the T- or G-matrix approximation one calculates the self energies from
\be
\underline{\Sigma}(\vec{1},\vec{1'}) \,=\, - {\rm i} \hbar \int\limits_C {\rm d}\vec{2} \,
\int\limits_C {\rm d}\vec{2'} \, \left(\,
\langle \vec{1} \, \vec{2} | \underline{T} | \vec{1'} \, \vec{2'} \rangle
- \langle \vec{1} \, \vec{2} | \underline{T} | \vec{2''} \, \vec{1'} \rangle \, \right) \, 
\underline{G}(\vec{2''},\vec{2}^+)
\label{eq11}
\ee
where the T-matrix is defined by the integral equation:
\bea
\langle\vec{1} \, \vec{2} | \underline{T} | \vec{1'} \, \vec{2'} \rangle &=&
\langle \vec{1} \, \vec{2} | \underline{V} | \vec{1'} \, \vec{2'} \rangle \nonumber \\
&+& {\rm i} \hbar
\int\limits_C {\rm d}\vec{1''} \, {\rm d}\vec{2''} {\rm d}\vec{1'''} \, {\rm d}\vec{2'''} \,
\langle \vec{1} \, \vec{2} | \underline{V} | \vec{1'''} \, \vec{2'''} \rangle \,
\underline{G}(\vec{1''},\vec{1'''}) \, \underline{G}(\vec{2''},\vec{2'''}) \,
\langle \vec{1'''} \, \vec{2'''} | \underline{T} | \vec{1'} \, \vec{2'} \rangle
\label{eq12}
\eea
For the effective interaction in nuclear matter (G-matrix), one has
additionally to incorporate the Pauli blocking of the intermediate states where
the two particles or holes propagate from $\vec{2'''}$ to $\vec{2''}$ and from 
$\vec{1'''}$ to $\vec{1''}$, respectively. For details, we refer the reader to 
\cite{botermans90a,brockmann90a}. In the dynamical situation of heavy-ion
collisions, the problem is more difficult. At the beginning of the
reaction, one
does not have a single Fermi sphere at zero temperature as for nuclear matter
in the ground state, but rather one has to deal with two Fermi spheres
separated by the beam momentum. Calculations of the G-matrix for colliding
nuclear matter have so far not been carried out. However, the effects of the
dynamical situation on the mean-fields (real part of the self-energy) have been
estimated by the Munich group \cite{sehn96a}. Only a few attempts have been
made to include realistic G-matrices into a transport description, see e.g. the
work of the T\"ubingen group \cite{jaenicke92a}. In many other cases, free or
in medium corrected cross sections are used together with parameterizations of
the mean-field (e.g. different versions of the QMD approach) or the latter is
calculated in local density approximation (see below). In order to avoid
confusion with the propagator $\underline{G}(\vec{1},\vec{1'})$, we use the
term T-matrix or T instead of G-matrix or G.

\begin{figure}[h]
\centerline{\psfig{figure=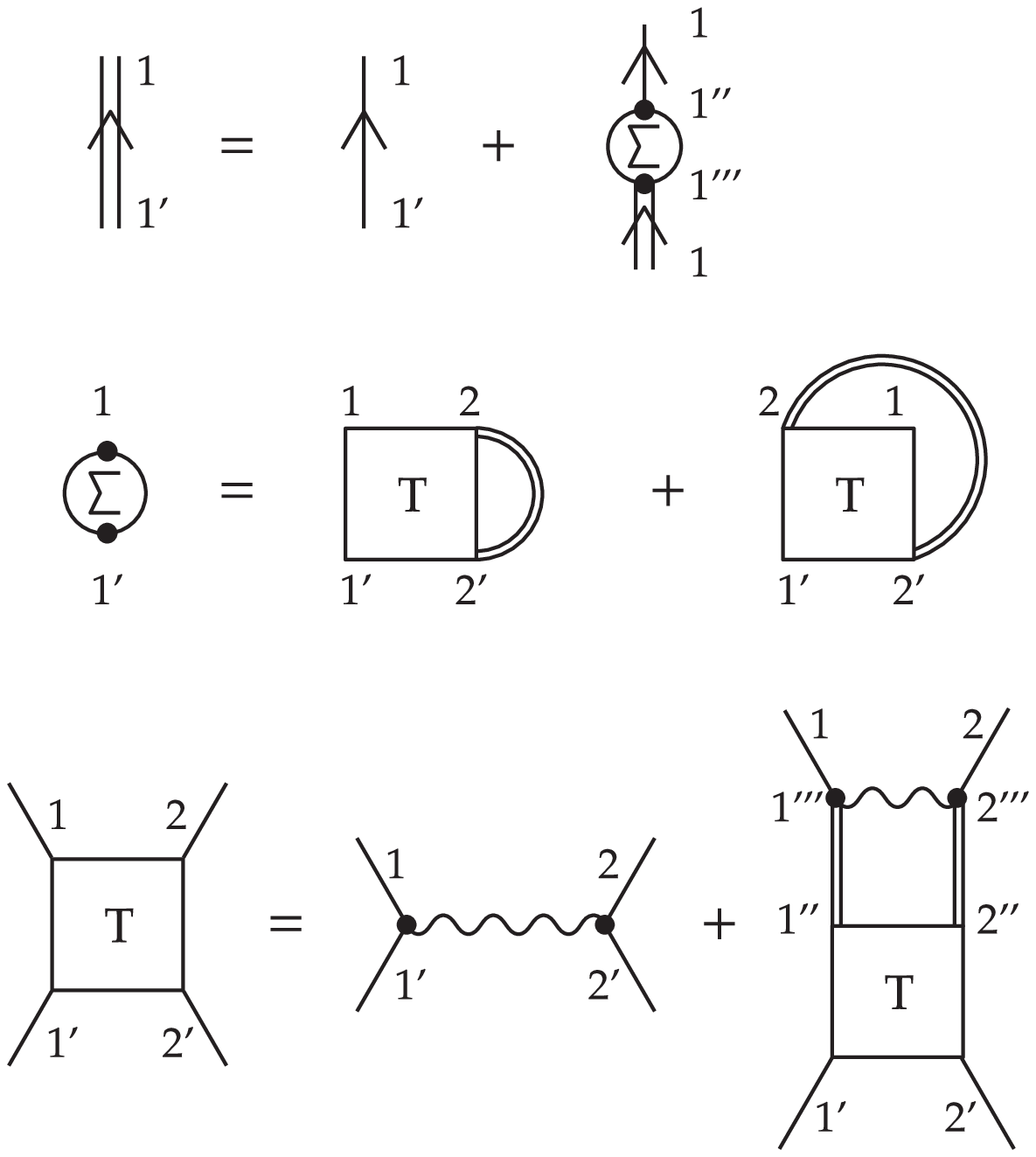,width=9cm}}
\caption{\label{diagramm1}
Feynman diagrams for the T-matrix approximation}
\end{figure}

Equation~(\ref{eq12}) corresponds to a resummation of all the ladder diagrams for the effective
interaction (T-matrix) in the medium, which is an appropriate approximation
for interactions with hard core, like the nucleon-nucleon interaction.
However, polarization insertions are neglected in this approximation. 
This will be of some importance for the relativistic generalization,
where pions (in general mesons), nucleons and resonances appear
(see below).
Unfortunately the coupled equations~(\ref{eq9}),~(\ref{eq11}) and~(\ref{eq12})
form a very complicated set of integral equations in an 8-dimensional
coordinate space. They might be visualized in a comprehensive way
by Feynman diagrams in figure~\ref{diagramm1}.

Here the double (single) line denotes the interacting (non-interacting)
Green's function in the T-matrix approximation, while the wiggled line
is the bare nucleon-nucleon interaction. $\Sigma$ and $T$ represent
the proper self-energy (two point function) and effective interaction
or T-matrix (four point function), respectively. Even in equilibrium
 nuclear matter calculations one usually has to make further 
approximations to come to solvable equations. E.g. one
approximates the two interacting propagators in the T-matrix equation
neglecting the hole contributions and/or one treats the self-consistency
only on the average etc.. Then one obtains e.g. the 
Brueckner-Hartree-Fock equations (in equilibrium!) or in the relativistic
version the Dirac-Brueckner equation (see e.g. \cite{brockmann90a}).

\begin{figure}[h]
\centerline{\psfig{figure=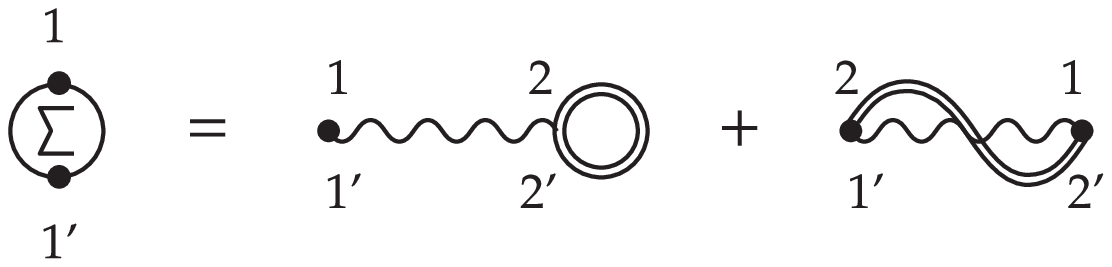,width=9cm}}
\caption{\label{diagramm2}
Feynman diagrams for the Hartree-Fock approximation}
\end{figure}

One important requirement for all kinds of approximations for the 
dynamical problem is the preservation of the basic conservation laws
of the underlying theory. It has been shown that the above mentioned
T-matrix approximation leads to energy, momentum and particle number
conserving transport equations \cite{botermans90a}. Furthermore it has
also been shown {\it{loc. cit.}} that the T-matrix approximation follows
from boundary conditions for the two body Green's function which 
correspond to Boltzmann's assumption of molecular chaos. This means
that the two body Green's function should factorize for times
in the infinite past, representing two non-interacting free particles.
In order to come to solvable approximations one can e.g. approximate
the T-matrix by the bare interaction alone ($T \approx V$). This
leads to the time dependent Hartree-Fock approach which will be
discussed below. In this approach the particles interact only via
a time dependent mean field generated by all particles. Direct
interactions (correlations) are neglected (see graphs in 
figure~\ref{diagramm2}).

\begin{figure}[h]
\centerline{\psfig{figure=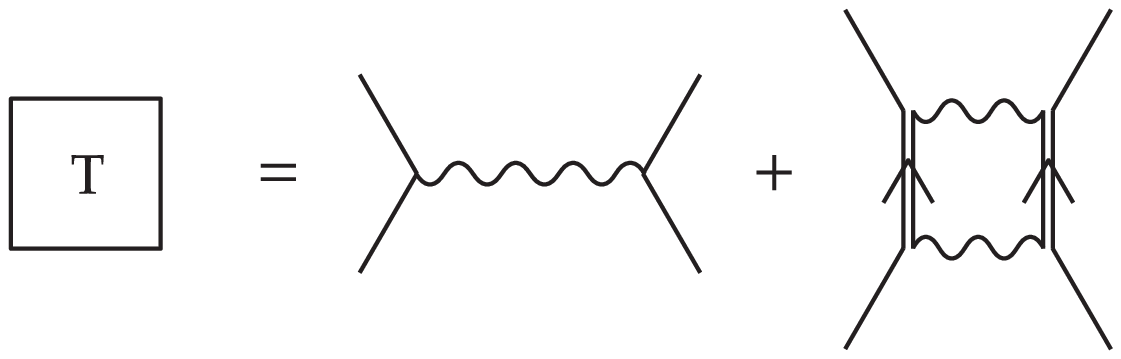,width=9cm}}
\caption{\label{diagramm3}
Feynman diagrams for the Born approximation}
\end{figure}

\begin{figure}[h]
\centerline{\psfig{figure=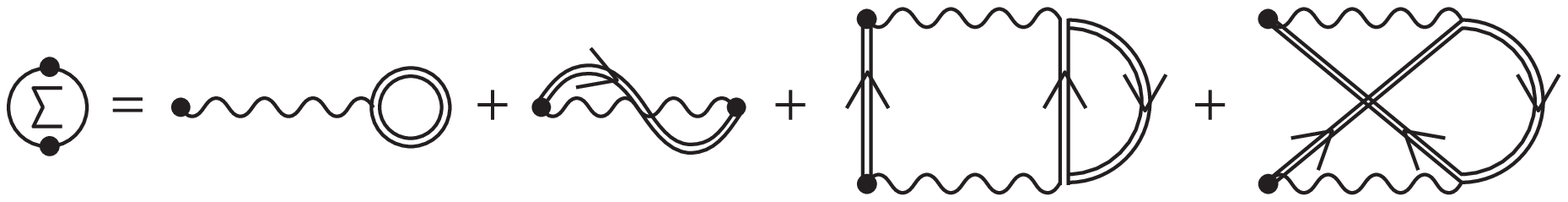,width=14cm}}
\caption{\label{diagramm4}
Feynman diagrams for the self-energy in the Born approximation}
\end{figure}

The self-consistency is still preserved, since the full Green's function
in Hartree-Fock approximation appears. One can give up the self-consistency 
to some extent but include higher orders in the interactions. The
Born approximation is obtained via the graphs in figure~\ref{diagramm3},
which leads to the self-energy as sketched in figure~\ref{diagramm4}.

However, in most cases the self consistent propagator has to be  replaced
by some further approximations for practical reasons. One can use e.g.
a propagator calculated for a local equilibrium situation, i.e. replacing
essentially the bare mass of the particle (here nucleons) by an effective
mass or effective dispersion relation, respectively, which is
calculated for an equilibrium system. Usually this is called 
dressing the propagator. In this way one can take also polarization
insertions into account, which are important for the relativistic
generalization. An analogous approximation can also be applied to the
self-energy itself. I.e. in the dynamical theory one uses the 
self-energy calculated from equilibrium theory (in T-matrix approximation)
\cite{jaenicke92a}. In this approach self-consistency is still preserved in the
equilibrium calculation of the effective interaction, but then this 
interaction is also used for the non equilibrium situation.

Only a few attempts have been made to take the non equilibrium situation
at the beginning of a high energy nucleus nucleus collision at least
in a stationary approximation into account, e.g. for two streams of 
nucleons inter-penetrating each other \cite{sehn96a,fuchs96a}.
It should be kept in mind, however, that in most works the approximations
mentioned above are combined with a gradient and quasi particle
approximation which then leads to transport equations for the
dynamical problem. In order to summarize the necessary steps to
derive a transport theory, we do not need to specify a certain
approximation scheme for the self-energy, but rather take this
quantity as calculable and therefore known.

We introduce the retarded and advanced Green's functions by
\begin{eqnarray}
g^+ &=& g^c- g^< \,=\, g^>-g^a \,,
\nonumber \\
g^-&=&g^c-g^> \,=\,g^<-g^a \,,
\label{eq13}
\end{eqnarray}
and analogous equations for the retarded and advanced self energies.
The advantage of these quantities is that the real part of the
retarded self-energy can be interpreted as the (in general non local)
mean field while the imaginary part describes the absorption or
finite life time of the quasi particles (dressed nucleons).

If the (anti)chronological quantities are eliminated from
equation~(\ref{eq8})
one obtains in obvious operator notation for the two point functions
\begin{eqnarray}
s(\vec{1})\,g^{\gk\ }& = &\Sigma^+ \,g^{\gk\ }\,+\,\Sigma^{\gk\ }\, g^- \,,
\label{e7} \vspace{3mm}
\\
s(\vec{1})\,g^{\pm }& =& 1\,+\,\Sigma^\pm\, g^{\pm } \,.
\label{e8}
\end{eqnarray}
If one adds or subtracts the corresponding hermitian conjugate 
equations one can separate the real and imaginary parts of these
equations, e.g.
\begin{eqnarray}
\left( s(\vec{1}) - s^*(\vec{1'}) \right)\, g^{\gk\ } &=&
 \left[    Re\,\Sigma^+\,,\,g^{\gk\ } \right]   _{-} \,+\,
 \left[    \Sigma^{\gk\ }\,,\,Re\,g^{+} \right]   _{-}
 \nonumber \\
& &+\,\frac{1}{2} \left[   \Sigma^>\,,\,g^< \right]   _{+} \,-\,
\frac{1}{2} \left[   \Sigma^<\,,\,g^> \right]   _{+}
\label{eq14}
\end{eqnarray}
\begin{eqnarray}
\left( s(\vec{1}) + s^*(\vec{1'}) \right)\, g^{\gk\ } &=&
 \left[    Re\,\Sigma^+\,,\,g^{\gk\ } \right]   _{+} \,+\,
 \left[    \Sigma^{\gk\ }\,,\,Re\,g^{+} \right]   _{+}
 \nonumber \\
& &+\,\frac{1}{2} \left[   \Sigma^>\,,\,g^< \right]   _{-} \,-\,
\frac{1}{2} \left[   \Sigma^<\,,\,g^> \right]   _{-}
\label{eq15}
\end{eqnarray}
where the brackets, $[   A\,,\,B]   _\pm$,
denote (anti) commutators of the operators.

Those equations for two point functions $A(\vec{1},\vec{1'})$ 
are now transformed into
phase space by the Wigner transform defined by
\begin{equation}
A_W(R\,,\,p)\,=\, \int d^4y \,e^{{\rm i}p\cdot y/\hbar}
\,A\left( R+\frac{1}{2}y
\,,\,R-\frac{1}{2}y\right) \quad.
\label{eq16}
\end{equation}
The main advantage of this transformation is that functions
sharply peaked with respect to the relative variable $y$
and smooth with respect to $R$ translate into smooth
phase space functions with respect to both arguments, $R$ and $p$.
In this way local operators have momentum independent
Wigner transforms etc..
To perform the Wigner transformations, equations~(\ref{eq14}) and~(\ref{eq15})
are rewritten in terms of the new variables $R$ and $y$ and the operation
$\int {\rm d}^4 y \exp({\rm i} p \cdot y/\hbar)$ 
is then applied to the equations. 
Derivatives with respect to $y$ can be expressed by multiplications
with $p$ by partial integration.
 \begin{eqnarray}
 {\rm i} \hbar \left[   \frac{\partial}{\partial T} + \frac{ \vec{p}}{m}
 \cdot \vec{\nabla}_R \right]   g^{\gk\ }_W \,&=&\,
 \left[   Re\,\Sigma^+\,,\,g^{\gk\ } \right]   _{-,W} \,+\,
 \left[   \Sigma^{\gk\ }\,,\,Re\,g^{+} \right]   _{-,W}
 \nonumber \\
 &  &
 \,+\,\frac{1}{2}\left[   \Sigma^>\,,\,g^{<} \right]   _{+,W}
 \,-\,\frac{1}{2}\left[   \Sigma^<\,,\,g^{>} \right]   _{+,W} \,,
 \label{e24}
 \\
 2\left[   \omega - \frac{ \vec{p}^{\,2} }{2m} + \frac{1}{4} \,
 \frac{\hbar^2}{2m} \Delta_R \right]   g^{\gk\ }_W \, &=&\,
 \left[   Re\,\Sigma^+\,,\,g^{\gk\ } \right]   _{+,W} \,+\,
 \left[   \Sigma^{\gk\ }\,,\,Re\,g^{+} \right]   _{+,W}
 \nonumber \\
 & & \,+\,
 \frac{1}{2}\left[   \Sigma^>\,,\,g^{<} \right]   _{-,W} \,-\,
 \frac{1}{2}\left[   \Sigma^<\,,\,g^{>} \right]   _{-,W} \,.
 \label{e25}
 \end{eqnarray}
Here the subscript $W$ at the (anti-) commutators denote the Wigner
transforms of the whole quantity.
In order to express the right hand side in terms of
Wigner transforms, we need the Wigner transform of Operator
products, which is in general given by
\begin{equation}
\left( AB \right)_W (R\,,\,p) \,=\, A_W(R\,,\,p)
\exp \left\{ \frac{{\rm i}\hbar}{2} \stackrel{\leftrightarrow}{\Lambda}
\right\} 
B_W (R\,,\,p) \quad.
\label{e20}
\end{equation}
The differential operator, $\stackrel{\leftrightarrow}{
\Lambda}$, in the exponent is given by $ \stackrel{ \leftarrow}{
\partial}_p \cdot \stackrel{ \rightarrow}{\partial}_R-$
 $ \stackrel{ \leftarrow}{
\partial}_R \cdot \stackrel{ \rightarrow}{\partial}_p$
in four vector notation and where the arrows indicate the direction
into which the partial derivatives act.

The essential simplification is now to neglect higher than first order
derivatives. 
More precisely: we require quantities of the form
\begin{equation}
\left|
\frac{{\rm i}\hbar}{2}\, \frac{\partial_p A_W \cdot \partial_R B_W }
{A_W B_W}\right|   \,\ll \,1 \hspace{1cm} \mbox{and} \hspace{1cm}
\left|
\frac{{\rm i}\hbar}{2}\, \frac{\partial_R A_W \cdot \partial_p B_W }
{A_W B_W}\right| \,\ll \,1
\label{eq17}
\end{equation}
to be small every where in phase space i.e. that the operators are sufficiently
diagonal (local) in space-time representation and slowly
varying with $R=(x+x')/2$, and that higher order derivatives can be omitted. 
It should be mentioned that the smoothness of the Wigner transformed quantities
in phase space can be enforced to some extent by averaging them over a
sufficiently large region in phase space  (of volume larger than $\hbar$), what
usually is called coarse graining.
Of course in doing this some of the information
contained in the exact Wigner transforms is lost, but
one gains in simplicity of the dynamical description and in the 
case of heavy ion collisions this is the only theoretical approximation which
leads to (numerically) solvable dynamical equations.
Other approaches like the Quantum Molecular Dynamics model 
are motivated by the form
of these equations but up to now these models are lacking a strict theoretical
derivation.
Now in a gradient expansion up to first order derivatives
the Wigner transforms of the
(anti) commutators, $[   A\,,\,B ]   _{\pm,W}$,
can be expressed by simple products or Poisson brackets of the Wigner
transformed operators by
\begin{equation}
[   A\,,\,B ]   _{+,W} =  2\,A_W\,B_W
\label{eq18}
\end{equation}
\begin{equation}
[   A\,,\,B ]   _{-,W}  =
 {\rm i}\hbar\{ A_W\,,\,B_W\} \equiv
{\rm i}\hbar( \partial_p A_W \cdot \partial_R
B_W-\partial_R A_W \cdot \partial_p B_W )
\label{eq19}
\end{equation}
Furthermore the left hand side of Eq. (\ref{e24}) can also be expressed
by a Poisson bracket namely ${\rm i}\hbar \{ \omega -\frac{p^2}{2m}\,,\,
g^{\gk\ }_W\}$, where $\omega$ denotes the zero component
of the momentum four vector, while the derivative on the left  hand side
of Eq. (\ref{e25})
is of second order and consequently should be neglected
in a gradient expansion up to first order. We arrive at
\begin{equation}
{\rm i}\hbar\left\{ \omega - \frac{\vec{p}\,^2}{2m}
 - Re\,\Sigma^+\,,\,g^{\gk\ }
\right\} -{\rm i}\hbar \left\{  \Sigma^{\gk\ }\,,\,Re\, g^+ \right\} \,=\,
\Sigma^> g^< \,-\, \Sigma^< g^>  \,,
\label{eq20}
\end{equation}
\begin{equation}
\left[   \omega - \frac{\vec{p}\,^2}{2m}
 -Re\,\Sigma^+ \right]   g^{\gk\ }
-\Sigma^{\gk\ }Re\,g^+ \,=\,
\frac{{\rm i}\hbar}{4} \left\{ \Sigma^>\,,\,g^< \right\} \,-\,
\frac{{\rm i}\hbar}{4} \left\{ \Sigma^<\,,\,g^> \right\} \,,
\label{eq21}
\end{equation}
where we dropped the index W for the Wigner transformed quantities.
These are of course just c-number functions in phase space and the
above equations of motion are therefore local, in contrast to the
exact equations. The analogous equations for the retarded Green's
functions are
\begin{equation}
{\rm i}\hbar\left\{ \omega - \frac{p^2}{2m} - \Sigma^\pm \,,\,g^\pm
\right\} \,=\, 0 \,,
\label{eq22}
\end{equation}
\begin{equation}
\left[   \omega - \frac{p^2}{2m} - \Sigma^\pm \right]
g^\pm\,=\,1 \,.
\label{eq23}
\end{equation}
The right hand side of Eq. (\ref{eq20}) is the collision term, with
the rate of loss minus the rate of gain (note an over all minus sign
implicitly contained in the Green's functions).

Especially the self energies $\Sigma^{>,<}$ can be interpreted as the rate of
scattering out (in) per occupied (empty) state respectively.
The correlation functions $g^{>,<}$ are connected to the local occupation number,
$f(R,p)$, and the local spectral function, $a(R,p)$, 
via the definitions of $a$ and $f$,
\begin{equation}
g^>\, = \, \frac{1}{{\rm i}\hbar} a (1-f) \,,
\hspace{3cm}
g^<\, = \,-\frac{1}{{\rm i}\hbar} a f \,.
\label{eq24}
\end{equation}
such that $a={\rm i} \hbar (g^> - g^< ) = {\rm i} \hbar ( g^+ - g^-)$.
The interpretation of $a$ and $f$ is motivated by the solutions for $g^{>,<}$ in a
non interacting equilibrium system of fermions, where $a(R,p) \sim
\delta(\omega-\frac{p^2}{2m})$ (independent of $R$) and $f(R,p)$ is the Fermi
distribution.
In this case the energy of the particles is fixed to their mass shell value.
On the left hand side of (\ref{eq20}), the first Poisson bracket describes
the motion of the phase space points propagated by a semi-classical
Hamiltonian
given by $p^2/2m+ \Re\,\Sigma^+$. 
The real part of the retarded self-energy can therefore be interpreted as a
space and momentum dependent potential.
The difference to a completely classical
motion is of course again the possibility of energies off the mass
shell. The second Poisson bracket on the left hand side of (\ref{eq20})
reflects the kinetic effects of the collisions, e.g. the effects
of the energy and momentum change in collisions on the effective mass
etc.
In contrast to the transport equations (\ref{eq20}), 
equations~(\ref{eq21})
which should be simultaneously fulfilled are not studied in great detail.
The reason is that the possibility of off shell propagation ($\omega \neq
\frac{p^2}{2m}+ \Re \Sigma^+$) in between collisions makes practical solutions
rather difficult.
Therefore one usually neglects all terms except the first one in equation
(\ref{eq21}), which is called the quasi particle approximation. In this
approximation the energy of the particles is fixed to their mass shell value
and therefore the energy is no independent variable anymore.
In order to see when this is justified we solve equation (\ref{eq23}) for the
retarded and advanced Green's function

\begin{equation}
g^\pm \,=\, \frac{1}{\omega-\frac{p^2}{2m}-\Sigma^\pm \pm
{\rm i} \epsilon} \,.
\label{eq25}
\end{equation}

The ${\rm i} \epsilon$ prescription is only needed if the imaginary part 
of the self
energy vanishes or is neglected. If this solution is put into (\ref{eq22}) the
latter equation is automatically fulfilled.
Subtracting the two equations (\ref{eq22}) from each other and doing the
same with (\ref{eq23}) we obtain two equations for the spectral function:
\begin{equation}
\{\omega-E\,,\,a\} \,=\,\{ \Gamma \,,\, \Re\,g^+\} \,,
\label{eq26}
\end{equation}
\begin{equation}
(\omega-E ) a \,=\, \Gamma \,\Re\,g^+  \,.
\label{eq27}
\end{equation}
Here we introduced
\begin{equation}
\Gamma \,=\, {\rm i}\hbar (\Sigma^>\,-\,\Sigma^<) = {\rm i}\hbar(\Sigma^+
\,-\,\Sigma^-)= -2\hbar \Im \,\Sigma^+
\label{eq28}
\end{equation}
in analogy to the spectral function and the abbreviation, $E$,
for $E(R,p)=\frac{p^2}{2m}+ \Re \,\Sigma^+(R,p)$.
Observing that from Eq. (\ref{eq25}) it follows that
\begin{equation}
\Re \,g^+ \,=\, \frac{ \omega-E}{(\omega-E)^2 + \left(
\frac{\Gamma}{2\hbar} \right)^2 }\,=\,
\frac{\omega -E}{ (\omega-E)^2 + \gamma^2}  \,,
\label{eq29}
\end{equation}
with $\gamma=\Gamma/2\hbar$, one can solve the algebraic
equation (\ref{eq27}) for $a$ with the result
\begin{equation}
a\,=\,\frac{\Gamma}{(\omega-E)^2 + \left( \frac{\Gamma}{2\hbar}
\right)^2} \,=\, 2\hbar \frac{\gamma}{(\omega-E)^2 +\gamma^2} \,.
\label{eq30}
\end{equation}
The spectral function is therefore (formally) a Lorentzian with the
width $\gamma=- \Im\,\Sigma^+$ around the mass shell $E=\frac{p^2}{2m}
+ \Re \,\Sigma^+$. It can easily be shown that the solution (\ref{eq30})
of (\ref{eq27}) is also a solution of (\ref{eq26}).
The quasi particle approximation is valid if the Lorentzian becomes a
$\delta$-function in the limit $\Gamma/E \ll 1$. The imaginary part of the
retarded self-energy varies from $-10$ MeV at $E=125$ MeV to $-25$ MeV
at $E=325$ MeV so that the fraction $\Gamma/E$ seems to be sufficiently
small for nonrelativistic energies. However, it should be kept in mind
that off shell propagation can cause changes in the order of more than
10\% in the scattering rates since in the collision term foldings
of four spectral functions occur, which increases the effective width
by a factor of $2-4$. However, if we adopt the quasi particle approximation
and if we neglect the momentum dependence of the real part of the 
retarded self-energy in the collision term, one immediately obtains
the so called Vlasov-Uehling-Uhlenbeck equation
(VUU, sometimes also called Boltzmann-Uehling-Uhlenbeck BUU):
\bea
\left[ \dpar{t}  \right. &+& \left. 
\left( \frac{\vec{p}_1}{m}+\grad_{p_1}\Re\Sigma^+ \right) \cdot \grad_{r_1} -
\grad_{r_1}\Re \Sigma^+ \cdot\grad_{p_1} \right] f_1(\vec{r}_1,\vec{p}_1,t)  
\nonumber \\
&=& \frac{2g}{m^2 (2\pi \hbar)^3}\int {\rm d}^3 \vec{p}_2 \int {\rm d}^3 
\vec{p}'_1 
\int {\rm d}^3 \vec{p}'_2 \, \delta^4(p_1+p_2-p'_1-p'_2)
 \frac{{\rm d}\sigma}{{\rm d}\Omega} 
\nonumber \\ 
&& \times \left[ f'_1 f'_2 (1-f_1)(1-f_2) -
 f_1 f_2 (1-f'_1)(1-f'_2) \right]
\label{eq31}
\eea
which especially in its relativistic version (see below) is a starting
point for many practical simulations of heavy ion collisions ($g=4$ denotes
the nucleon degeneracy factor). If the r.h.s of Eq.~(\ref{eq31}), the collision
integral, is neglected one obtains the Vlasov equation \cite{vlasov38a}. On the
other hand, if the real part of the retarded self-energy $\Re \Sigma^+$ is
neglected and if the so called Pauli blocking factors $(1-f)$ in the
collision integral are approximated by 1, one obtains the famous Boltzmann
equation. The modification of Boltzmann's collision integral by the Pauli
blocking factors, which take into account the Pauli principle for the final 
state of the two-body collisions, was first introduced by Nordheim
\cite{nordheim28a}, and latter worked out by Uehling and Uhlenbeck
\cite{uehling33a,uehling34a}.
Examples for the application of Eq.~(\ref{eq31}) to heavy ion collisions
can be found e.g. 
in \cite{molitoris84a,kruse85b,bertsch84a,aichelin85a,gregoire87a,wolf90a}.
It should be noted that the momentum
dependence of the self-energy causes some practical problems in the collision 
term, since scatterings change the momenta of the particles and therefore
also the potential energy. In principle this is correctly incorporated
in equation~(\ref{eq20}) where the energy $E=\vec{p}^2/2m + \Re \Sigma^+$ 
occurs in the spectral function ($\delta$-function in quasi particle limit) and
from this one can derive more general transport equations than (\ref{eq31}).
However, then additional factors (Z-factors) appear in the collision integral
which make the practical solution more difficult 
\cite{kadanoff62a,danielewicz84a}. Such collision terms are standard in other
fields \cite{landau56a,pines66a}.
In equation~(\ref{eq31}), only the free on shell energy
$\vec{p}^0=\vec{p}^2/2m$ appears in the
energy conserving $\delta$-function. Therefore in some practical computer
simulations the momentum or the energy respectively must be rescaled
to guarantee the conservation laws if the potential is momentum dependent.

The basic ingredients of transport equations like~(\ref{eq31}) are the 
real part of the retarded self-energy (the potential energy) and the
cross sections which both should be calculated consistently from the 
same basic interaction e.g. in G-matrix approximation, etc. Then the
cross section and the G-matrix are connected by
\be
\left| \langle \vec{p}_1 \, \vec{p}_2 | G | \vec{p}_1' \, \vec{p}_2' 
\rangle^{a.s.} \right|^2 \,=\, 4 g \, \frac{(2\pi)^5 \hbar^7}{m^2 V^3} \,
\frac{{\rm d}\sigma}{{\rm d}\Omega}(\vec{p}_1' - \vec{p}_1) 
\delta^3(\vec{p}_1 + \vec{p}_2 - \vec{p}_1' - \vec{p}_2')
\ee
where the plane waves are normalized to the volume $V$ \cite{jaenicke92a}.

In practice one solves equations like~(\ref{eq31}) by Monte-Carlo methods. 
First practical solutions of the Boltzmann equation (no self-energies, no
Pauli blocking) were done in the late 70's for hadron induced and heavy-ion
reactions
\cite{bondorf76a,gudima78a,yariv79a,cugnon80a,yariv81a,halbert81a}. Later, the
BUU simulations with test-particle methods were the first to include mean
potentials and Pauli blocking 
\cite{molitoris84a,kruse85b,bertsch84a,aichelin85a,gregoire87a}.
The phase space distributions can be represented by the density of test
particles. The latter are propagated classically with the Hamiltonian
$E(R,p)=\vec{p}^2/2m + \Re \Sigma^+(R,p)$ which yields a solution
of the Vlasov part (l.h.s. = 0) of the transport equation. At the
point of closest approach of two particles one performs a random scattering
process, if the distance at that point is smaller than
$\sqrt{\sigma_{tot}/\pi}$. The angular distribution for the scattering
is obtained from d$\sigma/$d$\Omega$. 

Up to now there is no stringent
proof that this procedure really yields 
a solution of the full transport equation, but it can be shown numerically
that this prescription yields
the correct asymptotic approach to
equilibrium, which can be deduced from the Boltzmann equation, i.e.
equation~(\ref{eq31}), without potential and without
the $(1-f)$ Pauli-blocking factors \cite{welke89a}.

However, a deficiency of all transport equations for the single particle phase
space distribution is that clusters or fragments in heavy ion collisions can
only be constructed in a statistical way, e.g. with the help of a coalescence
model. 
The reason is that in transport theory bound states of many nucleons cannot be
described directly.
This can be avoided if one combines the fully classical equations of motion
with the random scatterings of transport theory.
The classical many body theory with two body forces allows for bound states of
the particles but completely ignores any quantum effects. 
Especially the hard core of the nucleon-nucleon interaction would lead to a
completely different classical behavior as compared to quantum mechanical
scattering. Therefore in models like QMD, RQMD etc. one translates the soft
(long range) part of the effective interaction in the medium (Re$\Sigma^+$)
into a classical but density and possibly momentum dependent two body
interaction. However this must be supplemented by random scatterings to account
for the short range part of the interaction and the other quantum effects like
Pauli blocking. 
Unfortunately this procedure can only in part be justified from first
principles \cite{aichelin91a}, 
but it can be shown that this model yields quite
similar results as the transport model as long as one body observables are
concerned \cite{aichelin89b,hartnack90a,hartnack92b}. 
They are different though for many body
observables like fragments.

\section{Relativistic transport theory}

We now proceed to generalize the approximation scheme outlined above
to the relativistic case.
Here one should start from an effective Lagrangian with baryons and
mesons as elementary degrees of freedom. Since the well known
$\sigma - \omega -$model has been proven to describe static
properties of finite nuclei very well \cite{reinhard89a,rutz95a} it
seems to be reasonable to start with this model also for the
description of the dynamics of heavy ion collisions. It is known that
the linear energy dependence of the real part of the self-energy
is too strong as compared to the measured logarithmic behavior. 
However, this can be remedied by an effective 
(energy and density dependent) coupling strength \cite{maoG94a}.

We start from a Lagrangian density for nucleons, deltas and a nucleon
resonance like the $N^*_{1440}$ plus the $\sigma$ and $\omega$ and
$\pi$ meson, for details cf. \cite{maoG97a}. The free part reads:
\begin{eqnarray}
{\cal L}_{\rm F}&=&\bar{\psi}[{\rm i}\gamma_{\mu}\partial^{\mu}-M_{N}]\psi
+ \bar{\psi}^{*}[{\rm i}\gamma_{\mu}\partial^{\mu}-M_{N^{*}}]\psi^{*} \nonumber \\
&& +\bar{\psi}_{\Delta \nu}[{\rm i}\gamma_{\mu}\partial^{\mu}-M_{\Delta}]
\psi^{\nu}_{\Delta} \nonumber + \frac{1}{2}
\partial_{\mu}\sigma\partial^{\mu}\sigma-U(\sigma) \nonumber \\
&& -\frac{1}{4}\omega_{\mu\nu}\omega^{\mu\nu}+U(\omega) 
+ \frac{1}{2}(\partial_{\mu} \mbox{\boldmath $\pi$} \partial^{\mu}
\mbox{\boldmath $\pi$}
-m_{\pi}^{2}\mbox{\boldmath $\pi$}^{2})
    \end{eqnarray}
and U($\sigma$), U($\omega$) are the self-interaction part of the scalar field
\cite{boguta77a} and vector field \cite{bodmer91a}
\begin{eqnarray}
  && U(\sigma)=
   \frac{1}{2}m_{\sigma}^{2}\sigma^{2}+\frac{1}{3}b({\rm g}_{NN}^{\sigma}
\sigma)^{3}+\frac{1}{4}c({\rm g}_{NN}^{\sigma}\sigma)^{4}, \\
  && U(\omega)=\frac{1}{2}m_{\omega}^{2}\omega_{\mu}\omega^{\mu}
     (1+\frac{({\rm g}_{NN}^{\omega})^{2}}{2}\frac{\omega_{\mu}\omega^{\mu}}
     {Z^{2}}),
   \end{eqnarray}
respectively.  ${\cal L}_{I}$ is the interaction Lagrangian density
\begin{eqnarray}
 {\cal L}_{I}&=&{\cal L}_{NN}+{\cal L}_{N^{*}N^{*}}
 +{\cal L}_{\Delta \Delta}+{\cal L}_{NN^{*}}+{\cal L}_{\Delta N} 
 +{\cal L}_{\Delta N^{*}} \nonumber \\
     &=&{\rm g}^{\sigma}_{NN}\bar{\psi}(x)\psi(x)\sigma(x)
      - {\rm g}^{\omega}_{NN}\bar{\psi}(x)\gamma
  _{\mu}\psi(x)\omega^{\mu}(x)
   +{\rm g}_{NN}^{\pi}\bar{\psi}(x)\gamma_{\mu}\gamma_{5}\mbox{\boldmath $\tau$}
\cdot \psi(x)\partial^{\mu}\mbox{\boldmath $\pi$}(x) \nonumber \\
     &&+{\rm g}^{\sigma}_{N^{*}N^{*}}\bar{\psi}^{*}(x)\psi^{*}(x)\sigma(x)
      - {\rm g}^{\omega}_{N^{*}N^{*}}\bar{\psi}^{*}(x)\gamma
  _{\mu}\psi^{*}(x)\omega^{\mu}(x)
   +{\rm g}_{N^{*}N^{*}}^{\pi}\bar{\psi}^{*}(x)\gamma_{\mu}\gamma_{5}\mbox{\boldmath $\tau$}
\cdot \psi^{*}(x)\partial^{\mu}\mbox{\boldmath $\pi$}(x) \nonumber \\
 &&+{\rm g}^{\sigma}_{\Delta \Delta}
\bar{\psi}_{\Delta \nu}(x)\psi^{\nu}_{\Delta}(x)\sigma(x)
     -{\rm g}^{\omega}_{\Delta \Delta}\bar{\psi}_{\Delta \nu}(x)\gamma
  _{\mu}\psi^{\nu}_{\Delta}(x)\omega^{\mu}(x)
 +{\rm g}_{\Delta \Delta}^{\pi}\bar{\psi}_{\Delta\nu}(x)\gamma_{\mu}\gamma_{5} {\bf T}
 \cdot \psi^{\nu}_{\Delta}(x) \partial^{\mu}\mbox{\boldmath $\pi$}(x) \nonumber \\
     &&+\lbrack {\rm g}^{\sigma}_{NN^{*}}\bar{\psi}^{*}(x)\psi(x)\sigma(x)
      - {\rm g}^{\omega}_{NN^{*}}\bar{\psi}^{*}(x)\gamma
  _{\mu}\psi(x)\omega^{\mu}(x)
+{\rm g}_{NN^{*}}^{\pi}\bar{\psi}^{*}(x)\gamma_{\mu}\gamma_{5}\mbox{\boldmath $\tau$}
\cdot \psi(x)\partial^{\mu}\mbox{\boldmath $\pi$}(x) \nonumber \\
 &&- {\rm g}_{\Delta N}^{\pi}\bar{\psi}_{\Delta \mu}(x)
 \partial^{\mu}\mbox{\boldmath $\pi$}(x) \cdot {\bf S}^{+}\psi(x)
 - {\rm g}_{\Delta N^{*}}^{\pi}\bar{\psi}_{\Delta \mu}(x)
 \partial^{\mu}\mbox{\boldmath $\pi$}(x) \cdot {\bf S}^{+}\psi^{*}(x) +H.c. \rbrack
 \nonumber \\
 &=& {\rm g}_{NN}^{A}\bar{\psi}(x)\Gamma_{A}^{N} \psi(x) \Phi_{A}(x)
 +{\rm g}_{N^{*}N^{*}}^{A}\bar{\psi}^{*}(x)\Gamma_{A}^{N^{*}} \psi^{*}(x) \Phi_{A}(x)
 + {\rm g}_{\Delta \Delta}^{A} \bar{\psi}_{\Delta \nu}(x)\Gamma_{A}^{\Delta}\psi
 ^{\nu}_{\Delta}(x) \Phi_{A}(x) \nonumber \\
 && +\lbrack {\rm g}_{NN^{*}}^{A}\bar{\psi}^{*}(x)\Gamma_{A}^{N^{*}} \psi(x) \Phi_{A}(x)
 - {\rm g}_{\Delta N}^{\pi}\bar{\psi}_{\Delta \mu}(x)
 \partial^{\mu}\mbox{\boldmath $\pi$}(x) \cdot {\bf S}^{+}\psi(x) \nonumber \\
  && - {\rm g}_{\Delta N^{*}}^{\pi}\bar{\psi}_{\Delta \mu}(x)
 \partial^{\mu}\mbox{\boldmath $\pi$}(x) \cdot {\bf S}^{+}\psi^{*}(x) 
  + h.\, c. \rbrack
  \label{lagran1}
   \end{eqnarray}
Here, $\psi$, $\psi^{*}$ are the Dirac spinors of the nucleon and $N^{*}$(1440)
, and $\psi_{\Delta \mu}$ is the Rarita-Schwinger spinor of the
$\Delta$-baryon. $\mbox{\boldmath $\tau$}$ is the isospin operator of the nucleon and $N^{*}$
(1440),
${\bf T}$ is the isospin operator of the $\Delta$, and ${\bf S}^{+}$ is the
isospin transition operator between the isospin 1/2 and 3/2 fields. 
${\rm g}_{NN}^{\pi}=f_{\pi}/m_{\pi}$,
 ${\rm g}_{\Delta N}^{\pi}=f^{*}/m_{\pi}$;
 $\Gamma_{A}^{N}=\Gamma_{A}^{N^{*}}=
 \gamma_{A}\tau_{A}$, $\Gamma_{A}^{\Delta}=\gamma_{A}T_{A}$,
  A=$\sigma$, $\omega$, $\pi$,
the symbols and notation are defined in table~\ref{maotab1}.

In the closed time path formalism one obtains quite analogous Dyson
equations for the different particle species from which one obtains
in turn transport equations for nucleons, $N^*_{1440}$ and $\Delta_{1232}$ 
\cite{maoG94a,maoG97a,maoG94b}. 
However, in the relativistic case, even more difficulties arise as for the
nonrelativistic tranport equations. T- or G-matrix calculations starting from a
Lagrangian like (\ref{lagran1}) are very cumbersome even for the static
equilibrium situation. Also the non-locality of the Fock term in the mean-field
approximation, leads to a non-trivial momentum dependence of the self-energy.
The same is true for the Born approximation, therefore one usually takes only
the Hartree or the mean-field approximation for the Vlasov part in the
transport equation. Effective masses and momenta enter both the Vlasov part and
the collision term where the cross section has to be calculated in some
approximation. As a qualitative guideline, one can calculate cross sections
from Lagrangians as (\ref{lagran1}) in the Born approximation. However, in
medium dressed propagators should be used for the internal lines, taking into
account effective mass and momentum in Hartree approximation. 
For a more rigorous
derivation, we refer to the work of Cassing {\it et al.}
\cite{cassing90a,cassing90b}. 
How the cross sections
are determined in the UrQMD model can be found in section~\ref{res_sig}.

One then usually uses the propagators of 
non-interacting systems of particles with effective mass, e.g. for 
the nucleon and the $N^*_{1440}$ 
\begin{eqnarray}
g^{c,a}(x,p) &=& (\not\!p+m^{*}_{N^{*}}) \left[ \frac{\pm 1}{p^{2}-{m^{*}}^2_{N^{*}}\pm {\rm i} \epsilon}
 + \frac{\pi {\rm i}}{E(p)}\delta(p_{0}-E(p))f_{N^{*}}(x,p) \right] , \\
g^>(x,p)&=& -\frac{\pi {\rm i}}{E(p)}\delta(p_{0}-E(p))[1-f_{N^{*}}(x,p)](\not\! p+m^{*}_{N^{*}}) \\
g^<(x,p) &=& \frac{\pi {\rm i}}{E(p)}\delta(p_{0}-E(p))f_{N^{*}}(x,p)(\not\! p+m^{*}_{N^{*}}).
\end{eqnarray}
where $E(p)=\sqrt{\vec{p}^2+M_{N^*}^2}$.
The effective mass and momentum of e.g. the $N^*_{1440}$ are defined as
  \begin{eqnarray}
    &&m^{*}_{N^{*}}(x)=M_{N^{*}}+\Sigma^{S}_{N^{*}}(x) \\
    &&p^{\mu}(x)=P^{\mu} - \Sigma^{\mu}_{N^{*}}(x).
  \end{eqnarray}

The self-energies are taken here in Hartree Fock approximation only.
Herewith one assumes that the contribution of the Born diagrams to the
real part of the self-energy is small. Furthermore one often also
neglects the Fock part of the self-energy for the same reason. The
self energies in Hartree approximation for the 
$N^*_{1440}$ for example are given by
 \begin{eqnarray}
&&\Sigma_{N^{*}}^{S}(x)=-\frac{{\rm g}_{N^{*} N^{*}}^{\sigma}}{m_{\sigma}^{2}}
[ {\rm g}_{N N}^{\sigma}\rho_{S}(N)+{\rm g}_{N^{*} N^{*}}^{\sigma}\rho_{S}
(N^{*}) + {\rm g}_{\Delta \Delta}^{\sigma}\rho_{S}(\Delta)  ],\\
&&\Sigma_{N^{*}}^{\mu}(x)=\frac{{\rm g}_{N^{*} N^{*}}^{\omega}}{m_{\omega}^{2}}
 [{\rm g}_{N N}^{\omega}\rho_{V}^{\mu}(N) + {\rm g}_{N^{*} N^{*}}^{\omega}
 \rho_{V}^{\mu}(N^{*}) +
 {\rm g}_{\Delta \Delta}^{\omega}\rho_{V}^{\mu}(\Delta) ].
 \end{eqnarray}
where $\rho_S(i)$ and $\rho_V^\mu(i)$ are the scalar and vector densities
of particle species $i$ and analogous expressions for $N$ and $\Delta_{1232}$.
Furthermore one usually makes use of the so called local density
approximation. This means that the mean $\sigma$ and $\omega$ fields
are determined from the local scalar and vector densities alone,
neglecting retarded contributions from other space time regions, i.e. in
the Klein-Gordon equation for the $\sigma$ (and analogous for the $\omega$):
\be
(\Box + m_\sigma^2) \sigma + b (g^\sigma_{NN})^3 \, \sigma^2 
+ c (g^\sigma_{NN})^4 \, \sigma^3 \,=\, g^\sigma_{NN} \rho_S(N)
+ g^\sigma_{N^*N^*} \rho_S(N^*) + g^\sigma_{\Delta\Delta} \rho_S(\Delta)
\ee
One neglects the dynamical contributions of the $\Box$-operator.
This should be a good approximation as long as there are no
rapid field oscillations on the scale of the (small) inverse mass 
of $\sigma$ and
$\omega$ meson. 
However, the retardation of the interaction should not be 
neglected for all energies from non-relativistic up to the
ultra-relativistic energy regime. 
It is not clear up to now if one should take related phenomena
like $\sigma,\omega$ Bremsstrahlung seriously \cite{mishustin97a}.
Moreover, it has been shown that for
intermediate energies of several hundred MeV per nucleon the above
approximation is acceptable \cite{weber90a}.

As an example, the RBUU equation e.g. for the $N^*_{1440}$ 
distribution function now reads:
 \begin{eqnarray}
&&\lbrace p_{\mu} [
 \partial^{\mu}_{x}-\partial^{\mu}_{x}\Sigma_{N^{*}}
^{\nu}(x) \partial_{\nu}^{p}+\partial_{x}^{\nu}\Sigma_{N^{*}}^{\mu}(x)\partial
_{\nu}^{p} ] +m^{*}_{N^{*}}\partial^{\nu}_{x}\Sigma_{N^{*}}^{S}(x)\partial_{\nu}^{p}
\rbrace \frac{f_{N^{*}}({\bf x}, {\bf p}, t)}{E^{*}_{N^{*}}(p)}
 \nonumber \\
 && = C^{N^{*}}(x,p).
\label{rbuueqn}
 \end{eqnarray}
and analogous equations for the nucleon and the $\Delta_{1232}$.
A practical solution of the relativistic Vlasov equation (l.h.s. of
equation~(\ref{rbuueqn}) $=0$) can be found in \cite{koCM87a}.
Attempts have been made also to solve the corresponding quantum mechanical
problem in terms of the time-dependent Dirac equation with mean fields
\cite{cusson85a}. Realizations of the full relativistic Boltzmann
Uehling Uhlenbeck model (RBUU) can be found in \cite{weber90a,fuchs95a}.

\begin{table}
    \begin{center}
    \tabcolsep 0.10in
  \begin{tabular}{|c||c|c|c|c|c|c|c|c|c|c|c|}
      \hline
 {\rm A} & m$_{A}$ & g$^{A}_{NN}$ & g$^{A}_{N^{*}N^{*}}$ & 
 g$^{A}_{\Delta \Delta}$ & g$^{A}_{NN^{*}}$ & $\gamma_{A} $ & $\tau_{A}$ & $ T_{A} $ &
  $\Phi_{A} (x) $ & D$_{A}^{\mu}$ & D$_{A}^{i}$ \\
 \hline\hline
 $\sigma$ & m$_{\sigma}$ & g$^{\sigma}_{NN}$ & g$^{\sigma}_{N^{*}N^{*}}$ &
 g$_{\Delta \Delta}^{\sigma}$ & g$^{\sigma}_{NN^{*}}$ & 1 & 1 & 1 & $\sigma(x)$ &
 1 & 1   \\
 \hline
$\omega$ & m$_{\omega}$ & $-$ g$^{\omega}_{NN} $ & $-$ g$^{\omega}_{N^{*}N^{*}}$ & 
 $-$ g$_{\Delta \Delta}^{\omega}$ & $-$ g$^{\omega}_{NN^{*}}$ &  $\gamma_{\mu} $ &
 1 & 1 & $\omega^{\mu}(x)$ & $-$ g$^{\mu \nu}$ & 1 \\
 \hline
 $\pi$ & m$_{\pi}$ & g$_{NN}^{\pi}$& g$^{\pi}_{N^{*}N^{*}}$ & 
 g$_{\Delta \Delta}^{\pi}$ & g$^{\pi}_{NN^{*}} $ & 
 $\not\! k \gamma_{5} $
& \mbox{\boldmath $\tau$} &${\bf T}$
 & \mbox{\boldmath $\pi$}(x) & 1 & $\delta_{ij}$ \\
       \hline
      \end{tabular}
          \end{center}
\caption{\label{maotab1} Symbol and notation definitions.}
\end{table}

The collision term on the r.h.s. can be written as
   \begin{eqnarray}
  C^{N^{*}}(x,p)&=&\frac{1}{2}\int \frac{d^{3}p_{2}}{(2\pi)^{3}}
 \int\frac{d^{3}p_{3}}{(2\pi)^{3}} \int\frac{d^{3}p_{4}}{(2\pi)^{3}}
 (2\pi)^{4}\delta^{(4)}(p+p_{2}-p_{3}-p_{4})\nonumber \\
 && \times W^{N^{*}}(p,\,p_{2},\,p_{3},\,p_{4})
 (F_{2}-F_{1}),
    \end{eqnarray}
where $F_{2}$, $F_{1}$ are the Uehling-Uhlenbeck factors :
    \begin{eqnarray}
&& F_{2}=[1-f_{N^{*}}({\bf x},{\bf p},t)][1-f_{B_{2}}({\bf x},{\bf p}_{2},t)]
   f_{B_{3}}({\bf x},{\bf p}_{3},t)f_{B_{4}}({\bf x},{\bf p}_{4},t), \\
&&  F_{1}=f_{N^{*}}({\bf x},{\bf p},t)f_{B_{2}}({\bf x},{\bf p}_{2},t)
   [1-f_{B_{3}}({\bf x},{\bf p}_{3},t)]
   [1-f_{B_{4}}({\bf x},{\bf p}_{4},t)] , 
    \end{eqnarray}
$B_{2}$, $B_{3}$, $B_{4}$ can be $N$, $\Delta$ and $N^{*}$(1440) and
possibly many other resonances.
$W^{N^{*}}(p,\,p_{2},\,p_{3},\,p_{4})$  is the transition
 probability of different channels, which has the form (for details see 
ref. \cite{maoG97a})
  \begin{equation}
 W^{N^{*}}(p,\,p_{2},\,p_{3}\,,p_{4})=
  \frac{1}{16E^{*}_{N^{*}}(p)E^{*}_{B_{2}}(p_{2})E^{*}_{B_{3}}(p_{3})
  E^{*}_{B_{4}}(p_{4})}\sum_{AB} (T_{D}\Phi_{D} - T_{E}\Phi_{E})
   +p_{3} \longleftrightarrow p_{4}.
  \end{equation}
Here $T_{D}$, $T_{E}$ are the isospin matrices and $\Phi_{D}$, $\Phi_{E}$
are the spin matrices, respectively.
 D denotes the contribution of the direct diagrams and E is that of the exchange
diagrams. $A$, $B=\sigma$, $\omega$, $\pi$ represent the contributions of 
different mesons. The exchange of $p_{3}$ and $p_{4}$ is only for the 
case of identical particles in the final state. 
 The two-body scattering reactions relevant to the $N^{*}$(1440) 
in the $N$, $\Delta$ and $N^{*}$(1440) system are as follows: \\
 \indent (1) Elastic reactions: \\
\mbox{} \hspace{1.0cm} $NN^{*} \longrightarrow NN^{*}$, \hspace{1.0cm}
  $\Delta N^{*} \longrightarrow \Delta N^{*}$, \hspace{1.0cm} 
  $N^{*}N^{*} \longrightarrow N^{*} N^{*}$ .   \\
  \indent (2) Inelastic reactions:  \\
 \mbox{} \hspace{1.0cm} $N N \longleftrightarrow NN^{*}$, \hspace{1.0cm} 
  $N \Delta \longleftrightarrow N N^{*}$, \hspace{1.0cm} 
  $\Delta \Delta \longleftrightarrow N N^{*}$, \\
 \mbox{} \hspace{1.0cm} $N N^{*} \longleftrightarrow \Delta N^{*}$, \hspace{0.9cm} 
  $N N^{*} \longleftrightarrow N^{*} N^{*}$, \hspace{0.6cm} 
  $N N \longleftrightarrow \Delta N^{*}$, \\
 \mbox{} \hspace{1.0cm} $N \Delta \longleftrightarrow \Delta N^{*}$,
\hspace{1.0cm} $\Delta \Delta \longleftrightarrow \Delta N^{*}$, \hspace{1.0cm} 
  $N^{*} N^{*} \longleftrightarrow \Delta N^{*}$ , \\
 \mbox{} \hspace{1.0cm} $N N \longleftrightarrow N^{*}N^{*}$, \hspace{0.8cm} 
  $N \Delta \longleftrightarrow N^{*}N^{*}$, \hspace{0.8cm}
  $\Delta \Delta \longleftrightarrow N^{*}N^{*}$.

In principle one also obtains resonance formation and decay terms which are
not given here. From this it becomes clear that for higher energies 
(above several GeV per nucleon), where dozens of baryon and meson 
resonances come into play, the number of reaction channels increases
tremendously. Even for comparatively low energies higher resonances
contribute significantly to (subthreshold) particle production
(see section~\ref{subthreshold}). Most of the corresponding coupling
constants are not known and instead one directly parameterizes the
cross sections and potentials and decay widths. Furthermore the
resonances have a spectral function (mass distribution) of finite
width which is neglected in the quasi particle approximation. These
spectral functions might change (width and the position
of the pole) in the hot and dense nuclear medium, which has caused much
interest during the last years 
\cite{asakawa92a,herrmann93a,friman97a,rapp97a,klingl97a,peters97a}.

However, we would like to stress that a shift of the resonance mass
distribution to lower masses is not necessarily connected to 
chiral symmetry restoration ($m^* \to 0$), since such a shift could
also be obtained in a model without chiral symmetry in the high energy
limit. 

In a pragmatic way one can incorporate the mass distributions of 
resonances by folding the collision terms with appropriate
mass distributions of the involved resonances \cite{danielewicz91a}
(see also section~\ref{resonances}). In the actual numerical implementations
this is usually taken into account by choosing the actual resonance
mass (mass of the test-particles) according to a probability distribution,
which might depend on the local density and temperature. 

However, the 
basic concept of well established quasi particles becomes 
meaningless for energies beyond several GeV (baryons), where there
is no pronounced resonance structure on the corresponding cross section
anymore. For these energies one must incorporate a mechanism for 
continuous excitations of mesons and baryons. Since for such high
energies also the internal quark and gluon degrees of freedom become
more and more important one usually supplements the actual resonances
by continuous string excitations above a (model dependent) 
energy threshold. These excited strings can then fragment into
pre-hadrons by string fragmentation algorithms, which have been
proven to be successful as long as rescatterings of newly produced
hadrons can be neglected \cite{andersson83b}. A well known example is the 
FRITIOF model \cite{fritiof1.6,fritiof7.0}, which however does not incorporate 
a complete phase space picture but rather works only in momentum space so
that rescatterings are difficult to be included. 
Perturbative QCD effects like multiple minijets 
have been included in the PYTHIA model
\cite{sjoerstrand86a}.

In the UrQMD model,
to be discussed in more detail in the following chapter, a string excitation
and fragmentation scheme is combined with the transport theoretical
approach. Other models working along the same line are the RQMD
(Relativistic Quantum Molecular Dynamics) model \cite{sorge89a} or the
HSD (Hadron String Dynamics) \cite{ehehalt96a}. Some other models are based
mainly on the dynamics of strings like VENUS \cite{werner93a}
and QGSM \cite{bravina94a} or partons
like the parton cascade model PCM \cite{geiger92a,geiger95a} or the dual parton
 model \cite{capella80a,capella94a}. 

Since parton cascades are mainly based on the cross sections from
perturbative QCD, a basic requirement is that the typical momentum
transfer $\sqrt{Q^2}$ in parton scatterings is larger than 
$\sim 10$~GeV so that running coupling constants become sufficiently
small. Even if this is fulfilled in first collisions, secondary interactions
are much softer and require a non-perturbative treatment.
Therefore these models have to be supplemented by
hadronization prescriptions like the parton-hadron conversion model
HIJING \cite{wang91a} and prescriptions for rescattering of newly
produced hadrons like the hadron cascade HIJET \cite{shor89a}.

A common problem for all models involving a transition from the
string or parton picture to hadrons is the so called formation time.
This is the time after which newly produced quarks and anti-quarks can
be considered to be in a well established hadronic state (meson or
baryon) that can interact again with other particles. Not much is known
from the theoretical side how such a parameter can be calculated or how
an appropriate model prescription could be derived. For the prescription
used in the UrQMD model, we refer to section~\ref{formation-time}.

Even though models like RQMD or VENUS are impressively fine tuned for
CERN SPS data the price to be paid are additional mechanisms or 
model assumptions like string droplets in VENUS \cite{werner93b} or color ropes
in RQMD \cite{sorge95a}. Furthermore, many of these models are not designed to
work over a broad energy range (e.g. the relativistic cascade 
ARC \cite{pang92a} is only designed for AGS energies and models such as
QMD/IQMD \cite{hartnack90a,aichelin87b,aichelin86a,peilert89a,bass95c} 
or BUU \cite{bertsch84a,aichelin85a,teis97a} 
may only be applied to the  BEVALAC/SIS energy domain).
On the other hand the search for the quark gluon plasma in high energetic
heavy ion collisions will require a systematic study of the excitation
functions of observables over the broad energy range from SIS 
($\sim 1$~AGeV) up to SPS ($\sim 200$~AGeV) or even higher energies 
in colliders like RHIC ($\sqrt{s}=20$~AGeV) or LHC ($\sqrt{s}>1000$~AGeV).

A general problem of hadron or parton cascades or transport models is 
the geometric interpretation of the cross section in most of the actual
model implementations. Even though the underlying theory or model is
usually Lorentz invariant the practical realization might not be
\cite{kortemeyer95a}. This is due to the fact that the sequence of collisions
according to a specific collision criterion can be reference frame
dependent. The reason is that points of closest approach of several
particles in space and time might be interchanged in the time sequence
after a Lorentz transformation, such that in different reference frames
one obtains different results for the sequence of scatterings. In the
UrQMD model this problem has been minimized by the choice of the
collision criterion (see section~\ref{collcrit}) such that the 
frame dependence of the observables (going from the CM to the target
rest frame) is only in the order of $\sim 5$\% at CERN/SPS energies.

\section{The Quantum Molecular Dynamics approach}
In the following we discuss 
an approach which goes beyond the one-body descriptions as discussed  above,
which is the Quantum Molecular Dynamics (QMD) model
\cite{aichelin91a,aichelin87b,aichelin86a,peilert89a}. 
The QMD model is a $N$-body theory which simulates heavy ion reactions 
at intermediate energies on an event by event basis. 
Taking into account all fluctuations and correlations has basically two
advantages: i) many-body processes, in particular the formation of complex
fragments are explicitly treated and ii) the model allows for an 
event-by-event analysis of heavy ion reactions similar to the methods which
are used for the analysis of exclusive high acceptance data.

The major aspects of the formulation of QMD will now be discussed briefly.
For a more detailed description we refer to ref.~\cite{aichelin91a}.
The particular realization of the UrQMD model will be discussed later.

\subsection{Formal derivation of the transport equation}
In QMD each nucleon is represented
by a coherent state of the form (we set $\hbar,c =1$) 
\begin{equation}
\label{gaussians}
\phi_i (\vec{x}; \vec{q}_i,\vec{p}_i,t) = 
\left({\frac{2 }{L\pi}}\right)^{3/4}\, \exp \left\{
-\frac{2}{L}(\vec{x}-\vec{q}_i(t))^2 + \frac{1}{\hbar} {\rm i} \vec{p}_i(t)
\vec{x} \right\}
\end{equation}
which is characterized by 6 time-dependent parameters,
$\vec{q}_{i}$ and $\vec{p}_{i}$, respectively.
The parameter $L$, which is related  to the extension of the wave packet in
phase space, is fixed. 
The total $n$-body wave function is  assumed to be the direct
product of coherent states (\ref{gaussians})
\begin{equation}
\Phi = \prod_i \phi_i (\vec{x}, \vec{q_i}, \vec{p_i}, t)
\end{equation}
Note that we do not use a Slater determinant and
thus neglect antisymmetrization. The computational time scales like 
$(A_p + A_t)^4$ in the case of a Slater determinant, while in QMD it 
is like $(A_p + A_t)^2$. First successful attempts 
to simulate heavy ion reactions with antisymmetrized
states have been performed for small systems \cite{feldmeier90a,ono92a}, for a
recent review see Ref. \cite{feldmeier97a}. 
A consistent derivation of the QMD equations of motion for the wave function
under the influence of both, the real and the imaginary part of the 
G-matrix, however, has not yet been achieved. Therefore we will add the 
imaginary part as a cross section and treat them as in the cascade approach.
How to incorporate cross sections into an antisymmetrized molecular 
dynamics is not yet known. This limits its applicability to very low beam 
energies.

The initial values of the parameters
are chosen in a way that the ensemble of $A_T$ + $A_P$ 
nucleons gives a proper density distribution as well as a proper momentum
distribution of projectile and target nuclei. 

The equations of motion of the many-body 
system is calculated by means of a generalized variational principle: we
start out from the action 
\begin{equation}
S = \int\limits ^{t_2} _{t_1} {\cal L} [\Phi, \Phi^\ast] dt 
\end{equation}
with the Lagrange functional ${\cal L}$ 
\begin{equation}
{\cal L} = \left \langle\Phi \left\vert {\rm i}\hbar {\frac{d }{dt}} - H \right\vert
\Phi\right\rangle
\end{equation}
where the total time derivative includes the derivation with respect to the
parameters. 
The Hamiltonian $H$ contains a kinetic term and mutual 
interactions $V_{ij}$, which can be interpreted as the real part of the 
Brueckner G-matrix supplemented by the Coulomb interaction. We will later on 
describe the components of $H$ in detail.
The time evolution of the parameters is obtained by the
requirement that the action is stationary under the allowed variation of the
wave function. This yields an Euler-Lagrange equation for each parameter.

If the true solution of the Schr\"odinger equation is contained in the
restricted set of wave functions 
$\phi_i(\vec{x}; \vec{q}_i, \vec{p}_i, t)$
this variation of the action gives the exact solution of the Schr\"odinger
equation. If the parameter space is too restricted we obtain that wave
function in the restricted parameter space which comes closest to the
solution of the Schr\"odinger equation. Note that the set of wave functions
which can be covered with special parameterizations is not necessarily a
subspace of Hilbert-space, thus the superposition principle does not hold.

For the coherent states and a Hamiltonian of the form $H = \sum_i T_i + {\ 
\frac{1 }{2}} \sum_{ij} V_{ij}$ ($T_i$= kinetic energy, $V_{ij}$ = potential
energy) the Lagrangian and the variation can easily be calculated and we
obtain: 
\begin{equation}
{\cal L} = \sum_i \biggl[-\dot{\vec{q}_i} {\vec{p}_i}  - T_i -
{1\over 2}\sum _{j\neq i} \langle
V_{ik}\rangle - {\frac{3 }{2Lm}} \biggr]\end{equation}
\begin{equation}
\dot{\vec{q}_i} = {\frac{\vec{p}_i }{m}} + \nabla_{\vec{p}_i} \sum_j
\langle V_{ij}\rangle  = \nabla_{\vec{p}_i} \langle H \rangle
\end{equation}
\begin{equation}
\dot{\vec{p}_i} = - \nabla_{\vec{q}_i} \sum _{j\neq i} \langle
V_{ij}\rangle = -\nabla_{\vec{q}_i} \langle H \rangle
\end{equation}
with $\langle V_{ij}\rangle = \int d^3x_1\,d^3x_2\, 
\phi_i^* \phi_j^* V(x_1,x_2) \phi_i \phi_j$.
These are the time evolution equations which are solved numerically. 
Thus the variational principle reduces the time evolution of the 
$n$-body Schr\"odinger equation to the time evolution equations of 
$6 \cdot (A_P+A_T)$ parameters to which a physical meaning can be
attributed.
The equations of motion for the parameters $\vec{p}_i$ and $\vec{q}_i$ read
\begin{equation}\label{hamiltoneq}
\dot{\vec{p}}_i = - \frac{\partial \langle H \rangle}{\partial \vec{q}_i} 
\quad {\rm and} \quad
\dot{\vec{q}}_i = \frac{\partial \langle H \rangle}{\partial \vec{p}_i} \, ,
\end{equation}
and show the same structure as the classical Hamilton equations.
The numerical solution can be treated in a similar manner as it is done in
classical molecular dynamics 
\cite{molitoris84a,bodmer77a,wilets77a,bodmer80a,kiselev83a}.
Using trial wave functions other than Gaussians in Eq.\ (\ref{gaussians})
yields
more complex equations of motion and hence the analogy 
to classical molecular dynamics is lost. If $\langle H \rangle$ has no
explicit time dependence, QMD conserves energy and momentum by construction.

\subsection{Inclusion of collisions}
As stated above the imaginary part of the G-matrix acts like a collision 
term. In the QMD simulation we restrict ourselves to binary collisions 
(two-body level). The collisions are performed in a point-particle sense in a 
similar way as in VUU or cascade \cite{cugnon80a}:
Two particles collide if their minimum distance $d$, 
i.e.\ the minimum relative 
distance of the centroids of the Gaussians during their motion, 
in their CM frame
fulfills the requirement: 
\begin{equation}
 d \le d_0 = \sqrt{ \frac { \sigma_{\rm tot} } {\pi}  }  , \qquad
 \sigma_{\rm tot} = \sigma(\sqrt{s},\hbox{ type} ).
\end{equation}
where the cross section is assumed to be the free cross section of the
regarded collision type ($N-N$, $N-\Delta$, \ldots).

Beside the parameters describing the $N$--$N$ potential, the 
cross sections constitute another major part of the model. In principle,
both quantities are connected and can be deduced from 
Brueckner theory. QMD-calculations using consistently derived cross-sections
and potentials from the local phase space distributions have been discussed
e.g.\ in \cite{jaenicke92a}. Such simulations are time-consuming since the 
cross-sections and potentials do explicitly depend on the local phase space
population. 

Within the framework of using free cross section one may parameterize
the cross section of the processes to fit to the experimental data if 
available. 
For unknown cross sections isospin symmetry and time reversibility is assumed.

\subsection{Pauli blocking due to Fermi statistics}
The  cross section is reduced to an effective cross section by the
Pauli-blocking. For each collision 
the phase space densities in the final states are checked in order to assure
that the final distribution in phase space is in agreement with the Pauli 
principle ($f\le 1$).
Phase space in QMD is not discretized into elementary cells
as in one-body models like VUU. In order 
to obtain smooth distribution functions the following procedure is applied:
The phase space density $f_i'$ at the final states $1'$ and $2'$  is
measured and interpreted
as a blocking probability. Thus, the collision is only allowed with a 
probability of $(1-f_1')(1-f_2')$. If the collision is not allowed the 
particles remain at their original momenta.

\subsection{The Relativistic Quantum Molecular Dynamics approach}
The Relativistic Quantum Molecular Dynamics (RQMD) approach 
has been developed to extend the QMD model up to relativistic
energies (AGS and CERN/SPS domain) \cite{sorge89a}.
Its main improvements compared to the standard QMD model 
\cite{aichelin91a,hartnack92b,aichelin86a,aichelin87b,peilert89a,bass95c,hartnack89b}
are
\begin{enumerate}
\item covariant dynamics
\item an improved and extended collision term containing 
	heavy baryon-resonances,
	strange particles and string-excitation for high energy hadron-hadron
	interactions
\end{enumerate}

In this section we will focus on the first item -- the covariant dynamics --
since the description of the UrQMD collision term in the following chapter
will cover in great detail all the techniques which have also been
employed in the RQMD model.

The Relativistic Quantum Molecular Dynamics model describes the time-evolution
of a many-body system using classical covariant equations of motion.
The system propagates in a $8N$-dimensional phase space with $6N$ degrees
of freedom representing the classical configuration- and momentum-space.
The remaining $2N$ degrees of freedom contain the eigentime and energy
of each particle.

The necessity of employing an $8N$-dimensional phase-space is based
on the {\em no-interaction theorem} (NIT) by 
Curri, Sudarshan and Mukunda \cite{currie63a}:
In a $6N+1$-dimensional phase-space a Lagrangian for a Poincar\'e-invariant
dynamics can only be formulated in the case of non-interacting free 
particles.
In order to reduce the $8N$-dimensional phase-space to the 
commonly used $6N+1$ dimensions, Lorentz-covariant constraints have
to be introduced. These constraints have to yield reasonable equations
of motion in the non-relativistic limit of a dilute gas.
These constraints impose a time-ordering on the dynamics of the system.
They consist of $N$ mass-shell constraints:
\begin{equation}
H_i = \vec{p}_i^2 - m_i^2 - V_i \,=\, 0 \quad i=1,\ldots, N 
\end{equation}
and $N-1$ constraints which connect the relative times of the particles
\begin{equation}
\label{zfix}
\chi_i =  \sum\limits_{j \ne i} g_{ij} \vec{p}_{ij} \vec{q}_{ij} \,=\, 0 
\qquad i=1,\ldots,N
\end{equation}
with
\begin{equation}
\vec{q}_{ij}= \vec{q}_i - \vec{q}_j\,,\quad \vec{p}_{ij}=\vec{p}_i+\vec{p}_j \,, 
\quad
        g_{ij}= \frac{\exp(\vec{q}^2_{ij}/L)}{\vec{q}^2_{ij}}
\end{equation}
The $N$th constraint then serves to 
attach the individual eigentimes to a 
common time-evolution parameter $\tau$.

In the case of finite range interactions, acausalities may occur.
The constraints~(\ref{zfix}), however, suppress correlations
of particles with strongly differing eigentimes.

Furthermore a time-ordering for binary collisions must be imposed
(via the time-evolution parameter $\tau$).
This time-ordering, unfortunately, depends on the reference
frame. One has to assume that the number of collisions is large enough
for the dynamics of a single binary collision not to influence
macroscopic properties of the system.

The Hamiltonian is defined as linear combination of all 
Poincar\'e-invariant constraints:
\begin{equation}
H \,=\, \sum\limits_{i=1}^N \lambda_i H_i + \sum\limits_{i=1}^{N-1} 
        \delta\mu_i \chi_i 
\end{equation}
The canonical equations of motion are then given as
\begin{eqnarray}
\frac{d\vec{q}_j}{d\tau} &=& \frac{\partial H}{\partial \vec{p}_j} \,=\,
        2 \lambda_j \vec{p}_j - \sum\limits_{i=1}^N \lambda_i 
        \frac{\partial V_i}{\partial \vec{p}_j} \qquad j=1,\ldots,N \\
\frac{d\vec{p}_j}{d\tau} &=& - \frac{\partial H}{\partial \vec{q}_j} \,=\,
        \sum\limits_{i=1}^N  \lambda_i \frac{\partial V_i}{\partial \vec{q}_j}
\end{eqnarray}
with the coefficients $\lambda_i$:
\begin{equation}
\lambda_i \,\approx \, -\frac{\partial \chi_N}{\partial \tau} S_{Ni}
\qquad i=1,\ldots,N 
\end{equation}
The $N$th constraint $\chi_N$ 
connects the eigentimes of the individual particles to the 
time-evolution parameter $\tau$. Although this parameter is used
to describe the time-evolution of the system, it should, however, 
not be interpreted as the system time. 
The matrix  $S_{ij}$
is then given by:
\begin{equation}
(S^{-1})_{ij} \equiv \{ H_i, \chi_j \} \qquad i,j=1,\ldots,N 
\end{equation}
Here, the main drawback of the RQMD ansatz appears: In order to solve
the equations of motion one needs to calculate the coefficients
$\lambda_i$. For their calculation the matrix $S^{-1}$ must be
inverted. Since the number of elements of $S^{-1}$ is quadratic
in the number of particles $N$, the inversion of $S^{-1}$ (which
can only be done numerically) is very time-consuming, especially
for heavy collision systems at CERN/SPS energies.

\chapter{The UrQMD-Model}
\label{urqmdkap}
\section{Initialization}

This section describes the initialization 
of projectile and target
nuclei in the UrQMD model. Projectile and target are modeled
according to the Fermi-gas ansatz. 
The nucleons are represented by Gaussian   
shaped density distributions:
\begin{equation}
\label{uqmd_gaussian}
\varphi_j(\vec{x}_j,t) \,=\, \left(\frac{2 \alpha}{\pi}\right)^{\frac{3}{4}}
        \exp{\left\{ -\alpha (\vec{x}_j - \vec{r}_j(t))^2 
        + \frac{{\rm i}}{\hbar} \vec{p}_j(t) \vec{x}_j
        \right\} } \quad.
\end{equation}
The wave-function of the nucleus is defined as the product wave-function
of the single nucleon Gaussians:
\begin{equation}
\Phi \,=\,  \prod_j \varphi_j (\vec{x_i}, \vec{p_i}, t) \quad.
\end{equation}
 
Each initialized nucleus must meet the following constraints:
\begin{itemize}
\item $\sum_i \vec{q}_i = \vec{0}$, 
        i.e. it is centered in configuration space around $\vec{0}$,
\item $\sum_i \vec{v}_i = \vec{0}$, i.e. the nucleus is at rest
\item its binding energy should correspond to the value given by the
	Bethe-Weizs\"acker formula,
\item the radius should yield the following mass dependence
\begin{equation}
\label{kernradius}
        R(A)\,\sim \, r_0 \cdot A^{\frac{1}{3}}
\end{equation}
        and have a reasonable surface-thickness,
\item in its center, the nucleus should have nuclear matter ground-state
	density.
\end{itemize} 
In configuration space the centroids of the Gaussians are randomly
distributed within a sphere. The finite width of Gaussians
results in a surface region beyond the radius of that sphere. Therefore
its radius is reduced by half a layer of nucleons from the 
original nuclear radius of equation~(\ref{kernradius}):
\begin{equation}
R(A) \,=\, r_0 \left( \frac{1}{2} \left[ A + \left( A^{\frac{1}{3}} -1
        \right)^3 \right] \right)^{\frac{1}{3}} \, .
\end{equation}
The parameter $r_0$ is a function of the nuclear matter ground state
density $\rho_0$ used in the UrQMD model:
\begin{equation}
r_0 \,=\, \left( \frac{3}{4 \pi \rho_0} \right)^{\frac{1}{3}} \, .
\end{equation}
The relatively small number of nucleons to be distributed over
the volume of the nucleus may result in large fluctuations in
the mean density of the nucleus. Therefore  the phase-space
density at the location of each nucleon is evaluated after its placement.
If the phase-space density is too high (i.e. the respective area of
the nucleus is already occupied by other nucleons), then the location
of that nucleon is rejected and a new location is randomly chosen.

The initial momenta of the nucleons 
are randomly chosen between 0 and the local
Thomas-Fermi-momentum:
\begin{equation}
p_F^{max} \,=\, \hbar c 
        \left( 3 \pi^2 \rho \right)^{\frac{1}{3}} \, ,
\end{equation}
with $\rho$ being the corresponding local proton- or neutron-density.

A principal disadvantage of this type of initialization is
that the initialized nuclei are not really in their ground-state
with respect to the Hamiltonian used for their propagation.
The parameters of the Hamiltonian (see section~\ref{hamiltonian})
are tuned to the equation of state of infinite nuclear matter and 
to properties of finite nuclei (such as their binding energy and
their root mean square radius). If, however, the energy of the 
nucleons within the nucleus is minimized according to the Hamiltonian
in a self-consistent fashion, then the nucleus would collapse to 
a single point in momentum space because the Pauli-principle
has not been taken into account in the Hamiltonian. 

A viable solution to this problem is the inclusion of fermionic
properties of the nucleons via the antisymmetrization of the
wave-function of the nucleus. This ansatz has first been implemented
in the framework of the Fermionic Molecular Dynamics (FMD) \cite{feldmeier90a}.
The FMD equations of motion are computationally very expensive -- it
has therefore not been possible so far, to calculate systems heavier
than Ca+Ca in the framework of FMD \cite{feldmeier97a}.

An alternative recipe is the use of a so-called Pauli-Potential
\cite{wilets77a} in the Hamiltonian. This potential, 
which is repulsive in configuration and momentum space,
allows to maintain the product ansatz for the wave-function of the nucleus.
A self-consistent minimization (e.g. via a Metropolis-algorithm
\cite{metropolis53a})
of the energy of the nucleus results
in a reasonable ground-state due to the Pauli-Potential mimicking
the fermionic properties of the nucleons.
A drawback of the Pauli-Potential is, however, that the kinetic momenta
of the nucleons are not anymore equivalent to their canonic momenta,
i.e. the nucleons carry the correct Fermi-momentum, but their velocity
is zero. Furthermore, the Pauli-Potential leads to a wrong specific heat and
changes the dynamics of fragment formation. A big advantage of the 
Pauli-Potential is that the initialized
nuclei remain absolutely stable whereas in the {\em conventional}
initialization and propagation without the Pauli-Potential the
nuclei start evaporating single nucleons after approximately
20 - 30 fm/c.

\section{Equations of motion}
\label{hamiltonian}
This section describes the real part of the nucleon-nucleon interaction
as it is implemented into the UrQMD model \cite{konopka96a}.
The interaction is based on a non-relativistic density-dependent 
Skyrme-type equation of state
with additional Yukawa- and Coulomb potentials.
Momentum dependent potentials are not used -- a Pauli-potential, however,
may be included optionally.

The nucleon- or baryon-density can be obtained from 
the Gaussian~(\ref{uqmd_gaussian}):
\begin{equation}
\label{uqmd_dens1}
\varrho_j(\vec{x}_j,t) \,=\, \left(\frac{2 \alpha}{\pi}\right)^{\frac{3}{2}}
\exp{\left\{ - 2 \alpha (\vec{x}_j - \vec{r}_j(t))^2 
        \right\} } \, .
\end{equation}
where $\vec{x}_j$ denotes the quantum mechanical position variable, while 
$\vec{r}_j(t)$ is the classical parameter of the Gaussian. The Skyrme-Potential (momentum-dependence and spin-exchange has been
neglected) has the form:
\begin{equation}
V^{Sk} \,=\, \frac{1}{2!}t_1 \sum_{j,k}\mbox{}'\delta(\vec{x}_j -\vec{x}_k) +
        \frac{1}{3!} t_2 \sum_{j,k,l}\mbox{}' \delta(\vec{x}_j -\vec{x}_k) 
        \delta(\vec{x}_j -\vec{x}_l) \quad,
\end{equation}
where in order to avoid self-interactions, all terms where at least two indices
are identical are discarded in the primed sum. This potential consists of a 
sum of two- and a three-body interaction terms. The two-body
term, which has a linear density-dependence models the long range attractive
component of the nucleon-nucleon interaction, whereas the three-body
term with its quadratic density-dependence is responsible for the
short range repulsive part of the interaction. 
Using the Gaussian~(\ref{uqmd_gaussian}) as the wave-function of the nucleon
we obtain for the two-body Skyrme potential of particle $j$:
\begin{eqnarray}
\label{uqmd_sk2}
V_j^{\rm Sk2} &=& \sum\limits_{k}^{N}\mbox{}' \int {\rm d}\vec{x}_j \,{\rm d}
        \vec{x}_k \,\varphi^*_j(\vec{x}_j) \,\varphi^*_k(\vec{x}_k) \,
        t_1 \delta(\vec{x}_j -\vec{x}_k) \,
        \varphi_j(\vec{x}_j) \,\varphi_k(\vec{x}_k) \nonumber\\
        &=& t_1 \sum\limits_{k}^{N}\mbox{}'
        \left( \frac{\alpha}{\pi} \right)^{\frac{3}{2}}
        \exp{\left\{ - \alpha (\vec{r}_j - \vec{r}_k)^2 \right\} }\nonumber \\
        &=& t_1 \varrho_j^{\rm int}(\vec{r}_j) \quad.
\end{eqnarray}
In the last line the interaction density was introduced. This density
has the same form as the nucleon density~(\ref{uqmd_dens1}) obtained
from the Wigner-transform of the Gaussian~(\ref{uqmd_gaussian}), 
but omits the nucleon at the
location $j$ and its Gaussian has twice the width of that used
in equation~(\ref{uqmd_dens1}).
The three-body potential for particle $j$ can be derived in an analogous 
fashion: 
\begin{eqnarray}
V_j^{\rm Sk3} &=& \frac{1}{2!} \sum\limits_{kl}^{N}\mbox{}'  
        \int {\rm d}\vec{x}_j \,{\rm d} \vec{x}_k \,{\rm d} \vec{x}_l \, 
        \varphi^*_j(\vec{x}_j) \, \varphi^*_k(\vec{x}_k) \,
        \varphi^*_l(\vec{x}_l)
        \nonumber \\
&&\qquad \qquad \times t_2 \delta(\vec{x}_j -\vec{x}_k) 
        \delta(\vec{x}_j-\vec{x}_l) \,
        \varphi_j(\vec{x}_j) \, \varphi_k(\vec{x}_k) \, \varphi_l(\vec{x}_l) \\
&=& t_2 \frac{1}{2!} \sum\limits_{kl}^{N}\mbox{}' 
        \left( \frac{4 \alpha^2}{3 \pi^2} \right)^{\frac{3}{2}}
        \exp{\left\{ - \frac{2}{3} \alpha \left( 
        (\vec{r}_j - \vec{r}_k)^2 + (\vec{r}_k - \vec{r}_l)^2
        + (\vec{r}_l - \vec{r}_j)^2 \right) \right\} } \quad. \nonumber
\end{eqnarray}
In the case of infinite nuclear matter the individual relative distances
should be close to their average value. Therefore the relative distance
between particle $k$ and $l$ may be substituted by the average of the
other two relative distances:
\begin{equation}
\label{uqmd_sk3}
V_j^{\rm Sk3} \,\approx\, \frac{1}{2!}   
        t_2 \sum\limits_{kl}^{N}\mbox{}' 
        \left( \frac{4 \alpha^2}{3 \pi^2} \right)^{\frac{3}{2}}
        \exp{\left\{ - \alpha \left(
        (\vec{r}_j - \vec{r}_k)^2 + (\vec{r}_j - \vec{r}_l)^2
        \right) \right\} } \quad. \nonumber
\end{equation}
Using the definition of the interaction-density given in 
equation~(\ref{uqmd_sk2}), the quadratic density dependence of the
three particle term~(\ref{uqmd_sk3}) may be generalized to arbitrary
exponents for the density. This is of great importance for the
implementation of a so-called {\em soft} equation of state. Then,
however, the interpretation of $V_j^{\rm Sk3}$ as three particle
interaction is no longer valid:
\begin{equation}
\label{uqmd_sk3_approx}
V_j^{\rm Sk3} \,\approx\,  t_2 3^{-\frac{3}{2}} (\varrho_j^{\rm int})^2
        \rightarrow
        t_\gamma (\gamma+1)^{-\frac{3}{2}} (\varrho_j^{\rm int})^\gamma \quad.
\end{equation}
In the UrQMD model expression~(\ref{uqmd_sk3_approx}) is always used, even
for the case $\gamma=2$.

The Yukawa-, Coulomb- and (optional) Pauli-potentials may be written
in the form of two-particle interactions:
\begin{eqnarray}
 V^{ij}_{\rm Yuk}&=& V_0^{\rm Yuk}
 \frac{\hbox{exp}\{-|\vec{r}_i-\vec{r}_j|/\gamma_Y\}}
 {|\vec{r}_i-\vec{r}_j|}\\
 V^{ij}_{\rm Coul}&=&\frac{Z_i Z_j e^2}
 {|\vec{r}_i-\vec{r}_j|} \\
 V^{ij}_{\rm Pau}&=&
V^0_{\rm Pau} \, \left( \frac{\hbar}{q_0 p_0} \right)^3 
        \hbox{exp}\left\{ - \frac{|\vec{r}_i-\vec{r}_j|^2}{2 q_0^2}
        - \frac{|\vec{p}_i-\vec{p}_j|^2}{2 p_0^2}  \right\}
        \delta_{\tau_i \tau_j} \delta_{\sigma_i \sigma_j}
\end{eqnarray}
$\sigma_j$ and $\tau_j$ denote the spin and isospin of particle $j$ and
$Z_j$ represents its charge.

In infinite nuclear matter the contribution of the Yukawa-potential to
the total energy has a linear density-dependence, just like the two-body
Skyrme-contribution. Therefore all parameter sets which satisfy the
following relation for the parameter $t_1$ yield the same equation
of state in infinite nuclear matter:
\begin{equation}
\frac{1}{2} t_1 + 2 \pi V_0^{\rm Yuk} \gamma^2_Y \,=\, {\rm const.}
\end{equation}
In finite nuclei the usage of a Yukawa-potential has the advantage
that the parameters can be tuned to the proper  surface properties of 
the nuclei without changing the equation of state.

\begin{table}[ht]
\begin{center}
\renewcommand{\arraystretch}{1.3}
\begin{tabular}{|c||cc|} \hline
\bf parameter & \bf without Pauli-potential & \bf with Pauli-potential \\\hline \hline
 $\alpha$ (fm$^{-2}$) & $0.25$  & $0.1152$ \\
$t_1$ (MeV fm$^3$) & $-7264.04$ & $-84.5$ \\
$t_\gamma$ (MeV fm$^6$) & $87.65$ & $188.2$ \\
$\gamma$ & $1.676$ & $1.46$ \\
$V_0^{\rm Yuk}$ (MeV fm) & $-0.498$ & $-85.1$ \\
$\gamma_Y$ (fm) & $1.4$ & $1.0$ \\\hline
$V_0^{Pauli}$ (MeV) & --        &       $98.95$ \\
$q_0$ (fm)      & --    & $2.16$ \\
$p_0$ (MeV/c) & --      & $120$ \\
\hline
\end{tabular}
\end{center} 
\caption{\label{eos_tab}
Parameters of the hard equation of state implemented in the UrQMD model, 
with and
without Pauli-potential.}
\end{table}

The classical UrQMD Hamiltonian which governs the motion of the parameters 
$\vec{r}_j$ and $\vec{p}_j$ of the wave-functions is thus given by:
\begin{eqnarray}
\label{uqmdhamiltonian}
H_{\rm UrQMD}& =\, \sum\limits_{j=1}^N E_j^{\rm kin} & +
\frac{1}{2} \sum\limits_{j=1}^N \sum\limits_{k=1}^N \left( E_{jk}^{\rm Sk2} + 
E_{jk}^{\rm Yukawa} + E_{jk}^{\rm Coulomb} + E_{jk}^{\rm Pauli} \right)
\nonumber \\
&& + \frac{1}{6} \sum\limits_{j=1}^N \sum\limits_{k=1}^N
\sum\limits_{l=1}^N E_{jkl}^{\rm Sk3}
\end{eqnarray}
The individual contributions are defined as:
\begin{eqnarray}
E_j^{\rm kin} & = & \sqrt{p_j^2 + m_j^2} \quad, \\
E_{jk}^{\rm Sk2} & = & t_1 \left(\frac{\alpha}{\pi}\right)^{\frac{3}{2}}
\exp{ \left\{ - \alpha r_{jk}^2 \right\} } \quad, \\
E_{jkl}^{\rm Sk3} &=& t_\gamma 
\left(\frac{4 \alpha^2}{3 \pi^2}\right)^{\frac{3}{2}}
\exp{\left\{-\alpha (r_{jk}^2+r_{jl}^2) \right\}} \quad, \\
E_{jk}^{\rm Yukawa} & = & V_0^{\rm Yuk} \frac{1}{2 r_{jk}}
\exp{\left\{\frac{1}{4 \alpha \gamma_Y^2}\right\}}
\left[\exp{\left\{-\frac{r_{jk}}{\gamma_Y}\right\}} \left( 1 - {\rm erf}\left(
        \frac{1}{2 \gamma_Y \sqrt{\alpha}} - \sqrt{\alpha} r_{jk}
                \right) \right) \right.  \nonumber \\
&& \qquad \qquad \qquad
        \left. - \exp{ \left\{ \frac{r_{jk}}{\gamma_Y} \right\} } 
        \left( 1 - {\rm erf}\left(
        \frac{1}{2 \gamma_Y \sqrt{\alpha}} + \sqrt{\alpha} r_{jk}
                \right) \right) \right] \quad, \\
E_{jk}^{\rm Coulomb} & = & \frac{Z_i Z_j \rm e^2}{r_{jk}} 
        {\rm erf}\left( \sqrt{\alpha} r_{jk} \right) \quad, \\
E_{jk}^{\rm Pauli} & = &
        V_0^{\rm Pau} \left(\frac{\hbar}{p_0 q_0}\right)^3
        \left(1+\frac{1}{2 \alpha q_0^2} \right)^{-\frac{3}{2}} \nonumber \\
&&  \qquad \qquad \qquad \times
        \exp{ \left\{ - \frac{\alpha r_{jk}^2}{2 \alpha q_0^2 +1} 
        - \frac{p_{jk}^2}{2 p_0^2} \right\}} 
        \delta_{\tau_j \tau_k} \delta_{\sigma_j \sigma_k} \quad,
\end{eqnarray}
with
\begin{equation}
r_{jk} = | \vec{r}_j - \vec{r}_k | \qquad \mbox{and} \quad
p_{jk} = | \vec{p}_j - \vec{p}_k | \quad.
\end{equation}
So far, only a hard equation of state has been implemented into the 
UrQMD model. The respective parameters are listed in table~\ref{eos_tab}.

Unfortunately, the generalization of such two body forces to the relativistic
region is not a simple task and is not incorporated in the present
UrQMD model.
In principle, the interaction must be mediated by fields, which are
propagated according to wave equations or one must make use of the
so called constraint Hamiltonian dynamics \cite{currie63a,sorge89a}.
Many other models avoid the propagation of fields using local
density approximation. Above 2 AGeV, we therefore resort to a cluster
decomposition in phase space, i.e. potential interactions are only enforced for
particles with relative momenta smaller than 2 GeV/c.

\section{The collision term}

The UrQMD collision term contains 55 different baryon species
(including nucleon, delta and hyperon resonances with masses up to 2.25 GeV/$c^2$) 
and 32 different meson species (including strange meson resonances), which
are supplemented by their corresponding anti-particle 
and all isospin-projected states.
The baryons and baryon-resonances which can be populated in UrQMD are listed
in table~\ref{bartab}, the respective mesons in table~\ref{mestab}. 
The states listed can either be produced in string decays, s-channel
collisions or resonance decays.
For excitations with higher masses than 2 GeV/$c^2$ a string picture is used.
Full baryon/antibaryon symmetry is included:
The number of implemented baryons therefore defines the number
of antibaryons in the model and the antibaryon-antibaryon interaction 
is defined via the baryon-baryon interaction cross sections.

\begin{table}
\begin{center}
\begin{tabular}{cccccc}
\hline \hline
nucleon&delta&lambda&sigma&xi&omega\\  \hline \hline
$N_{938} $&$\Delta_{1232}$&$\Lambda_{1116}$&$\Sigma_{1192}$
&$\Xi_{1315}$&$\Omega_{1672}$\\
$N_{1440}$&$\Delta_{1600}$&$\Lambda_{1405}$&$\Sigma_{1385}$&$\Xi_{1530}$&\\
$N_{1520}$&$\Delta_{1620}$&$\Lambda_{1520}$&$\Sigma_{1660}$&$\Xi_{1690}$&\\
$N_{1535}$&$\Delta_{1700}$&$\Lambda_{1600}$&$\Sigma_{1670}$&$\Xi_{1820}$&\\
$N_{1650}$&$\Delta_{1900}$&$\Lambda_{1670}$&$\Sigma_{1750}$&$\Xi_{1950}$&\\
$N_{1675}$&$\Delta_{1905}$&$\Lambda_{1690}$&$\Sigma_{1775}$&$\Xi_{2030}$&\\
$N_{1680}$&$\Delta_{1910}$&$\Lambda_{1800}$&$\Sigma_{1915}$&&\\
$N_{1700}$&$\Delta_{1920}$&$\Lambda_{1810}$&$\Sigma_{1940}$&&\\
$N_{1710}$&$\Delta_{1930}$&$\Lambda_{1820}$&$\Sigma_{2030}$&&\\
$N_{1720}$&$\Delta_{1950}$&$\Lambda_{1830}$&&&\\
$N_{1900}$&&$\Lambda_{1890}$&&&\\
$N_{1990}$&&$\Lambda_{2100}$&&&\\
$N_{2080}$&&$\Lambda_{2110}$&&&\\
$N_{2190}$ &&&&&\\
$N_{2200}$ &&&&&\\
$N_{2250}$ &&&&&\\
\hline \hline
\end{tabular}
\caption{\label{bartab} Baryons and baryon-resonances included into
the UrQMD model. Through baryon-antibaryon symmetry the respective
antibaryon states are included as well.}
\end{center}
\end{table}

Elementary cross sections are fitted to available proton-proton or
pion-proton data. Isospin symmetry is used when possible in order
to reduce the number of individual cross sections which have to
be parameterized or tabulated.

\begin{table}
\begin{center}
\begin{tabular}{cccccccc}
$0^{-+}$  & $1^{--}$   & $ 0^{++}$ & $ 1^{++}$ & $1^{+-}$ &$ 2^{++}$ & $(1^{--})^*$    & $(1^{--})^{**}$\\ \hline \hline
 $\pi$    & $ \rho$    & $ a_0$    &  $ a_1$   & $ b_1 $  &  $ a_2$  & $ \rho_{1450}$  & $\rho_{1700}$\\ 
 $K  $    & $   K^*$   & $ K_0^*$  &  $ K_1^*$ & $ K_1 $  & $ K_2^*$ & $ K^*_{1410}$   & $K^*_{1680}$\\
 $\eta$   & $  \omega$ & $ f_0 $   &  $f_1$    & $ h_1 $  & $ f_2 $  & $ \omega_{1420}$& $\omega_{1662}$\\
 $\eta'$  & $\phi $    & $f_0^*$   &  $ f_1'$  & $ h_1'$  & $ f_2'$  & $ \phi_{1680}$  & $\phi_{1900}$\\
\end{tabular}
\caption{\label{mestab} Mesons and meson-resonances, sorted with
respect to spin and parity, included into
the UrQMD model.}
\end{center}
\end{table}

\subsection{The collision criterion}
\label{collcrit}

In the UrQMD model hadron-hadron collisions are performed
stochastically, in a similar way as in the original cascade 
models \cite{cugnon80a}. The cross section is
interpreted geometrically as an area.
Two particles collide if their distance $d_{\rm trans}$ 
fulfills the relation:
\begin{equation}
\label{collcrit1}
 d_{\rm trans} \le d_0 = \sqrt{ \frac { \sigma_{\rm tot} } {\pi}  }  , \qquad
 \sigma_{\rm tot} = \sigma(\sqrt{s},{\rm type} )\quad.
\end{equation}
The total cross section $\sigma_{\rm tot}$ depends on the c.m. energy
$\sqrt{s}$ and on the species and quantum numbers 
of the incoming particles. $d_{\rm trans}$ is the relative distance between
the two colliding particles (in three dimensional configuration space).
At the point of closest approach this distance is purely transverse
with regard to the relative velocity vector of the particles.
 
During the calculation each particle is checked at the beginning of each
time step whether it will collide according to criterion~(\ref{collcrit1})
within that time step. After each binary collision
or decay the outgoing particles are checked for further collisions
within the respective time step. This procedure
assumes that all particles have the same clock, i.e. that a 
$6N+1$ dimensional phase-space is used.

In relativistic heavy ion collisions, however, the relative distance
between the particles depends on the reference-frame of the calculation.
Therefore the time-order of the binary collisions and their cross sections
also depend on the reference frame. One possible ansatz to overcome 
this unphysical frame-dependence is the use of a $8N$ dimensional
phase-space \cite{sorge89a}, although even here the time-order of the
collisions is not unique.  
In $8N$ dimensional phase space 
each particle has its own eigentime. The connection
between a time-evolution parameter $\tau$ (the system clock) 
and the individual eigentimes is constructed via $2N-1$ Lorentz-covariant
constraints, which reduce the $8N$ dimensional phase-space to a
$6N+1$ dimensional phase-space. The particles move in Minkowsky-space
along 4-dimensional curved trajectories. At the beginning of each time step
(in units of $\tau$) criterion~(\ref{collcrit1}) is applied to scan
for collisions. Now, however, $d_{\rm trans}$ is defined as the
covariant relative distance:
\begin{equation}
d_{\rm trans}=\sqrt{\left(\frac{(q_n-q_m)^\nu (p_m+p_n)_\nu}{(p_n+p_m)^2}
(p_n+p_m)^\mu-(q_n-q_m)^\mu\right)^2} \quad,
\end{equation}
using 4-vectors for the locations $q_\nu$ and momenta $p_\nu$ of
the particles.

In the UrQMD model a $6N+1$ dimensional phase-space has been
used. We therefore use a different ansatz to  minimize the frame-dependence 
of the collisions:
The impact parameter of two colliding particles -- here also called
$d_{\rm trans}$ -- is well defined in the local rest-frame of the
two particles. It corresponds to the relative distance between the
two particles at the time of closest approach. 
In order to compute the time of closest approach of the two particles
one always transforms into the local rest-frame of the two particles
(using a $6N+1$ dimensional phase space -- this implies that no time-coordinates
are transformed). With $\vec{q}_i$ being the locations and $\vec{p'}_i$
the momenta in the local rest-frame one obtains for the squared
impact parameter $d^2_{\rm trans}$:
\begin{equation}
d^2_{\rm trans}\,=\, \vec{d}^2 - \vec{d}^2_{\|} \,=\,
        (\vec{q}_1 - \vec{q}_2)^2 - 
        \frac{\left((\vec{q}_1 - 
        \vec{q}_2)\cdot(\vec{p'}_1 - \vec{p'}_2)\right)^2}
             {(\vec{p'}_1 - \vec{p'}_2)^2} \quad,
\end{equation}
The constraint of always using the local rest-frame of the colliding
particles ensures that the cross section of the two particles is always
calculated in the same fashion and does not depend on the reference frame
of the nucleus-nucleus collision.

The time of closest approach $\tau_{\rm coll}$ 
(i.e. the collision time), however, still
depends on the reference-frame of the nucleus-nucleus reaction.
This dependence cannot be avoided since the system clock is linked to
that reference-frame. Therefore the time-order of the individual binary
collisions strongly varies with the respective reference-frame 
\cite{kodama83a}.
Using the locations $\vec{r}_i$ and the momenta $\vec{p}_i$ in
the reference frame of the nucleus-nucleus collision one obtains for
the time of closest approach for the two colliding particles:
\begin{equation}
\tau_{\rm coll} \, = \, - 
        \frac{(\vec{r}_1 - \vec{r}_2)\cdot(\vec{p}_1/E_1 - \vec{p}_2/E_2)}
             {(\vec{p}_1/E_1 - \vec{p}_2/E_2)^2} \quad.
\end{equation}

We have studied the computational frame dependence in the system
S+S at 200 GeV/nucleon: using the above described algorithm, particle
multiplicities and collision numbers vary by less than 3\% between
the laboratory frame and the CM frame. 

The frame-independent definition of the cross section (via the 
impact parameter in the two-particle rest frame) is an important
factor in ensuring the approximate reference-frame independence.

\subsection{Cross sections}

\subsubsection{Nucleon-nucleon interactions}
\label{res_sig}
\label{resonances}

In UrQMD cross sections are a function of the incoming and outgoing particle
types, their isospins and their c.m. energy. They may either be tabulated,
parameterized according to an algebraic function or extracted from
other cross sections via general principles, such as detailed balance or
the additive quark model.

\begin{figure}[tb]
\begin{minipage}[t]{9cm}
\centerline{\psfig{figure=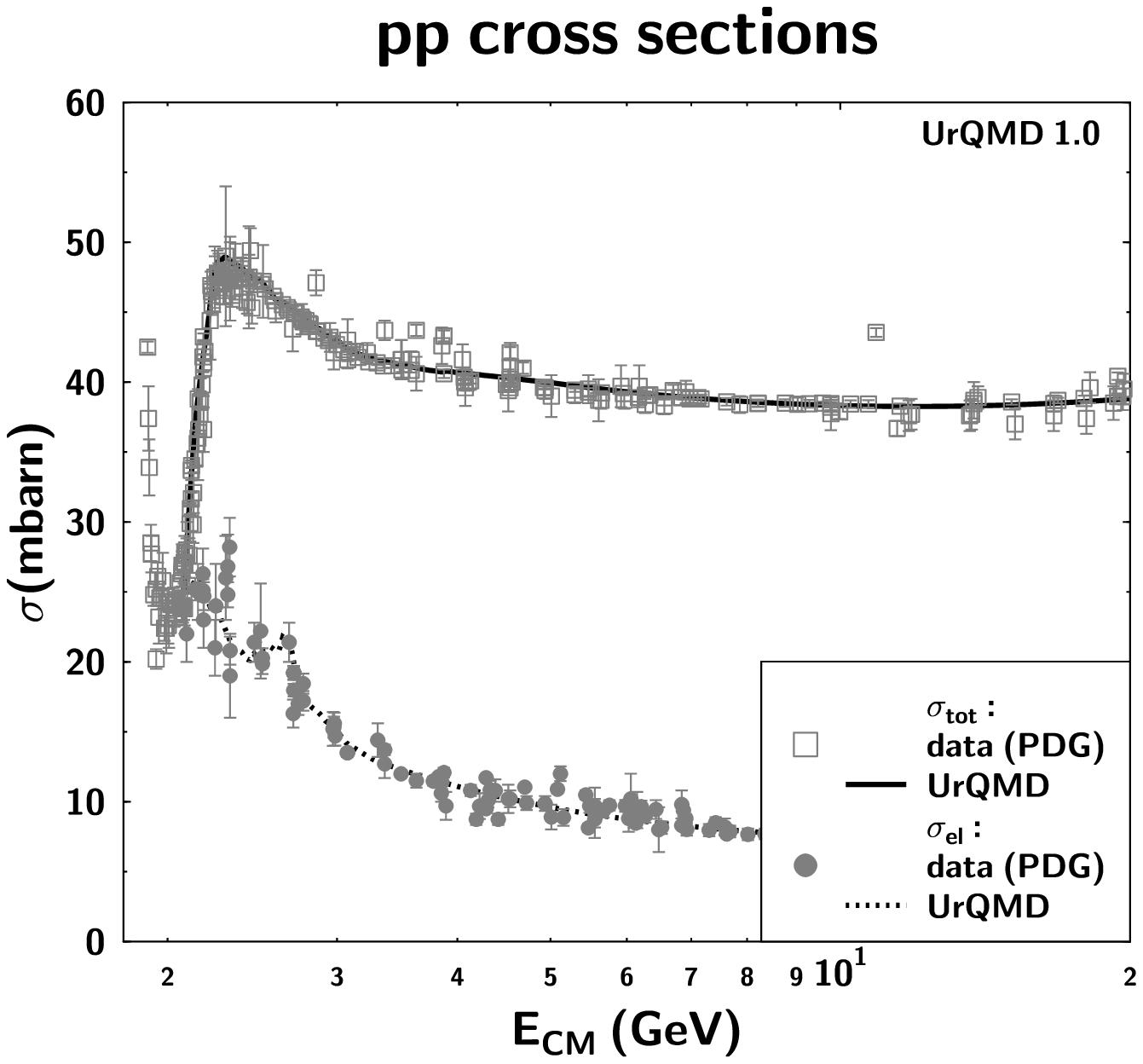,width=9cm}}
\caption{\label{sigpp}
UrQMD parameterization of the total and elastic proton-proton cross section.
The data has been taken from \protect \cite{PDG96}. A table-lookup
has been used at low energies to properly describe the data.}
\end{minipage}
\hfill
\begin{minipage}[t]{9cm}
\centerline{\psfig{figure=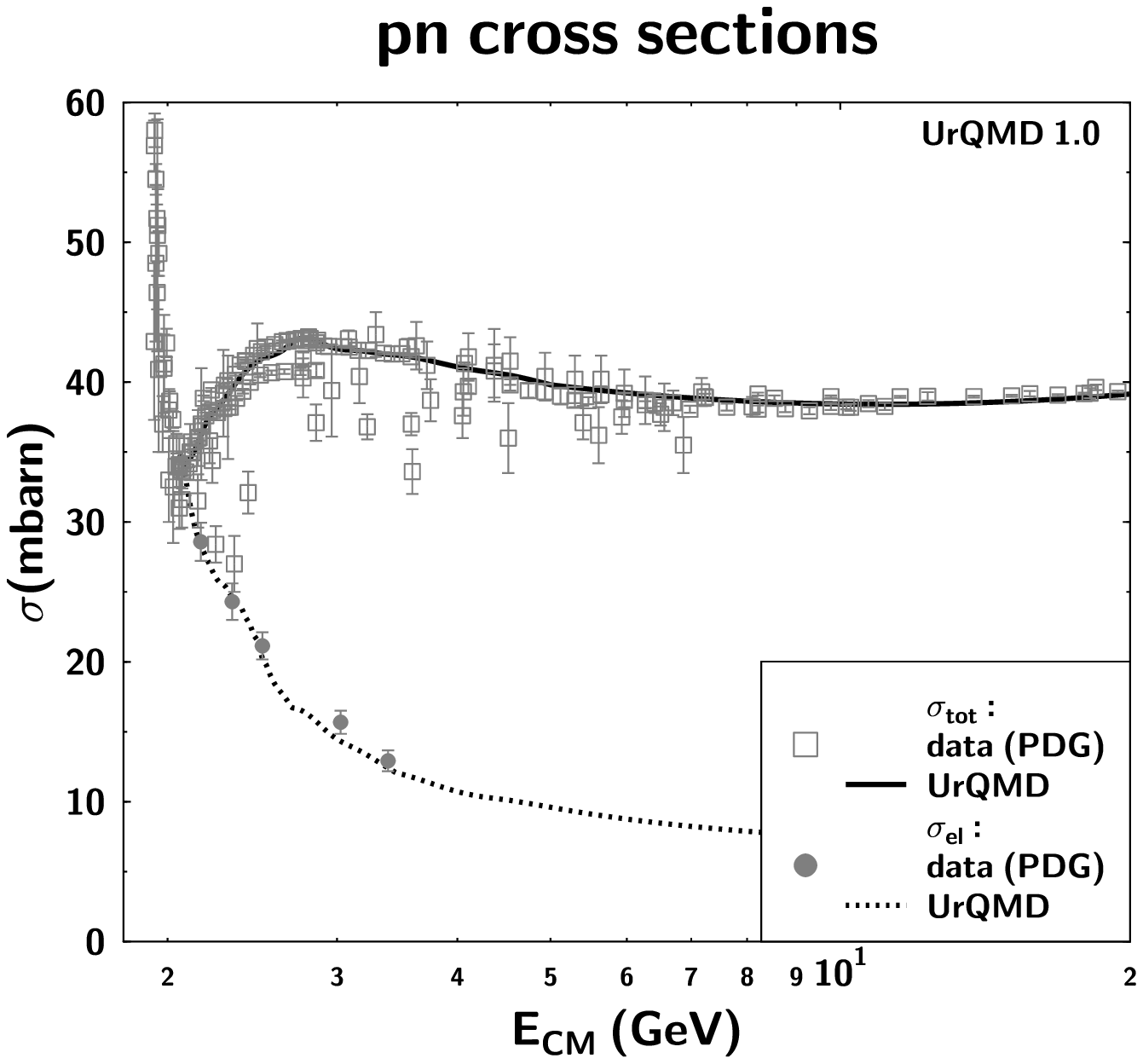,width=9cm}}
\caption{\label{sigpn}
UrQMD parameterization of the total and elastic proton-neutron cross section.
The data has been taken from \protect \cite{PDG96}. A table-lookup
has been used at low energies to properly describe the data.
Note the large differences between the proton-proton and the proton-neutron
data which emphasize the importance of correct isospin treatment}
\end{minipage}
\end{figure}

The total and elastic proton-proton and proton-neutron cross sections
are well known \cite{PDG96}. 
Since their functional dependence on $\sqrt{s}$ shows
at low energies a complicated shape, UrQMD uses a table-lookup for those
cross sections. Figure~\ref{sigpp} shows the comparison between 
measurements for the proton-proton elastic and total cross sections and
the respective UrQMD table lookup. The same comparison for the proton-neutron
case is shown in figure~\ref{sigpn}. 
At low energies large differences
are visible between the proton-proton and the proton-neutron cross sections.
Therefore the proper treatment of isospin -- especially at low energies --
is of major importance; simple averaged nucleon-nucleon cross sections
should not be used.
The neutron-neutron cross section is treated as equal to the proton-proton
cross section (isospin-symmetry).
In the high energy limit
($\sqrt{s} \ge 5$~GeV) the CERN/HERA parameterization for the proton-proton
cross section is used \cite{PDG96}.

Particle production in UrQMD either takes place via the decay of a
meson- or baryon-resonance or via a string excitation and fragmentation.
Up to incident beam energies of 8--10 GeV/nucleon particle production
is dominated by resonance decays. Production cross sections for 
the excitation of individual resonances can be calculated in the framework of
OPE or OBE models \cite{berenguer93a}.
Regarding the number of implemented resonances in UrQMD and considering 
the limited applicable energy-range for cross sections calculated within
OPE and OBE models the 
calculation of all implemented resonance excitation cross sections 
in the framework of these models is not practical. 
We therefore  employ in UrQMD 
an effective parameterization based on simple phase space considerations;
free parameters are tuned to experimental measurements.
The cross section has the general form:
\begin{equation}
\label{ppinelform}
\sigma_{1,2 \to 3,4}(\sqrt{s}) \,\sim (2 S_3 + 1) (2 S_4 +1)
\frac{\langle p_{3,4} \rangle}{\langle p_{1,2} \rangle} \,
\frac{1}{(\sqrt{s})^2} \, | {\cal M}(m_3,m_4) |^2 \, .
\end{equation}
The matrix element $| {\cal M}(m_3,m_4) |^2$ is assumed to have
no spin-dependence but may depend on the masses of the outgoing
particles. The cross section depends also on the momenta of the
in- and outgoing particles in the two-particle rest-frame
$\langle p_{i,j} \rangle$.
If the outgoing particles are stable particles with a well-defined mass, 
$\langle p_{3,4} \rangle$ is defined as:
\begin{equation}
\label{pcmsfixed}
  \langle p_{3,4}(\sqrt{s}) \rangle \,=\, p_{CMS}(\sqrt{s}) \,=\, {1 \over 2\sqrt{s}  }\sqrt{
   \left( s-(m_3 + m_4)^2 \right)   \left( s-(m_3-m_4)^2 \right)} \quad.
\end{equation}
($\langle p_{1,2} \rangle$ is defined as above exchanging $m_3$, $m_4$ by
$m_1$, $m_2$, respectively).
If, however, 3 or 4 are resonances, the width of their mass distribution
must be taken into account. Then  relation~(\ref{pcmsfixed}) must be
modified by an integral over the mass distribution $A_r(m)$  of the 
respective resonance (in the case both are resonances):
\begin{equation}
\label{pcmsmd}
  \langle p_{3,4}(\sqrt{s}) \rangle \,=\,
 \int  \int 
  p_{CMS}(\sqrt{s},m_3,m_4)\, A_3(m_3) \, A_4(m_4) \, 
  {\rm d} m_3\; {\rm d} m_4 \, .
\end{equation}
The mass distribution $A_r(m)$ is normalized to unity.
The upper and lower boundaries for the integration in equation~(\ref{pcmsmd})
are determined by the production thresholds for the respective resonances
(e.g. in UrQMD this threshold is 
$m_N + m_{\pi} = 1.07$ GeV for non-strange baryon resonances) and the 
maximum available energy (e.g. in the case of $p p \to p \Delta$ 
the maximum available energy is $\sqrt{s} - m_N$).
For the solution of this integral we assume the mass distribution to have
the shape of a free Breit-Wigner distribution with a mass-dependent
width:
\begin{equation}
\label{bwnorm}
A_r(m) \, = \, \frac{1}{N} 
        \frac{\Gamma(m)}{(m_r - m)^2 + \Gamma(m)^2/4} \qquad \mbox{with} \quad
        \lim_{\Gamma \rightarrow 0} A_r(m) = \delta(m_r - m) \, ,
\end{equation}
with the normalization constant
\begin{equation}
\label{bwnorm_norm}
N \,=\, \int\limits_{-\infty}^{\infty} 
\frac{\Gamma(m)}{(m_r - m)^2 + \Gamma(m)^2/4} \, {\rm d}m \, .
\end{equation}
Alternatively one can also choose a Breit-Wigner distribution with a fixed
width, the normalization constant then has the value $N=2\pi$.

\begin{figure}[tb]
\begin{minipage}[t]{9cm}
\centerline{\psfig{figure=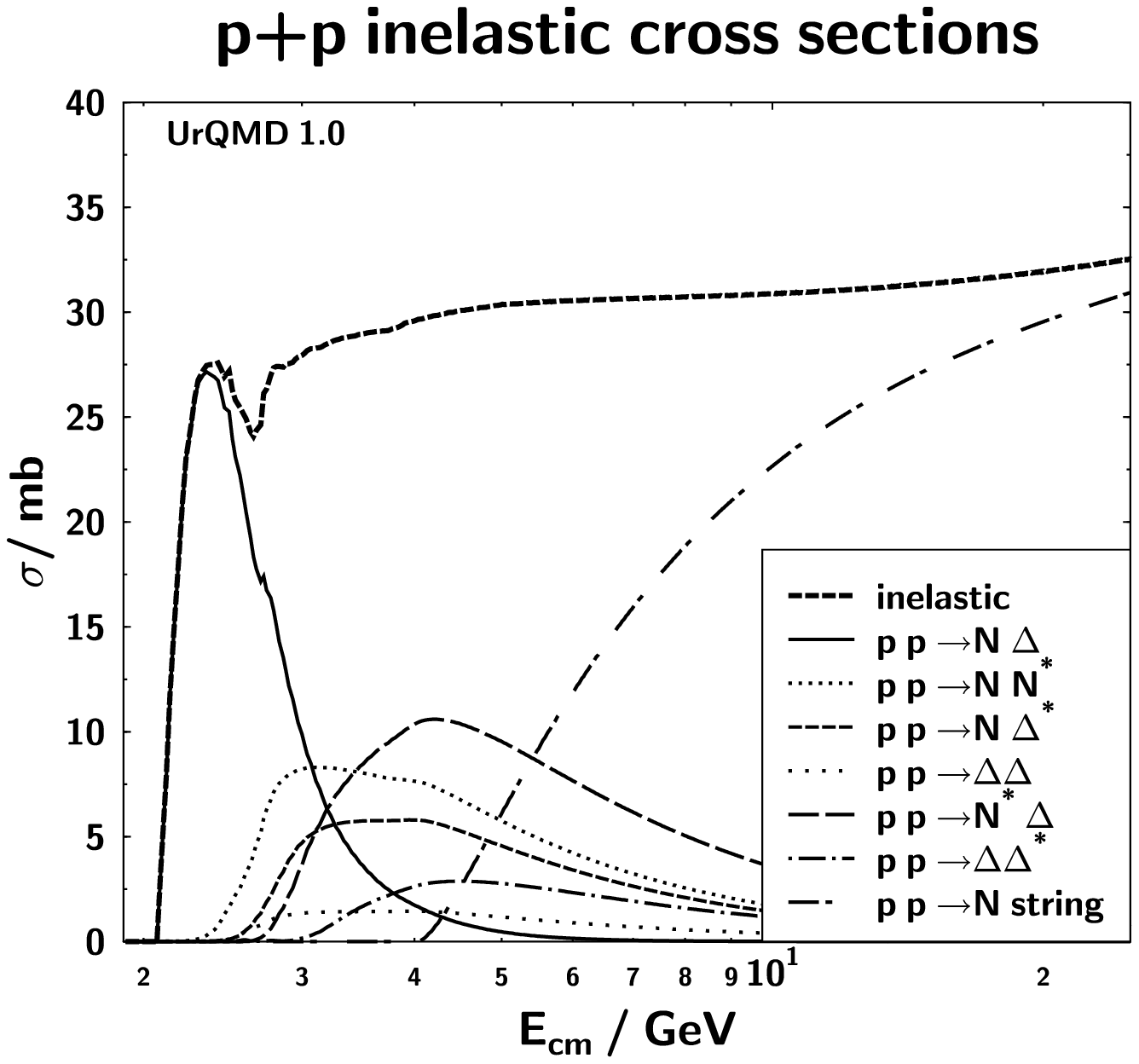,width=9cm}}
\caption{\label{pp-inel}
Inelastic proton-proton cross section in UrQMD, subdivided into the 
classes
$N N \to N \Delta_{1232}$, $N N \to N N^*$, $N N \to N \Delta^*$,
$N N \to \Delta_{1232} \Delta_{1232}$, $N N \to \Delta_{1232} N^*$, 
$N N \to \Delta_{1232} \Delta^*$ and $N N \to N string$.
For high energies, string-excitation is dominant.}
\end{minipage}
\hfill
\begin{minipage}[t]{9cm}
\centerline{\psfig{figure=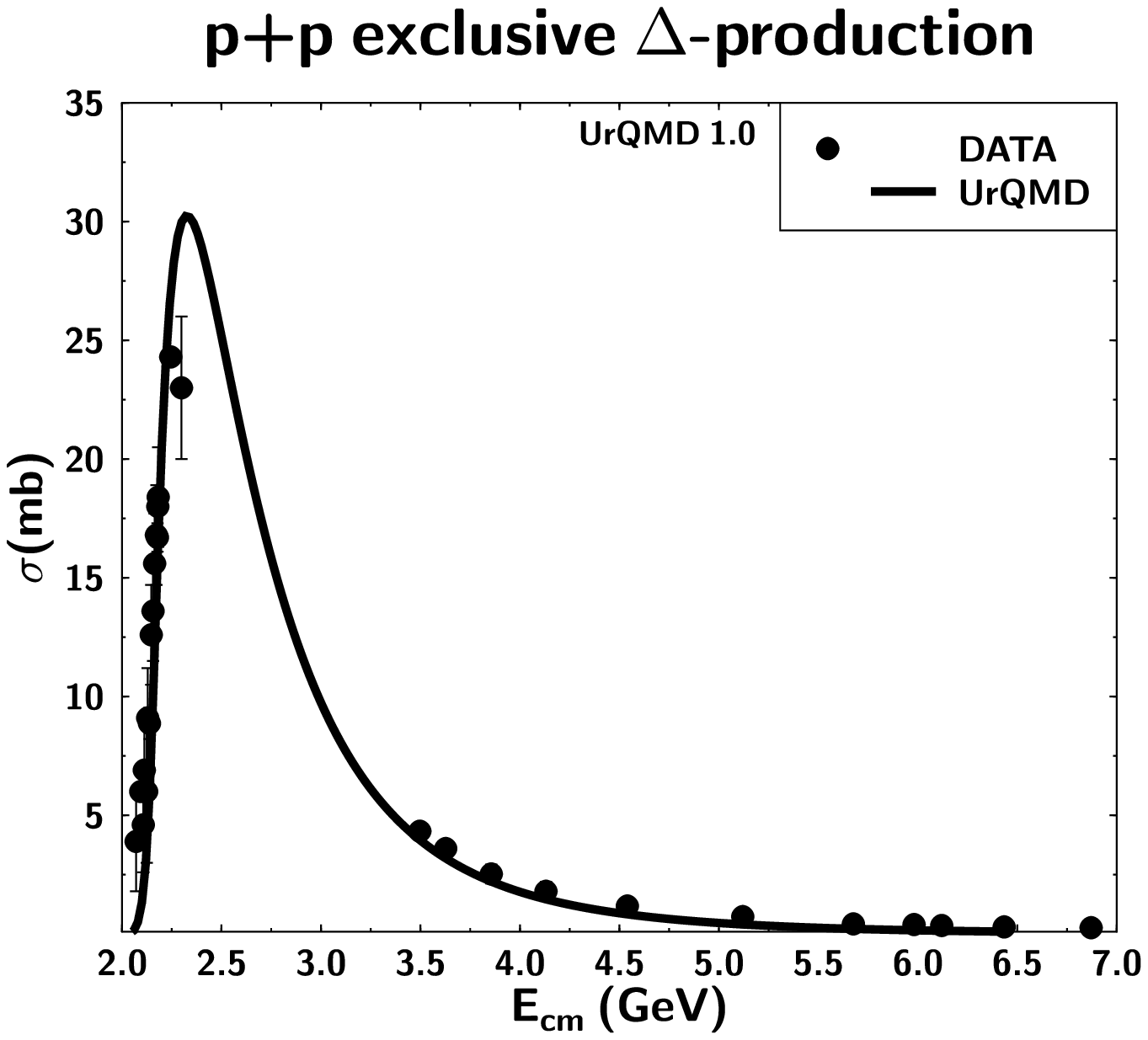,width=9cm}}
\caption{\label{pp-nd1232}
UrQMD fit for the exclusive $\Delta_{1232}$ production in proton-proton
reactions  compared to data
\protect\cite{flaminio84a}. }
\end{minipage}
\end{figure}

In UrQMD, the excitation of non-strange resonances is 
subdivided into 6 classes:
$N N \to N \Delta_{1232}$, $N N \to N N^*$, 
$N N \to N \Delta^*$, $N N \to \Delta_{1232} \Delta_{1232}$, 
$N N \to \Delta_{1232} N^*$ and $N N \to \Delta_{1232} \Delta^*$.
Here the $\Delta_{1232}$ is explicitly listed,
whereas higher excitations of the $\Delta$ resonance
have been denoted as $\Delta^*$.
For each of these classes specific assumptions are made with regard
to the form of the matrix-element $| {\cal M}(m_3,m_4) |^2$.
Free parameters are tuned to experimental measurements, when available.
We make the following assumptions for the matrix elements:
\begin{enumerate}
\item $N N \to N \Delta_{1232}$ excitation:
\begin{equation}
| {\cal M}(\sqrt{s},m_3,m_4) |^2 \,=\, A\, \frac{m_\Delta^2 \Gamma_\Delta^2}
{\left( ( \sqrt{s})^2 - m_\Delta^2 \right)^2 + m_\Delta^2 \Gamma_\Delta^2}
\quad,
\end{equation}
with $m_\Delta=1.232$ GeV, $\Gamma_\Delta=0.115$ GeV and $A=40000$. Note that
this form of the matrix element has been adjusted to fit the data shown in
figure \ref{pp-nd1232}.
\item $N N \to N N^*$, $N N \to N \Delta^*$, 
$N N \to \Delta_{1232} N^*$ and $N N \to \Delta_{1232} \Delta^*$ excitation:
\begin{equation}
| {\cal M}(m_3,m_4) |^2 \,=\, A\, \frac{1}
{(m_4-m_3)^2 \, (m_4+m_3)^2}
\quad,
\end{equation}
with $A=6.3$ for $N N \to N N^*$, $A=12$ for $N N \to N \Delta^*$
and $A=3.5$ for  $N N \to \Delta_{1232} N^*$. In the case of
$N N \to \Delta_{1232} \Delta^*$ there are insufficient data available,
therefore we use the same matrix element and parameters as in the
case of $N N \to \Delta_{1232} N^*$.
Since $m_3 \ne m_4$ is valid for all above cases, 
the matrix element cannot diverge.
\item $N N \to \Delta \Delta$ excitation:
\begin{equation}
| {\cal M}(m_3,m_4) |^2 \,=\, A
\quad,
\end{equation}
with $A=2.8$.
\end{enumerate}

E. g. the total cross section for the excitation of a $N^*$ resonance
can be obtained by summing equation~(\ref{ppinelform}) over all 
$N^*$ resonances implemented in the UrQMD model. Figure~\ref{pp-inel}
shows the total cross sections for the excitation of the 
different resonance classes $N N \to N \Delta_{1232}$, $N N \to N N^*$, 
$N N \to N \Delta^*$,
$N N \to \Delta_{1232} \Delta_{1232}$, $N N \to \Delta_{1232} N^*$ and
$N N \to \Delta_{1232} \Delta^*$.
Also plotted is the cross section for the excitation of a string
which is defined as the difference between total inelastic
cross section and the sum of all the resonance excitation cross sections:
$\sigma_{\rm string} = \sigma_{\rm tot} - 
\sigma_{\rm el} - \sum \sigma^i_{\rm inel}$.
For $\sqrt{s} \ge 6$ GeV the string excitation cross section yields
the strongest contribution to the total inelastic cross section.

Figure~\ref{pp-nd1232} shows  the fit of the UrQMD
$ p p \to N \Delta_{1232}$ cross section to experimental measurements
\cite{flaminio84a}. The measurements refer to the $\Delta^++ n$ 
exit channel and have been rescaled to match the full isospin-summed
cross section. In the case of the exclusive $\Delta_{1232}$
cross section the quality of the data and thus also the quality
of the resulting fit is very good. For higher resonance excitations
this is unfortunately no longer the case and additional measurements 
are needed to clarify the situation.
One has to keep in mind, however, that the experimental extraction
of exclusive resonance production cross sections is only possible
via two- or three-particle correlations (e.g. a pion-nucleon correlation
in the case of the $\Delta$) which introduces large systematic errors,
especially for broad resonances.

\begin{figure}[bth]
\centerline{\psfig{figure=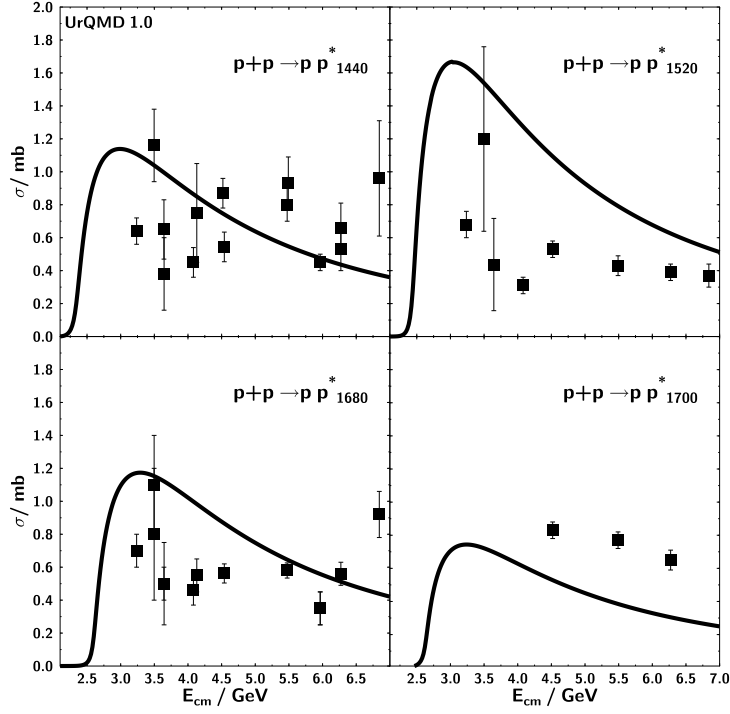,width=11cm}}
\caption{\label{pp-ppstar}
UrQMD parameterization for exclusive $p p^*$ cross sections.
Only one parameter was used to describe all  available 
cross section data \protect\cite{flaminio84a}. 
}
\end{figure}

In figure~\ref{pp-ppstar} the UrQMD cross sections for the processes
$p p \to p p^*_{1440}$, $p p \to p p^*_{1520}$
$p p \to p p^*_{1680}$ and $p p \to p p^*_{1700}$
are compared to data \cite{flaminio84a}.
One single parameter has been used to fit all four cross sections.
The data situation is not as good as in the case of the
$\Delta_{1232}$ resonance, some ambiguities are visible which results in
the quality of the fit being not as good as in the previous case. 
The parameters for the other classes are fitted in the same fashion.

After the species of the excited resonance has been assigned,
its mass must be determined: This happens stochastically
between the minimum allowed mass for the resonance and the maximum
available free energy according to the mass distribution~(\ref{bwnorm}).

In the UrQMD model, the full decay width $\Gamma_{tot}(M)$ of a resonance is 
defined as the sum of all partial decay widths and depends on the
mass of the excited resonance:
\begin{equation}
\Gamma_{tot}(M)  
\label{gammatot}
       \,=\, \sum  \limits_{br= \{i,j\}}^{N_{br}} \Gamma_{i,j}(M) \quad.
\end{equation}
The partial decay widths $\Gamma_{i,j}(M)$ for the decay into the 
exit channel with particles $i$ and $j$ is given by:
\begin{equation}
\label{gammapart}
\Gamma_{i,j}(M)
        \,=\,
       \Gamma^{i,j}_{R} \frac{M_{R}}{M}
        \left( \frac{\langle p_{i,j}(M) \rangle}
                    {\langle p_{i,j}(M_{R}) \rangle} \right)^{2l+1}
         \frac{1.2}{1+ 0.2 
        \left( \frac{\langle p_{i,j}(M) \rangle}
                    {\langle p_{i,j}(M_{R}) \rangle} \right)^{2l} } \quad,
\end{equation}
here $M_R$ denotes the pole mass of the resonance, $\Gamma^{i,j}_{R}$
its partial decay width into the channel $i$ and $j$ at the pole and
$l$ the decay angular momentum of the exit channel.
All pole masses and partial decay widths at the pole are taken from the Review
of Particle Properties \cite{PDG96}. 
$\Gamma_{i,j}(M)$ is constructed in such a way that 
$\Gamma_{i,j}(M_R)=\Gamma^{i,j}_R$ is fulfilled at the pole.
In many cases only crude estimates for $\Gamma^{i,j}_R$ are given
in \cite{PDG96} -- the partial decay widths must then be fixed by
studying exclusive particle production in elementary proton-proton
and pion-proton reactions. 

All masses, full widths and decay probabilities used in UrQMD are
listed in table~\ref{respars1} for non-strange baryon-resonances,
in table~\ref{respars2} for single-strange baryon-resonances,
 in table~\ref{respars3} for double-strange baryon-resonances, 
and in tables~\ref{mespars1} and~\ref{mespars2} for meson-resonances.
Due to experimental uncertainties there frequently exists a large interval
in which the respective parameters may be tuned. 
Therefore the given set of parameters is not unique. However, the known
uncertainties have been exploited to gain an optimized fit to the available
inclusive and exclusive meson production data.

An important component in the computation of mass-dependent decay widths
is the momentum of the decay products in the rest frame of the
resonance, $\langle p_{i,j}(M) \rangle$, which has been defined
in equations~(\ref{pcmsfixed}) and~(\ref{pcmsmd}).
In order to avoid recursion (the full decay width as defined in
(\ref{gammatot}) is needed for the evaluation of 
(\ref{pcmsmd})),
the full and partial decay widths are evaluated at the beginning
of the calculation using a mass distribution $A_r(M)$ with a fixed
mass-independent width  and then tabulated for
further use.

\pagebreak

\begin{table}
\renewcommand{\arraystretch}{1.3}
\begin{center}
\scriptsize
\begin{tabular}{|l||rr|rrrrrrrrr|} \hline
\bf resonance & \bf mass & \bf width & 
$N \gamma$ & $N \pi$ & $N \eta$ & $N \omega$ & $N \varrho$ &
$N \pi \pi$ & $\Delta_{1232} \pi$ & $N^*_{1440} \pi$ & $\Lambda K$ \\\hline\hline
$N^*_{1440}$ & 1.440 & 200  &      & 0.70 &      &      &      & 0.05 & 0.25 &      &      \\
$N^*_{1520}$ & 1.520 & 125  &      & 0.60 &      &      &      & 0.15 & 0.25 &      &      \\
$N^*_{1535}$ & 1.535 & 150  & 0.001& 0.55 & 0.35 &      &      & 0.05 &      & 0.05 &      \\
$N^*_{1650}$ & 1.650 & 150  &      & 0.65 & 0.05 &      &      & 0.05 & 0.10 & 0.05 & 0.10 \\
$N^*_{1675}$ & 1.675 & 140  &      & 0.45 &      &      &      &      & 0.55 &      &      \\
$N^*_{1680}$ & 1.680 & 120  &      & 0.65 &      &      &      & 0.20 & 0.15 &      &      \\
$N^*_{1700}$ & 1.700 & 100  &      & 0.10 & 0.05 &      & 0.05 & 0.45 & 0.35 &      &      \\
$N^*_{1710}$ & 1.710 & 110  &      & 0.15 & 0.20 &      & 0.05 & 0.20 & 0.20 & 0.10 & 0.10 \\
$N^*_{1720}$ & 1.720 & 150  &      & 0.15 &      &      & 0.25 & 0.45 & 0.10 &      & 0.05 \\
$N^*_{1900}$ & 1.870 & 500  &      & 0.35 &      & 0.55 & 0.05 &      & 0.05 &      &      \\
$N^*_{1990}$ & 1.990 & 550  &      & 0.05 &      &      & 0.15 & 0.25 & 0.30 & 0.15 & 0.10 \\
$N^*_{2080}$ & 2.040 & 250  &      & 0.60 & 0.05 &      & 0.25 & 0.05 & 0.05 &      &      \\
$N^*_{2190}$ & 2.190 & 550  &      & 0.35 &      &      & 0.30 & 0.15 & 0.15 & 0.05 &      \\
$N^*_{2220}$ & 2.220 & 550  &      & 0.35 &      &      & 0.25 & 0.20 & 0.20 &      &      \\
$N^*_{2250}$ & 2.250 & 470  &      & 0.30 &      &      & 0.25 & 0.20 & 0.20 & 0.05 &      \\
$\Delta_{1232}$ & 1.232 & 115. & 0.01 & 1.00 &      &      &      &      &      &      &      \\
$\Delta^*_{1600}$ & 1.700 & 200  &      & 0.15 &      &      &      &      & 0.55 & 0.30 &      \\
$\Delta^*_{1620}$ & 1.675 & 180  &      & 0.25 &      &      &      &      & 0.60 & 0.15 &      \\
$\Delta^*_{1700}$ & 1.750 & 300  &      & 0.20 &      &      & 0.10 &      & 0.55 & 0.15 &      \\
$\Delta^*_{1900}$ & 1.850 & 240  &      & 0.30 &      &      & 0.15 &      & 0.30 & 0.25 &      \\
$\Delta^*_{1905}$ & 1.880 & 280  &      & 0.20 &      &      & 0.60 &      & 0.10 & 0.10 &      \\
$\Delta^*_{1910}$ & 1.900 & 250  &      & 0.35 &      &      & 0.40 &      & 0.15 & 0.10 &      \\
$\Delta^*_{1920}$ & 1.920 & 150  &      & 0.15 &      &      & 0.30 &      & 0.30 & 0.25 &      \\
$\Delta^*_{1930}$ & 1.930 & 250  &      & 0.20 &      &      & 0.25 &      & 0.25 & 0.30 &      \\
$\Delta^*_{1950}$ & 1.950 & 250  & 0.01 & 0.45 &      &      & 0.15 &      & 0.20 & 0.20 &      \\
\hline
\end{tabular} 
\end{center}
\caption{\label{respars1}
Masses, widths and branching ratios for non-strange baryon-resonances
in UrQMD. Masses are given in GeV and the widths in MeV. All parameters
are within the range given by the Review of Particle Properties 
\protect\cite{PDG96} and have been tuned to exclusive particle production
channels.}
\end{table}

\begin{table}
\renewcommand{\arraystretch}{1.3}
\begin{center}
\scriptsize
\begin{tabular}{|l||rr|rrrrrrrrrrr|} \hline
\bf resonance & \bf mass & \bf width & 
$N \bar K$ & $N \bar K^*_{892}$ & $\Sigma \pi$ & $\Sigma^* \pi$ & 
$\Lambda \gamma$ & $\Lambda \eta$ &  $\Lambda \omega$ & $\Lambda \pi$ & 
$\Sigma \eta$ & $\Lambda^* \pi$ & $\Delta \bar K$ \\\hline\hline
$\Lambda^*_{1405}$ & 1.407 &  50  &      &      & 1.00 &      &      &      &      &      &      &      &      \\
$\Lambda^*_{1520}$ & 1.520 &  16  & 0.45 &      & 0.43 & 0.11 & 0.01 &      &      &      &      &      &      \\
$\Lambda^*_{1600}$ & 1.600 & 150  & 0.35 &      & 0.65 &      &      &      &      &      &      &      &      \\
$\Lambda^*_{1670}$ & 1.670 &  35  & 0.20 &      & 0.50 &      &      & 0.30 &      &      &      &      &      \\
$\Lambda^*_{1690}$ & 1.690 &  60  & 0.25 &      & 0.45 & 0.30 &      &      &      &      &      &      &      \\
$\Lambda^*_{1800}$ & 1.800 & 300  & 0.40 & 0.20 & 0.20 & 0.20 &      &      &      &      &      &      &      \\
$\Lambda^*_{1810}$ & 1.810 & 150  & 0.35 & 0.45 & 0.15 & 0.05 &      &      &      &      &      &      &      \\
$\Lambda^*_{1820}$ & 1.820 &  80  & 0.73 &      & 0.16 & 0.11 &      &      &      &      &      &      &      \\
$\Lambda^*_{1830}$ & 1.830 &  95  & 0.10 &      & 0.70 & 0.20 &      &      &      &      &      &      &      \\
$\Lambda^*_{1890}$ & 1.890 & 100  & 0.37 & 0.21 & 0.11 & 0.31 &      &      &      &      &      &      &      \\
$\Lambda^*_{2100}$ & 2.100 & 200  & 0.35 & 0.20 & 0.05 & 0.30 &      & 0.02 & 0.08 &      &      &      &      \\
$\Lambda^*_{2110}$ & 2.110 & 200  & 0.25 & 0.45 & 0.30 &      &      &      &      &      &      &      &      \\
$\Sigma^*_{1385}$ & 1.384 &  36  &      &      & 0.12 &      &      &      &      & 0.88 &      &      &      \\
$\Sigma^*_{1660}$ & 1.660 & 100  & 0.30 &      & 0.35 &      &      &      &      & 0.35 &      &      &      \\
$\Sigma^*_{1670}$ & 1.670 &  60  & 0.15 &      & 0.70 &      &      &      &      & 0.15 &      &      &      \\
$\Sigma^*_{1750}$ & 1.750 &  90  & 0.40 &      & 0.05 &      &      &      &      &      & 0.55 &      &      \\
$\Sigma^*_{1775}$ & 1.775 & 120  & 0.40 &      & 0.04 & 0.10 &      &      &      & 0.23 &      & 0.23 &      \\
$\Sigma^*_{1915}$ & 1.915 & 120  & 0.15 &      & 0.40 & 0.05 &      &      &      & 0.40 &      &      &      \\
$\Sigma^*_{1940}$ & 1.940 & 220  & 0.10 & 0.15 & 0.15 & 0.15 &      &      &      & 0.15 &      & 0.15 & 0.15 \\
$\Sigma^*_{2030}$ & 2.030 & 180  & 0.20 & 0.04 & 0.10 & 0.10 &      &      &      & 0.20 &      & 0.18 & 0.18 \\
\hline
\end{tabular} 
\end{center}
\caption{\label{respars2}
Masses, widths and branching ratios for single-strange baryon-resonances
in UrQMD. Masses are given in GeV and the widths in MeV. All parameters
are within the range given by the Review of Particle Properties 
\protect\cite{PDG96} and have been tuned to exclusive particle production
channels and to the total kaon-nucleon cross section.}
\end{table}

\begin{table}
\renewcommand{\arraystretch}{1.3}
\begin{center}
\scriptsize
\begin{tabular}{|l||rr|rrrr|} \hline
\bf resonance & \bf mass & \bf width & 
$\Xi \pi$ & $\Xi \gamma$ & $\Lambda \bar K$ & $\Sigma \bar K$ \\\hline\hline
$\Xi^*_{1530}$ & 1.532 &   9  & 0.98 & 0.02 &      &      \\
$\Xi^*_{1690}$ & 1.700 &  50  & 0.10 &      & 0.70 & 0.20 \\
$\Xi^*_{1820}$ & 1.823 &  24  & 0.15 &      & 0.70 & 0.15 \\
$\Xi^*_{1950}$ & 1.950 &  60  & 0.25 &      & 0.50 & 0.25 \\
$\Xi^*_{2030}$ & 2.025 &  20  & 0.10 &      & 0.20 & 0.70 \\
\hline
\end{tabular}
\end{center} 
\caption{\label{respars3}
Masses, widths and branching ratios for double-strange baryon-resonances
in UrQMD. Masses are given in GeV and the widths in MeV. All parameters
are within the range given by the Review of Particle Properties 
\protect\cite{PDG96} and have been tuned to exclusive particle production
channels.}
\end{table}

\begin{table}
\renewcommand{\arraystretch}{1.3}
\begin{center}
\scriptsize
\begin{tabular}{|l|rr|rrrrrrrrrrrr|} \hline
\bf meson & \bf mass & \bf width &  
$\gamma \pi$ & $\gamma \rho$ & $\gamma \omega$ & $\gamma \eta$ &
$\gamma K$ & $\pi \pi$ & $\pi \rho$ & $ 3 \pi$ & $\pi \eta$ &$ 4 \pi$&
$K \bar K^*$ & $\bar K K^*$
\\\hline\hline
$\omega$     & 0.782 &   8  & 0.09 &      &      &      &      & 0.02 &      & 0.89 &      &      &      &      \\
$\rho$       & 0.769 & 151  &      &      &      &      &      & 1.00 &      &      &      &      &      &      \\
$f_0(980)$   & 0.980 & 100  &      &      &      &      &      & 0.70 &      &      &      &      &      &      \\
$\eta'$      & 0.958 &   0.2&      & 0.30 & 0.05 &      &      &      &      &      &      &      &      &      \\
$K^*$        & 0.893 &  50  &      &      &      &      &      &      &      &      &      &      &      &      \\
$\phi$       & 1.019 &   4  &      &      &      & 0.01 &      &      & 0.13 & 0.02 &      &      &      &      \\
$K_0^*$      & 1.429 & 287  &      &      &      &      &      &      &      &      &      &      &      &      \\
$a_0$        & 0.984 & 100  &      &      &      &      &      &      &      &      & 0.90 &      &      &      \\
$f_0(1370)$  & 1.370 & 200  &      &      &      &      &      & 0.10 &      &      &      & 0.70 &      &      \\
$K_1(1270)$  & 1.273 &  90  &      &      &      &      &      &      &      &      &      &      &      &      \\
$a_1$        & 1.230 & 400  & 0.10 &      &      &      &      &      & 0.90 &      &      &      &      &      \\
$f_1$        & 1.282 &  24  &      & 0.07 &      &      &      &      &      &      &      & 0.20 &      &      \\
$f_1(1510)$  & 1.512 & 350  &      &      &      &      &      &      &      &      &      &      & 0.50 & 0.50 \\
$K_2(1430)$  & 1.430 & 100  &      &      &      &      &      &      &      &      &      &      &      &      \\
$a_2(1320)$  & 1.318 & 107  &      &      &      &      &      &      & 0.70 &      & 0.14 &      &      &      \\
$f_2(1270)$  & 1.275 & 185  &      &      &      &      &      & 0.50 &      &      &      & 0.30 &      &      \\
$f_2'(1525)$ & 1.525 &  76  &      &      &      &      &      & 0.01 &      &      &      &      &      &      \\
$K_1(1400)$  & 1.400 & 174  &      &      &      &      &      &      &      &      &      &      &      &      \\
$b_1(1235)$  & 1.235 & 142  &      &      &      &      &      &      &      &      &      &      &      &      \\
$h_1(1170)$  & 1.170 & 360  &      &      &      &      &      &      & 0.90 & 0.10 &      &      &      &      \\
$h_1'(1380)$ & 1.380 &  80  &      &      &      &      &      &      &      &      &      &      & 0.50 & 0.50 \\
$K^*(1410)$  & 1.410 & 227  &      &      &      &      &      &      &      &      &      &      &      &      \\
$\rho(1465)$ & 1.465 & 310  &      &      &      &      &      & 0.50 &      &      &      & 0.50 &      &      \\
$\omega(1419)$& 1.419 & 174  &      &      &      &      &      &      & 1.00 &      &      &      &      &      \\
$\phi(1680)$ & 1.680 & 150  &      &      &      &      &      &      &      &      &      &      & 0.40 & 0.40 \\
$K^*(1680)$  & 1.680 & 323  &      &      &      &      &      &      &      &      &      &      &      &      \\
$\rho(1700)$ & 1.700 & 235  &      &      &      &      &      & 0.10 &      &      &      & 0.20 &      &      \\
$\omega(1662)$& 1.662 & 280  &      &      &      &      &      &      & 0.50 &      &      &      &      &      \\
$\phi(1900)$ & 1.900 & 400  &      &      &      &      &      &      &      &      &      &      & 0.40 & 0.40 \\
\hline
\end{tabular} 
\end{center}
\caption{\label{mespars1}
Masses, widths and branching ratios for meson-resonances
in UrQMD, part~I. Masses are given in GeV and the widths in MeV. All parameters
are within the range given by the Review of Particle Properties 
\protect\cite{PDG96}. Additional branching ratios can be found
in table~\protect\ref{mespars2}.}
\end{table}

\begin{table}
\renewcommand{\arraystretch}{1.3}
\begin{center}
\scriptsize
\begin{tabular}{|l|rr|rrrrrrrrrrrrr|} \hline
\bf meson & \bf mass & \bf width &  
$\eta \pi \pi$ & $\eta \rho$ & $\rho \pi \pi$ & $\omega \pi \pi$ &
$\eta \eta$ & $K \bar K$ & $K \bar K \pi$ & $K \pi$ & $K^* \pi$ &
$K \rho$ & $K \omega$ & $K^* \pi \pi$ & $\omega \pi$ 
\\\hline\hline
$\omega$      & 0.782 &   8  &      &      &      &      &      &      &      &      &      &      &      &      &      \\
$\rho$        & 0.769 & 151  &      &      &      &      &      &      &      &      &      &      &      &      &      \\
$f_0(980)$    & 0.980 & 100  &      &      &      &      &      & 0.30 &      &      &      &      &      &      &      \\
$\eta'$       & 0.958 &   0  & 0.65 &      &      &      &      &      &      &      &      &      &      &      &      \\
$K^*$         & 0.893 &  50  &      &      &      &      &      &      &      & 1.00 &      &      &      &      &      \\
$\phi$        & 1.019 &   4  &      &      &      &      &      & 0.84 &      &      &      &      &      &      &      \\
$K_0^*$       & 1.429 & 287  &      &      &      &      &      &      &      & 1.00 &      &      &      &      &      \\
$a_0$         & 0.984 & 100  &      &      &      &      &      & 0.10 &      &      &      &      &      &      &      \\
$f_0(1370)$   & 1.370 & 200  &      &      &      &      &      & 0.20 &      &      &      &      &      &      &      \\
$K_1(1270)$   & 1.273 &  90  &      &      &      &      &      &      &      &      & 0.47 & 0.42 & 0.11 &      &      \\
$a_1$         & 1.230 & 400  &      &      &      &      &      &      &      &      &      &      &      &      &      \\
$f_1$         & 1.282 &  24  & 0.54 &      & 0.10 &      &      &      & 0.09 &      &      &      &      &      &      \\
$f_1(1510)$   & 1.512 & 350  &      &      &      &      &      &      &      &      &      &      &      &      &      \\
$K_2(1430$    & 1.430 & 100  &      &      &      &      &      &      &      & 0.50 & 0.25 & 0.09 & 0.03 & 0.13 &      \\
$a_2(1320)$   & 1.318 & 107  &      &      &      & 0.11 &      & 0.05 &      &      &      &      &      &      &      \\
$f_2(1270$    & 1.275 & 185  &      &      &      &      &      & 0.20 &      &      &      &      &      &      &      \\
$f_2'(1525)$  & 1.525 &  76  &      &      &      &      & 0.10 & 0.89 &      &      &      &      &      &      &      \\
$K_1(1400)$  & 1.400 & 174   &      &      &      &      &      &      &      &      & 0.96 & 0.03 & 0.01 &      &      \\
$b_1(1235)$   & 1.235 & 142  &      & 0.10 &      &      &      &      &      &      &      &      &      &      & 0.90 \\
$h_1(1170)$   & 1.170 & 360  &      &      &      &      &      &      &      &      &      &      &      &      &      \\
$h_1'(1380)$  & 1.380 &  80  &      &      &      &      &      &      &      &      &      &      &      &      &      \\
$K^*(1410)$   & 1.410 & 227  &      &      &      &      &      &      &      & 0.30 & 0.65 & 0.05 &      &      &      \\
$\rho(1465)$  & 1.465 & 310  &      &      &      &      &      &      &      &      &      &      &      &      &      \\
$\omega(1419)$ & 1.419 & 174 &      &      &      &      &      &      &      &      &      &      &      &      &      \\
 $\phi(1680)$ & 1.680 & 150  &      &      &      &      &      & 0.10 & 0.10 &      &      &      &      &      &      \\
$K^*(1680)$   & 1.680 & 323  &      &      &      &      &      &      &      & 0.40 & 0.30 & 0.30 &      &      &      \\
$\rho(1700)$  & 1.700 & 235  &      &      & 0.70 &      &      &      &      &      &      &      &      &      &      \\
$\omega(1662)$ & 1.662 & 280 &      &      &      & 0.50 &      &      &      &      &      &      &      &      &      \\
$\phi(1900)$ & 1.900 & 400   &      &      &      &      &      & 0.10 & 0.10 &      &      &      &      &      &      \\
\hline
\end{tabular}
\end{center} 
\caption{\label{mespars2}
Masses, widths and branching ratios for meson-resonances
in UrQMD, part~II. Masses are given in GeV and the widths in MeV. 
All parameters
are within the range given by the Review of Particle Properties 
\protect\cite{PDG96}.}
\end{table}

\pagebreak

\begin{figure}[tb]
\begin{minipage}[t]{9cm}
\centerline{\psfig{figure=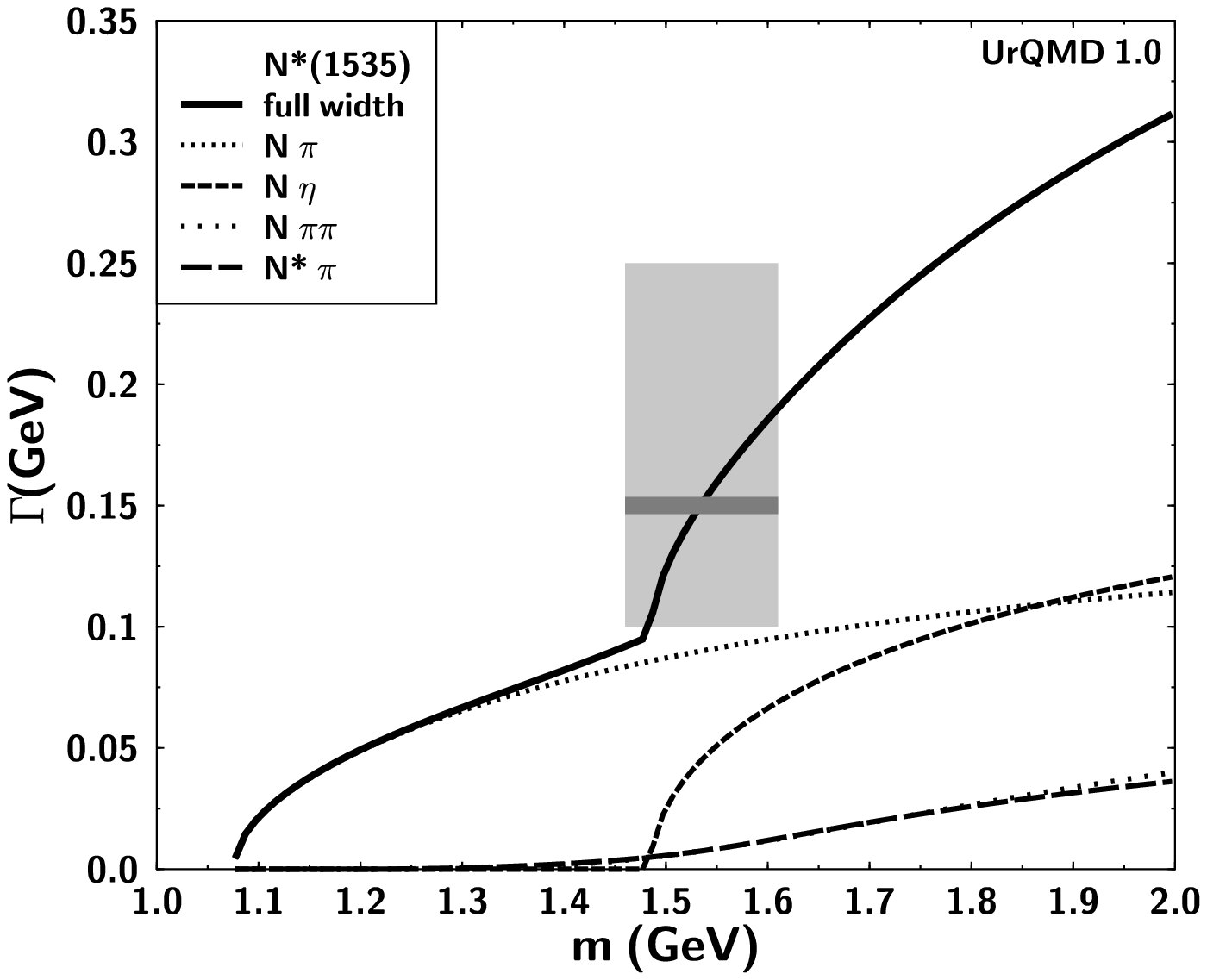,width=9cm}}
\caption{\label{pwn1535}
Total and partial mass-dependent decay widths of the $N^*_{1535}$
resonance. The grey-shaded area marks the size of the error-bar
for the total decay-widths according to \protect\cite{PDG96}.
The opening of the $N \eta$ decay-branch above its energetic 
threshold is clearly visible. }
\end{minipage}
\hfill
\begin{minipage}[t]{9cm}
\centerline{\psfig{figure=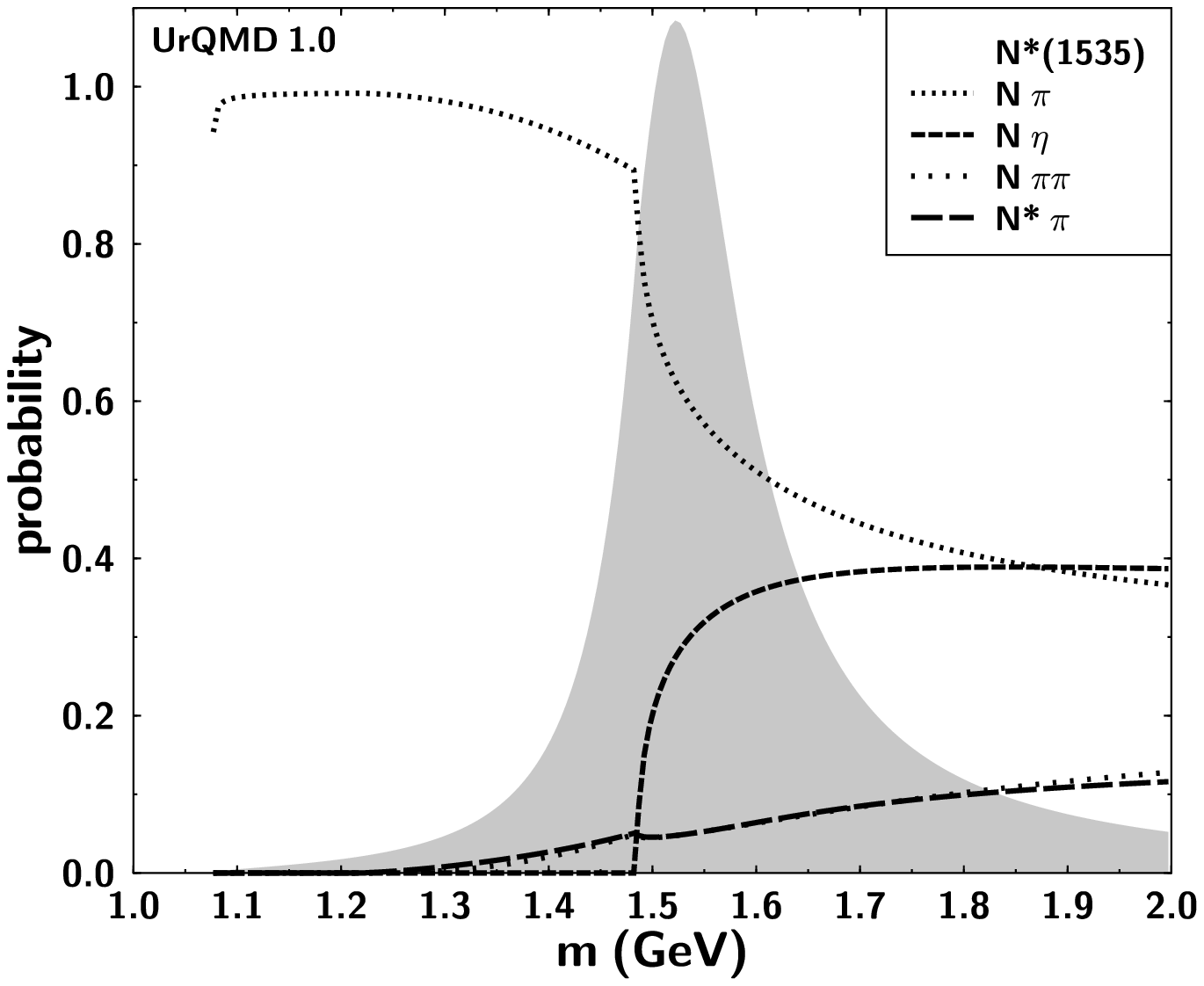,width=9cm}}
\caption{\label{ppn1535}
Mass-dependent branching ratios for the 
different exit-channels of a $N^*_{1535}$ decay.
The grey-shaded area symbolizes the Breit-Wigner mass-distribution
of the resonance. The resonance may also be populated in a mass-range
in which the $\eta$-production is strongly suppressed or even
energetically impossible.}
\end{minipage}
\end{figure}

Figure~\ref{pwn1535} shows the total and partial decay-widths of
the $N^*_{1535}$ resonance as a function of its mass. 
This resonance is particularly interesting since it dominates the
production of the $\eta$ meson at SIS-energies.
The grey-shaded
area represents the experimental uncertainty of the
full width at the resonance pole \cite{PDG96}.
The opening of the $N \eta$ decay-channel at its threshold energy is
clearly visible. Figure~\ref{ppn1535} shows the respective probabilities
for the different decay channels. Here, the grey-shaded area depicts
the Breit-Wigner mass-distribution of the $N^*_{1535}$ resonance.
Obviously the resonance can also be easily populated below the
$\eta$ production threshold -- due to limited phase-space in a heavy-ion
reaction the integrated decay-probability of a $N^*_{1535}$ into a
nucleon and an $\eta$ meson may lie well below the free branching ratio
given in the Review of Particle Properties \cite{PDG96}.

Unfortunately, equation~(\ref{gammapart}) cannot be easily 
extended to include three- or four-body decay channels.
In order to treat all decay channels on an equal footing in UrQMD,
the outgoing particles of a three- or four-body decay are combined
into two ``effective'' particles which are used to compute
the respective partial decay-widths. $N$-body phase-space, however, is treated
explicitly. 

All resonances decay isotropically in their rest frame.
For a two-particle exit channel the momenta are given by
equation~(\ref{pcmsfixed}). If a resonance is among the outgoing particles,
its mass must first be determined  according to a Breit-Wigner 
mass-distribution. If the exit channel contains three or four particles,
then the respective N-body phase-space must be taken into account
for their momenta \cite{block80a}.

\begin{figure}[tb]
\begin{minipage}[t]{9cm}
\centerline{\psfig{figure=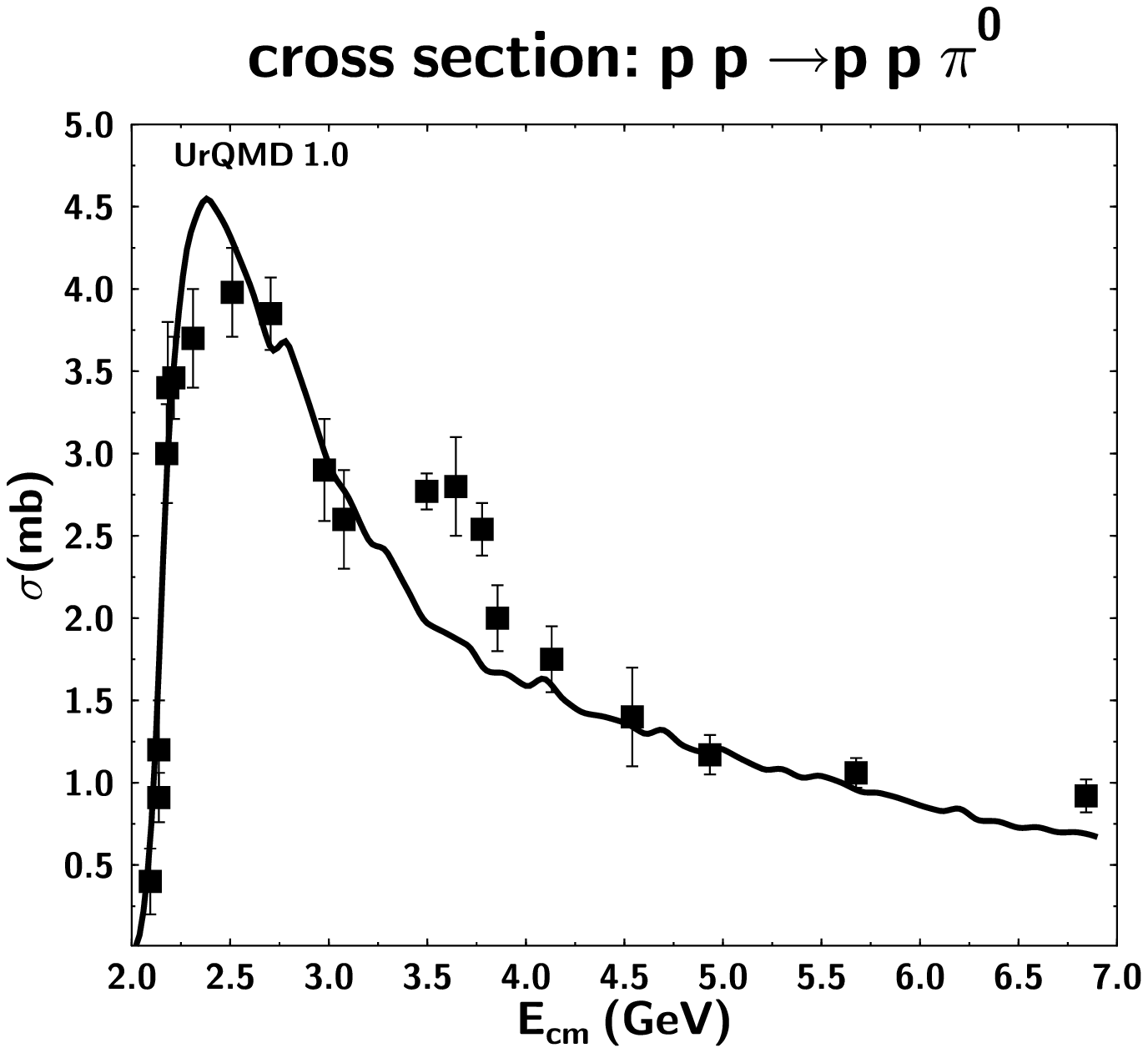,width=9cm}}
\caption{\label{ppexcl_pi}
Exclusive $\pi^0$ production cross section in proton-proton
reactions within the UrQMD model (solid line) compared to data (solid squares)
\protect\cite{flaminio84a}. 
The comparison allows to evaluate the implementation of 
baryon-resonances into the UrQMD model.}
\end{minipage}
\hfill
\begin{minipage}[t]{9cm}
\centerline{\psfig{figure=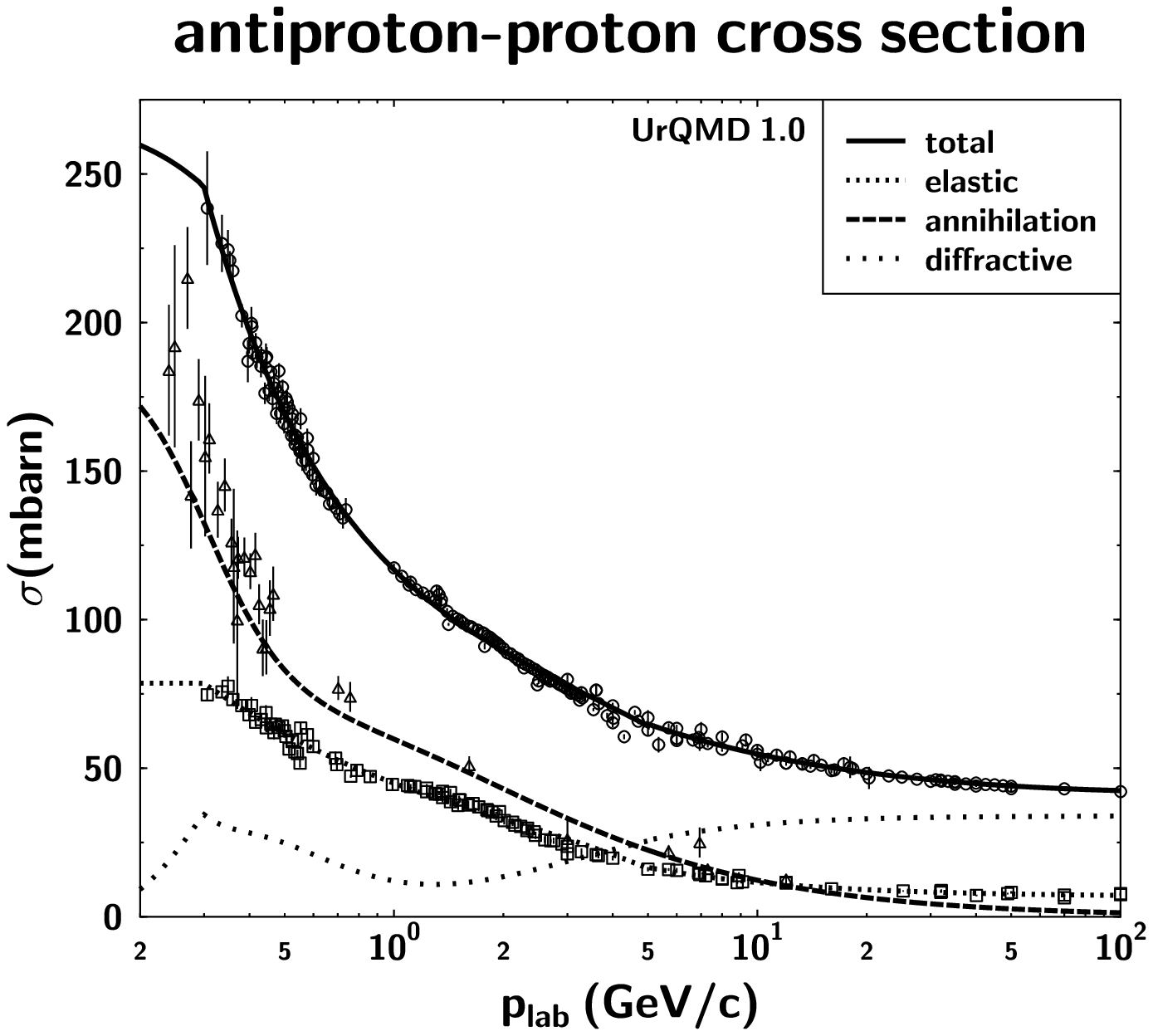,width=9cm}}
\caption{\label{bbar}
Parameterization of antiproton-proton cross sections in UrQMD
compared to data \protect \cite{PDG96}.
The ``diffractive'' cross section is the difference between
the total cross section 
and the sum of elastic and annihilation cross sections.
}
\end{minipage}
\end{figure}

The implementation of the resonances and their inelastic production
cross sections can be evaluated by studying exclusive meson production
in proton-proton reactions: Figure~\ref{ppexcl_pi} shows a comparison for
exclusive $\pi^0$ production in proton-proton reactions between
the UrQMD model and data \cite{flaminio84a}.
Especially at low energies close to the production threshold, such
exclusive cross sections are very sensitive to the partial decay width
and pole mass of the resonances from which the respective particle
emerges. Above $\sqrt{s} \approx 5$~GeV all inclusive particle
production cross sections are dominated by string decays, which,
however, have no influence on the exclusive $p + p \to p + p + \pi^0$
channel.

\subsubsection{Baryon-antibaryon annihilation}

The physics of baryon-antibaryon annihilation is still a subject
of active research and not yet fully understood. A number of
different models exist -- for an overview we refer to
\cite{dover92a,loiseau92a}.
In UrQMD the elementary annihilation cross section is fitted to
the available data: Figure~\ref{bbar} shows the UrQMD parameterization
in comparison to data for the antiproton-proton total, elastic and 
annihilation cross section \cite{PDG96,flaminio84a}.

For the annihilation cross section we employ a parameterization
first suggested by Koch and Dover \cite{kochP89a}:

\begin{equation}
\sigma _{\rm ann}^{\bar p p} \,=\, \sigma _{0}^{N} \, \frac {s_0}{s} \left [
\frac {{\rm A}^2 s_0}{(s - s_0)^2 + {\rm A}^2 s_0} + {\rm B} \right ] \quad ,
\end{equation}
with $\sigma _{0}^{N} = 120$ mb, $s_0 = 4 m_{N}^2$, 
A = 50 MeV and B= $0.6$. 

The antiproton-neutron annihilation cross section is very similar to 
the antiproton-proton annihilation cross section \cite{elioff62a}
and is therefore treated identically.

The total and elastic proton-antiproton cross sections are treated
according to the CERN/HERA parameterization:
\begin{equation}
\sigma(p)\,=\,{\rm A} + {\rm B}\, p^n + {\rm C}\,{\rm ln}^2(p) 
        + {\rm D}\,{\rm ln}(p)\, ,
\end{equation}
with the laboratory-momentum $p$ in GeV/c.
The respective parameters are listed in table~\ref{ppbarfit1}.

\begin{table}[ht]
\begin{center}
\renewcommand{\arraystretch}{1.3}
\begin{tabular}{|l|lllll|} \hline
& A & B & n & C & D \\\hline\hline
$\sigma_{\rm tot}$ & $38.4$ & $77.6$ & $-0.64$ & $0.26$ & $-1.2$ \\
$\sigma_{\rm el}$ & $10.2$ & $52.7$ & $-1.16$ & $0.125$ & $-1.28$ \\
\hline
\end{tabular} 
\caption{\label{ppbarfit1}
Parameters for the CERN/HERA parameterization for the total and
elastic proton-antiproton cross sections.
This parameterization is used in UrQMD for momenta
$p_{lab}>5$~GeV/c.}
\end{center}
\end{table}

For momenta $p_{lab}<5$~GeV/c, UrQMD uses another parameterization
to obtain a good fit to the data:
\begin{equation}
\sigma_{\rm tot}(p) =\left\{ \begin{array}{ll}
75.0+43.1 p^{-1} + 2.6 p^{-2} -3.9 p &\quad :\quad 0.3<p<5\\
271.6\exp (-1.1\, p^2)  & \quad:\quad p< 0.3
\end{array} \right. 
\end{equation}

\begin{equation}
\sigma_{\rm el}=\left\{ \begin{array}{ll}
31.6+18.3 p^{-1} -1.1 p^{-2} - 3.8 p &\quad :\quad 0.3<p< 5\\
78.6 &\quad :\quad p< 0.3\;
\end{array} \right. 
\end{equation}

For low lab-momenta the annihilation cross section is dominant.
The sum of annihilation and elastic cross section, however, is
smaller than the total cross section:
\begin{equation}
 \Delta\sigma = \sigma_{\rm tot}-\sigma_{\rm el}-\sigma_{\rm ann}
\end{equation}
This difference is called ``diffractive'' cross section in UrQMD,
$\sigma_{\rm diff}=\Delta \sigma$, and is used to excite one
(or both) of the collision partners to a resonance or to
a string. In the string case the same excitation scheme as for
proton-proton reactions is used. For high energies  the ``diffractive''
cross section is the dominant contribution to the total
antiproton-proton cross section.

The extrapolation from nucleon-antinucleon annihilation to generic
baryon-antibaryon annihilation poses a large problem since the
respective cross sections have not been measured and the overall
uncertainties in the understanding of the basic annihilation mechanism
leave ample room for speculation.
Due to different masses and quark-contents, modifications to the
nucleon-antinucleon cross sections are inevitable.

Assuming a weak $t$-dependence of the matrix-element, a phase-space
correction analogous to that for exchange reactions was suggested
in \cite{kochP86a}, which leads to the following rescaling of
the nucleon-antinucleon cross section:
\begin{equation}
\sigma_{\rm \bar B B}\sim  \frac{p_{\rm \bar NN}}{p_{\rm \bar
BB}}\sigma_{\rm \bar N N} \, .
\end{equation}
A disadvantage of this ansatz is that the cross section diverges
for low energies. 

The same $\sqrt{s}$-dependence is
used in UrQMD for all baryon-antibaryon cross sections (i.e. 
the parameterization
of the proton-antiproton cross section). The different quark-content
of the colliding (anti-)baryons is taken into account by rescaling
the cross section with the ratio of the total baryon-baryon and
nucleon-nucleon cross sections as obtained from the Additive Quark Model
(AQM, see also the following section):
\begin{equation}
\sigma_{\rm \bar B B} (\sqrt{s}) = 
\frac{\sigma^{\rm AQM}_{\rm B B}}{\sigma^{\rm AQM}_{\rm N N}} ~ 
\sigma_{\rm \bar N N} (\sqrt{s}) \, .
\end{equation}
By scaling the cross section with the ratio of the respective
AQM cross sections, the strangeness content of the colliding
baryons is taken into account.

The final state of a baryon-antibaryon annihilation is generated
via the formation of two meson-strings. The available c.m. energy
of the reaction is distributed in equal parts to the two strings
which decay in the rest frame of the reaction.
On the quark level this procedure implies the annihilation of a
quark-antiquark pair and the reordering of the remaining
constituent quarks into newly produced hadrons (additionally taking
sea-quarks into account). This model for the baryon-antibaryon
annihilation thus follows the topology of a {\em rearrangement}-graph.
Another possibility would be the annihilation of a diquark-antidiquark
with the subsequent fragmentation of a single meson-string.
This ansatz, which is being used in RQMD \cite{sorge89a} follows
the topology of an {\em annihilation}-graph.

Details in the treatment of the baryon-antibaryon annihilation cross section 
may have a large influence on the final yield of antiprotons and antihyperons:
If the baryon-antibaryon annihilation cross section is parameterized as
the proton-antiproton annihilation cross section but then rescaled
to equivalent relative momenta in the incoming channel, 
changes in the order of 50\% -- 300\% may occur.
At CERN/SPS energies
the $\bar \Xi$ yield in central Pb+Pb reactions at 160 GeV/nucleon 
would decrease by a factor of 3. The $\bar p$ and $\bar Y$
yields would then be diminished by 50\% and 25\%, respectively.

\subsubsection{Additive Quark Model}
For all interactions for which no experimental data exist
(e.g. hyperon-baryon resonance scattering) {\em ad hoc}
assumptions must be made. Ignoring all unknown cross sections and
setting them to zero would contradict the experimentally observed
hadron universality, which states that particle production in
highly energetic jets does not strongly depend on the incoming
collision partners (taking all conservation laws into account).
A relatively simple prescription for obtaining unknown hadron-hadron
cross sections is the Additive Quark Model: In this model the 
cross section only depends on the quark-content of the colliding
hadrons \cite{goulianos83a}:
\begin{eqnarray}
\sigma_{tot} & = & 40 \cdot
\left(\frac{2}{3}\right)^{n_{\rm M}}\cdot(1- 0.4 x_1^s)\cdot(1- 0.4 x_2^s) \\
\sigma_{el} & = & 0.039 \cdot \sigma_{tot}^{2/3} \, .
\end{eqnarray}
Here $n_{\rm M}$ is the number of colliding mesons and 
$x_i^s$ is the ratio of strange quarks to non-strange quarks in 
the $i$-th hadron.
Both relations are phenomenological and do not contain
any energy- or momentum-dependence.
For high energies they agree well with experimentally known hadron-hadron
cross sections. The missing momentum- or energy-dependence, however,
leads to a breakdown of this prescription for cross sections
close to threshold.

\subsubsection{Detailed Balance}
The principle of detailed balance is based on the time-reversal invariance 
 of the matrix element of the reaction. It is most commonly found
in text books in the form:
\begin{equation}
\label{dbgl3}
\sigma_{f \rightarrow i } \,=\, \frac{\vec{p}_i^2}{\vec{p}_f^2} \frac{g_i}{g_f}
\sigma_{i \rightarrow f} \quad ,
\end{equation}
with $g$ denoting the spin-isospin degeneracy factors.
UrQMD applies the general principle of detailed balance to the 
following two process classes:
\begin{enumerate}
\item 
Resonant meson-meson and meson-baryon interactions: Each resonance created
via a meson-baryon or a meson-meson annihilation may again decay into
the two hadron species which originally formed it. This symmetry is only
violated in the case of three- or four-body decays and string fragmentations, 
since N-body collisions with (N$>2$) are not implemented in UrQMD. 
\item
Resonance-nucleon or resonance-resonance interactions: the excitation
of baryon-resonances in UrQMD is handled via parameterized cross sections
which have been tuned to data. The reverse reactions usually have not
been measured - here the principle of detailed balance is applied.
Inelastic baryon-resonance deexcitation is the only method in UrQMD
to absorb mesons (which are {\em bound} in the resonance). Therefore
the application of the detailed balance principle is of crucial
importance for heavy nucleus-nucleus collisions.
\end{enumerate}
Equation~(\ref{dbgl3}), however, is only valid in the case of stable
particles with well-defined masses. Since in UrQMD detailed balance
is applied to reactions involving resonances with finite lifetimes
and broad mass distributions, equation~(\ref{dbgl3}) has to be 
modified accordingly. For the case of one incoming resonance the
respective modified detailed balance relation has been derived
in \cite{danielewicz91a}. Here, we generalize this expression for
up to two resonances in both, the incoming and the outgoing channels.

The differential cross section for the reaction 
$(1\,,\,2) \rightarrow (3\,,\,4)$ is given by:
\begin{equation}
\label{diffcx1}
{\rm d} \sigma_{12}^{34} \,=\,
    \frac{| {\cal M} |^2}{64 \pi^2 s} \, \frac{p_{34}}{p_{12}} 
        \,{\rm d \Omega}\,
     \prod_{i=3}^4    \delta(p_i^2 -M_i^2) {\rm d}p_i^2  \quad,
\end{equation}
here the $p_i$ in the $\delta$-function denote four-momenta.
The $\delta$-function ensures that the particles are on mass-shell,
i.e. their masses are well-defined. If the particle, however, has 
a broad mass distribution, then the $\delta$-function
must be substituted by the respective mass distribution (including
an integration over the mass):
\begin{equation}
\label{diffcx2}
{\rm d} \sigma_{12}^{34} \,=\,
    \frac{| {\cal M} |^2}{64 \pi^2 s} \, \frac{1}{p_{12}} 
        \,{\rm d \Omega}\,
     \prod_{i=3}^4  p_{34} \cdot
   \frac{ \Gamma}{\left(m-M_i\right)^2+ \Gamma^2/4} \frac{{\rm d} m}{2\pi}
\, .
\end{equation}
Incorporating these modifications into equation~(\ref{dbgl3}) and
neglecting a possible mass-dependence of the matrix element we
obtain:
\begin{equation}
\label{uqmddetbal}
     \frac{ {\rm d} \sigma_{34}^{12} }{{\rm d} \Omega }   
       = \frac{\langle p_{12 }^2 \rangle   }
       {\langle p_{34 }^2 \rangle  } \,
       \frac{(2 S_1 + 1) (2 S_1 + 1)}
            {(2 S_3 + 1) (2 S_4 + 1)}\,
      \sum_{J=J_-}^{J_+} 
      \langle j_1 m_1 j_2 m_2 \| J M \rangle \, 
        \frac{ {\rm d} \sigma_{12}^{34} }{{\rm d} \Omega } \, .
\end{equation}
Here, $S_i$ indicates the spin of particle $i$ and 
the summation of the Clebsch-Gordan-coefficients is for the isospin of the
outgoing channel only. For the incoming channel, isospin is 
treated explicitly. The summation limits are given by:
\begin{eqnarray}
  J_- &=& \max \left( |j_1-j_2|,  |j_3-j_4| \right) \\  
  J_+ &=& \min \left( j_1+j_2,  j_3+j_4     \right)  \quad.
\end{eqnarray}
The integration over the mass distributions of the resonances  
in equation~(\ref{uqmddetbal}) has been denoted by the brackets $\langle\rangle$,
e.g.
\begin{displaymath}
p_{3,4}^2 \,\Rightarrow \, \langle p_{3,4}^2 \rangle \, = \,
 \int  \int 
  p_{CMS}^2(\sqrt{s},m_3,m_4)\, A_3(m_3) \, A_4(m_4) \, 
  {\rm d} m_3\; {\rm d} m_4 \quad,
\end{displaymath}
which is identical to equation~(\ref{pcmsmd}). Correspondingly, the 
mass distribution $A_r(m)$ is given by equation~(\ref{bwnorm}).

The most frequent applications of equation~(\ref{uqmddetbal}) in UrQMD
are the processes $\Delta_{1232}\, N \to N\, N$ and
$\Delta_{1232}\, \Delta_{1232} \to N\, N$.

\subsubsection{Meson-baryon and meson-meson interactions}

Due to the large pion-nucleon cross section at low
c.m. energies, resonant meson-baryon and meson-meson cross sections
are among the most important in UrQMD.
Up to c.m. energies of 2.2~GeV meson-baryon and meson-meson interactions
in UrQMD are dominated by resonance scattering, i.e. the formation
of an intermediate resonance.
For example, the total meson-baryon cross section for
non-strange particles is  given by
\begin{eqnarray}
\label{mbbreitwig}
\sigma^{MB}_{tot}(\sqrt{s}) &=& \sum\limits_{R=\Delta,N^*}
       \langle j_B, m_B, j_{M}, m_{M} \| J_R, M_R \rangle \,
        \frac{2 S_R +1}{(2 S_B +1) (2 S_{M} +1 )} \nonumber \\
&&\times        \frac{\pi}{p^2_{CMS}}\, 
        \frac{\Gamma_{R \rightarrow MB} \Gamma_{tot}}
             {(M_R - \sqrt{s})^2 + \frac{\Gamma_{tot}^2}{4}}
\end{eqnarray}
with the total and partial $\sqrt{s}$-dependent decay widths $\Gamma_{tot}$ and
$\Gamma_{R \rightarrow MB}$ 
(see equations~(\ref{gammatot}) and (\ref{gammapart})).
Therefore, the total pion-nucleon cross section depends on all 
pole masses, widths and branching ratios of all $N^*$ and $\Delta^*$
resonances listed in table~\ref{respars1}.

\begin{figure}[tb]
\begin{minipage}[t]{9cm}
\centerline{\psfig{figure=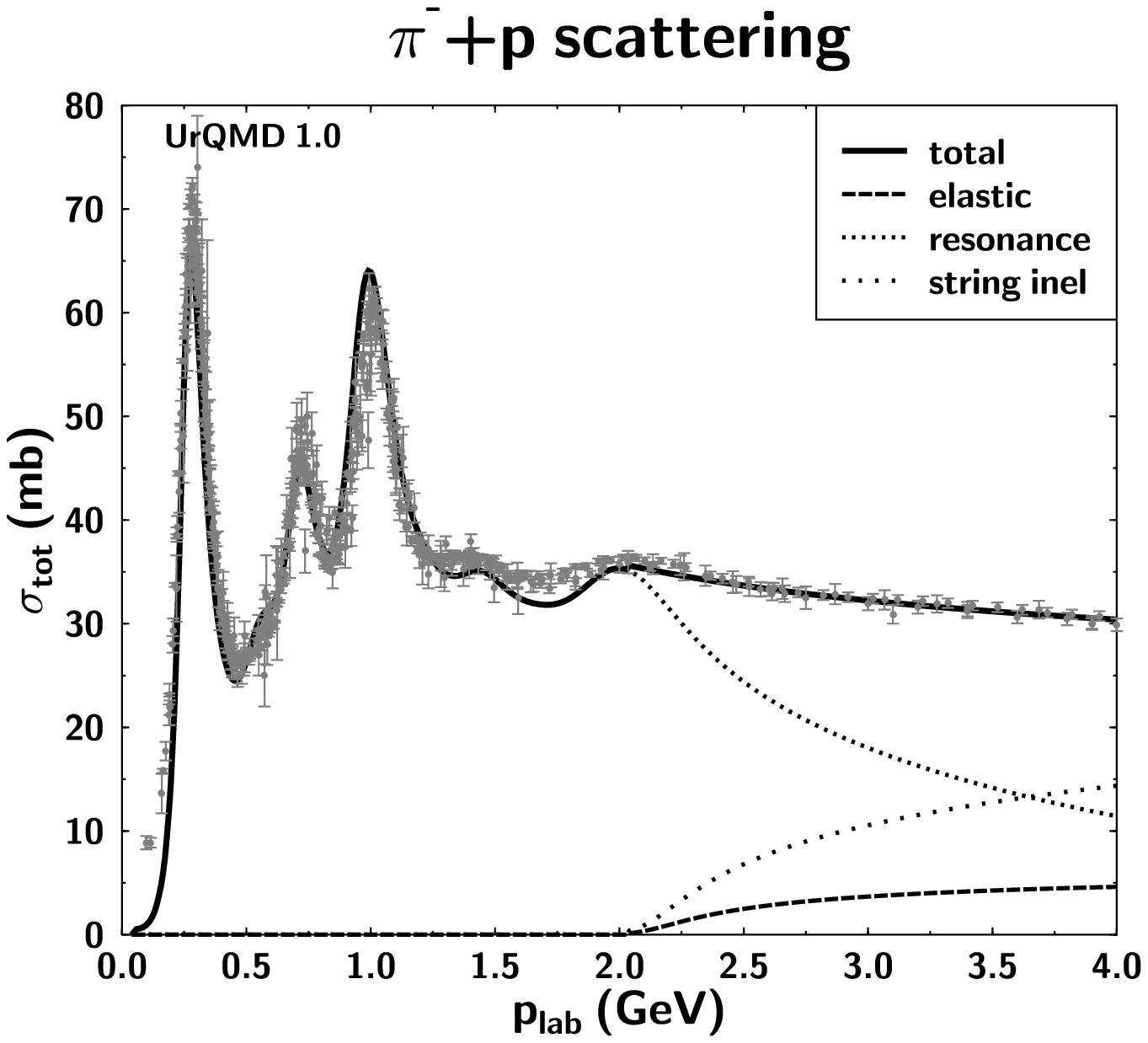,width=9cm}}
\caption{\label{pim-p_total}
Total $\pi^- p$ cross section in UrQMD, compared to data \protect\cite{PDG96}. Up
to $p_{lab} = 2$~GeV the total cross section consists solely of
resonance scattering. For higher momenta elastic scattering
and string excitation are employed to fill the total cross 
section.
}
\end{minipage}
\hfill
\begin{minipage}[t]{9cm}
\centerline{\psfig{figure=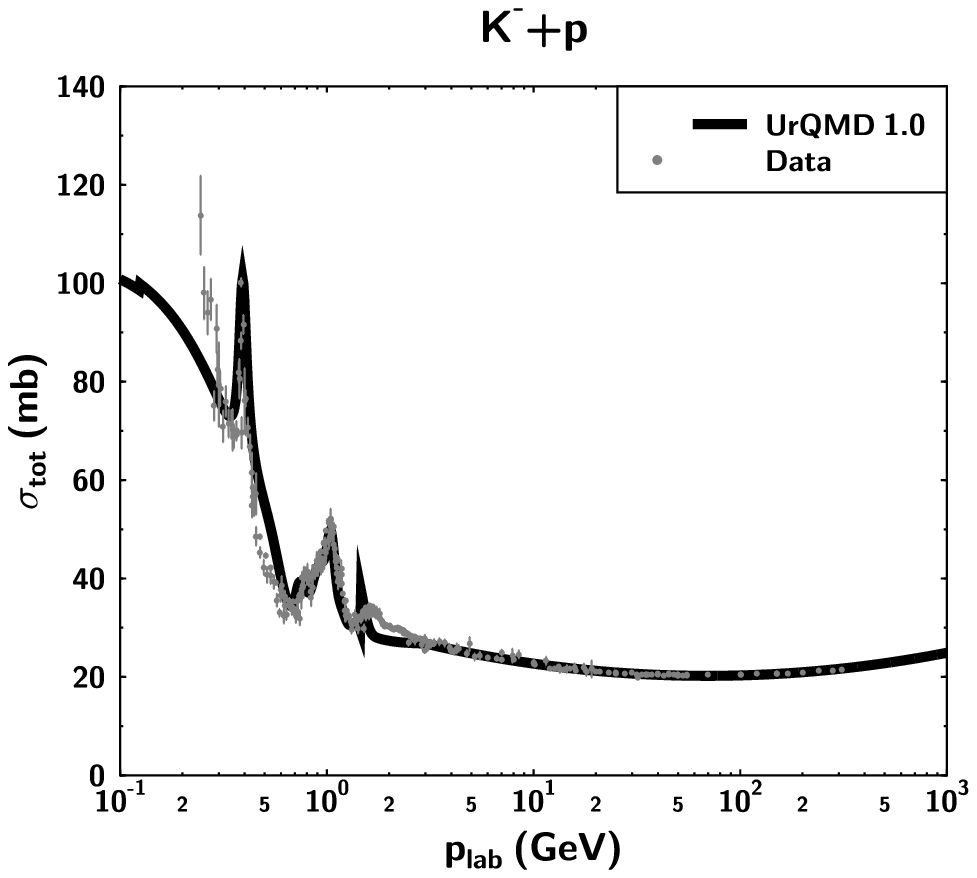,width=10cm}}
\caption{\label{kmp}
Total $K^- p$ cross section in UrQMD. The data is from \protect\cite{PDG96}. 
Resonant $K^- p$
scattering is only possible via $\Lambda$ and $\Sigma$ resonances, whereas
in the case $\pi^- p$ only non-strange resonances contribute.
}
\end{minipage}
\end{figure}

Figure~\ref{pim-p_total} compares the total $\pi^- p$ cross section
of UrQMD to data \cite{PDG96}. Up to $p_{lab}=2$~GeV the total
cross section is very well described by resonance scattering. For
higher momenta the resonance description breaks down and the total
cross section is described as a superposition of elastic scattering
and inelastic string excitation. A comparison between the 
total $\pi^- p$ and the total $K^- p$ cross sections 
(see figures~\ref{pim-p_total} and~\ref{kmp}) is of particular
interest: In the first case only $N^*$- and $\Delta$-resonances
contribute to the cross section, whereas in the latter case only 
hyperon resonances may be excited.

\begin{figure}[tb]
\begin{minipage}[t]{9cm}
\centerline{\psfig{figure=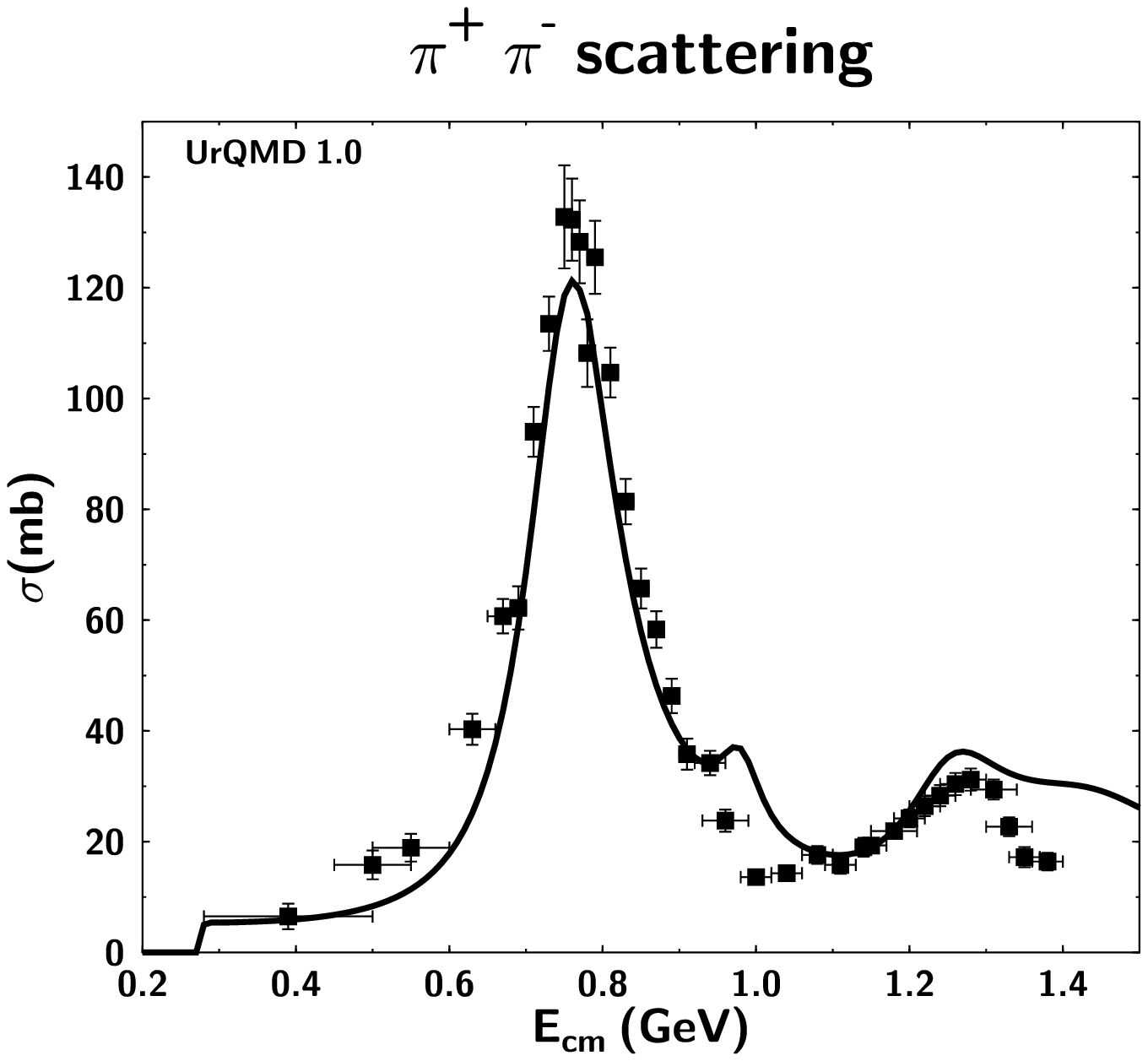,width=9cm}}
\caption{\label{pipi}
Total $\pi^+ \pi^-$ cross section in UrQMD compared to data
\protect\cite{flaminio84a}. 
The cross section is dominated by the $\rho$-resonance.
}
\end{minipage}
\hfill
\begin{minipage}[t]{9cm}
\centerline{\psfig{figure=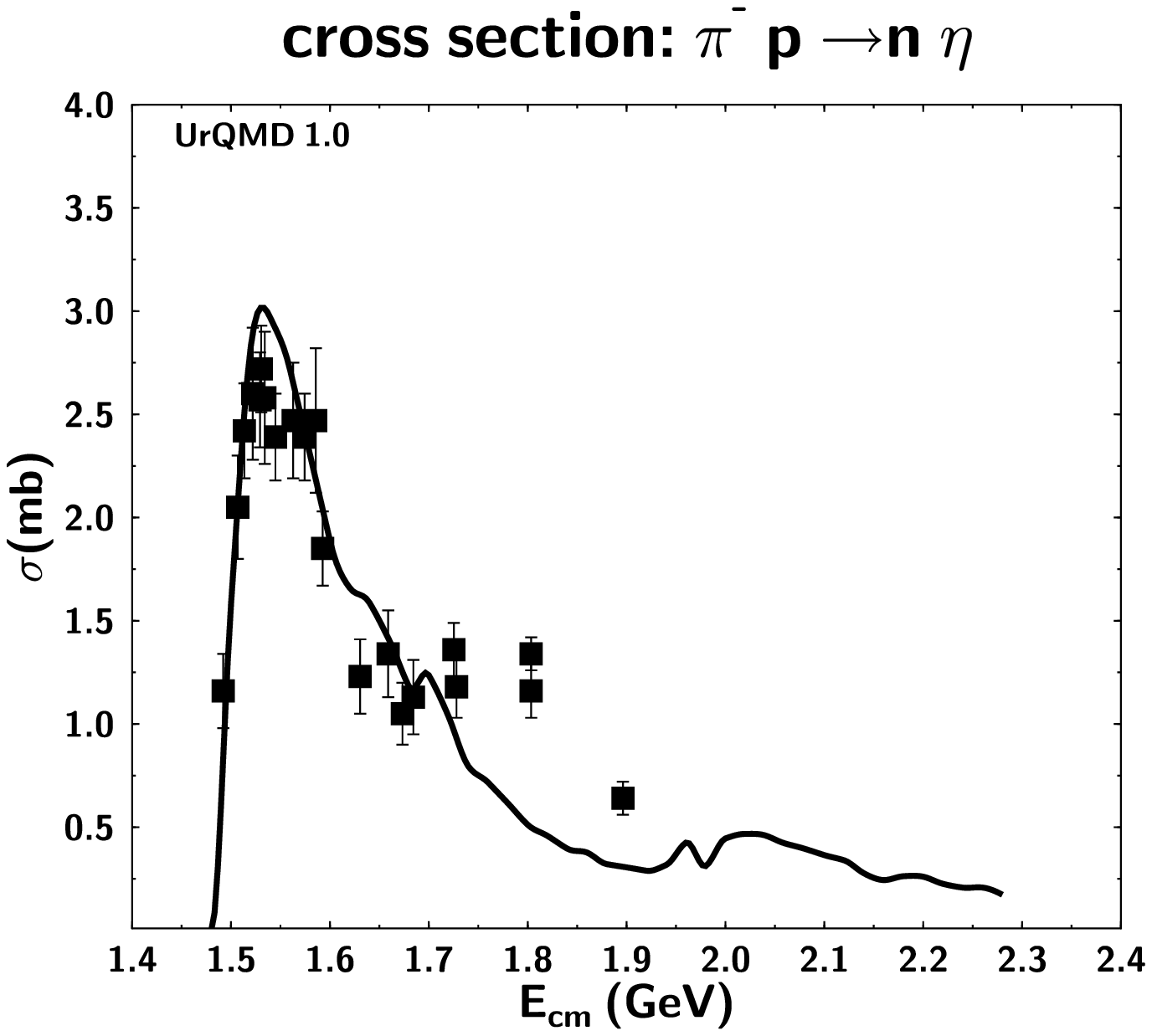,width=9cm}}
\caption{\label{mbexcl_eta}
Exclusive pion-induced $\eta$ production in UrQMD compared to data
\protect\cite{flaminio84a}: 
Apart from the $N^*_{1535}$ resonance, also the $N^*_{1650}$, the $N^*_{1700}$, 
the $N^*_{1710}$  and the $N^*_{2080}$ contribute to the cross section. 
}
\end{minipage}
\end{figure}

The same basic principles apply in the meson-meson case: Figure~\ref{pipi}
shows the total $\pi^+ \pi^-$ cross section, which is dominated by
the $\rho$-resonance. However, in the case of meson-meson scattering
a constant elastic cross section of 5 mb is added in order to fully
reproduce the measured cross section.

The cross section for an exclusive exit channel 
$M \, B \rightarrow R \rightarrow M' \, B'$ 
may be obtained by 
replacing the total width $\Gamma_{tot}$ in equation~(\ref{mbbreitwig})
by the respective partial decay-width $\Gamma_{R \rightarrow M' B'}$.
Figure~\ref{mbexcl_eta} shows the cross section for the process
$\pi^- p \to \eta n$ compared to data. Such exclusive processes
allow for the fine-tuning of the resonance parameters listed in
table~\ref{respars1}, which is very important since the parameters
given in \cite{PDG96} often are only crude estimates. $\eta$-production
is a rather selective probe:
Apart from
the $N^*_{1535}$ resonance, also the $N^*_{1650}$, the $N^*_{1710}$, 
the $N^*_{1710}$ and the $N^*_{2080}$ contribute to this cross section.

\subsection{Angular distributions}

The UrQMD approach uses an analytical expression for the differential
cross-section of in-medium $NN$ elastic scattering derived from the 
collision term
of the RBUU equation \cite{maoG96a,maoG97a}
to determine the scattering angles between the outgoing partners
of all elementary hadron-hadron collisions. 
Here it is assumed that the angular distributions
for all relevant two-body processes are similar and 
can be described approximately
by the differential cross-section of in-medium $NN$ elastic scattering.
This cross section is 
\begin{equation}
\label{sigdiff}
  \sigma_{NN \rightarrow NN}(s,t) = \frac{1}{(2 \pi)^{2} s} \lbrack D(s,t)
  + E(s,t) + (s, t \longleftrightarrow u) \rbrack, 
  \end{equation}
 \begin{eqnarray}
  D(s,t)&=& \frac{({\rm g}_{NN}^{\sigma})^{4}}{2(t-m_{\sigma}^{2})^{2}}
 (t- 4 m^{*2})^{2} + \frac{({\rm g}_{NN}^{\omega})^{4}}{(t-m_{\omega}^{2})^{2}}
 (2 s^{2} + 2st +t^{2} -8m^{*2}s +8m^{*4}) \nonumber \\
  &&  + \frac{24({\rm g}_{NN}^{\pi})^{4}}{(t- m_{\pi}^{2})^{2}} m^{*4}t^{2}
 - \frac{4({\rm g}_{NN}^{\sigma}{\rm g}_{NN}^{\omega})^{2}}
{(t - m_{\sigma}^{2})(t-m_{\omega}^{2})} (2s + t -4m^{*2})m^{*2}, \\
  E(s,t)&=& -\frac{({\rm g}_{NN}^{\sigma})^{4}}{8(t-m_{\sigma}^{2})
 (u-m_{\sigma}^{2})} \lbrack t (t+s) + 4m^{*2}(s-t) \rbrack
  + \frac{({\rm g}_{NN}^{\omega})^{4}}{2(t-m_{\omega}^{2})(u-m_{\omega}^{2})}
 (s- 2m^{*2}) \nonumber \\ && \times (s-6m^{*2})  
    - \frac{6({\rm g}_{NN}^{\pi})^{4}}{(t- m_{\pi}^{2})(u-m_{\pi}^{2})}
  (4m^{*2}-s -t ) m^{*4}t \nonumber \\
&& + ({\rm g}_{NN}^{\sigma}{\rm g}_{NN}^{\omega})^{2}
 \lbrack \frac{t^{2} - 4m^{*2}s -10m^{*2}t +24m^{*4}}{4(t-m_{\sigma}^{2})
(u-m_{\omega}^{2})} + \frac{(t+s)^{2} - 2m^{*2}s + 2m^{*2}t }{4 (t-m_{\omega}
^{2})(u-m_{\sigma}^{2})}  \rbrack \nonumber \\
  && + ({\rm g}_{NN}^{\sigma}{\rm g}_{NN}^{\pi})^{2}
 \lbrack \frac{3m^{*2}(4m^{*2}-s-t)(4m^{*2}-t)}{2(t-m_{\sigma}^{2})
(u-m_{\pi}^{2})} + \frac{3t(t+s)m^{*2}}{2 (t-m_{\pi}
^{2})(u-m_{\sigma}^{2})}  \rbrack \nonumber \\
  && + ({\rm g}_{NN}^{\omega}{\rm g}_{NN}^{\pi})^{2}
 \lbrack \frac{3m^{*2}(t+s-4m^{*2})(t+s-2m^{*2})}{(t-m_{\omega}^{2})
(u-m_{\pi}^{2})} + \frac{3m^{*2}(t^{2}-2m^{*2}t)}{ (t-m_{\pi}
^{2})(u-m_{\omega}^{2})}  \rbrack, 
  \end{eqnarray}
where the function $D$ represents the contribution of the 
direct term and $E$ is the
exchange term. The coupling strengths are ${\rm g}_{NN}^{\sigma}=6.9$,
${\rm g}_{NN}^{\omega}=7.54$ and ${\rm g}_{NN}^{\pi}=1.434$ and 
Mandelstam variables are given by:
\begin{eqnarray}
  &&s=(p+p_{2})^{2}= \lbrack E^{*}(p)+E^{*}(p_{2}) \rbrack ^{2}
   - ( {\bf p} + {\bf p}_{2})^{2}, \\
  &&t=(p-p_{3})^{2}=\frac{1}{2}(s - 4m^{*2}) ( \cos \theta -1 ), \\
  &&u=(p-p_{4})^{2}=4m^{*2}-s -t,
  \end{eqnarray}
with $\theta$ denoting the scattering angle in the c.m. system. 
The in-medium 
single-particle energy is given by
\begin{equation}
E^{*}(p)=\sqrt{{\bf p}^{2}+m^{*2}} \; .
\end{equation}

The formula for the differential cross section of in-medium $NN$ 
elastic scattering
can be extended to all elementary hadron-hadron collisions if it is 
scaled by the replacement
\begin{equation}
s \rightarrow s-(m_1^{*}+m_2^{*})^2+4m^{*^2} \; ,
\end{equation}
where $m_1^{*}$ and $m_2^{*}$ denote the effective masses of the 
incoming hadrons.
Furthermore, we take into account the effects stemming from the 
finite size of hadrons and
a part of the short range correlation by introducing 
a phenomenological form factor at
each vertex. For the nucleon-nucleon-meson vertex we take the commonly 
used form
\begin{equation}
F_{NNA} = \frac{\Lambda_A^2}{\Lambda_A^2-t} \; .
\end{equation}
Here $\Lambda_A$ is the cut-off mass of the meson $A$. These cut-off masses are
$\Lambda_{\sigma}=$1200 MeV, $\Lambda_{\omega}=$808 MeV and 
$\Lambda_{\pi}=$500 MeV. \\
The total energy and the masses of the incoming hadrons serve as an input
for calculating the angular distribution.

Since at the current stage UrQMD only uses free cross sections and
free on-shell particles the effective in-medium quantities $E^*$ and
$m^*$ are replaced by the respective free quantities $E$ and $m$.

It is worth to stress again that equation~(\ref{sigdiff}) 
is only used for the angular distributions
of all elementary two-body processes but  not for the 
corresponding total cross sections.

\subsection{Resonance lifetimes}

\begin{figure}[t]
\begin{minipage}[t]{9cm}
\centerline{\psfig{figure=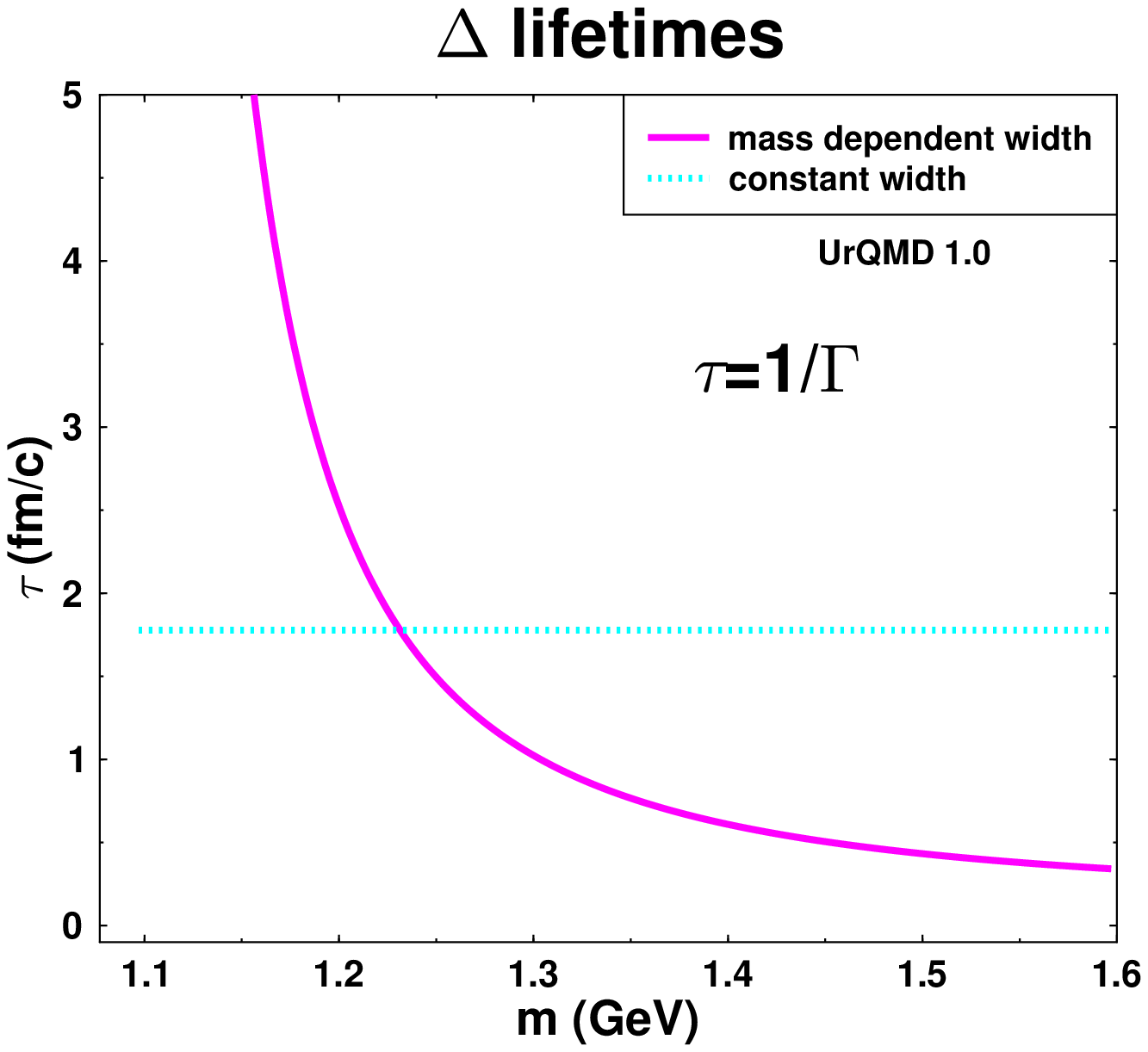,width=9cm}}
\end{minipage}
\hfill
\begin{minipage}[t]{9cm}
\centerline{\psfig{figure=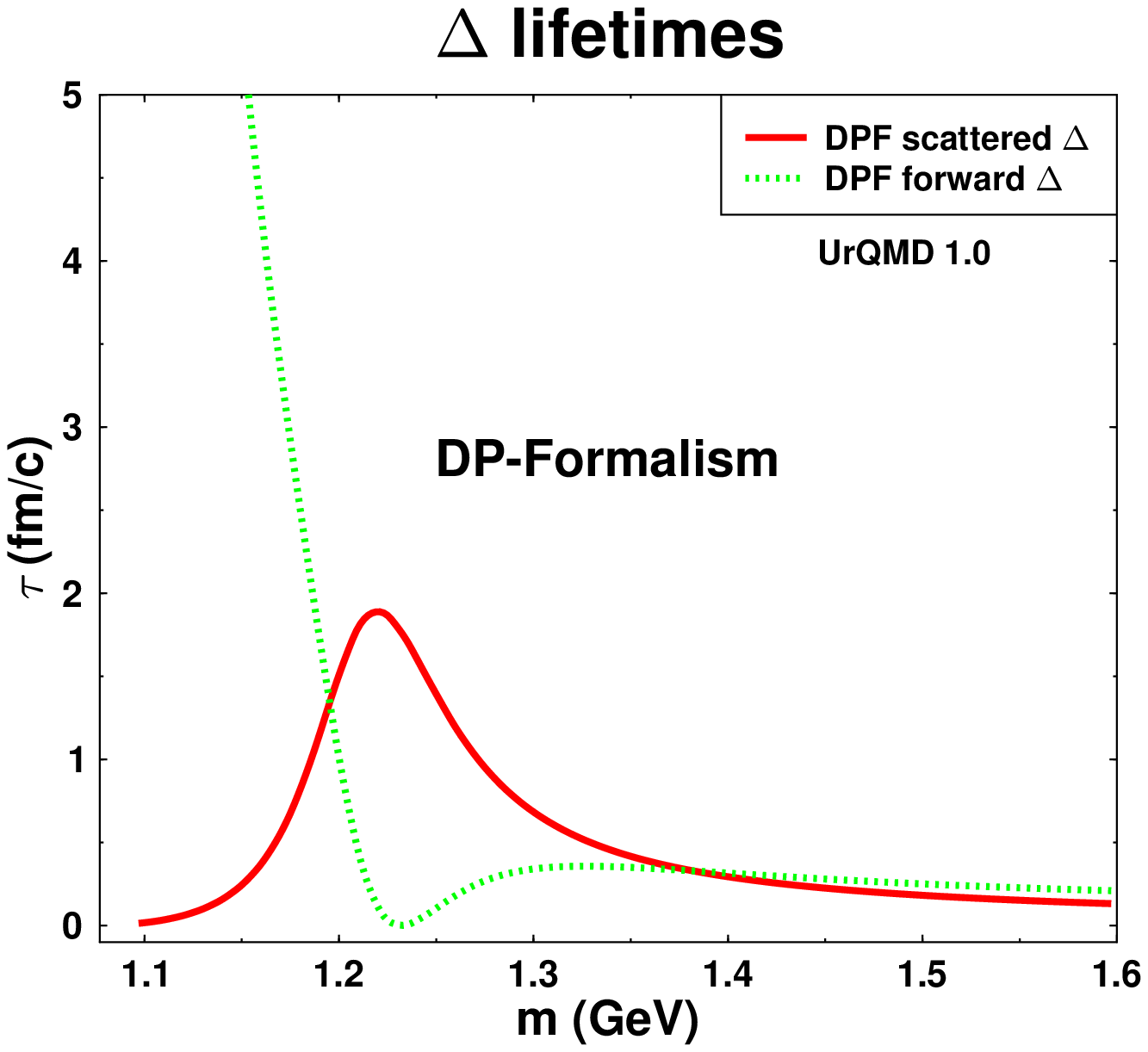,width=9cm}}
\end{minipage}
\caption{\label{lifetimes} Mass-dependent lifetime of the $\Delta(1232)$ 
resonance as used in transport model calculations. The current prescription 
of the lifetime computed as inverse (optionally mass-dependent) width is 
shown on the left. The r.h.s. displays lifetimes according to 
Danielewicz and Pratt
\protect\cite{danielewicz96a}.}
\end{figure}

Recently, the treatment of resonance lifetimes in microscopic transport
calculations has received much attention \cite{danielewicz96a}.
The standard approach to resonance lifetimes $\tau_R$ in transport calculations
is the application of $\tau_R = 1/\Gamma_R$ in conjunction with a 
Monte-Carlo sampling of the exponential decay law. The total decay width
of the resonance $\Gamma_R$ is either taken to be constant or 
mass dependent (see e.g. figure \ref{pwn1535}). 
The resulting lifetimes in the case of the $\Delta_{1232}$ resonance
can be seen in the left frame of figure~\ref{lifetimes}.

However, this treatment is highly questionable. Time delays and resonance
lifetimes in the context of semi-classical transport calculations can be
investigated via the scattering of an incident wave packet on a spherical
volume. The details of this investigation
are presented in the work of Danielewicz and Pratt (DP) \cite{danielewicz96a}. 
Here, 
only a brief summary with emphasis on its application in a microscopic
transport model is given. 

As a result of the scattering, both, the
forward going and the scattered part of the wave packet suffer
time delays. In terms of phase shifts they are given by:
\begin{equation}
\Delta \tau_S = \frac{{\rm d} \delta_l}{{\rm d} E} \quad ; \qquad
\Delta \tau_F = 
        \frac{(2l+1)\, 2 \cos(2\delta_l)}{\sum_l (2l+1) (1-4 \sin^2(\delta_l))}
\frac{{\rm d} \delta_l}{{\rm d} E} 
\end{equation}

If one is only interested in the thermodynamic quantities of the system,
one can define an ergodic constraint which relates the time delays to
its densities of state.
The investigation of this system, which in the context of transport 
model calculations may be thought of as a ``thermal'' or infinite matter
limit yields the following results:
\begin{itemize} 
\item[-] $\Delta \tau_F$ is consistent with the motion of particle through
        a mean field: \\$ U = \rho \cdot  {\rm Re}\,{\cal F}(0)$  
        (impulse approximation)
\item[-] all interaction effects (to lowest non-trivial order in density) 
        can either be absorbed into a  mean field
        or into a modified scattering delay $\Delta \tau_S'$ 
        \begin{equation}
        \Delta \tau_S' = \frac{1}{2 \sin^2(\delta_l)} 
        \frac{{\rm d} \delta_l}{{\rm d} E} 
        \end{equation}
\end{itemize}
Transport model calculations dominantly treat resonances as Breit-Winger
resonances.
In that case the form of the phase shift $\delta_l$ is well known and
the application of the formalism yields:
\begin{equation}
\Delta \tau_S = \frac{\Gamma/2}{(E-E_R)^2 + \Gamma^2/4 } 
\end{equation}
\begin{equation}
\Delta \tau_F = \frac{\sigma_R}{\sum_l (2l+1) \frac{\pi}{k^2}} 
\frac{(E-E_R)^2 - \Gamma^2/4}{\Gamma \left( (E-E_R)^2 + \Gamma^2/4 \right)}
\end{equation}
\begin{equation}
\Delta \tau_S' = 1/\Gamma_R \quad 
        \mbox{(``thermal'' or infinite matter limit)}
\end{equation}
During these delay-times, particle identity should change to ensure the correct
density of states. However, 
$\Delta \tau_F$ is negative near $E_R$. When treating the forward
time delay as a mean field interaction this negative time delay corresponds
to an acceleration of the respective particle. If, on the other hand,
the time delay is treated as such with the scattered particle in a resonance
state during that time-span, then
$\Delta \tau_F$ and (in order to remain consistent) $\Delta \tau_S$
must be rewritten in a way that no negative delays occur:
\begin{equation}
\Delta \tau_S^{\rm UrQMD} = \frac{\Gamma/4}{(E-E_R)^2 + \Gamma^2/4 } 
\end{equation}
\begin{equation}
\Delta \tau_F^{\rm UrQMD} = 
\frac{\sigma_{R}}{\sigma_F}
        \frac{(E-E_R)^2 }{\Gamma \left( (E-E_R)^2 + \Gamma^2/4 \right)}
\end{equation}
These time delays are plotted in the right frame of
figure~\ref{lifetimes}.
The $1/\Gamma(E)$ divergence of $\Delta \tau_F$ should not pose a 
serious problem since it appears at an energy far off the pole 
at which the cross section
for resonance production usually vanishes. It can therefore 
be omitted through the
proper energy dependence of the ratio $\sigma_R / \sigma_F$.
However, the determination of the cross section $\sigma_F$ is not
straightforward\footnote{Note that when following the mean field
approach for the forward time delay,
the form
of the mean field can directly be determined from the experimentally 
accessible forward scattering amplitude ${\cal F}(0)$},
since it is experimentally not directly observable 
via~$\frac{\rm d \sigma}{\rm d \Omega}$. 
For the application in a transport model one can attempt to determine
the strength of $\sigma_F$ via the infinite matter limit: an infinite
matter calculation employing $\Delta \tau_S$ and $\Delta \tau_F$ with
their respective cross sections must yield the same result as a calculation
with $\Delta \tau^{\prime}_S$ and the scattering cross section alone --
if the forward cross section has been chosen properly. This approach
does of course not yield a unique choice for $\sigma_F$ but provides
a range of $\sigma_F$ parameterizations which are consistent with
the well known ``thermal'' or infinite matter limit.

Different ansatzes for $\sigma_F$ are currently under investigation in
the framework of UrQMD. Apart from this novel approach it is also possible
to perform calculations with $\tau_R=1/\Gamma_{R}$, both, for fixed
and mass-dependent widths.

\subsection{String-excitation and -fragmentation}

The strong interactions of hadrons and nuclei are described by
Quantum Chromodynamics. Due to asymptotic freedom of QCD 
the coupling constant $\alpha _s(Q^2)$ becomes small at small distances or 
large $Q^2$. It gives a possibility to apply the perturbation theory 
to the processes with large momentum transfer. 
On the other hand at large distances ($\approx 1/\Lambda _{QCD}$) 
the coupling constant is not small and non perturbative effects,
which are responsible for the confinement should be important. This large
distance dynamics is very essential for understanding the processes with
small momentum transfer, which give a dominant contribution to 
the high energy hadronic interaction. 

If one considers the hadrons as  bubbles in QCD vacuum liquid  
with the chromodynamical fields of quarks which do not penetrate 
in the vacuum medium, 
then the interactions of such bubbles at high energies leads 
to the production of new objects - color tubes, or strings. 
The interaction between such objects can take place with color transfer
when quarks from different hadrons are interacting and without color transfer
when hadrons are interacting only by exchange of the momenta. 
In both cases it leads to increase of energy and at some moment 
it will be energetically favorable to break the color tube by producing 
$q\bar q$-pair from the vacuum. This process repeats until many white bubbles
- hadrons will be produced. According to the uncertainty principle 
the time needed for production of a hadron with momentum 
$p$ is $\tau\approx p/m^2$, 
so in the c.m. system the last will be produced the fastest 
hadrons containing the spectator quarks (inside-outside cascade).
Each produced $q$ and $\bar q$
have small relative momenta in their rest frame or small rapidity distances.
As a result due to Lorentz invariance of the picture, finally produced
hadrons at high energies will be uniformly distributed in rapidity and
will have limited transverse momenta.

Hereby for high energy reactions 
we use a similar picture corresponding to the case when two hadrons
are interacting only by momentum transfer. It imitates the processes 
with the double and single diffraction. 
In baryon-baryon (meson-meson) 
interaction
strings between quark $q_v$ and diquark $qq_v$ (antiquark $\bar q_v$) 
from the same hadron are produced.  
The hadron-hadron interactions at high energies are simulated in 3 stages.
According to the cross sections the type of interaction is  defined:
elastic, inelastic, antibaryon-baryon annihilation etc.
In the case of inelastic collision with string excitation the kinematical
characteristics of strings are  modeled in the following way: 
The hadron momentum transfer $p_T$
is simulated according to a Gaussian distribution
\begin{equation}
f(\vec{p_{T}}) \propto \frac{1}{\sqrt{\pi \sigma^2}}
\exp(- p_{T}^{2}/\sigma ^2)
\label{gausspt}
\end{equation}
where  $\sigma =$ 1.6 GeV/c. 
The other interacting hadron gets the same momentum transfer but in
the opposite direction. The excited strings have the continuous 
mass distribution $f(M) \propto 1/M^2$
with the masses $M$, limited by the total collision energy $\sqrt{s}$: 
$M_1+M_2 \le \sqrt{s}$. The rest of the $\sqrt{s}$ is equally distributed
between the longitudinal momenta of two produced strings:
$\vec{p_{1\parallel}}=-\vec{p_{2\parallel}}$. 
The energy of the $i$th string $E_i$ 
is defined by 
the longitudinal momentum of the string $\vec{p_{i\parallel}}$, 
momentum transfer
$p_T$ and the mass of the string $M_i$:
\begin{eqnarray}
E_i^2 &=& p_{i\parallel}^2 + p_T^2 + M_i^2, \ \ i=1,2  \\
E_1 + E_2 &=& \sqrt{s}.                                      
\end{eqnarray}

The longitudinal momenta
of the  constituent quarks are  chosen according to the structure functions
of hadrons:
\begin{equation}
f(x_{q}) = f_{0} (x_{q})^{\alpha -1} (1-x_{q})^{\beta-1}
\end{equation}
with $\alpha =$ 0.5 and $\beta =$ 2.5 for valence quark in nucleons.

The transverse distribution of the constituent quarks  
was generated according to the same Gaussian distribution as
for the momentum transfer Eq. (\ref{gausspt}).
The diquark transverse momentum is equal in magnitude, but of
opposite direction.

The second stage of h-h interactions is connected with 
{\bf string fragmentation}.
UrQMD uses the Field-Feynman fragmentation procedure 
when the strings decay independently 
from both ends. It includes energy, momentum and
quantum number conservation laws and
the possibility of converting diquarks
into mesons via diquark breaking.

\begin{figure}[tb]
\centerline{\psfig{figure=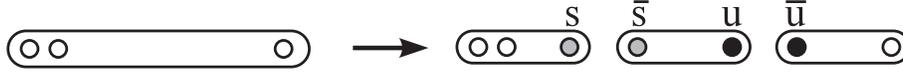,width=12cm}}
\caption{\label{strfrag}
Particle production in a  string-excitation. Two quark-antiquark
pairs are spontaneously created in the color flux-tube between
the constituent diquark and the constituent quark.
}
\end{figure}

Figure~\ref{strfrag} schematically sketches the fragmentation excitation
of a baryon-string: Two quark-antiquark pairs are created (a $\bar u u$ and
an $\bar s s$ pair). The leading diquark combines with a newly produced 
$s$ quark to form a hyperon, the newly produced $\bar s$ quark combines
with a newly produced $u$ quark to form a kaon and the newly produced
$\bar u$ quark forms together with the leading quark a pion.
In order to produce baryon-antibaryon pairs in a string-excitation, also
the creation of diquark-antidiquark pairs from the Dirac-sea must be
possible. 
The suppression factor for the flavors
of the quark-antiquark pair produced in string decays 
can be  defined by a Schwinger-like formula \cite{schwinger51a}: 
\begin{equation}
\label{qmassenwichtung}
|{\cal M}|^2 \sim  \exp  \left(-\frac {\pi m^2 }{\kappa }\right) \quad ,
\end{equation}
where ${\cal M}$ gives the matrix-element for the production
of a quark-antiquark pair. 
The string-tension $\kappa$ is ($\kappa \approx 1$~GeV/fm) 
and $m$ is the mass of the quark-antiquark
pair \cite{andersson83b,andersson82a,andersson83a}.

The suppression factors can also
be tuned to the production probabilities for certain meson- and baryon-species
in elementary proton-proton collisions:  The $s$-quark suppression factor
is very sensitive with regard to kaon-production, whereas the 
diquark suppression factor plays an important role for the antinucleon
production. The strange diquark suppression factor can be tuned by studying
antihyperon production. The standard values for the different suppression
factors in UrQMD are:
\begin{equation}
u:d:s:qq \,=\, 1:1:0.35:0.1 \quad.
\end{equation}

In the UrQMD string-excitation scheme it is not only  possible to excite 
the ground-states of baryons
and mesons, but also their excited states, which have the
same quark-content.  If for baryons, the quark-content does not determine
whether the state belongs to the lowest octet or to the 
lowest decuplet, then octet or decuplet are chosen with equal probability.
The probabilities for the higher excited multiplets are tuned to 
multiplicities measured in proton-proton reactions.
In the case of mesons the multiplet must also be determined before a
species can be assigned. The probability of choosing a certain multiplet
depends on the spin of the multiplet and its average mass:
\begin{equation}
  P_{J^{PC}} \sim \frac{ 2S+1}{ \langle m \rangle_{J^{PC}} } \quad.
\end{equation}
The respective values for the meson multiplets implemented in UrQMD are
listed in table~\ref{pnonett}.

\begin{table}[ht]
\begin{center}
\renewcommand{\arraystretch}{1.3}
\begin{tabular}{|l||llllllll|} \hline
{\bf multiplet} $J^{PC}$ & $0^{-+}$ & $1^{--}$ & $0^{++}$ & $1^{++}$ & 
                $2^{++}$ & $1^{+-}$ & $1^{--*}$ & $1^{--**}$ \\\hline
{\bf probability} & $0.102$ & $0.190$ & $0.056$ & $0.124$ &
                      $0.197$ & $0.127$ & $0.110$ & $0.095$ \\
\hline
\end{tabular} 
\caption{\label{pnonett}
Excitation probabilities for the different meson-multiplets in UrQMD.
The probabilities depend on spin and average mass of the multiplet. }
\end{center}
\end{table}


\noindent
The excitation and fragmentation of a string follows an iterative
scheme:
\begin{displaymath}
\mbox{\rm string} \, \Longrightarrow \mbox{\rm hadron} \,+\, 
        \mbox{\rm new string} \qquad,
\end{displaymath}
i.e. a quark-antiquark (or a diquark-antidiquark) pair is created and
placed between leading constituent quark-antiquark (or diquark-quark)
pair.  Then a hadron is formed randomly on one of the end-points of the
string. The quark content of the hadron determines its species
and charge. 
In case of resonances the mass is determined according to
a Breit-Wigner distribution. Finally, the energy-fraction of the string
which is assigned to the newly created hadron must be determined:
After the hadron has been stochastically assigned a transverse momentum,
the fraction of longitudinal momentum transferred from the string
to the hadron is determined by the fragmentation function.

For the following discussion, it is convenient to introduce 
light-cone variables
in configuration space ($z^{\pm } $) and in momentum space ($p^{\pm } $). 
They depend on space-time ($t$,$z$) and energy-momentum ($E$,$p$) coordinates
and are defined as:
\begin{equation}
z^{\pm } = t \pm z \quad \mbox{\rm and} \quad p^{\pm } = E \pm  p \quad .
\end{equation}
where 
For the stochastic fragmentation of a string one defines the
Lorentz-invariant quantities:
\begin{equation}
x^{\pm } =\frac {p^{\pm}_{\rm hadron}}{p^{\pm}_{\rm total}} \quad
(0 \leq x^{\pm} \leq 1) \quad ,
\end{equation}
which represent the longitudinal momentum-fraction of the string which is
transferred to the  new hadron.
As examples we show a meson string fragmenting from the r.h.s.:
\begin{equation}
\label{fragmesstr}
       p^+ \underbrace{(q\, \bar q q \,  \bar q )}_{\mbox{string}}
      =   x^+ p^+\underbrace{( q \bar q  )}_{\mbox{meson}}
                              +  \, (p^+-x^+p^+) 
        \underbrace{q \bar q}_{\mbox{string}} \qquad,
\end{equation}
and a baryon string fragmenting from the l.h.s:
\begin{equation}
\label{fragbarstr}
       p^- \underbrace{(qq\, q \bar q \,  q )}_{\mbox{string}}
      =   x^- p^-\underbrace{( qqq  )}_{\mbox{baryon}}
                              +\,  (p^--x^-p^-) 
        \underbrace{\bar q q}_{\mbox{string}} \qquad.
\end{equation}
This iterative fragmentation process is repeated until the remaining
energy of the string gets too small for a further fragmentation.

\begin{figure}[tb]
\begin{minipage}[t]{9cm}
\centerline{\psfig{figure=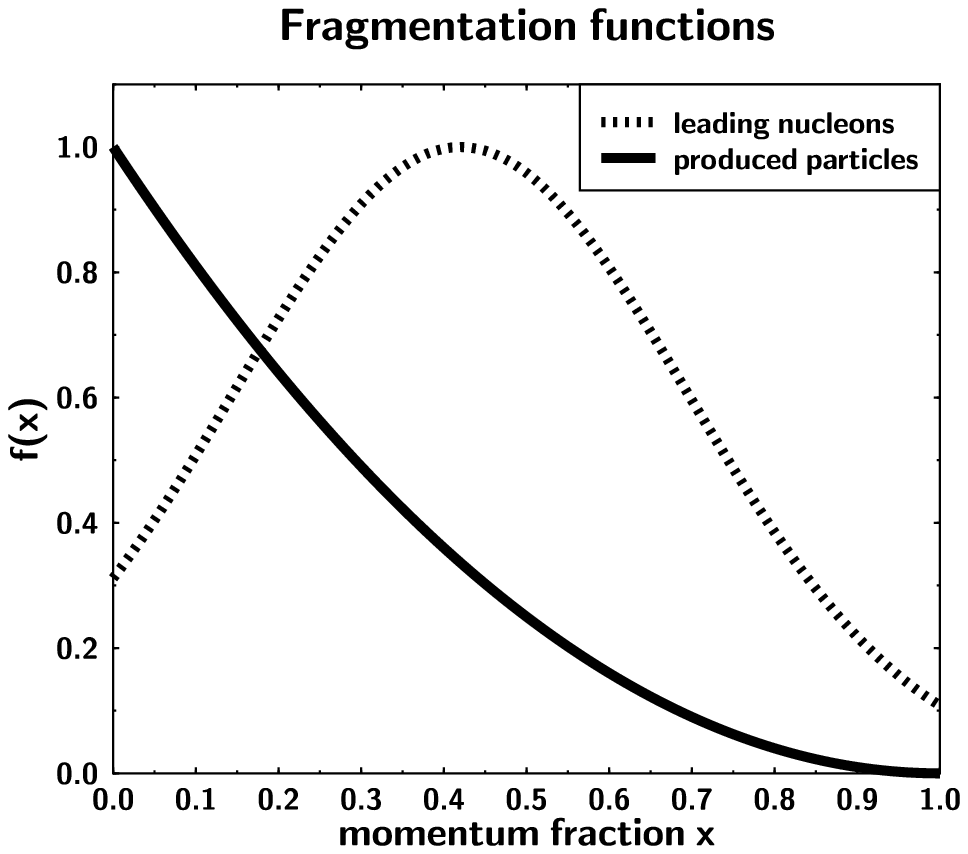,width=10cm}}
\caption{\label{frag}
UrQMD fragmentation functions: nucleons containing leading constituent
quarks have a Gaussian fragmentation function whereas  a Field-Feynman
fragmentation function is used for newly produced particles.
}
\end{minipage}
\hfill
\begin{minipage}[t]{9cm}
\centerline{\psfig{figure=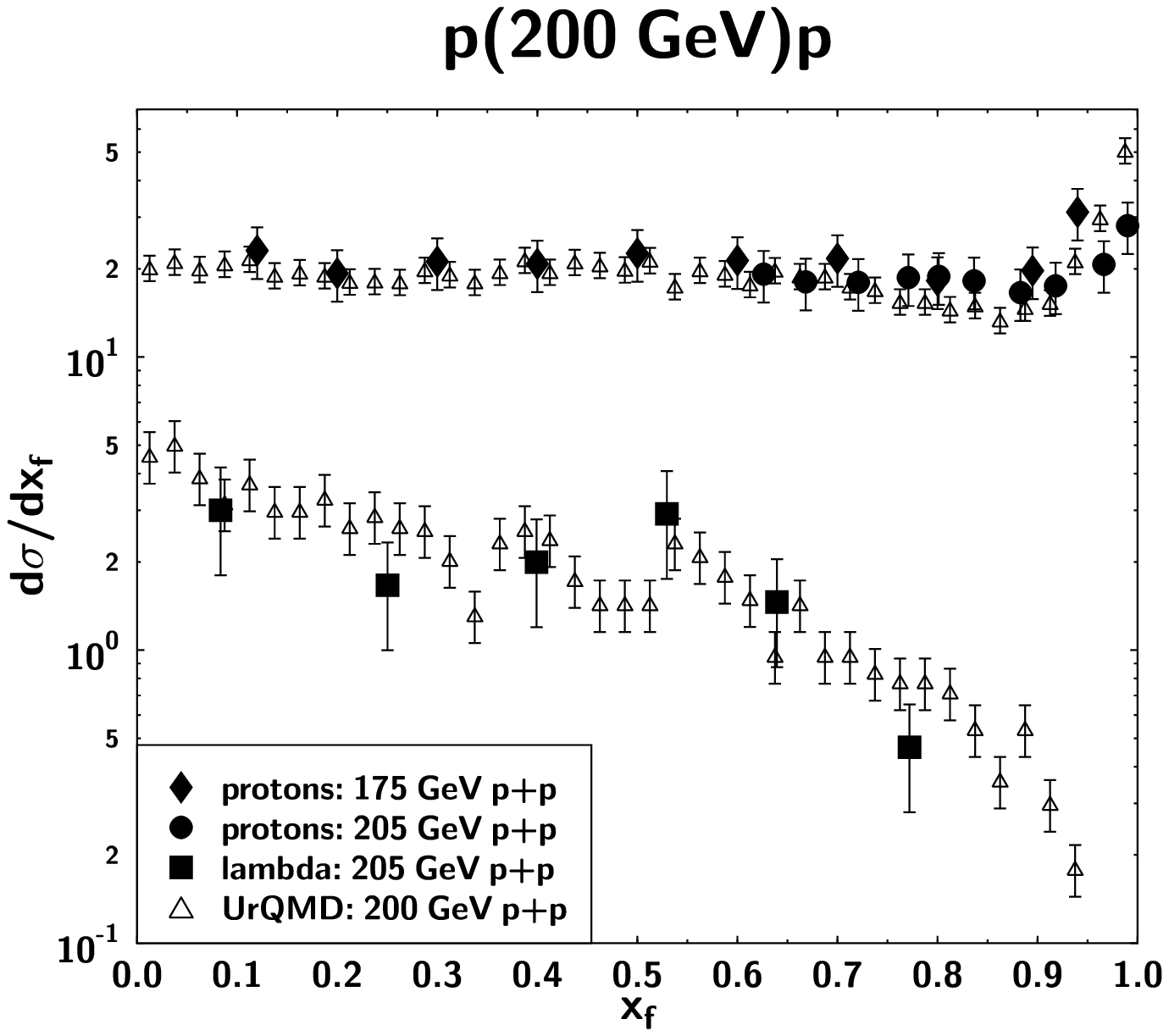,width=9cm}}
\caption{\label{xf_p175p}
Comparison of the Feynman-$x$ distribution for lambdas and protons between
UrQMD and data \protect\cite{charlton73a} in proton-proton reactions
at $\approx 200$~GeV.
}
\end{minipage}
\end{figure}

The fragmentation function $f(x,m_t)$ represents the probability distribution 
for hadrons with the transverse mass $m_t$ to acquire the longitudinal 
momentum fraction
$x$ from the fragmenting string. One of the most common fragmentation
functions is the one used in the LUND model \cite{andersson83a}:
\begin{equation}
\label{flund}
     f(x,m_t)  \sim  \frac{1}{x}  
         (1-x)^a  \exp{\left(-b \cdot m_t^2/x\right)} \, .
\end{equation}
In UrQMD, different fragmentation functions are used for leading nucleons
and newly produced particles, respectively (see figure~\ref{frag}):
\begin{eqnarray}
\label{fuqmd1}
f(x)_{\rm nuc} &=& \exp\left(-\frac{(x-B)^2}{2\,A^2}\right)  
                        \quad \mbox{for leading nucleons} \\
\label{fuqmd2}
f(x)_{\rm prod} &=& (1-x)^2  
                        \quad \mbox{for produced particles} 
\end{eqnarray}
with $A=0.275$ and $B=0.42$. 
The fragmentation function $f(x)_{\rm prod}$, used for newly produced
particles, is the well-known Field-Feynman 
fragmentation function \cite{field77a,field78a}. 
At string break-up,  $q\bar{q}$-pairs have zero transverse
momentum in the string frame, but  the transverse momenta of a single quark,
$\vec{p_{t}}$, and the corresponding antiquark, $-\vec{p_{t}}$, is taken 
according to eq. (\ref{gausspt}).

The main criteria for the selection of the fragmentation functions and the
fine-tuning of their parameters is that the particle multiplicities
and momentum distributions in elementary proton-proton reactions must
match the data.

\begin{figure}[tb]
\begin{minipage}[t]{9cm}
\centerline{\psfig{figure=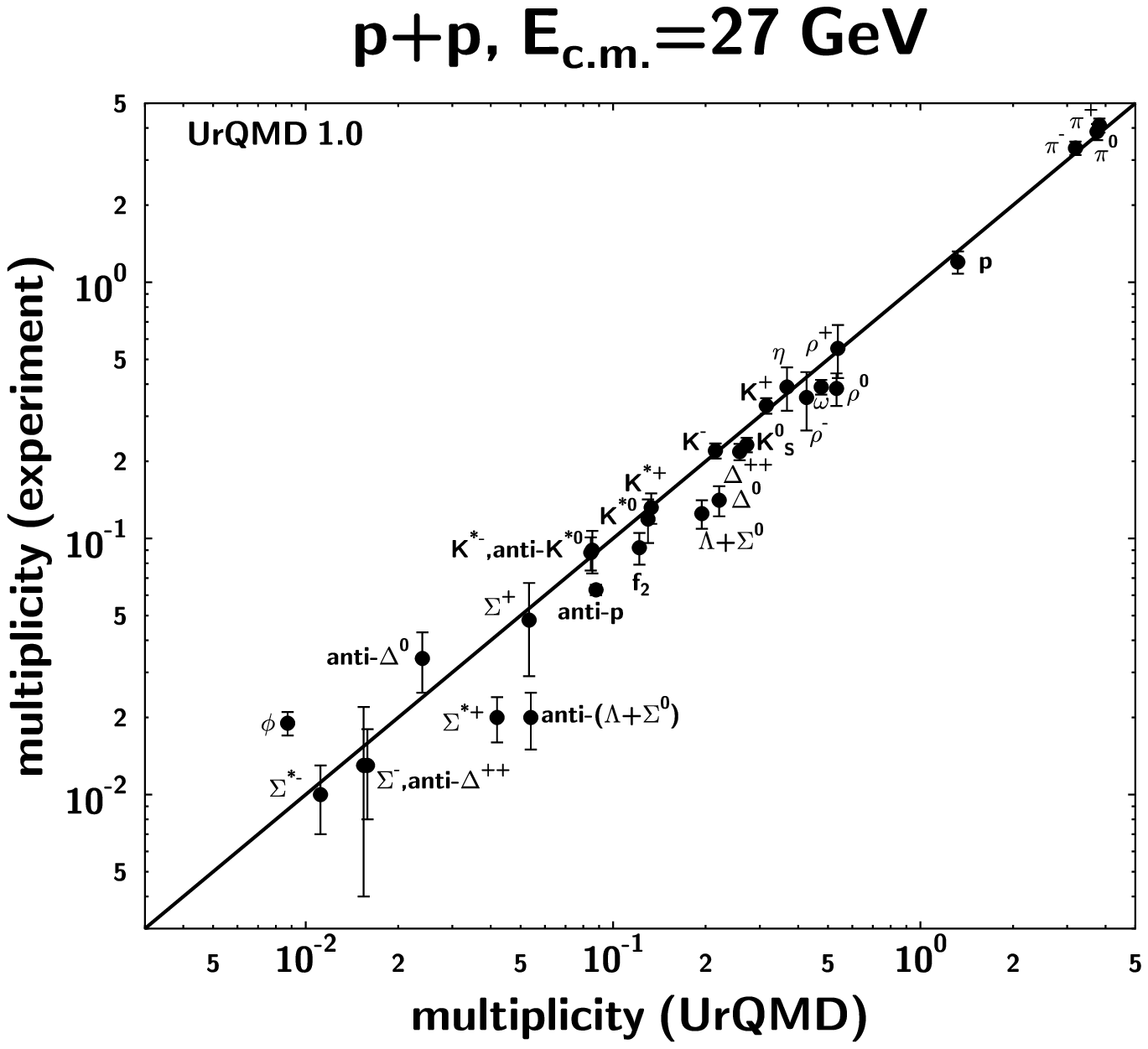,width=9.5cm}}
\caption{\label{becattini}
Data \protect\cite{becattini97a,aguilar91a,kichimi79a} vs. UrQMD particle multiplicities in 
proton-proton reactions
at a c.m. energy of 27 GeV. 
}
\end{minipage}
\hfill
\begin{minipage}[t]{9cm}
\centerline{\psfig{figure=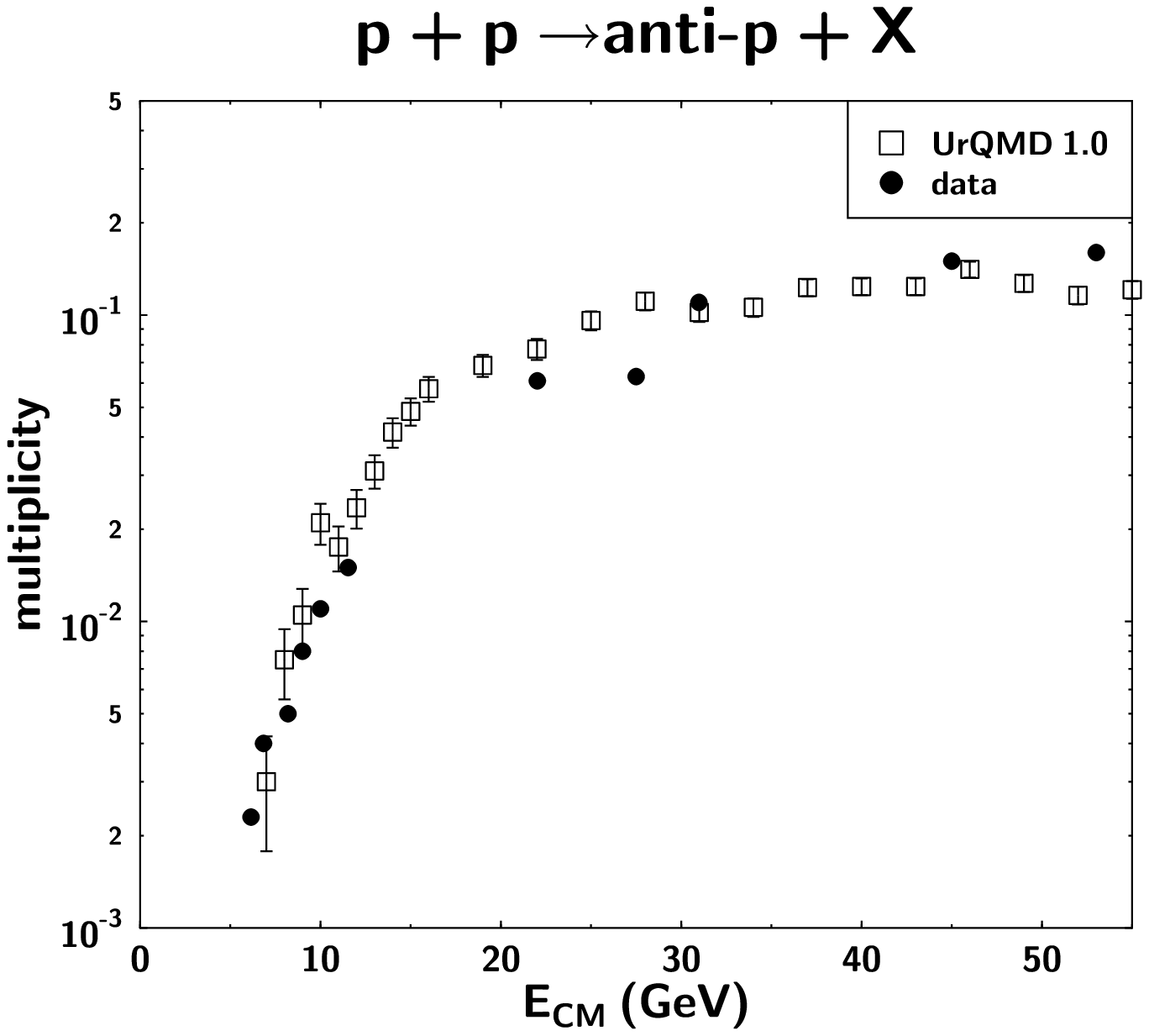,width=9.5cm}}
\caption{\label{pbar}
Proton-proton excitation function of anti-proton multiplicities as obtained
in UrQMD. The experimental data are taken from \cite{antinucci73a}
}
\end{minipage}
\end{figure}

Figure \ref{xf_p175p} shows the Feynman-$x$ ($x_f = 2 p_{lab}/\sqrt{s}$) 
distributions for emitted protons
and lambdas in proton-proton reactions at $E_{lab} \approx 200$~GeV.
UrQMD fits quite well the data \cite{charlton73a}, which is important
for the predictive power with regard to stopping and particle production 
in nucleus-nucleus collisions.
The UrQMD estimate of particle multiplicities in comparison to data is shown 
for proton-proton reactions at 200~GeV in table~\ref{pp200mult}, 
and in figure~\ref{becattini}
at $\sqrt{s}=27$~GeV. The overall agreement is very good, however,
currently the $\phi$-meson production is underestimated by almost
a factor of 2 and (anti-)lambda and $\Sigma^0$ production is overestimated
by 50\%. Also shown in Fig. \ref{pbar} is the excitation function of
anti-proton multiplicities in inelastic proton-proton collisions.

\begin{table}
\begin{center}
\renewcommand{\arraystretch}{1.3}
\begin{tabular}{|l||r|r|} \hline
\bf particle  & \bf data & \bf UrQMD  \\\hline\hline
$\pi^-$ & $2.62 \pm  0.06$ & $ 2.47 $ \\
$\pi^+$ & $3.22 \pm  0.12$ & $ 3.07 $ \\
$\pi^0$ & $3.34 \pm  0.24$ & $ 3.01 $ \\
$K^+$   & $0.28 \pm  0.06$ & $ 0.23 $ \\
$K^-$   & $0.18 \pm  0.05$ & $ 0.13 $ \\
$K^0_S$ & $0.17 \pm  0.01$ & $ 0.18 $ \\
$ \Lambda + \Sigma^0$ & $0.096 \pm 0.01$ & $0.17$ \\
$\bar \Lambda + \bar \Sigma^0$ & $0.013 \pm 0.004$ & $0.03$ \\
$p$ &    $1.34 \pm  0.15$ & $1.32$ \\
$\bar p$ & $0.05 \pm  0.02$ & $0.05$ \\
\hline
\end{tabular} 
\caption{\label{pp200mult}
UrQMD particle multiplicities compared to data  \protect\cite{kafka77a}
for proton-proton reactions at 200 GeV.}
\end{center}
\end{table}

\label{formation-time}
The last central issue to be discussed in this section is the definition
of the formation time:
The formation time of a hadron is the time the constituent quarks of 
the respective hadron need in order to bind together and form the hadron
and to tunnel out of the vacuum. For composed particles, 
there are two possibilities to define the formation
point: {\it i}) ``constituent'' formation point has the coordinates
of the string break-up points; {\it ii}) ``yo-yo'' formation point 
has the coordinates of the quark trajectories intersection.
In UrQMD, the ``yo-yo'' formation time definition is used. 
The formation time depends on the momentum and energy 
of the created hadron.

\begin{figure}[tb]
\centerline{\psfig{figure=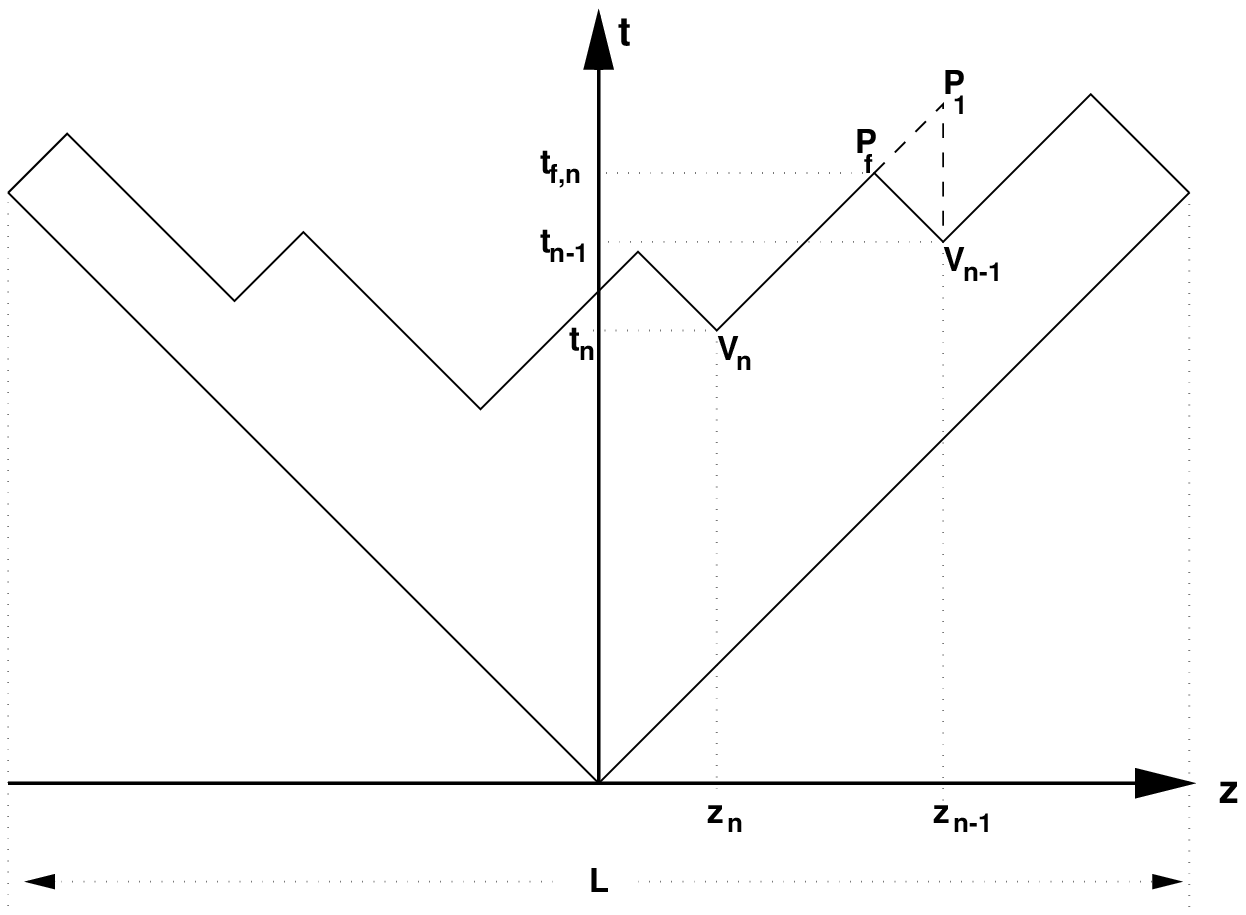,width=9cm}}
\caption{\label{string2}
Schematic picture of a fragmenting meson string in configuration
space. The first newly produced meson on the r.h.s. forms at the 
point $P_f$ at which the quark lines first meet.}
\end{figure}

Figure~\ref{string2} schematically shows the fragmentation of a 
meson string into several mesons. At the vertices $V_n$ and $V_{n-1}$
quark-antiquark pairs are created. The antiquark of vertex $V_n$
then forms a new meson together with the quark of vertex $V_{n-1}$.
Configuration space (as depicted in figure \ref{string2}) and momentum
space are linked by the following relations
\cite{andersson83b}:
\begin{eqnarray}
\label{e2z}
        E_n &=& \kappa \,(z_{n-1} - z_n) \\
\label{p2t}
        p_n &=& \kappa \,(t_{n-1} - t_n) \quad. 
\end{eqnarray}
and for the light-cone variables:
\begin{eqnarray}
\label{pp2zp}
p_n^+ &=& \kappa\,(z_{n-1}^+ - z_n^+) \\
\label{pm2zm}
p_n^- &=& \kappa\, (z_n^- - z_{n-1}^-) \quad. 
\end{eqnarray}
The formation point and time of the hadron is defined as the point in
space-time where the quark trajectories of the quark-antiquark pair forming
the hadron meet for the first time. In figure~\ref{string2} this
point is denoted as $P_f$. 
The formation time is then given by:
\begin{equation}
\label{tform_ort}
t_{f,n} \,=\, \frac{1}{2} ( t_{n-1} \,+\, t_n +z_{n-1} - z_n) \,=\, 
        \frac{1}{2}(z_{n-1}^+ + z_n^-)
\end{equation}
If the momenta of the produced particles are numbered consecutively from
right  to left, one obtains in momentum space:
\begin{equation}
\label{tform_imp}
t_{f,n} \,=\, \frac{1}{2 \kappa} \left( M + E_n - p_n - 2 
        \sum\limits_{j=0}^{n-1} p_j \right)  \quad,
\end{equation}
with $ L = M/\kappa $ and $ T= 2L = 2M/ \kappa $.

Depending on the hadron-species, formation times are in the order
of 1 -- 2 fm/c. During the formation time, the cross sections of
the leading hadrons containing leading constituent quarks are reduced
since prior to the formation of the hadron only the constituent
quark part of the cross section is present:
\begin{eqnarray}
\sigma _{\rm qh} &\approx & \frac {1}{3}\sigma_{\rm Bh}  
   \quad \mbox{\rm for baryons}\quad, \\  \nonumber
\sigma _{\rm qqh} &\approx & \frac {2}{3}\sigma_{\rm Bh} 
   \quad \mbox{\rm for baryons}\quad, \\ \nonumber
\sigma _{\rm qh} &\approx & \frac {1}{2} \sigma_{\rm Mh} 
   \quad \mbox{\rm for mesons}\quad .
\end{eqnarray}
Newly created hadrons without any leading constituent quarks of the initial
hadron have zero interaction cross section
during their formation time.

\subsection{Color fluctuations, color opacity and transparency}
Quantum Chromodynamics (QCD) has important - and so far not considered -
implications on high energy processes \cite{frankfurt94a}: 
\begin{itemize}
\item{Hadrons consist of configurations of very different spatial size;}
\item{At high energies, hadronic quark-gluon configurations can be
considered frozen. Due to the long coherence length at these energies
geometrical color optics can be applied;}
\item{Small object have reduced interactions;}
\item{Small objects will expand at high energies;}
\item{Large-size objects can be captured in the central zone of a heavy
ion collision.}
\end{itemize}

These effects have diverse consequences. One of them is color transparency:
Suppose a small sized object is created; this object does not interact very
much due to color screening. At sufficiently high energies, it can pass
through an entire nucleus without fluctuating into a large
configuration. Thus it does not interact - the nucleus is transparent.
Nucleus-Nucleus collisions provide a tool to investigate the effects of
color transparency e.g. in the production of $J/\Psi$ \cite{gerland98a}. 
However,
the complementary color opacity effect (large size configurations) can
cause stronger stopping and significant fluctuations in the transverse
energy for central reactions.

\begin{figure}[tb]
\centerline{\psfig{figure=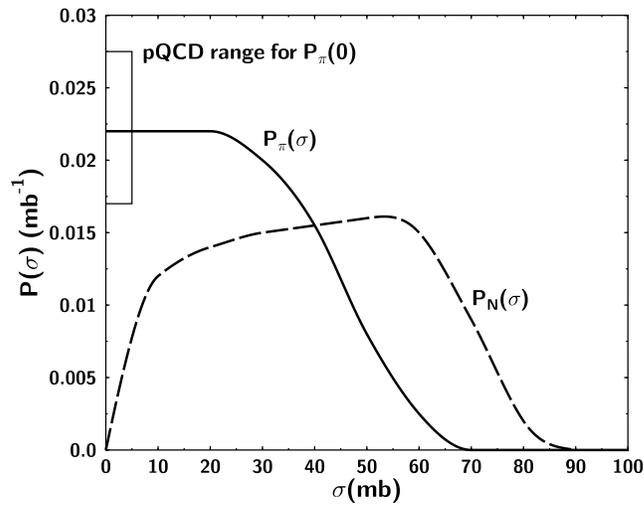,width=10cm}}
\caption{\label{colfluc} 
Cross-section probability for pions $P_\pi(\sigma)$ and nucleons
$P_N(\sigma)$ as extracted from experimental data. $P_\pi(\sigma=0)$ is
compared with the perturbative QCD prediction. (Figure taken from
\cite{frankfurt94a}).}
\end{figure}

A first step to investigate these QCD effects within a microscopic
transport model can be made by using fluctuations in the elementary
hadron hadron reactions. Thus one needs the probability $P(\sigma)$ 
for different sized configurations.  
It is convenient to consider moments of the distribution:
\begin{eqnarray}
<\sigma^0>=\int{\rm d}\sigma \overline{\sigma}^0 P(\sigma) =1\quad, \\
<\sigma^1>=\int{\rm d}\sigma \overline{\sigma}^1 P(\sigma)
=\overline{\sigma}\quad,\\
\vdots
\end{eqnarray}
where $\overline{\sigma}$ denotes the average cross section, and
$P(\sigma)$ the probability of a certain cross section.
The second moment $<\sigma^2>$can be determined from diffractive dissociation
experiments. In addition further information can be obtained from QCD,
which implies:
\be
P(\sigma)\propto\sigma^{N_q-2}\quad,
\ee
for $\sigma\rightarrow 0$, with $N_q$ being the number of valence
quarks.
Thus, we get for the nucleon and pion distributions:
\begin{eqnarray}
P_N(\sigma)\propto \sigma \quad\mbox{nucleon}\quad,\\
P_N(\sigma)\propto {\rm const.}\quad\mbox{pion}\quad.
\end{eqnarray}

from these arguments, we can construct $P(\sigma)$. As shown in
Fig.\ref{colfluc}, one gets a broad distribution for proton projectiles
and an even broader one for pions.

By exploring these color coherent phenomena, we have an effective new
method to investigate QCD and the nuclear structure. This helps
to disentangle and select distinct properties of interacting hadronic
systems. This has been so far neglected in microscopic calculations 
of heavy ion collisions.

\chapter{Probing hot and dense nuclear matter with relativistic heavy ion collisions}
\label{hotdense}
		\section{Baryon stopping}

Head-on collisions of two nuclei can be used to create highly excited
nuclear matter \cite{chapline73a,scheid68a}. 
In the course of the reaction, a part of the respective longitudinal momenta
is converted into transverse momentum and secondary particles. This
causes the creation of a zone of high energy density. Nuclear shock
waves have been suggested as a primary mechanism of creating high
energy densities in such collisions \cite{chapline73a,scheid68a,scheid74a}.

The term {\em nuclear stopping power} \cite{busza84a} 
characterizes the degree of stopping which
an incident nucleon suffers when it collides with the nuclear matter
of another nucleus. It can be studied by measuring
the rapidity distribution of the net-baryons present
in the reaction. Obviously, the average collision number per baryon
increases with the mass number of the colliding nuclei. Thus, the heaviest
systems available, such as Pb+Pb or Au+Au, are best suited
for the creation of strongly stopped matter and high energy densities.

The shape of the baryon rapidity distribution can give clear indications
on the onset of critical phenomena: Due to the strong dependence
of the baryon rapidity distribution on the baryon--baryon cross section
\cite{hartnack89a,berenguer92a,schmidt93a},
a rapid change in the shape of the scaled $dN/d(y/y_p)$
distribution with varying incident
beam energy is a clear signal for new degrees of freedom entering
the reaction (i.e. deconfinement) or for phenomena such as critical
scattering \cite{gyulassy77a}. The width of the $dN/d(y/y_p)$
distribution for baryons is inversely proportional to their cross section. 
In \cite{frankfurt91a} it was pointed out that at energies $E\ge 200$~GeV/nucleon 
large fluctuations of size in the nucleon wave function, leading to the
color-transparency or color-opacity effect, respectively, 
should be observed.

The degree of  stopping can furthermore be used to estimate the 
achieved energy density in the course of the collision. 
One often assumes that particles produced at $y=y_{CM}$ originate
from the central reaction zone at $z=0$ and the initial
proper time $\tau_0$.
The rapidity distribution of these produced particles can then be used
to estimate the initial energy density in the central reaction
zone \cite{bjorken83a}:
\begin{equation}
\epsilon_0 \,=\, 
\frac{m_T}{\tau_0 A} \,\left. \frac{dN}{dy} \right|_{y=y_{CM}}\quad. 
\end{equation}
Here $A$ is the transverse overlapping region area in the collision and
$m_T$ the transverse mass of the produced particles.
The proper production time is estimated to be of the order of 1 fm/c.
Estimates for the CERN/SPS energy region were in the order
of 1 to 10 GeV/fm$^3$ \protect\cite{bjorken83a}.

\begin{figure}[htb]
\begin{minipage}[t]{9cm}
\centerline{\epsfig{figure=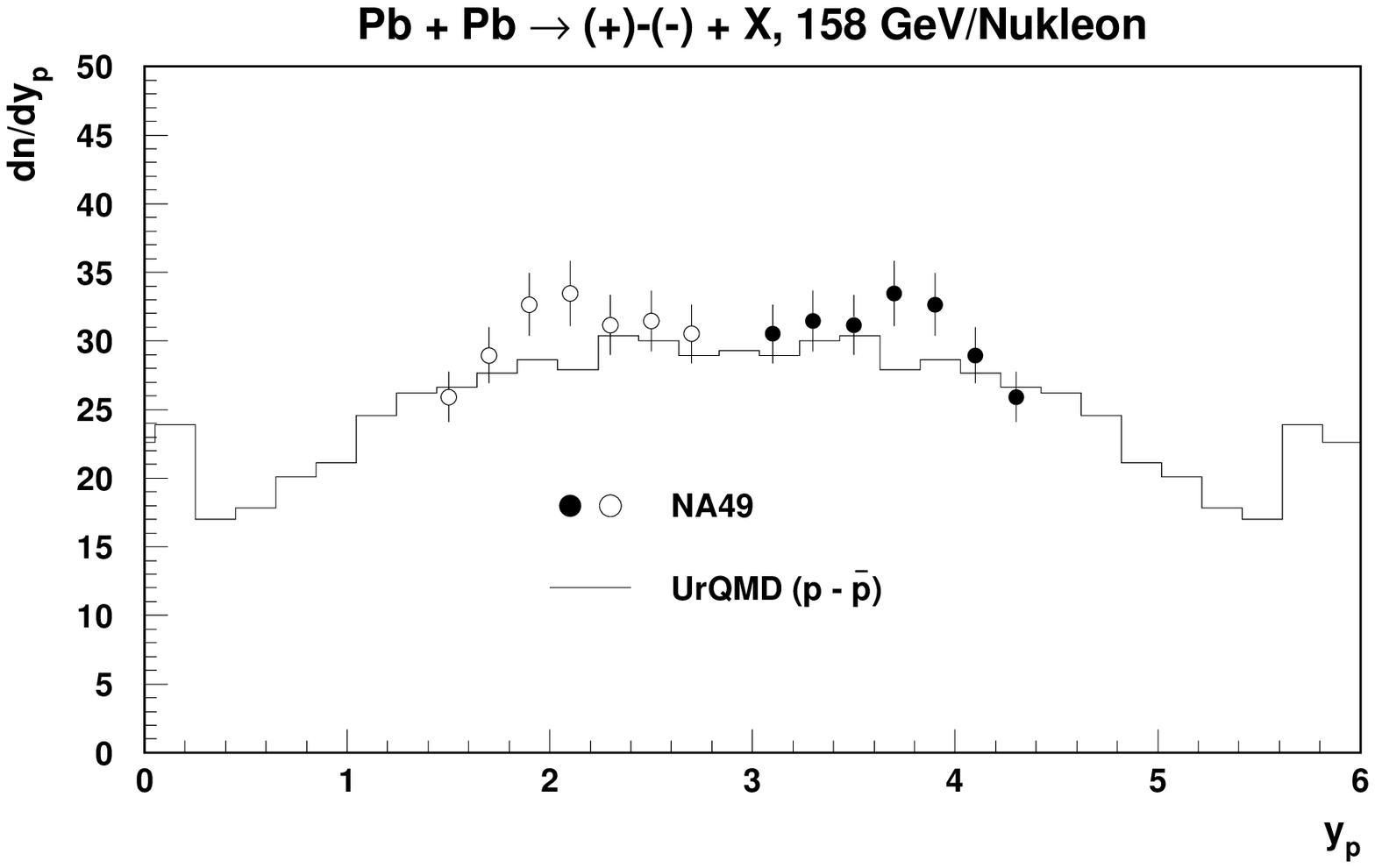,width=9.0cm}}
\caption{\label{pmm}Rapidity distributions of net-protons for Pb+Pb 
collisions at the SPS (160 AGeV) in comparison to preliminary NA49 data
\cite{guenther98a}.}
\end{minipage}
\hfill
\begin{minipage}[t]{9cm}
\centerline{\epsfig{figure=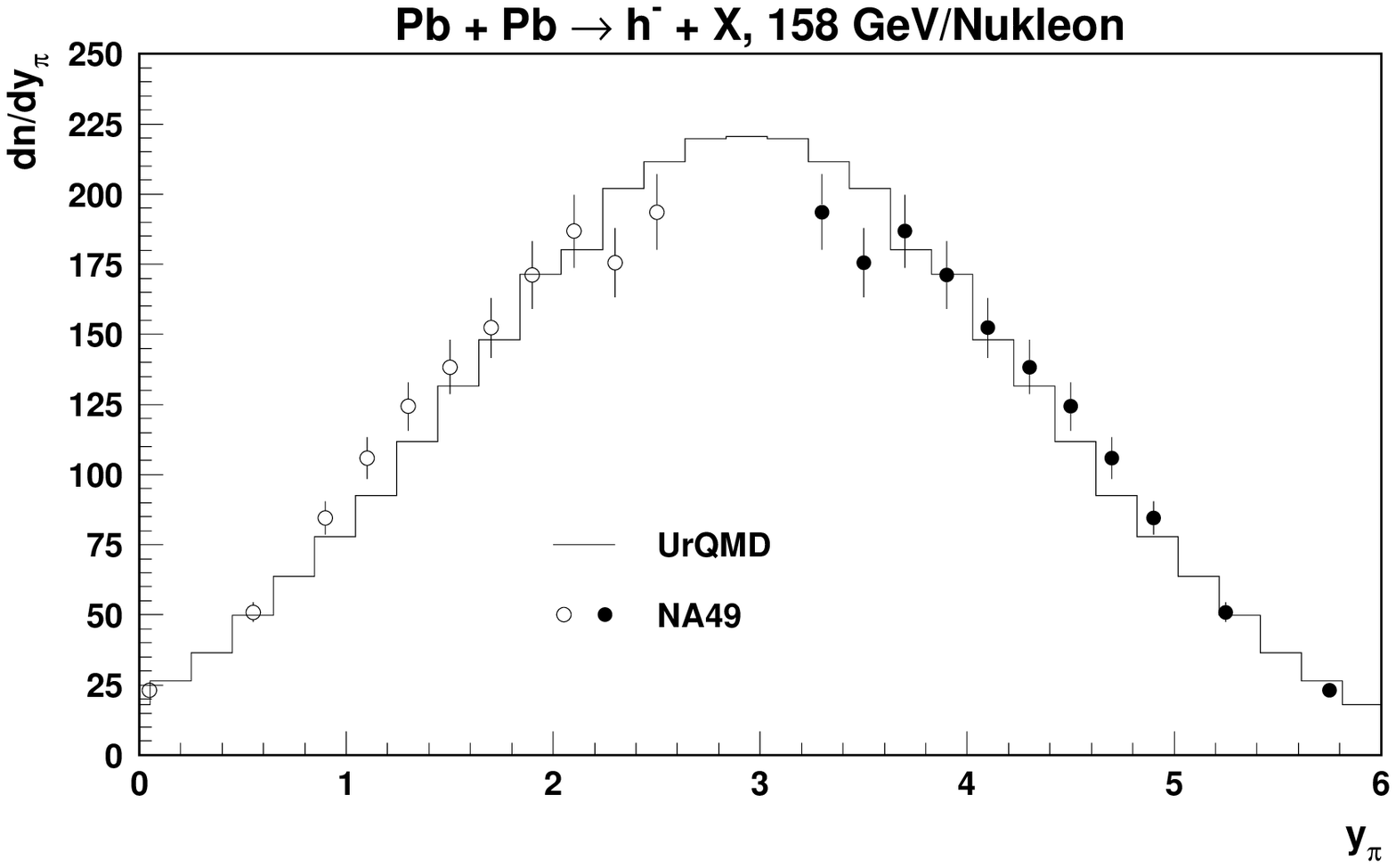,width=9.0cm}}
\caption{\label{hm}Rapidity distributions of negatively charged hadrons (i.e.
$\pi^-$, $K^-$ and $\overline p$)
for Pb+Pb collisions at the SPS (160 AGeV) in comparison to preliminary 
NA49 data \protect\cite{guenther98a}.}
\end{minipage}
\end{figure}

\begin{figure}[htb]
\begin{minipage}[t]{9cm}
\centerline{\epsfig{figure=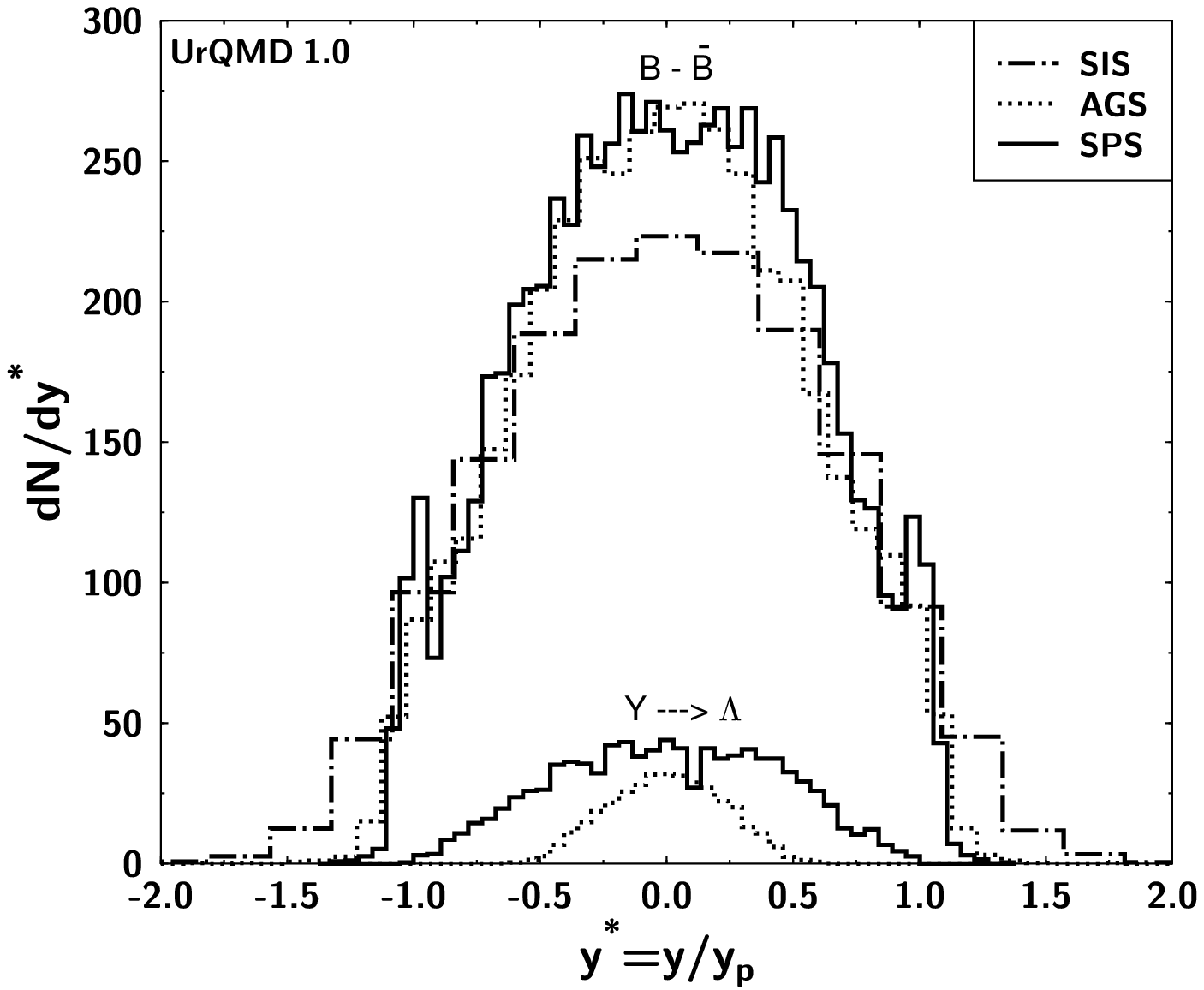,width=9.0cm}}
\caption{\label{uqmd_stopp}Rapidity distributions of net-baryons for Au+Au 
collisions at SIS (1 GeV/nucleon),
AGS (10.6 GeV/nucleon) and CERN/SPS energies (160 GeV/nucleon) 
calculated with the UrQMD model (higher curves). All 
distributions have been normalized to the projectile rapidity in the 
center of mass frame. We have also plotted rapidity distributions of strange
baryons decaying into $\Lambda$ (lower curves).}
\end{minipage}
\hfill
\begin{minipage}[t]{9cm}
\centerline{\epsfig{figure=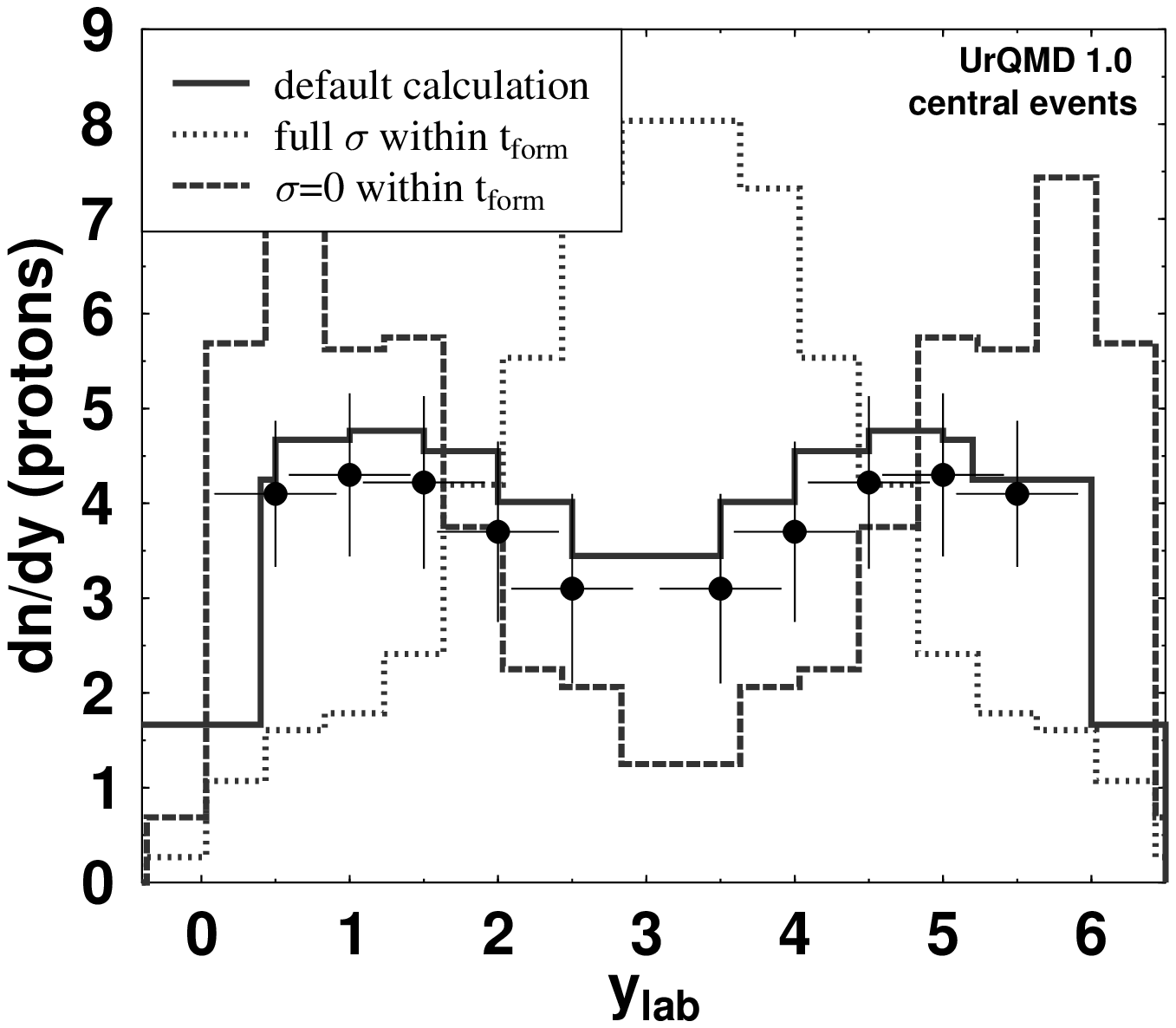,width=9.0cm}}
\caption{\label{fig:dndyform}
Formation time dependence of stopping power for the reaction S+S at 200
GeV/nucleon. Solid dots represent experimental data from \protect\cite{baechler94a}.
 The cross sections of particles
within their formation time introduce large uncertainties for their rapidity
distribution.}
\end{minipage}
\end{figure}

Hadronic transport models are well able to predict or reproduce the
measured rapidity distributions over a vast energy range, if baryon and meson
rescattering and particle production via string decay 
\cite{pang92a,keitz91a,sorge92a,bass96b} are incorporated. 
The VENUS model including possible formation and decay of
multi--quark clusters \cite{aichelin93a} seems to give similar results. 
A simple ``first collision model'' without rescattering 
\cite{werner93a,andersson87a,andersson87b,sjoestrand94a} does not suffice
to reproduce the data. However, it has been shown recently that a simple 
geometrical model of nucleus-nucleus collisions (LEXUS)
\cite{jeon97a} describes rather well the rapidity and transverse momentum
distributions of protons at SPS energies. It is based on linear
extrapolations of nucleon-nucleon collisions, following the concepts of
Glauber \cite{glauber59a} and the rows-on-rows model of H\"ufner and Knoll
\cite{huefner77a}. 
LEXUS ignores rescattering of secondary particles or any other
collective effect completely. This is visible in the resulting numbers
of produced (anti-)strange particles which differ strongly from experimental values.
Similar in spirit are investigations, where the final state of
nucleus-nucleus collisions is described as a superposition of isotropically
decaying fireballs \cite{leonidov97a}. 
Here, the longitudinal motion is extrapolated
from p+p data while the transverse motion is due to random walk collisions,
fitted to p+A reactions. Like in \cite{jeon97a} the model is designed to serve
as a baseline for further analyses. Unfortunately some genuine collective
effects which arise in AA collisions may counteract themselves, leaving the
discussed observables unchanged.
With respect to the net-baryon distribution, these studies show that 
gross features alone are not necessarily decisive signatures of
new physics.

Let us first explore the stopping behavior of UrQMD in central Pb+Pb
collisions at the SPS (160 AGeV) as depicted in Fig. \ref{pmm} in comparison to recently
measured data by the NA49 collaboration \cite{guenther98a}. Overall good
agreement is observed. A large amount of longitudinal momentum has been
deposited, leading to excessive particle production at central rapidities in
line with preliminary data from NA49 \cite{guenther98a} (cf. Fig. \ref{hm}).

Figure \ref{uqmd_stopp} shows a UrQMD calculation
of the proton rapidity distribution for three presently used
heavy ion accelerators. In all cases, a system as heavy as Au+Au or Pb+Pb 
exhibits a central pile-up at mid-rapidity. However,
the physical processes associated are different: The average longitudinal
momentum loss in the SIS energy regime is mainly due to the creation of 
transverse momentum whereas at the AGS/SPS energies abundant particle 
production eats up a considerable amount of the incident beam energy.
The form of the distributions change from a Gaussian at SIS energies to
a plateau at AGS and finally to a slight two-bump structure at CERN energies. 

At CERN/SPS energies baryon stopping is dominated not only by rescattering
effects but also by the formation time after hard collisions in which strings 
are excited. Within their formation time, baryons originating from a 
leading (di)quark interact with (2/3) 1/3 of their free cross section  and 
mesons with 1/2 of their free cross sections.
The influence of the formation time
is shown in figure~\ref{fig:dndyform}
for the system S+S at 200 GeV/nucleon. The default calculation (including
formation time) reproduces the data \cite{baechler94a} 
fairly well whereas the calculation
with zero formation time (dotted line) exhibits total stopping.
A calculation with zero cross section within the formation time, however,
exhibits transparency.

\begin{figure}[htb]
\begin{minipage}[t]{9cm}
\centerline{\epsfig{figure=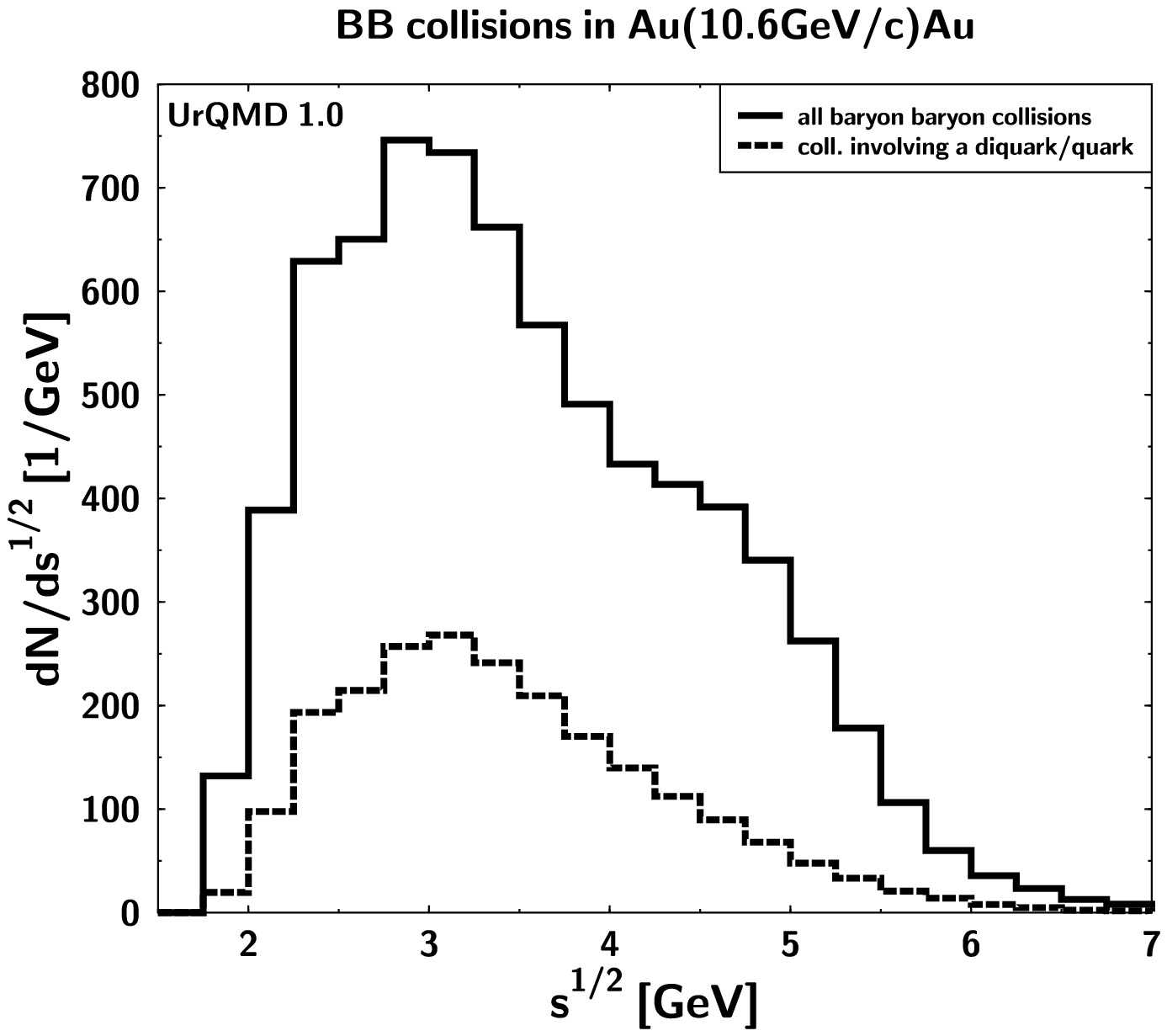,width=9.0cm}}
\caption{\label{fig:srtags} $E^{coll}_{CM}$ distribution for 
baryon baryon collisions in a central Au+Au reaction at the AGS. }
\end{minipage}
\hfill
\begin{minipage}[t]{9cm}
\centerline{\epsfig{figure=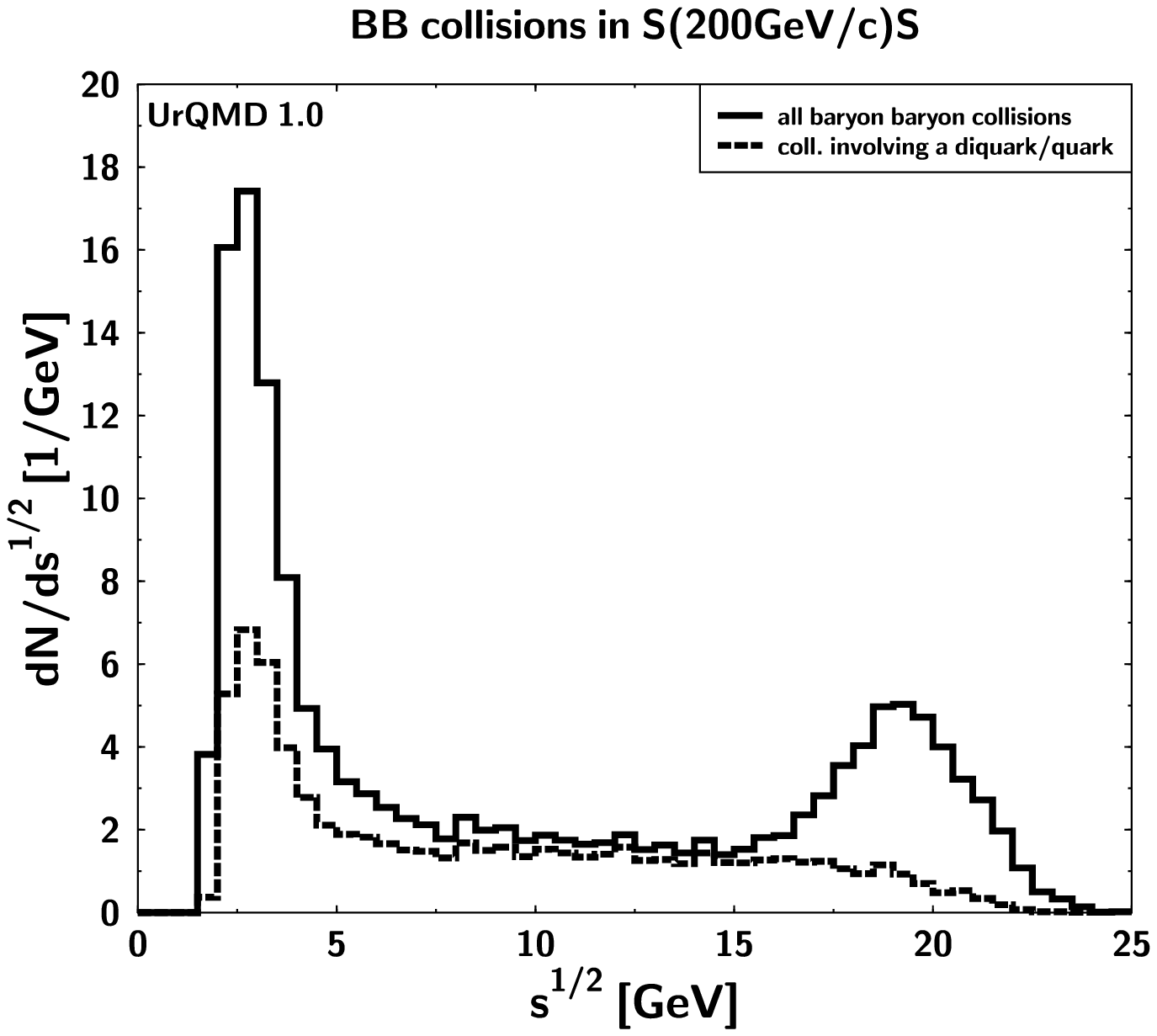,width=9.0cm}}
\caption{\label{fig:srtsps} $E^{coll}_{CM}$ distribution for baryon baryon 
collisions in a central S+S reaction at SPS energies. }
\end{minipage}
\end{figure}

In order to study this effect more closely, the $\sqrt{s}$ distributions
of elementary baryon-baryon collisions  are 
analyzed for Au+Au collisions at AGS and S+S 
collisions at SPS energies. 
Figure~\ref{fig:srtags} shows the respective distribution for 
Au+Au. The collision spectrum is
dominated by BB collisions with full cross sections and exhibits a
maximum at low energies. Approximately 20\% of the collisions
involve a diquark, i.e. a baryon originating from a string decay whose
cross section is reduced to 2/3  of its full cross section.

In figure~\ref{fig:srtsps}, the same analysis is performed for S+S at
200 GeV/nucleon. In contrast to the heavy system at AGS 
the collision spectrum 
exhibits two pronounced peaks dominated by full BB collisions, 
one in the beam energy range and one in the low (thermal) energy  range.
Approximately 50\% of the collisions, most of them at intermediate
$\sqrt{s}$ values, involve baryons stemming from string 
excitations whose
cross sections are reduced by factors of 2/3  (referred to as 
{\em diquarks}) or 1/3  (referred to as {\em quarks}).
The peak at high $\sqrt{s}$ values stems from the initial hard collisions
whereas the peak at low energies is related to the late, thermal stages
of the reaction.

\begin{figure}
\centerline{\epsfig{figure=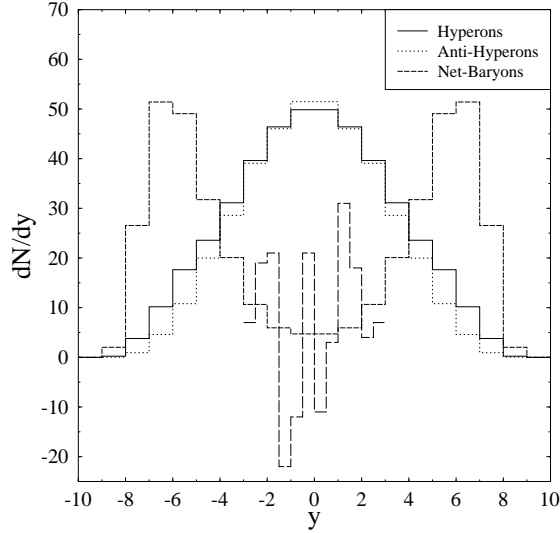,width=9.0cm}}
\caption{\label{dndy}
Event-averaged rapidity density of net-baryons, 
hyperons and anti-hyperons 
in collisions of 
$Pb+Pb$ at 6.5~TeV and $b=0$, calculated with Fritiof~7.02. For the
mid-rapidity range, the net-baryon distribution of one single event is also
depicted (figure taken from \cite{spieles96b}).}
\end{figure}

Figure~\ref{dndy} shows the importance of fluctuations of the net-baryon
distribution in a single event for highest energies. Although the average
net-baryon ${\rm d}N/{\rm d}y$ is less than 5 at mid-rapidity, regions of
high positive or negative net-baryon density occur stochastically in a limited
region of phase space.

%

	\section{Particle production}
		\subsection{Subthreshold particle production from resonance
		matter}
	\label{subthreshold}

Subthreshold particle production is 
kinematically forbidden in first nu\-cle\-on-nu\-cle\-on collisions.  
Therefore  it provides a good tool to extract information about 
collective effects and the high density phase.

As an example, we consider in the following the subthreshold production of
antiprotons: The process has first been 
observed experimentally for Si+Si at BEVALAC \cite{carroll89a}. 
Experimental studies have also been done recently with a
$2.0$~GeV/u and $1.8$~GeV/u Ne beam from the heavy ion synchrotron
(SIS) at GSI \cite{schroeter93a,schroeter94a}. 

Since the antiproton yields cannot be described in terms of
a first collision model \cite{shor90a}, 
various
mechanisms of antibaryon enrichment have been proposed,
e.g. hadronic multi-step
processes \cite{kochP89a,jahns92a,jahns92b,jahns93a}, strong mean 
field effects \cite{schaffner91a} and meson-meson interaction like 
$\rho\rho\rightarrow p\bar p$ \cite{koCM89a}.

Subthreshold antiproton production has been investigated with respect to
a reduction of the effective nucleon mass in the framework of
RBUU \cite{teis94a,sibirtsev97a}, QMD \cite{batko94a} and RVUU \cite{liGQ94a}. 
The important contribution of high lying baryon resonances without 
consideration of in-medium masses has also been studied\cite{spieles93a}.
This we will discuss in the following.
Note that the threshold behavior of $\bar{p}$-production 
in $NN$-reactions for
$\sqrt{s}-4m>1$~GeV is not experimentally determined, so any
model calculation depends on extrapolations.

Schematically there are three different energy regions in hadron-hadron
interactions: the (quasi-)\-elas\-tic scattering, the region of resonance
production and formation and the high energy region characterized by
abundant production of particles. In the resonance region the kinetic
energy of the reaction partners is sufficient to form excited states
of the ingoing hadrons which subsequently decay again into the stable
hadrons. The most prominent examples are
\begin{equation}
\begin{array}{ccc}
\pi N &\longrightarrow &\Delta_{1232} \\
\pi \pi & \longrightarrow & \varrho (770) \\
N N & \longrightarrow & N \Delta_{1232}
\end{array}
\end{equation}
\\
The formation of resonances influences the particle production rates
and their momentum distributions, e.g. the $p_t$ spectra. The particle
chemistry may change, too, distorting signatures from earlier
reaction stages, e.g. due to reactions like
\begin{equation}
\begin{array}{ccccc}
\pi Y & \longleftrightarrow & Y^{\star} & \longleftrightarrow & \bar{K}N \\
\pi \pi & \longleftrightarrow & f_0(980) &
\longleftrightarrow & K \bar{K}
\end{array}
\end{equation}
Any of these secondary interactions can be of utmost importance  
for particle production as discussed in 
\cite{sorge92a,jahns92a,jahns92b,mattiello89a},
the dominant  process being
the annihilation of produced mesons on
baryons which leads to the formation
of $s$ channel
resonances or strings.
Those resonances formed are not only
responsible for strangeness enrichment at the AGS
\cite{mattiello89a,sorge91a},
but they may be further excited in a subsequent collision
to a mass larger than 3$m_N$ which allows for
$\bar{N}N$ creation.

If the projectile nucleon gets excited in the first collision, but
does not lose momentum for target excitation
(which may happen in a subset of cases),
a further excitation in a subsequent collision
then may help to bring the projectile beyond the $\bar{B}$ creation
threshold.

On the one hand, the larger the mass of the incoming hadrons, 
the better the chance to
overcome the
$\bar{B}$ creation
threshold in a subsequent collision, because the second step
of excitation can be smaller.
On the other hand, the life time goes down with increasing mass
reducing the chance to hit another target nucleon.
Therefore the multi-step process of $\bar{p}$ production has to be viewed
as a complicated interplay between excitation and decay
processes.

Calculations in the framework of RQMD \cite{spieles94a} exhibit a scenario 
of cumulative
scattering of participants and secondaries, a multi-step process of
successive binary collisions like $\pi N$, $\rho N$, $\eta N^\ast$, etc, 
as well as $\Delta \Delta$, etc.  Baryons and mesons get more and more
excited and decay or rescatter subsequently.  Possibly, the invariant
mass of a single resonance-resonance collision exceeds the threshold
for antiproton production. The collision spectra of $AA$-reactions 
at 2~GeV/u show
 that meson-baryon-interactions play a decisive role for the
excitation of these high masses.
We do not find evidence for any antiproton production via the meson-meson
channel (especially $\rho\rho$) at SIS-energies.

\begin{figure}
\centerline{\epsfig{figure=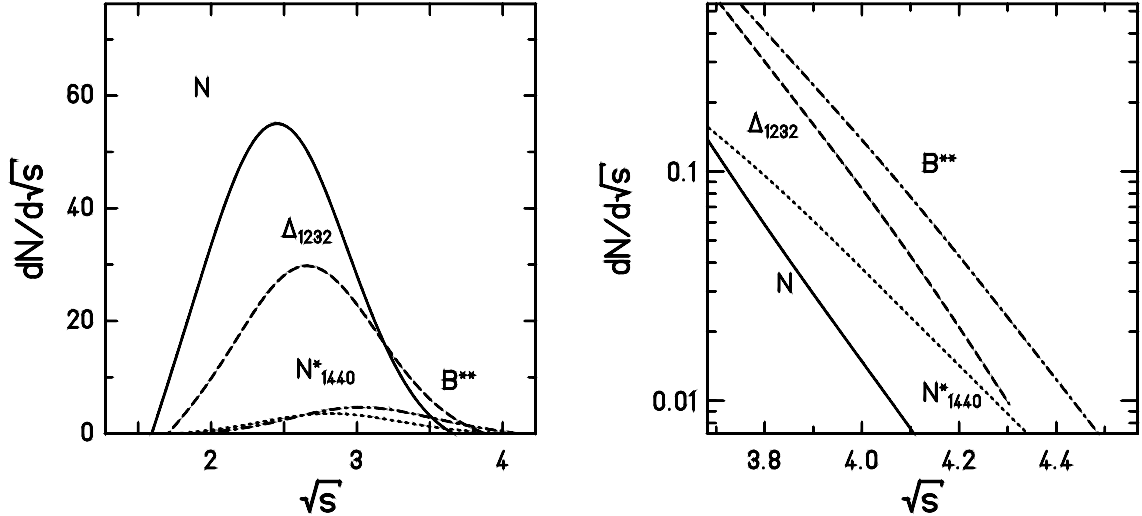,width=15cm}}
\caption[]{
RQMD calculation of the collision spectrum for the reaction
$Ni+Ni$ at 2~GeV/u (min. bias). The contributions are classified
according to the heaviest baryon involved: $N$ (solid line), $\Delta_{1232}$
(dashed line), $N^{\ast}_{1440}$ (dotted line) and all higher resonances
(dashed-dotted line). Shown is the total spectrum (left) and the tails
above the $\bar{p}$-production threshold (right).
\label{collspec}} 
\end{figure}

According to RQMD, BUU and IQMD calculations, the enrichment of
nuclear matter with $\Delta$ resonances is only about $15\%$ for medium
systems at SIS energies \cite{mattiello93a,ehehalt93b,bass94b}, higher
resonances are even less populated. Nevertheless, they turn out to
be most important for the production of heavy particles, since the
collisions at high center of mass energy
preferentially involve these highly excited baryons.
Fig. \ref{collspec} shows the baryon-baryon collision spectrum of the
reaction Ni+Ni at 2~GeV/u. 
Above
$\sqrt{s}=4\cdot m_{\rm N}$ only $1/3$ of the $BB$-reactions are $NN$-, $N\Delta_{1232}$- 
or $\Delta_{1232}\Delta_{1232}$-collisions. $50\%$ involve resonances 
higher than $N^*_{1440}$. They act as energy depots and dominate
the $\bar{p}$-production.

In the other microscopic models mentioned above the contribution of 
high mass resonances
are neglected, but the antiproton production is
strongly enhanced due to a reduced effective mass. 
In particular the $\bar{p}$ annihilation probability differs drastically
between the models \cite{spieles93a}. 
This question might be resolved by the measurements of
anti-flow, as was suggested in \cite{jahns94a}.
Thus, different microscopic models give totally incompatible scenarios
( see also section \ref{composite} ).
		\subsection{Temperatures and single particle spectra}

The hot and dense reaction zone consists of incident nucleons and 
produced particles. The {\em fireball} model considers the hadrons
as a gas in thermodynamic equilibrium.
For temperatures above 50 MeV, Fermi-Dirac and Bose-Einstein distribution 
functions for baryons and mesons may 
be approximated by a Maxwell-Boltzmann distribution 
\cite{gosset77a,hagedorn80a}, with the temperature $T$ as parameter.
Equilibration is thought to be visible predominantly in the transverse
degrees of freedom;
therefore, transverse momentum or transverse mass distributions are 
used to extract temperatures.

\begin{figure}[thb]
\centerline{\psfig{figure=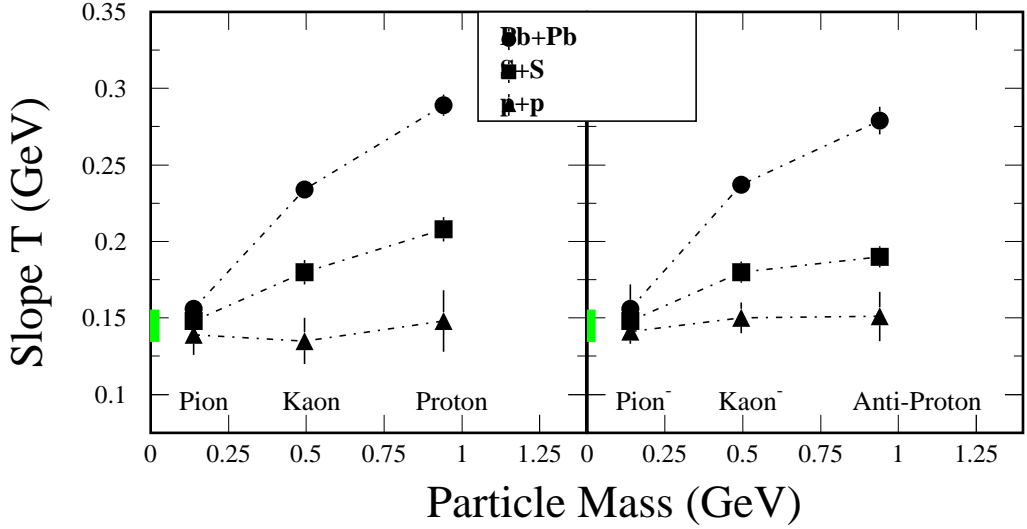,width=16cm}}
\caption{ \label{xufig} Mass and collision system dependence of the
inverse slope parameter $T$  measured by the NA44 collaboration 
\cite{xun96a}. }
\end{figure}

The inverse slope $T$ is obtained from a fit to the transverse mass
$m_T$ spectrum using 
\be
\frac{1}{m_T}\frac{dN}{dm_T}\propto {\rm e}^{m_T/T} \quad.
\ee

Heavy ion collisions at the AGS and, even more at the SPS, exhibit an 
intricate scenario. Here, typical MB collision energies are in the
range of 2-10~GeV in a baryon rich regime. 
Thus two problems have to be addressed:  The influence of 
the modeling of meson baryon scattering as well as the creation of
collective transverse flow (first measured at the SPS from the NA44
collaboration, see Fig. \ref{xufig}) which lead to increasing apparent 
temperatures ''$T$''. To study these aspects, the UrQMD model was used in 
two different modes: 
\begin{enumerate}
\item A complete downward extrapolation of the high energy longitudinal
      color flux picture to $\sqrt s = 2$~GeV (referred to as
      forward-backward peaked (f-b) variant).
\item An upward extrapolation of the resonance behavior, i.e. an 
      isotropic angular distribution of all outgoing particles in MB reactions
      (indicated as (iso)).
\end{enumerate}

\begin{figure}[htb]
\begin{minipage}[t]{9cm}
\centerline{\epsfig{figure=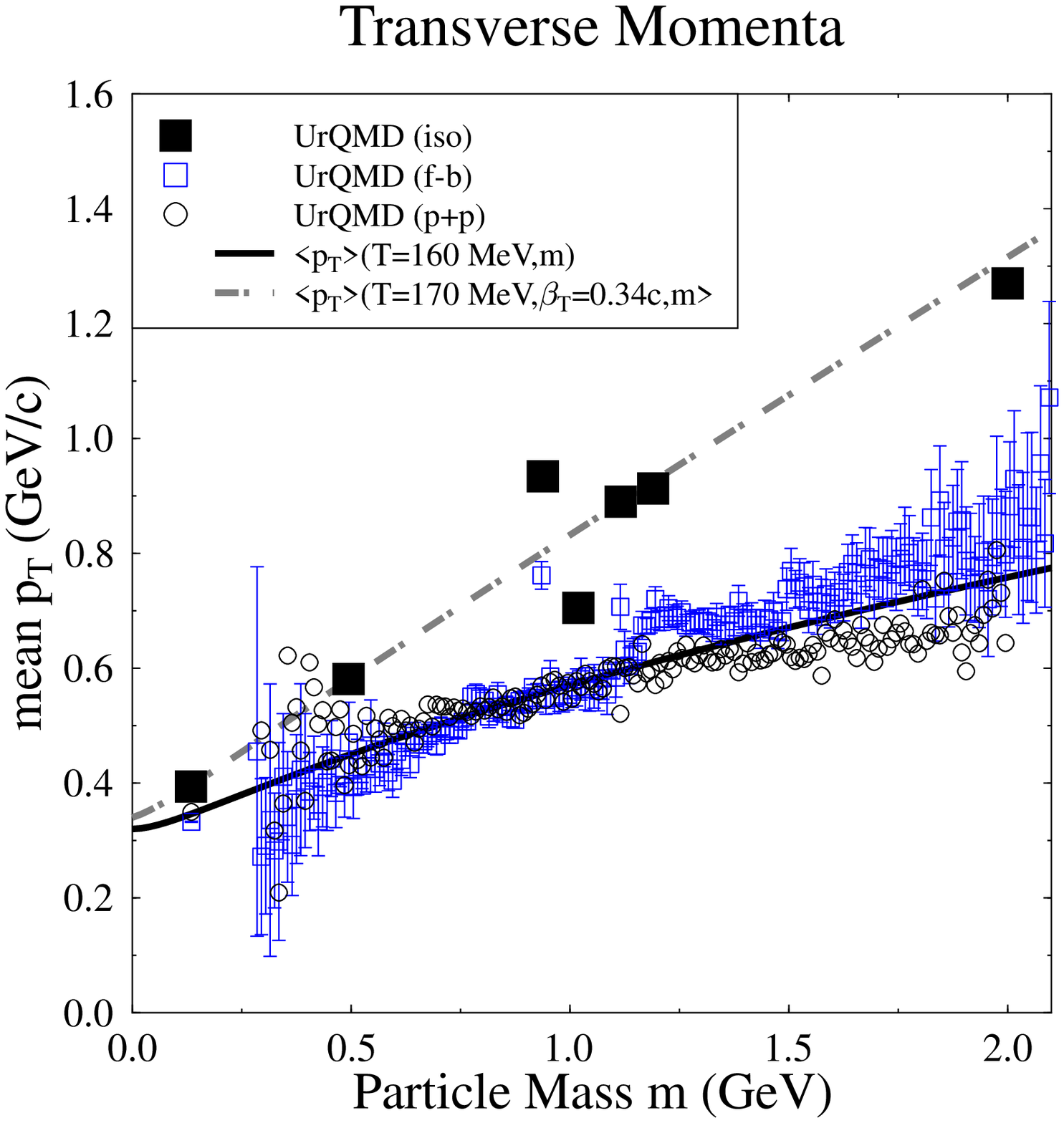,width=9.0cm}}
\caption{\label{mass} Mean $p_t$ at midrapidity ($|y|<0.5$) as a function 
of particle mass
in central Pb(160GeV)+Pb reactions. Figure taken from \protect 
\cite{bleicher98b}.}
\end{minipage}
\hfill
\begin{minipage}[t]{9cm}
\centerline{\epsfig{figure=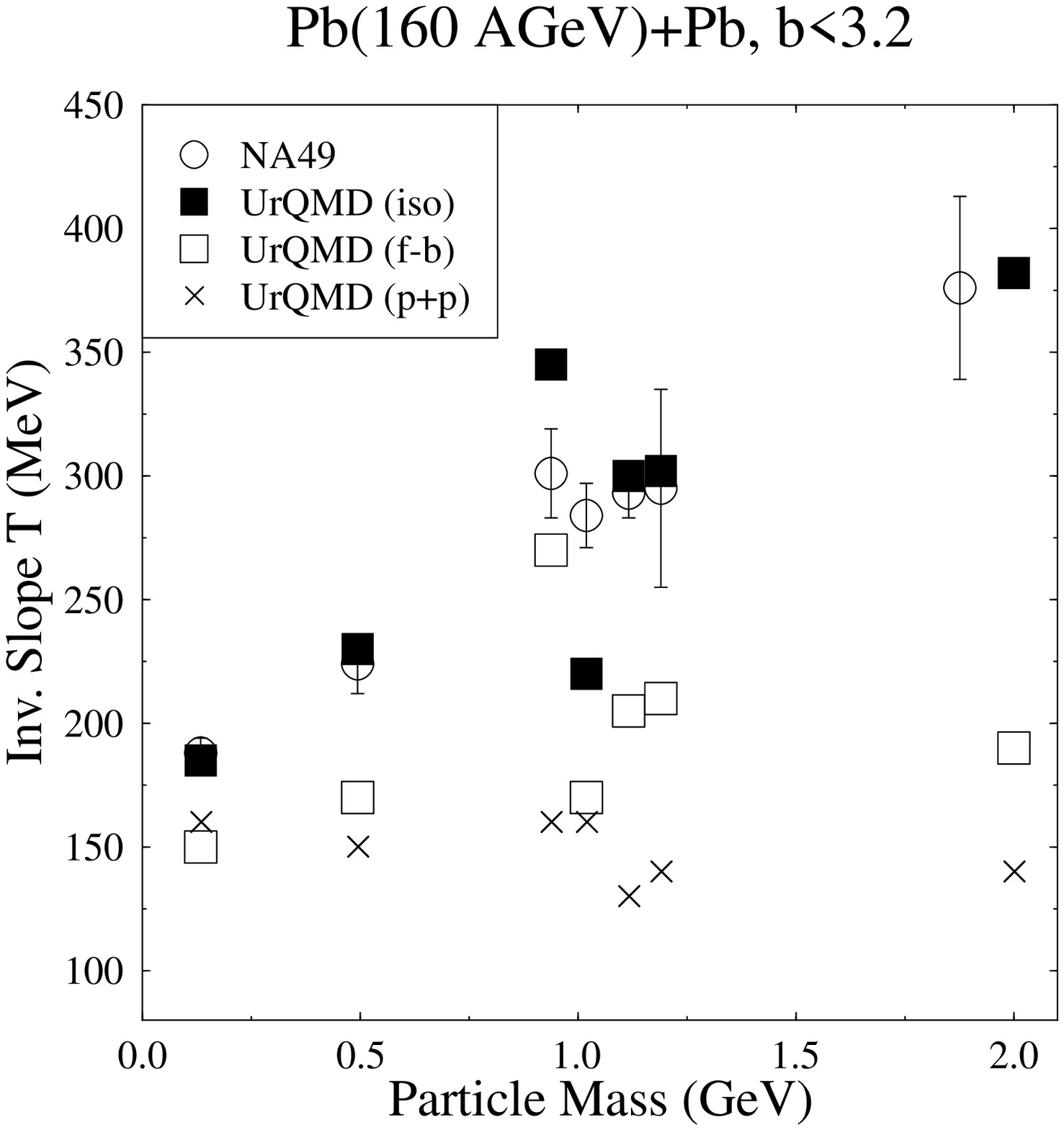,width=9.0cm}}
\caption{\label{slope}Mass and collision system dependence of the
inverse slope parameter $T$ in comparison to measured NA49 data 
\cite{roland97a}.
Figure taken from \protect \cite{bleicher98b}.}
\end{minipage}
\end{figure}

To gain insight into the creation of transverse flow, 
Fig. \ref{mass} discusses mean transverse momenta of particles 
for different mass bins at $|y|<0.5$ in central Pb+Pb collisions 
at the SPS. Circles show the p(160 AGeV)+p events, open squares indicate 
the (f-b) Pb+Pb events and the full black squares show the isotropic
model (iso). In addition, we fit the resulting mass spectra with a simple
fireball plus flow model. The full black line is the mean transverse
momentum $<p_T(m,T_0,\beta_T)>$ for 
different masses $m$ calculated from a thermal distribution with
temperature $T_0=160$~MeV without any additional flow ($\beta_T=0$). 
The grey dashed 
line shows the $<p_T(m,T_0,\beta_T)>$ from an expanding thermal 
source ($T_0=170$~MeV) with an additional transverse flow velocity 
of $\beta_T=0.34c$. This yields in the saddle point approximation a
rough value for the {\em apparent} temperature $T$\cite{leeKS90a}: 
\be
T\approx T_0+m\beta_T^2 \quad.
\ee
The proton proton reactions and the f-b scenario show no
significant difference in mean $p_T$ to a non expanding thermal source.
Only the isotropic model produces a transversally expanding
hadronic source with significant additional transverse flow velocity of
$0.34c$. Note that the statistical error bars are only stated for 
the (f-b) Pb+Pb reaction, in the (iso) and p+p case they are of 
approximately the same order.

This behavior of the (iso) prescription is also seen if we study  
the apparent temperatures (inverse slopes) at midrapidity ($|y|<0.5$) 
as a function of particle mass $m$ as depicted in Fig.\ref{slope}. 
Proton proton collisions (crosses) show about the same freeze-out
temperature for all particles from 130~MeV mass (pions) to 2~GeV mass. 
This is
mostly the same in the forward backward scenario (high energy limit
extrapolated downwards) in Pb+Pb. The inverse slopes increase only
slightly with particle mass. Only the isotropic treatment of 
intermediate energy 
meson baryon collisions in line with the resonance picture
yields an increasing apparent temperature with the particle mass as 
indicated by the preliminary NA49 data\cite{roland97a}.

The mass dependence of the apparent
temperature is sensitive to the detailed modeling of the meson baryon
rescattering process above the resonance region and below the high energy 
domain. Strong transverse expansion only builds up in the 'resonance
model' in contrast to the high energy model which creates much less transverse 
flow.

\begin{figure}[thb]
\centerline{\epsfig{figure=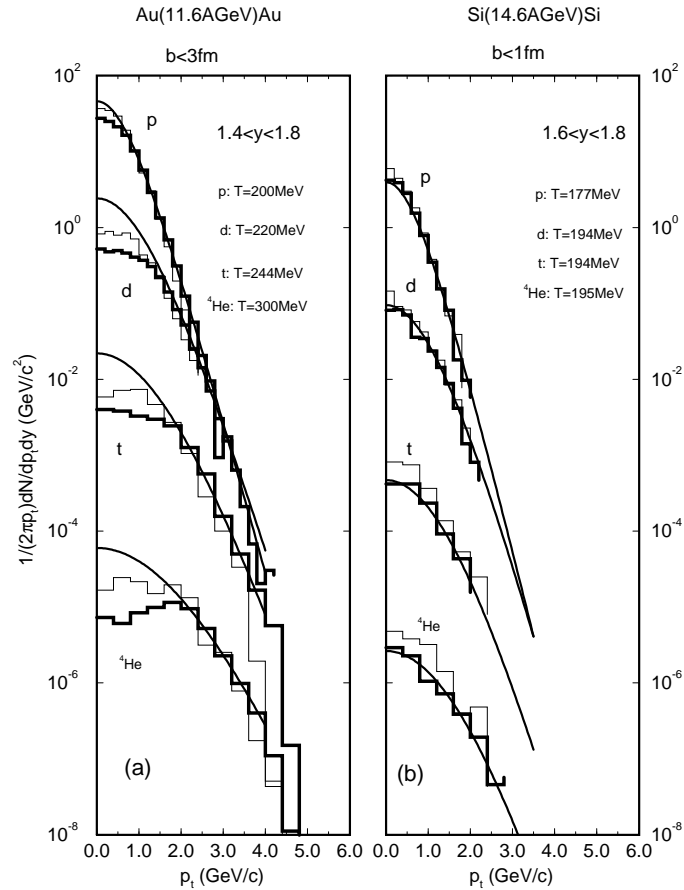,width=10cm}}
\caption{ \label{mattfig1} RQMD prediction of transverse momentum 
spectra at mid-rapidity for protons
and light composite particles in central Si and Au collisions
at the AGS. The bold solid histogram is a calculation with potentials,
the thin histogram the respective calculation without potentials and
the smooth solid line depicts  a Boltzmann parameterization adjusted
to the high momentum part of the spectra. The strong shoulder arm structure
is due to collective flow.
The figure
has been taken from \protect \cite{mattiello96a}.}
\end{figure}

Thus, collective (radial) flow \cite{siemens79a,stoecker81b} as well as 
feeding from  resonances, strongly influences
the shape of particle spectra \cite{sollfrank90a,mattiello95a}.
For light composite particles, the influence of collective flow is
visible in a strong shoulder-arm shape of transverse momentum spectra.
This can be seen in figure \ref{mattfig1}.
In order to account for flow effects, the spectra have to be fitted with
a thermal distribution including collective flow. The temperature $T$ and
the transverse flow velocity $\beta_t$ are the fit-parameters.
The shapes of the flow velocity profile and density profile at freeze out 
enter as additional degrees of freedom in the analysis.
Usually a box shaped density profile and a linearly increasing transverse
velocity profile are assumed.

In microscopic calculations, temperatures and flow velocities can also be
extracted by subdividing the system into cells and analyzing the local
transverse and longitudinal velocity distributions. 
Temperatures extracted via a global  two parameter fit are more than 
a factor of two higher than
temperatures obtained from such a microscopic analysis, at least at beam
energies in the 100 MeV/nucleon to 10 GeV/nucleon regime \cite{konopka96a}.

The reason for this discrepancy lies in the assumed shape of the 
freeze out density and velocity profiles. 
Analyses based on microscopic transport calculations show an extreme
sensitivity of the extracted values of $\beta_t$ and $T$ on the shapes
of the profiles \cite{konopka96a,mattiello96a}. 
Whereas a linearly increasing transverse freeze-out velocity
profile seems a tolerable assumption the shape of the freeze-out density
profile has a rather Gaussian shape (centered at $r_t=0$), far from the
usually assumed box-like distribution.
When using realistic density and velocity profiles one finds the high $m_t$
components of the particle spectra dominated by large collective effects
(i.e. the high expansion velocity). This results in a lower value for 
the temperature $T$.
A microscopic analysis of spectra of protons and light composite particles
at AGS energies showed $\beta_t$ and $T$ depending on the mass of the
particle \cite{mattiello96a}.

\begin{figure}
\begin{minipage}[t]{9cm}
\centerline{\epsfig{figure=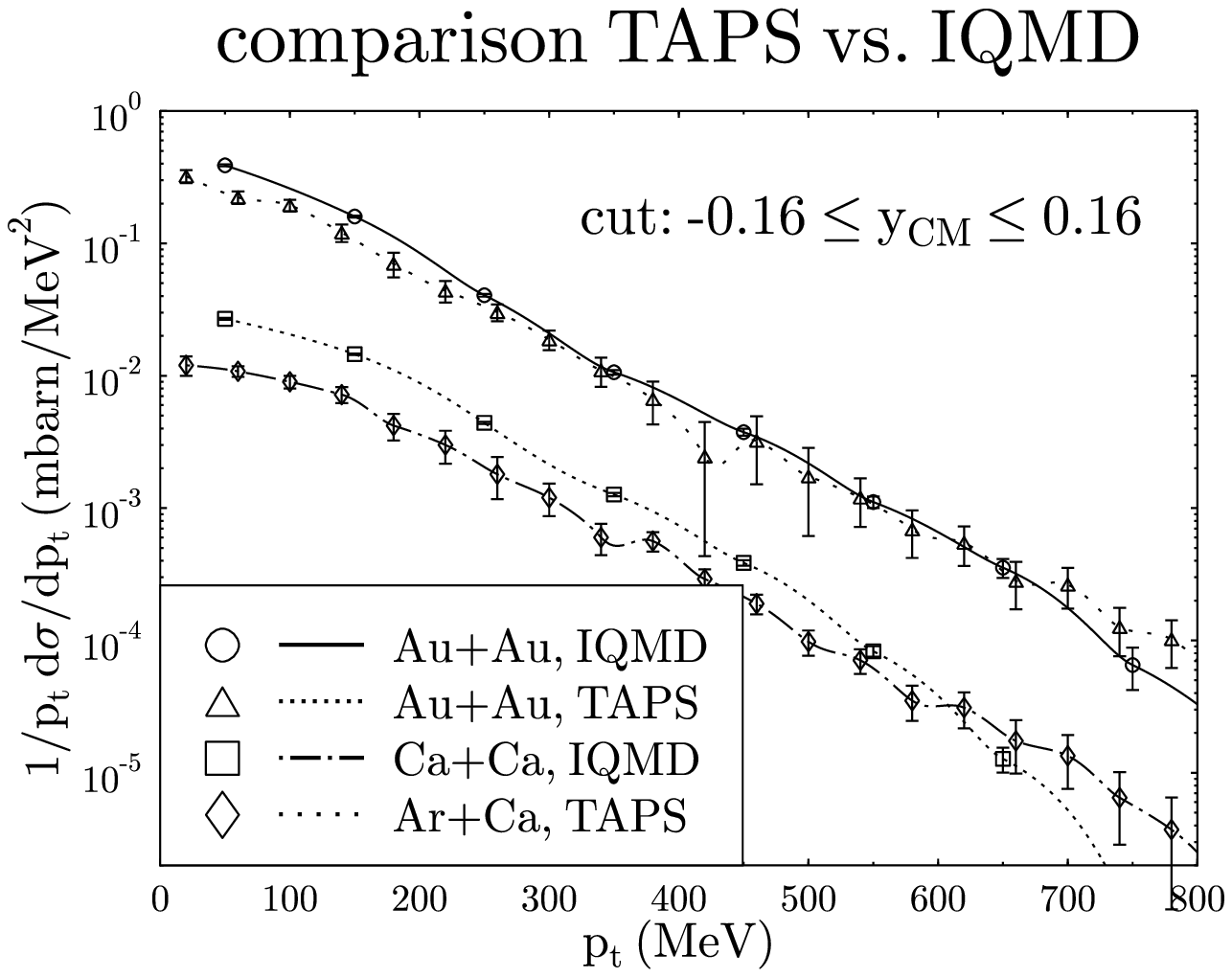,width=9cm}}
\caption{ \label{tapsspek} Comparison of inclusive $\pi^0$ spectra 
$\frac{d\sigma}{p_t dp_t}$ for Au+Au and Ca+Ca (minimum bias) collisions 
between the IQMD model and data measured by the TAPS collaboration. 
A hard EoS including momentum dependence is used. Figure taken
from \cite{bass95c}.}
\end{minipage}
\hfill
\begin{minipage}[t]{9cm}
\centerline{\epsfig{figure=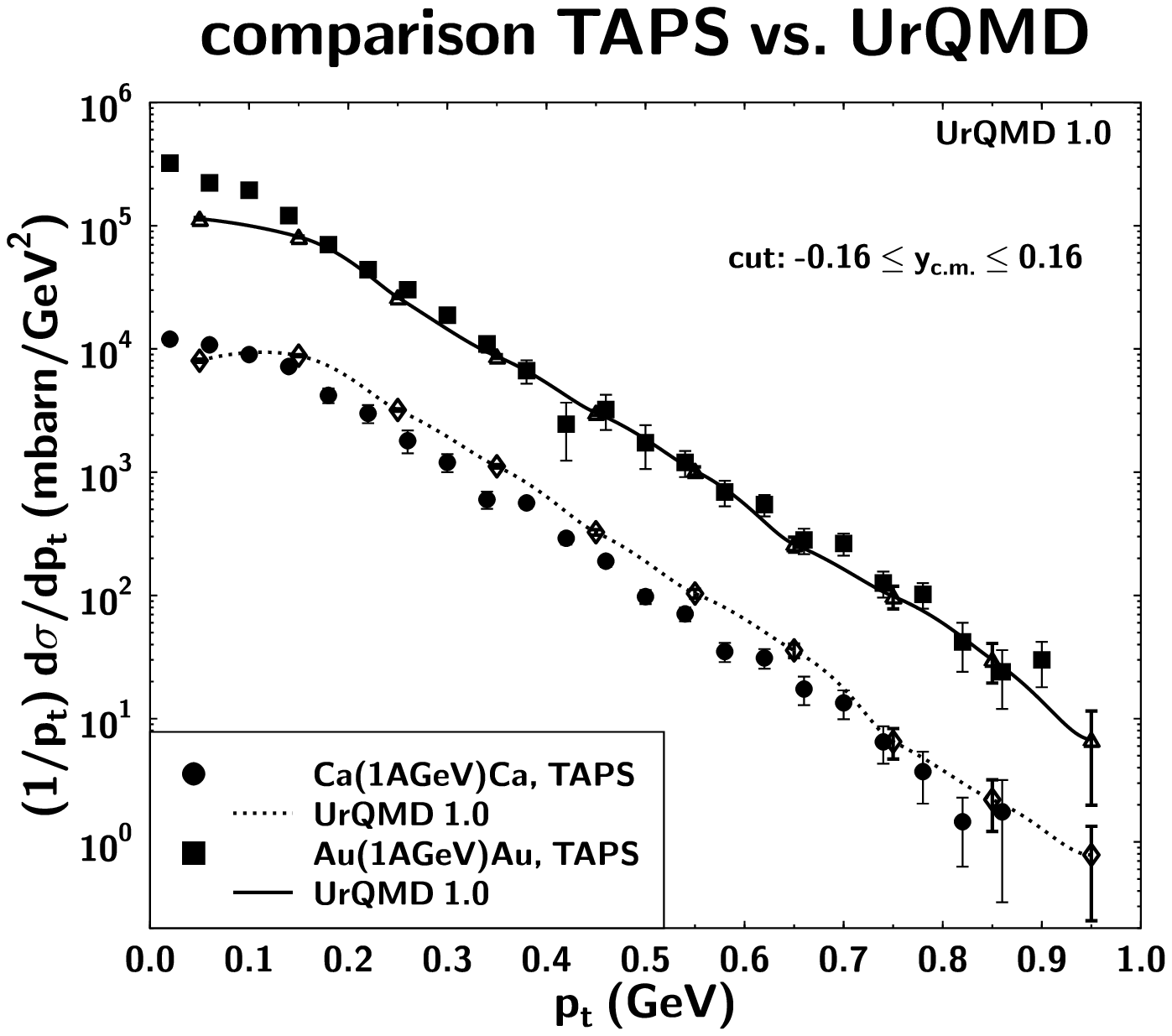,width=9cm}}
\caption{ \label{tapsuqmd} Same analysis as in 
figure~\protect\ref{tapsspek}, but
with the UrQMD model: The influence of higher resonances on the high
energy part of the spectrum is clearly visible.
}
\end{minipage}
\end{figure}

Spectra of light secondary particles, such as pions, at SIS and low
AGS energies are dominated by resonance decays: Here we will use
pion spectra in Au+Au reactions at 1 GeV/nucleon as example
for the importance of resonances. They show how the
shape of the spectra and the origin of the spectra can be explained
in a fully dynamical and non-thermal scenario:

Figure \ref{tapsspek} shows a comparison of inclusive $\pi^0$ spectra
for Au+Au and Ca+Ca (minimum bias and $y_{c.m.}\pm 0.16$) between the IQMD
model \cite{bass95c,hartnack92a}
and data published by the TAPS collaboration \cite{schwalb94a}. 
The data have been rescaled by a factor of 0.6 in oder to obtain a
minimum bias distribution from the measurement. 
The IQMD model overpredicts the yield for
the heavy system Au+Au by approximately 20\% which is consistent
with the findings of the KaoS collaboration for charged pions (a comparison
has been published in ref. \cite{bass95c}). However,
the $\pi^0$ yield of the light system Ca+Ca is overpredicted 
by almost a factor of 2. 
For both systems IQMD calculations fail to reproduce the high transverse
momentum tails.

Newer transport models perform much better when compared to the data
\cite{teis97a,teis97b}. Figure~\ref{tapsuqmd} shows a comparison to the same
data with the UrQMD model. Note that the slopes agree extremely well with
the experiment and the yield for the system Ca+Ca deviates less than
20\% from the data. However, the yield for the heavy Au+Au system is
underpredicted for low transverse momenta. A reason for this may lie
in the strength of the $\Delta_{1232}-N$ absorption cross section, which
is calculated via the principle detailed balance from the 
$N N \to N \Delta_{1232}$ cross section (see figure~\ref{pp-nd1232}).
The UrQMD fit to the  $\Delta_{1232}$ is very good. The absorption
channel, however, diverges for low energies (due to the detailed balance principle) 
and is therefore extremely sensitive to even the minutest changes in the
cross section parameterization. A reparameterization of the exclusive
$\Delta_{1232}$ production cross section (within the constraints given
by the data) may therefore improve the pion yield in heavy collision systems.

The importance of higher baryon resonances can be seen when comparing
the IQMD\footnote{Note that the IQMD model includes only the $\Delta_{1232}$ 
and no other resonances are considered.} 
pion spectrum at 1 GeV/nucleon, figure~\ref{tapsspek}, 
with the respective spectrum
generated by the UrQMD model in figure~\ref{tapsuqmd}.
The enhancement at high $p_t$ in the UrQMD
calculation -- which is due to the incorporation of higher 
baryon resonances \cite{liBA94a} --
is clearly visible and UrQMD agrees with the data even at the highest
momenta.

\begin{figure}[thb]
\begin{minipage}[t]{9cm}
\centerline{\epsfig{figure=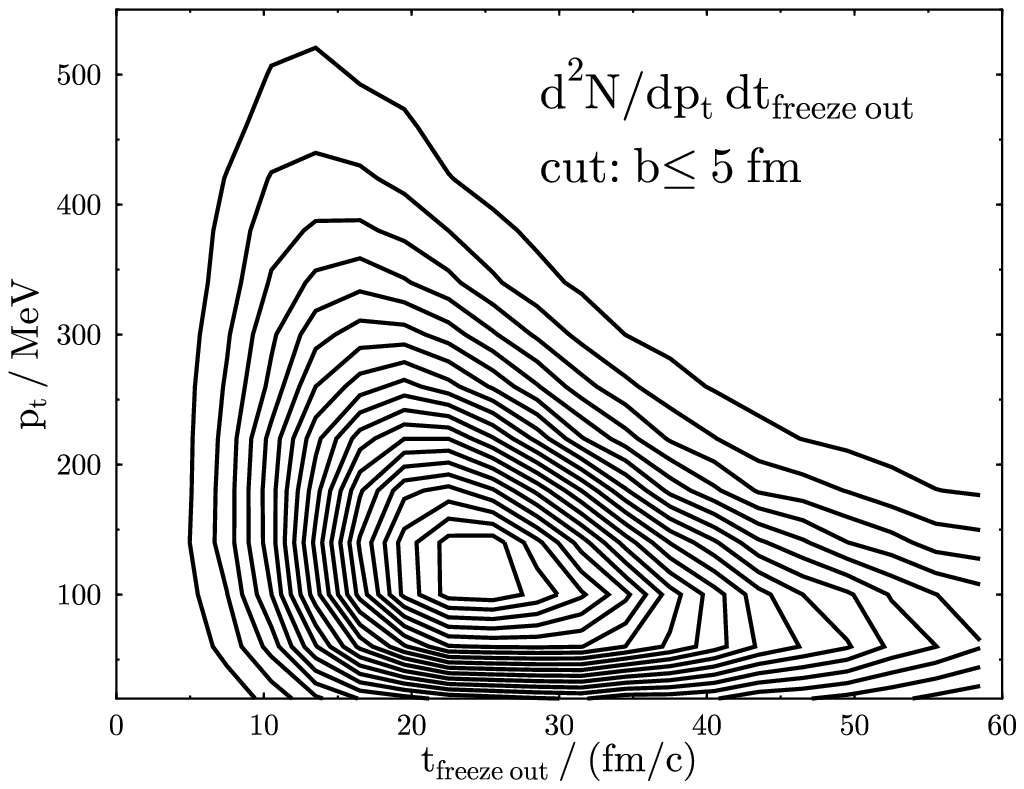,width=9cm}}
\caption{\label{pipttf} Transverse momentum versus freeze out time for pions
in semi-central Au+Au collisions at 1 GeV/nucl. in the IQMD model
\protect\cite{bass94c}. High $p_t$
pions freeze out predominantly in the hot and dense reaction phase.
The scaling of the contour lines is linear; far higher transverse momenta
occur than is depicted by the contour lines.}
\end{minipage}
\hfill
\begin{minipage}[t]{9cm}
\centerline{\epsfig{figure=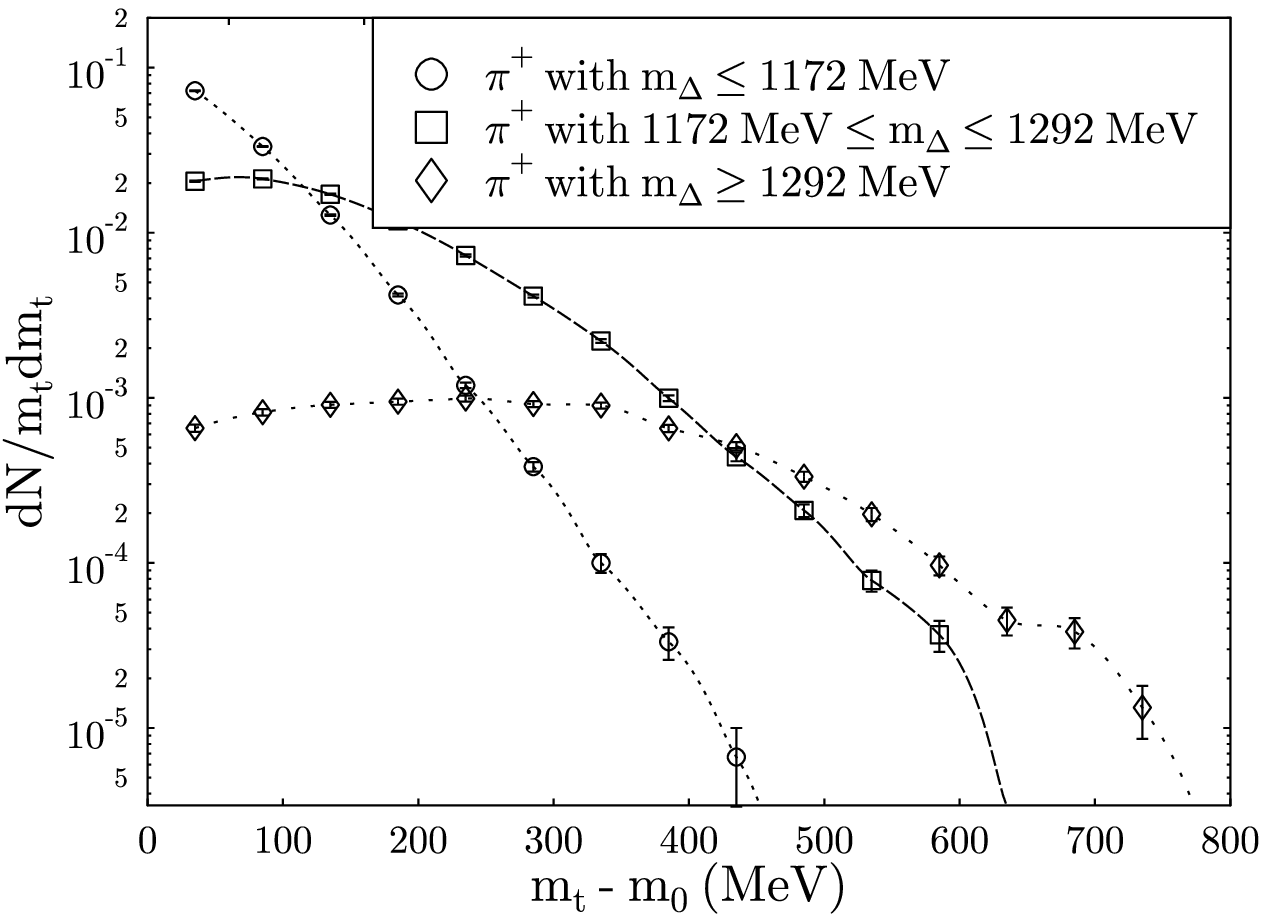,width=9cm}}
\caption{\label{deltadecomp} Transverse mass $m_t$ distribution 
of freeze out $\pi^+$ 
with cuts on the mass of the emitting $\Delta$ resonance. The calculation are
done for Au+Au collisions at 1 GeV/nucleon within the IQMD model.
The connection between high $m_t$ and large resonance mass is obvious: 
The production of freeze out pions with high $m_t$ is
dominated by the decay of heavy $\Delta$ resonances.}
\end{minipage}
\end{figure}

Especially the measurement of those high energy
pions is of great interest. They correlate directly to early freeze out
times and heavy resonances \cite{bass94c}:

Transverse momenta $p_t$ of pions in semi-central Au+Au collisions at 1
GeV/nucleon are shown in
figure \ref{pipttf} as a contour plot versus their freeze out time
(the time of their final interaction in the heavy ion collision): 
High $p_t$ pions are produced almost
exclusively in the early reaction phase with freeze out times smaller
than 20 fm/c, well in the high density phase of the reaction.  
The scaling of the contour lines in figure \ref{pipttf} 
is linear, far higher transverse
momenta are obtained than depicted by the contour lines. 
We can establish a correlation between high $p_t$ pions, early freeze out
times and the hot and dense reaction phase. 
A correlation between  high energy pions and early freeze out
times has also been observed in BUU calculations of La+La collisions at 1.35
GeV/nucleon \cite{liBA91a,liBA91b}. However, the global, fixed event averaged
time cut of 20 fm/c for La+La at 1.35 GeV/nucl. employed 
in ref. \cite{liBA91a,liBA91b} is
a rough estimate of the {\bf nucleonic} freeze out time
(while in our approach the actual time (event by event) for {\bf pion} freeze
out is employed): 
The densities in the collision center are
then already far lower than ground state density. 

The average freeze out density for all pions lies well below nuclear 
ground state
density; this is well known and has severely diminished the usefulness of pions
as probes for the hot and dense reaction phase 
\cite{wolf90a,hartnack88a,berenguer89a}. 
High $p_t$ pions, however, freeze
out at far higher densities: Their average freeze out density lies 
between 1.2 and
1.5 $\rho/\rho_0$ with some of them even freezing out at densities near the 
maximum density obtained in the collision. 

Figure \ref{deltadecomp} shows the transverse mass spectrum of $\pi^+$
in Au+Au collisions at 1 GeV/nucleon within the IQMD model \cite{bass94c}. 
The spectrum
has been decomposed into contributions of light ($m_{\Delta} \le 1172$ MeV),
medium (1172 MeV $\le m_{\Delta} \le 1292$ MeV) and heavy 
($m_{\Delta} \ge 1292$ MeV) $\Delta$ resonances.
The correlation between high $m_t$ pions and heavy resonances is obvious.
Note that since the IQMD model is limited to the $\Delta_{1232}$ resonance
and the inelastic nucleon nucleon cross section is projected 
onto the excitation of this particular resonance
this statement has to be interpreted in a more general manner: at high
transverse masses heavy resonances 
such as the $N^*_{1440}, N^*_{1520}$ and $N^*_{1535}$ should be substituted
for the heavy $\Delta(1232)$ resonances of the IQMD model and contribute
strongly to the transverse momentum or mass spectrum. 

		\subsection{Hadrochemistry: multiplicities and ratios}

\begin{figure}[thb]
\begin{minipage}[t]{9cm}
\centerline{\epsfig{figure=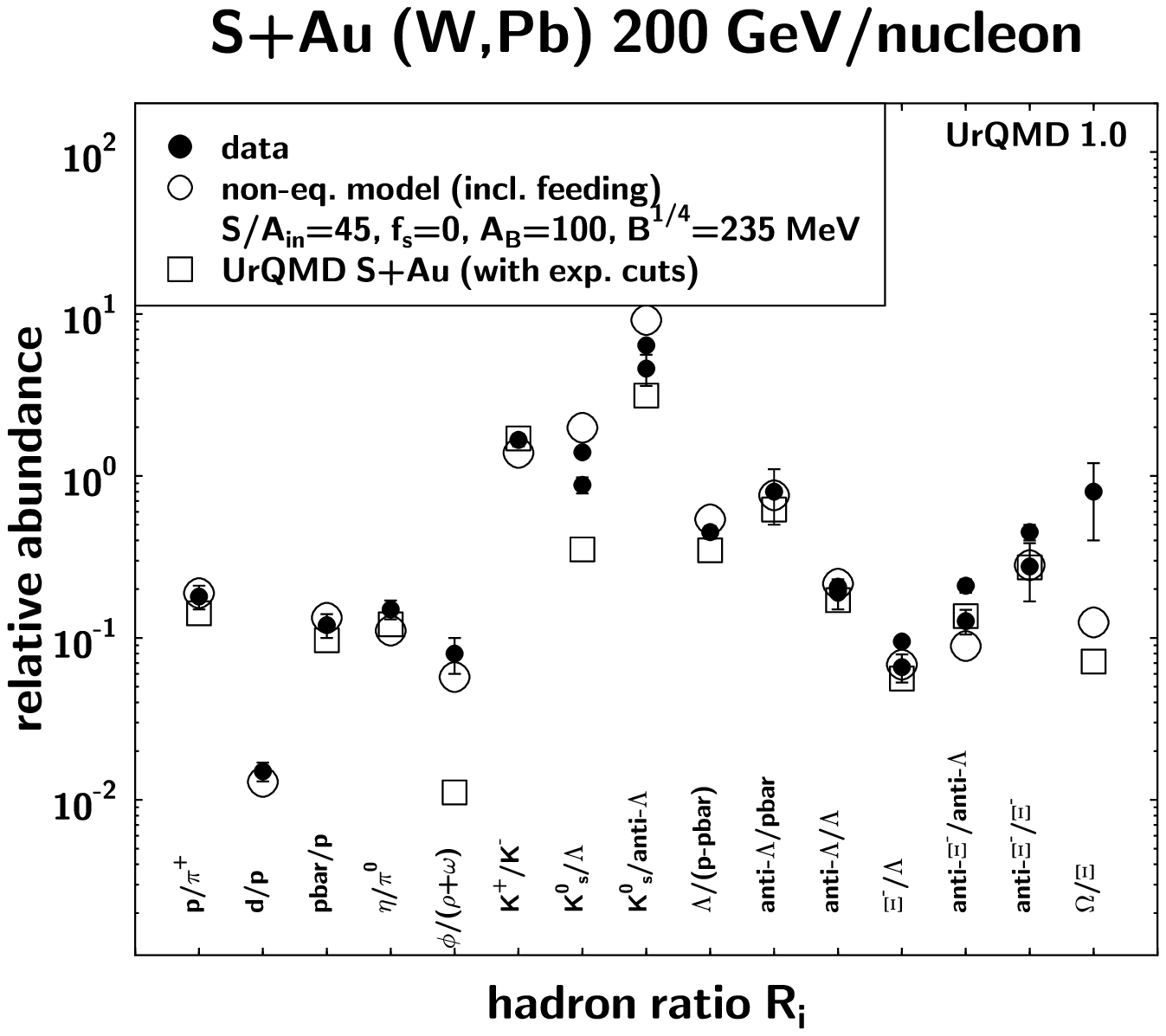,width=9cm}}
\caption{ \label{ratiosfig}
comparison between the UrQMD model
and data for the system S+Au(W,Pb) at 200 GeV/nucleon. Also shown
is a fit by a simple hadronization model \protect\cite{spieles97a}. 
Both models agree quite well with the data compilation 
\protect\cite{braun-munzinger96a}. The original data are from
\cite{murray94a,gazdzicki95a,mitchell94a,iyono92a,simon95a,masera95a,
dibari95a,alber94a,roehrich94a,guenther95a,abatzis94a,andersen94a}}.
\end{minipage}
\hfill
\begin{minipage}[t]{9cm}
\centerline{\epsfig{figure=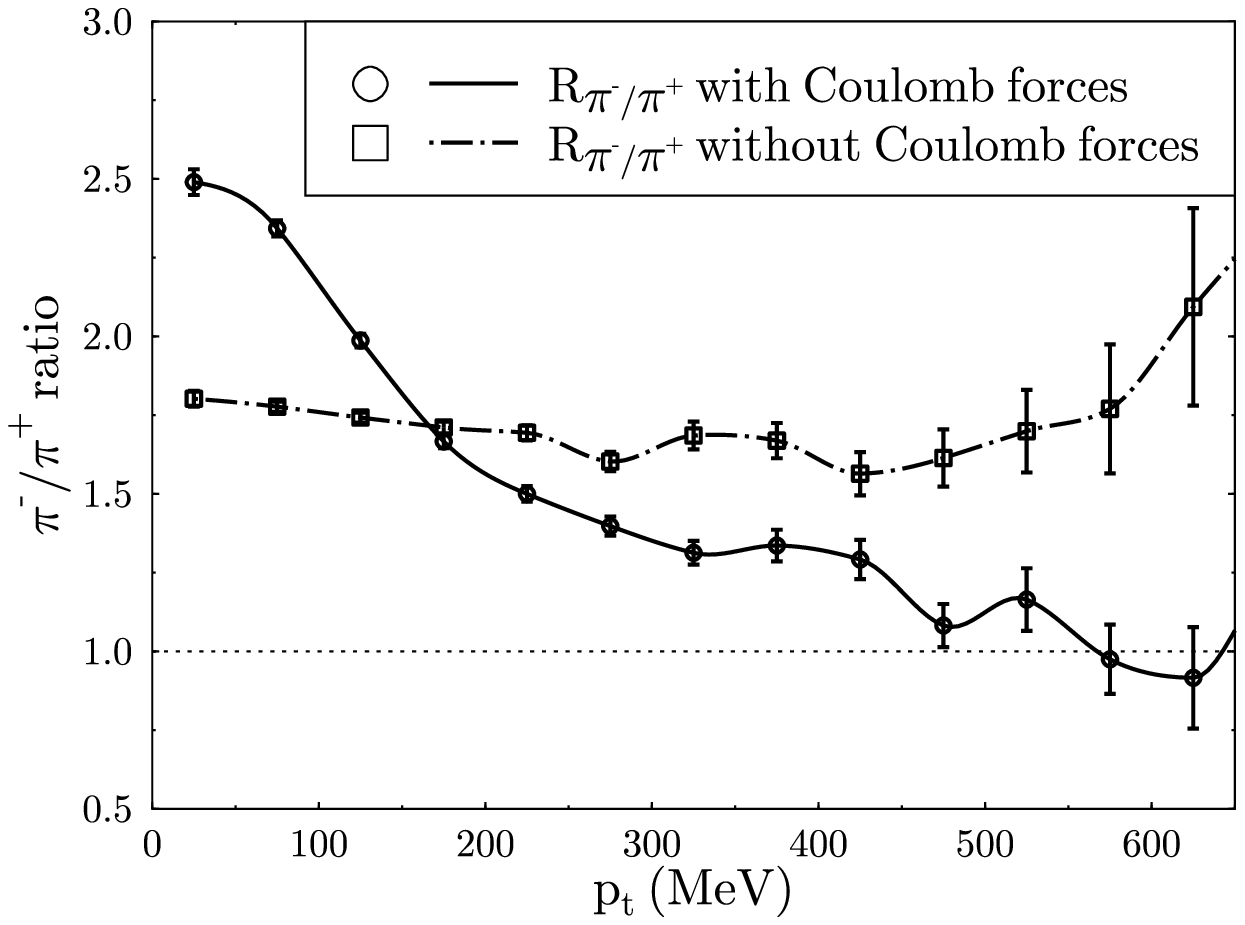,width=9cm}}
\caption{ \label{picoulomb} IQMD prediction for the $\pi^-$ to $\pi^+$ 
ratio versus transverse momentum $p_t$ in Au+Au 
collisions (minimum bias) at 1 GeV/nucl. with a hard EoS and momentum
dependence \protect\cite{bass95c}. The solid line
shows the full calculation including Coulomb forces. 
The dashed-dotted line shows a calculation  without Coulomb forces
acting upon the pions.}
\end{minipage}
\end{figure}

Hadron abundances and ratios
have been suggested as possible signatures  
for exotic states and phase transitions in dense
nuclear matter.
In addition they have been applied
to study the degree of chemical equilibration in a relativistic
heavy-ion reaction.
Bulk properties like temperatures, entropies and chemical potentials
of highly excited hadronic matter have been extracted
assuming thermal and chemical equilibrium
\cite{stoecker78a,stoecker86a,becattini97a,spieles97a,hahn88a,
hahn86a,stock86a,rafelski82a,rafelski82b,letessier93a,letessier94a,
sollfrank94a,cleymans93a,braun-munzinger95a,braun-munzinger96a}.
Recent SPS data on hadron yields and ratios have been fit
either in the framework of a hadronizing QGP droplet 
\cite{spieles97a,barz88a,barz90a,barz91a}
or of a hadron gas in
thermal and chemical equilibrium \cite{braun-munzinger96a}
(including elementary proton-proton interactions \cite{becattini97a}).
It has been shown that the thermodynamic parameters $T$ and $\mu_B$
imply that these systems have been either very close to 
or even above the critical
$T$, $\mu_B$ line for QGP formation \cite{braun-munzinger96b,spieles97a}.
However, a UrQMD calculation \cite{bass97b} agrees with many of the data
($\pi/p$, $d/p$, $\bar p/p$, $\bar \Lambda/\Lambda$, 
$\bar \Xi/ \bar \Lambda$ etc.) without assuming
thermal and chemical equilibrium. Large discrepancies
to the data ($> 50$ \%) are only found for the
$K^0_S/\Lambda$ and $\Omega/\Xi$  ratios (see figure \ref{ratiosfig}).
The microscopic transport model calculation shows that secondary 
interactions significantly change  the primordial 
hadronic cocktail of the system. The ratios exhibit a strong dependence
on rapidity \cite{bass97b}.

The assumption of global thermal and chemical equilibrium -- as being
employed by the static thermal models -- is not
justified:
Both, the discovery of directed collective flow of baryons and 
anti-flow of mesons 
in Au+Au reactions at 10.6 GeV/nucleon \cite{barrette94c} 
and Pb+Pb reactions at 160 GeV/nucleon
energies \cite{peitzmann97a} (see also section~\ref{flow})
as well as transport model analyses presented in section~\ref{freezeout}, 
which show distinctly different
freeze-out times and radii for different hadron species,
indicate that
the yields and ratios result from a complex non-equilibrium
time evolution of the hadronic system. 
A thermal model fit may therefore not seem meaningful.

On the other hand, particle ratios also contain valuable information about the
time-evolution and the dynamics of the collision system:
The inclusion of Coulomb forces and energy dependent
cross sections is also important in the domain of particle production. 
Figure \ref{picoulomb} demonstrates this 
by showing the IQMD prediction \cite{bass95c} for the  $\pi^-$ to $\pi^+$
ratio versus transverse momentum $p_t$ for Au+Au at 1 GeV/nucleon.
The solid line
shows a full calculation including Coulomb forces. For high $p_t$ the
$\pi^-/\pi^+$ ratio decreases towards 1, whereas for low $p_t$ 
it increases to 2.5 -- 
considerably higher than the value of 1.9 predicted by 
the $\Delta$-isobar model.  
The dashed line shows a calculation  without Coulomb forces.
This ratio remains constant at $\approx 1.9$.
The (small) remaining variations might be
due to the different energy dependence of the $\pi^+ - p$ and $\pi^- - p$
inelastic cross sections. The IQMD prediction is compatible to
recent measurements by the KaoS collaboration \cite{wagner97a}. A detailed
comparison of the data to the UrQMD model has yet to be performed.

Note that the $p_t$-integrated values agree very well with
the $\Delta$-isobar model and thus hide any further information possibly
contained in the $p_t$ dependent ratio.
The $\Delta$-isobar model \cite{barshay74a} 
was originally developed to describe p-wave
pion-nucleon scattering and is based on considering the $\Delta_{1232}$ as
an independent baryon species. Here it is applied in a statistical limit,
i.e. it is the basis for the assumption that all pions are produced via
the $\Delta_{1232}$ and the resulting multiplicity ratios only depend
on the number of incident protons and neutrons and can be calculated via
combinatorics applying the respective Clebsch-Gordan coefficients for the 
$\Delta_{1232}$ excitation and the subsequent decay into pions.

\begin{figure}[thb]
\centerline{\epsfig{figure=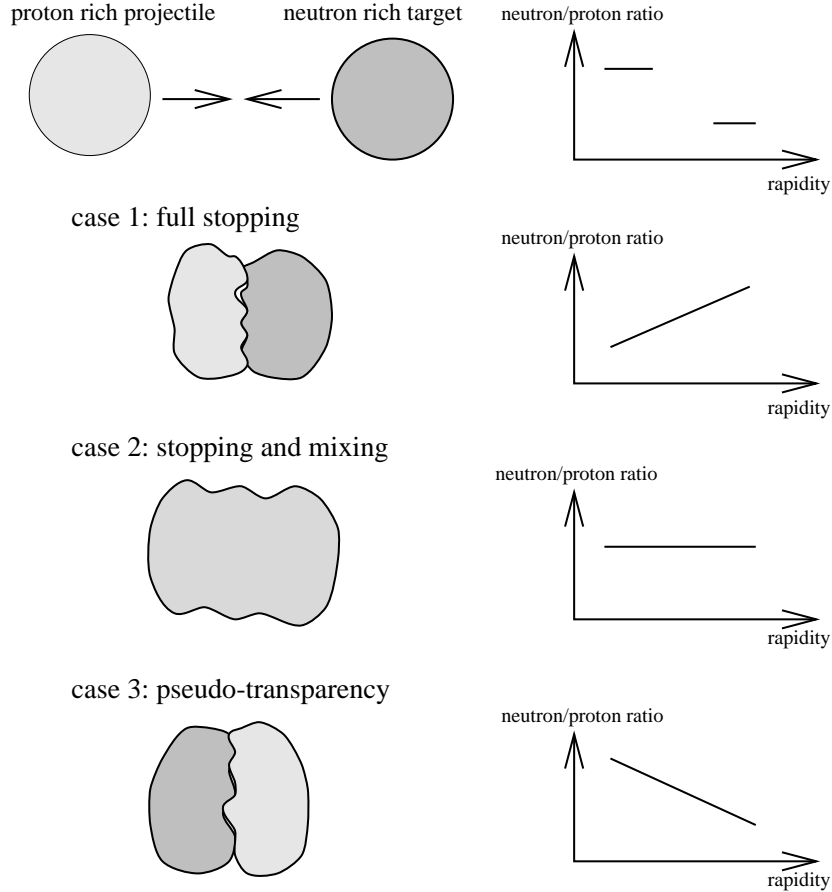,width=11cm}}
\caption{\label{stopnat1} By using projectile and target with different
isospin content, the mechanism of stopping, i.e. the amount of penetration
between projectile and target nucleons, can be measured by plotting
the neutron to proton ratio (or related quantities) versus the rapidity.
Three possible scenarios are schematically depicted in the figure 
together with their respective signature. }
\end{figure}

Particle ratios like the proton to neutron or the experimentally
accessible $^3$He to triton ratio may also be used to gain experimental
information on the mechanism of stopping 
(i. e. the amount of inter-penetration between projectile and target 
nucleons). So far the analysis of the stopping mechanism has only been
accessible to transport model calculations.
We propose an experiment with
which regions in rapidity dominated by projectile or target nucleons can
be identified (even at central collisions!) and with which therefore
the question of the stopping mechanism can be investigated.
If projectile and target have a large difference in their total isospin
(i.e. they differ strongly in their neutron to proton ratio) then the
comparison of the neutron to proton ratio vs. rapidity {\bf before}
the reaction with the  respective ratio {\bf after} the reaction
yields direct information on the degree of penetration between projectile
and target nucleons. Figure \ref{stopnat1} 
depicts three extreme cases: (a) full stopping,
(b) complete mixing of projectile and target nucleons and (c) transparency.
Note that the configuration in momentum space is opposite
to that in coordinate space.

\begin{figure}[thb]
\centerline{\epsfig{figure=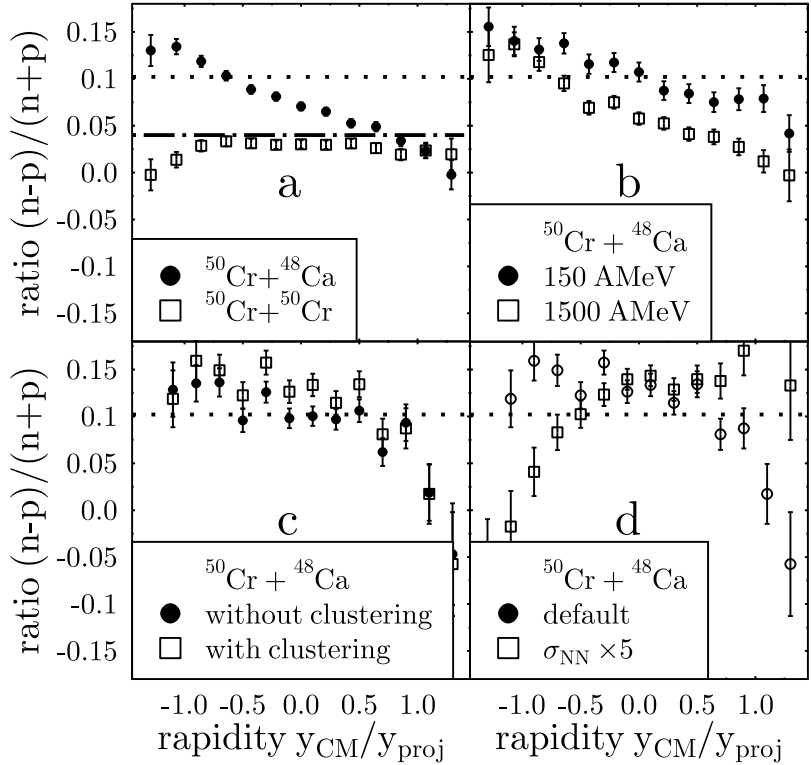,width=11cm}}
\caption{\label{jb94-1} Neutron to proton asymmetry as obtained in IQMD model 
for $^{50}$Cr+$^{48}$Ca and
$^{50}$Cr+$^{50}$Cr at 1 GeV/nucl. (a), for  $^{50}$Cr+$^{48}$Ca at
150 MeV/nucl. vs. 1.5 GeV/nucl. (b), for the same system without vs. with
cluster subtraction (c) and with an enhanced (factor 5) cross section
(d). }
\end{figure}

We have used the QMD/IQMD model in order to test the
feasibility of the suggested observable. As a test system we selected
$^{50}$Cr+$^{48}$Ca which has a large difference
in the neutron to proton ratio at almost identical mass.  
The calculations were performed with impact parameters b$\le 2$ fm.

Figure \ref{jb94-1}a shows the neutron to proton asymmetry vs. rapidity for 
$^{50}$Cr+$^{48}$Ca and the symmetric system $^{50}$Cr+$^{50}$Cr,
both at 1 GeV/nucleon.
The horizontal lines mark the values expected for case (2) of 
fig.~\ref{stopnat1}
for the isospin asymmetric (dotted) and symmetric system
(dash-dotted). The calculations clearly indicate that projectile and target
move through each other although the nucleon rapidity distribution
(which usually  characterizes stopping) shows a peak at c.m. rapidity.
Figure \ref{jb94-1}b shows the energy dependence of the asymmetry. 
At 1.5 GeV/nucl. the transparency is far more pronounced than at 
150 MeV/nucl. (which is within error bars almost compatible
with case (2) of Fig. \ref{stopnat1}). 
However, Figs. (a) and (b) only show nucleons without
taking clusterization into account. Fig. \ref{jb94-1}c compares an analysis
based on (all) nucleons with the respective analysis in which clusters
were subtracted (250 MeV/nucl. incident beam energy).
The qualitative trends remain unchanged. In order to investigate the 
sensitivity of the asymmetry to the stopping power we have performed 
a calculation with a 5 times higher nucleon-nucleon cross section.
Fig. \ref{jb94-1}d shows the respective calculation
at 250 MeV/nucl.  and compares it with the default calculation. A sign 
reversal in the slope of the asymmetry is visible.  

Experimentally it may be easier to study the $^3$He to triton ratio
instead of the neutron to proton ratio. At beam energies above 1.5 GeV/nucl.
the isospin content of the system is transferred to pionic degrees of freedom
and the neutron to proton ratio looses its 
effectiveness.

\clearpage
\subsection{Equilibrium properties of hadronic matter}\label{boxcalc}

\begin{figure}[tb]
\begin{minipage}[t]{9cm}
\centerline{\psfig{figure=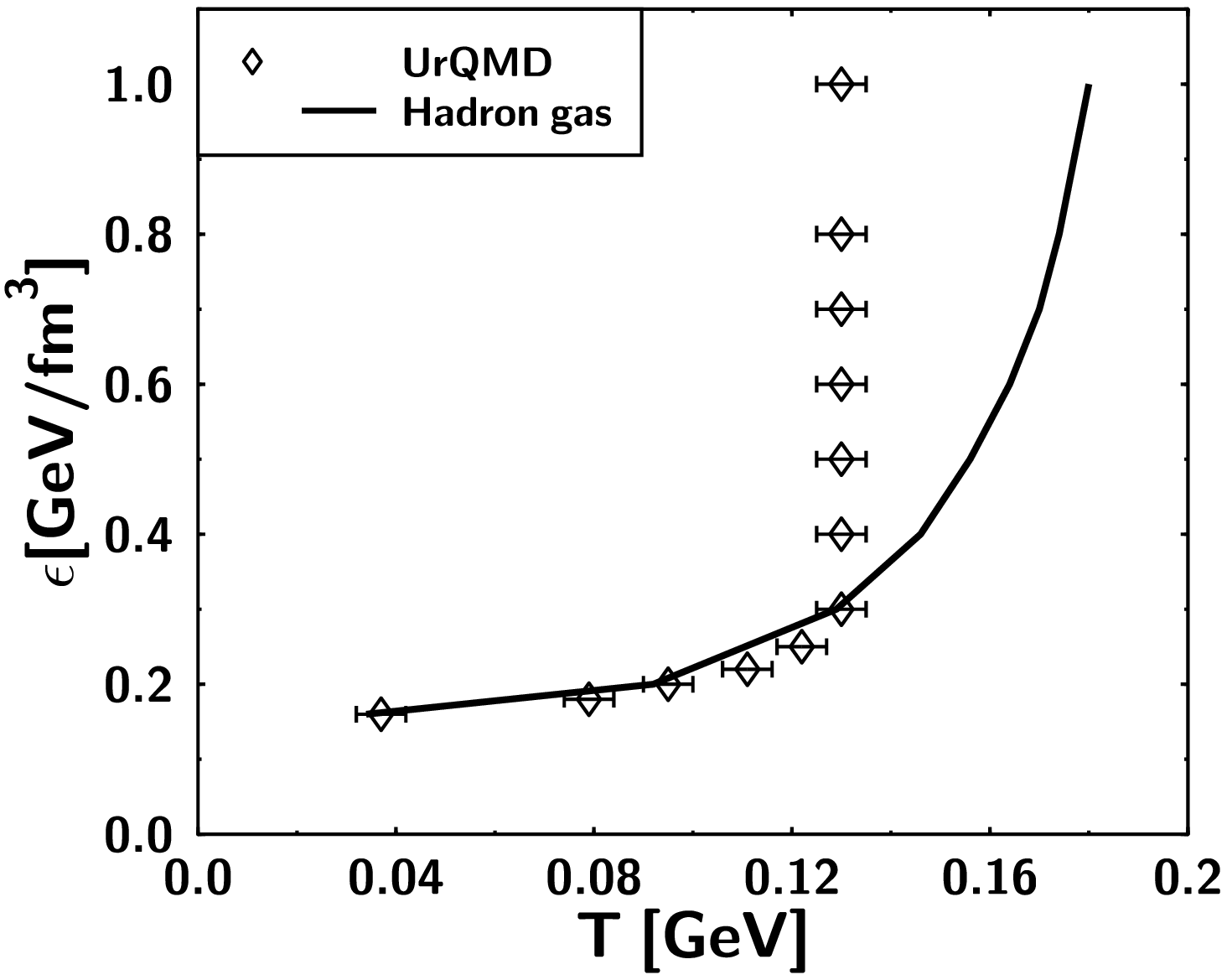,width=9cm}}
\caption{\label{eos}
Equation of State $\epsilon=\epsilon(T)$ obtained 
in UrQMD box calculations 
(squares) and in ideal hadron gas (solid line) 
at normal nuclear matter density and zero strangeness.}
\end{minipage}
\hfill
\begin{minipage}[t]{9cm}
\centerline{\psfig{figure=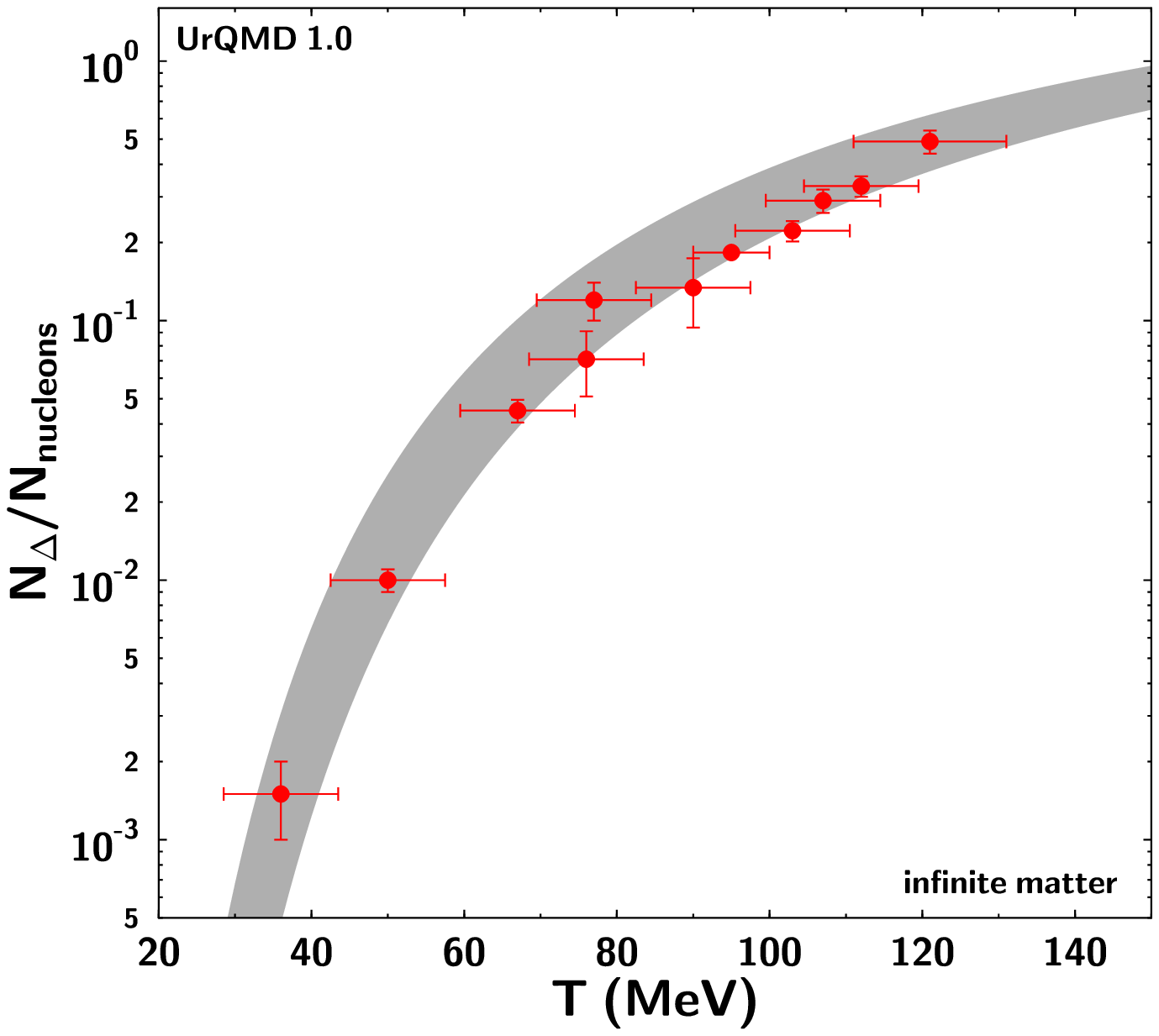,width=9cm}}
\caption{\label{box} The delta to
nucleon ratio vs. temperature as obtained in UrQMD infinite matter calculation.
The gray dashed area shows the delta to nucleon ratio calculated from the law 
of mass action in a Boltzmann approximation, 
taking fluctuations in the delta mass into account.}
\end{minipage}
\end{figure}

The possibility of thermal and chemical equilibration of hadronic matter 
is studied using UrQMD  box
calculations with periodic boundary conditions for normal baryonic density 
and zero strangeness at 
different energy densities from $\epsilon=0.2$ to $2.0$ GeV/fm$^3$. 
Starting from a state of pure nucleons with random  momentum  distribution, 
one observes that
particle multiplicities equilibrate after some time.
At low energy density $\epsilon$,
equilibration times for strange particles are much larger than
for non-strange particles, but this difference decreases with growing $\epsilon$.
In equilibrium, the energy spectra 
of the different hadronic species are nicely fitted by a  Boltzmann 
distribution $\exp{(-E/T)}$
with the same temperature $T$. Figure~\ref{eos} shows the variation of the 
temperature $T$ versus
energy density $\epsilon$ at $\rho_B=\rho_0$ and $\rho_S=0$, obtained in UrQMD.
The solid line shows the results of an ideal gas model including the same 
hadrons as UrQMD (tables \ref{bartab} and \ref{mestab}).
The UrQMD Equation of State shows a Hagedorn-like shape with a
limiting temperature $T\approx 130$ MeV, while the hadron gas EoS 
shows  a continuous rise of the temperature with $\epsilon$. Both models are
however in good agreement at low energy densities $\epsilon \le$ 0.3-0.35
GeV/fm$^3$. Because the UrQMD model uses a stochastic collision term, and no
hard core repulsion is considered for the different particles (cascade mode),
excluded volume correction is not included in the ideal gas formulation. This
correction, however, could be significant \cite{yenGD97a} and could be treated
in the Chappman-Enskog approximation.

In particular, the delta to nucleon ratio is consistent with the
theoretical expectation of a hadron gas. This can be seen in figure~\ref{box}
which displays the delta to nucleon ratio 
obtained at various energy densities ranging from 0.16 to 0.3 GeV/fm$^3$. 
The gray shaded
area shows the delta to nucleon ratio calculated from the law of mass action 
in a Boltzmann approximation, 
taking fluctuations in the delta mass into account.
At higher energy densities, despite the smaller temperature in UrQMD compared
to ideal gas, the number of pions and
other mesons is larger than in the Ideal Gas.
The difference between the two models resides in  
the string degrees of freedom included in UrQMD \cite{belkacem98a}.
At high energy densities, they keep a large 
fraction of the system energy. Moreover, the inclusion of strings in UrQMD
leads to a violation of the detailed balance at high energies, assumed
in statistical model.
Box calculations in a simplified version
where no strings and no many-body decays were included, show a very good 
agreement with the ideal gas. They 
nicely reproduce both temperature slopes and different hadronic 
multiplicities \cite{belkacem98a}.

\begin{figure}[bth]
\centerline{\psfig{figure=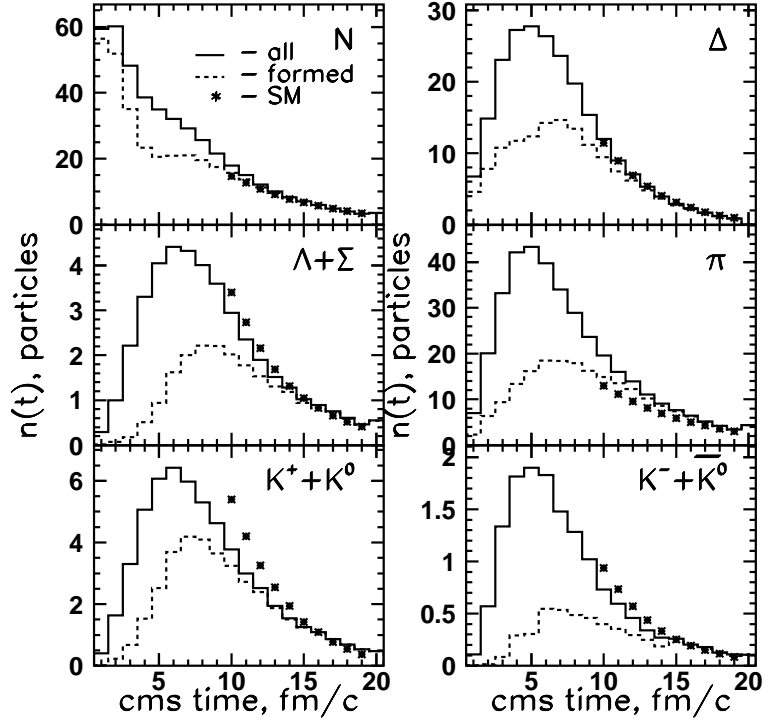,width=11cm}}
\caption{\label{n_t}
Number of particles as a function of cms time, obtained  in the central 
zone ($5\times 5\times 5$ fm$^3$)
of central ($b=$0) Au+Au collisions 
at 10.7 AGeV. Solid lines correspond to all hadrons in the zone,
dashed lines - to only formed hadrons. Points are the predictions
of ideal gas calculated at the same energy, baryonic and
strangeness densities as in UrQMD full sample in the zone. }
\end{figure}

The possibility of local equilibration 
in Au+Au collisions at AGS energies (10.7 AGeV) is studied by 
analyzing 
the time evolution of hadronic matter in the central reaction zone
\cite{bravina98a}.
It is found that between $t_{cm}=10$ and 20 fm/c, there is an 
isotropic velocity distribution
of hadrons in the central region. In this time interval, there are practically 
no strings left in this zone. Figure~\ref{n_t} shows the time evolution of
different particle multiplicities in the central region. The solid lines
correspond to the total number of particles for the different species, whereas
the dashed lines show the multiplicities of only already formed particles 
(cf. section ~\ref{formation-time}). The dots depict the results of the 
statistical model, obtained at the same $\epsilon$, $\rho_B $ and $\rho_S$ as 
in the central zone of UrQMD calculation.

\begin{figure}[bth]
\centerline{\psfig{figure=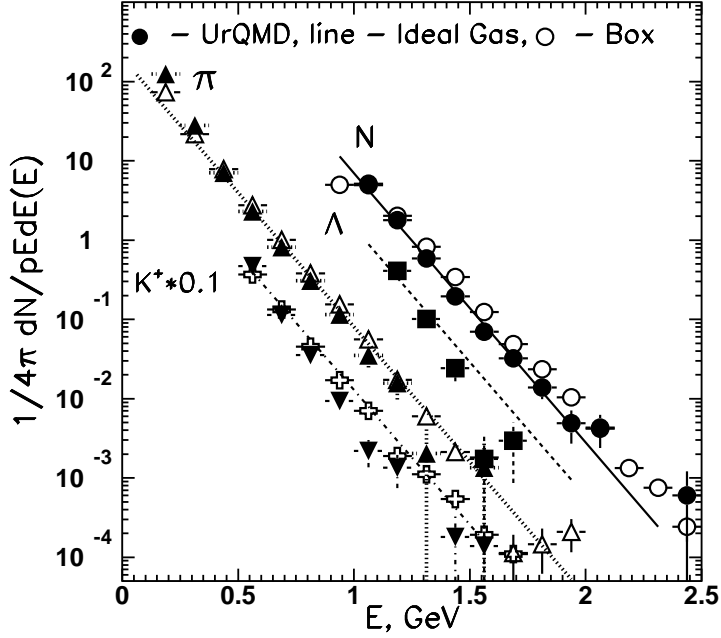,width=11cm}}
\caption{\label{dnde_13}
Energy spectra of N ($\bullet $), $\Lambda $ ($\Box $), 
$\pi $ ($\bigtriangleup $) and $K^+$ ($\bigtriangledown $) 
in central zone of Au+Au collisions at 10.7 AGeV at $t_{cm}$=13 fm/c fitted 
by Boltzmann distributions
with parameters $T$=128 MeV, $\mu _B$=534 MeV, $\mu _S=112$ MeV,
predicted by the statistical model (lines)
and obtained in UrQMD box calculations (open symbols). }
\end{figure}

\begin{figure}[bth]
\centerline{\psfig{figure=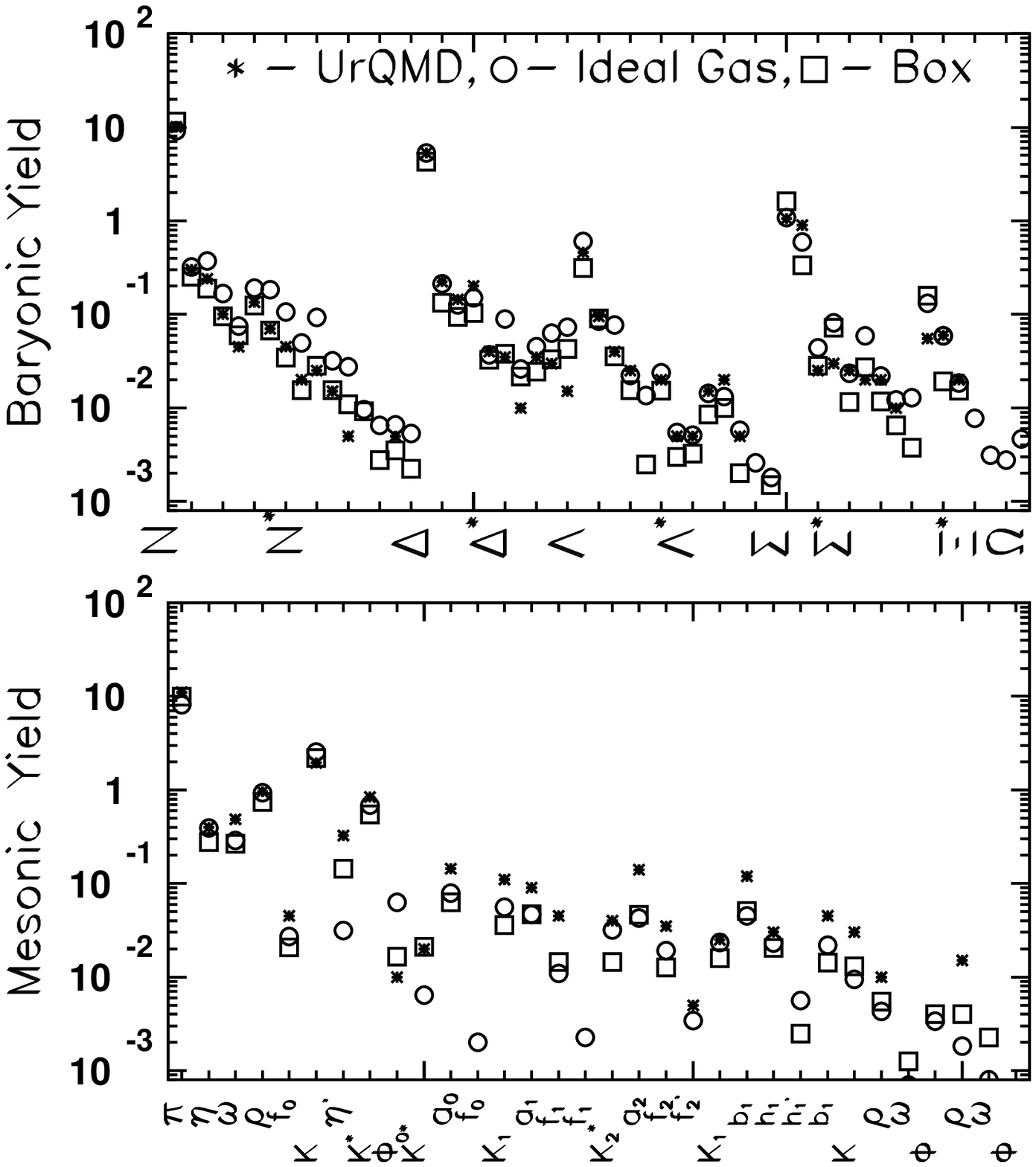,width=11cm}}
\caption{\label{bm_13}
Yields of hadrons obtained in central zone of Au+Au collisions at $t_{cm}$=13 
fm/c (*), 
in statistical model ($\bigcirc $) and in UrQMD box calculations ($\Box $).}
\end{figure}

Baryonic density 
$\rho_B$ drops from $\approx 2 \rho _0$ at $t_{cm}=$10 fm/c  to 0.5 $\rho_0$
at $t_{cm}=$15 fm/c.
Energy spectra of the different species
are nicely reproduced by Boltzmann distributions with the temperature
predicted by the statistical model (fig.~\ref{dnde_13}). The statistical model 
reproduces also the 
multiplicities of baryons and mesons obtained in UrQMD (fig.~\ref{bm_13}). 
One can see from fig.~\ref{eos}
that for the energy densities reached in the central zone of Au+Au collisions
at AGS at these times (e.g. $\epsilon = 293$ MeV/fm$^3$, $\rho_B=\rho_0$ and 
$\rho_S \approx 0$ for 
$t_{cm}=$13 fm/c)
the equation of state of the UrQMD model agrees quite well with that 
of the ideal resonance gas. The comparison with the equilibrated
state obtained in box calculations at the same $\epsilon$, 
$\rho_B$ and 
$\rho_S$ (fig.~\ref{dnde_13}-\ref{bm_13}) also indicates that in the central 
zone of Au+Au 
collisions, 
one probably reaches local chemical and thermal equilibrium.
%
However, the existence of collective flow \cite{barrette94c,barrette96a} as
well the analysis of the freeze-out properties of nucleons, hyperons
and various meson species in section~\ref{freezeout} 
indicate that the final state does not originate from an equilibrium 
configuration. One therefore has to be very careful when trying to draw
conclusions from the existence of a local zone of hadrochemical equilibrium
to the final state of the reaction.


\clearpage
		\subsection{Resonance matter}

The possibility of producing 
{\em $\Delta$-matter} (or in more general terms: {\em resonance matter})
and density isomers has been already
discussed early on \cite{stoecker78a,chapline73a}. Recently
this topic has received renewed attention 
\cite{boguta83a,boguta82a,waldhauser87a,metag93a,ehehalt93a,hofmann94a,liZ97a}: 
At beam energies above a few hundred MeV/nucleon, 
the nucleons can be excited into 
$\Delta$-resonances. If the density of these resonances 
is as high  as the nuclear
matter ground state density, then a new state of matter, 
{\em $\Delta$-matter}, has
been created. 
One of the potential signals for the presence of {\em $\Delta$-matter}
is the creation of pions as decay products of the $\Delta$-resonance.

How can {\em $\Delta$-matter} be produced? 
In order to address this question we first limit ourselves 
to Au+Au collisions at 1 GeV/nucleon. In that case the hadrochemistry
is well described by a simple closed system of nucleons, pions 
and the $\Delta_{1232}$ resonance:

\begin{figure}[thb]
\centerline{\epsfig{figure=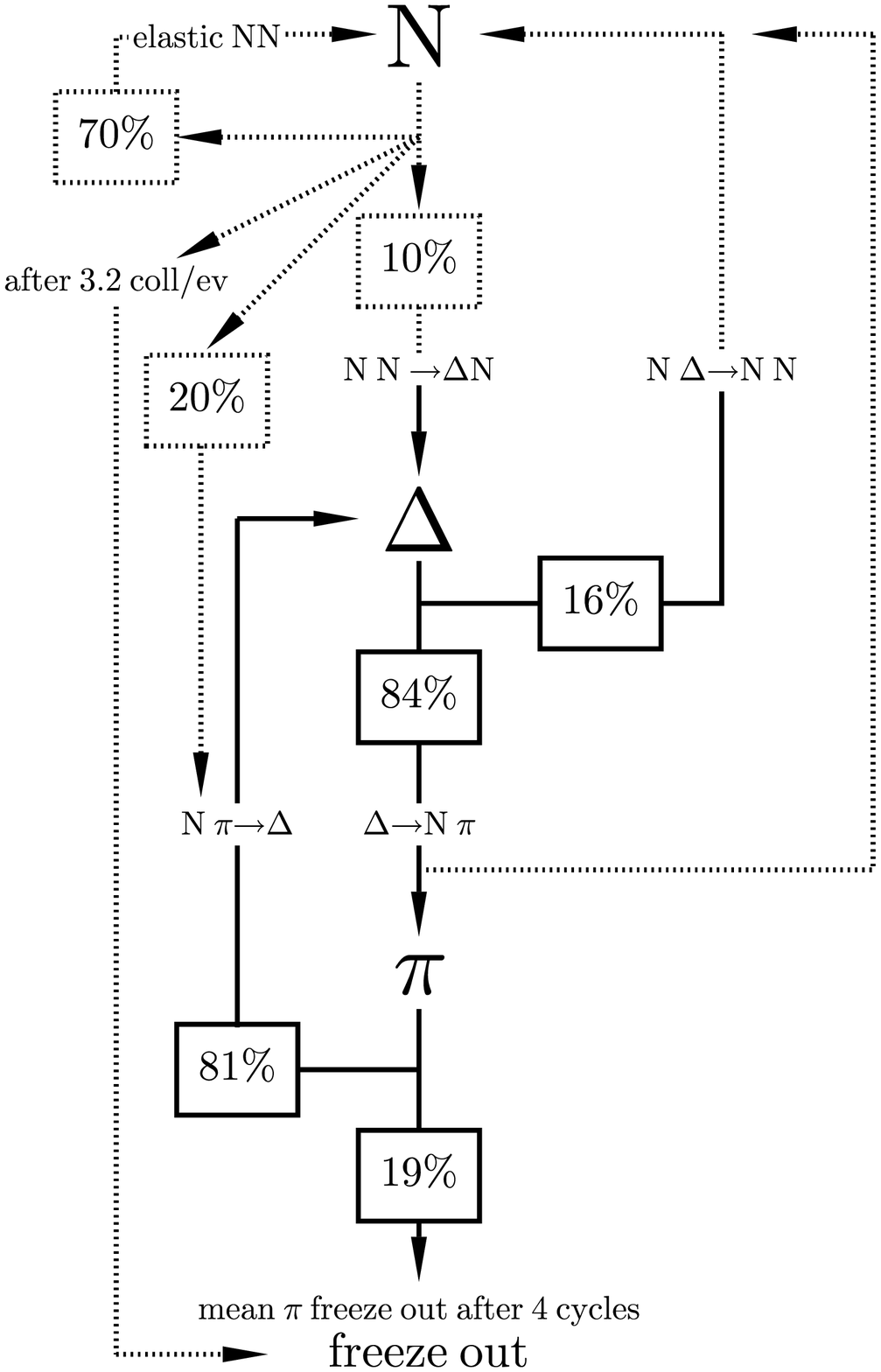,width=10cm}}
\caption{\label{pumpe} Pion - nucleon cycle in the IQMD model for Au+Au 
collisions at 1 GeV/nucleon. 
The scheme describes (for b
$\le 5$ fm and time-averaged) all possible processes (in the IQMD model) 
linked to the
creation of {\em $\Delta$-matter}. The probabilities in the boxes always 
refer to the
vertices they are directly connected with. The main process for sustaining
{\em $\Delta$-matter} is the $\Delta \to N \pi \to \Delta$ loop, which, 
however,first has to be fueled by the $N N \to \Delta N$ process.}
\end{figure}

Figure \ref{pumpe} shows the 
pion -- nucleon cycle in the IQMD model, which is limited to nucleons,
pions and the $\Delta_{1232}$ resonance. 
The scheme describes (for impact parameters b
$\le 5$ fm and averaged over 60 fm/c
possible processes linked to the
creation of {\em $\Delta$-matter}. 
The probabilities in the boxes always refer to the
vertices they are directly connected with.
$\Delta$-resonances are initially produced via inelastic nucleon 
nucleon scattering. The produced resonances can either be reabsorbed  via 
inelastic scattering or decay by
emitting a pion. The pion can then either {\em freeze out} 
or interact with a nucleon to  form a 
$\Delta$ again. In case the $\Delta$ has been 
absorbed the corresponding high energetic  
nucleon might have a second chance of becoming a $\Delta$ by inelastic 
scattering. It could also transfer energy via elastic scattering onto another 
nucleon which then could scatter inelastically and form a new $\Delta$.
A nucleon interacts in the average about three times before it freezes out. 
This value fluctuates considerably, depending on whether the
nucleon was in the participant zone (geometrical overlap of the colliding heavy
ions) or in the spectator zone of the collision.

\begin{figure}[thb]
\centerline{\epsfig{figure=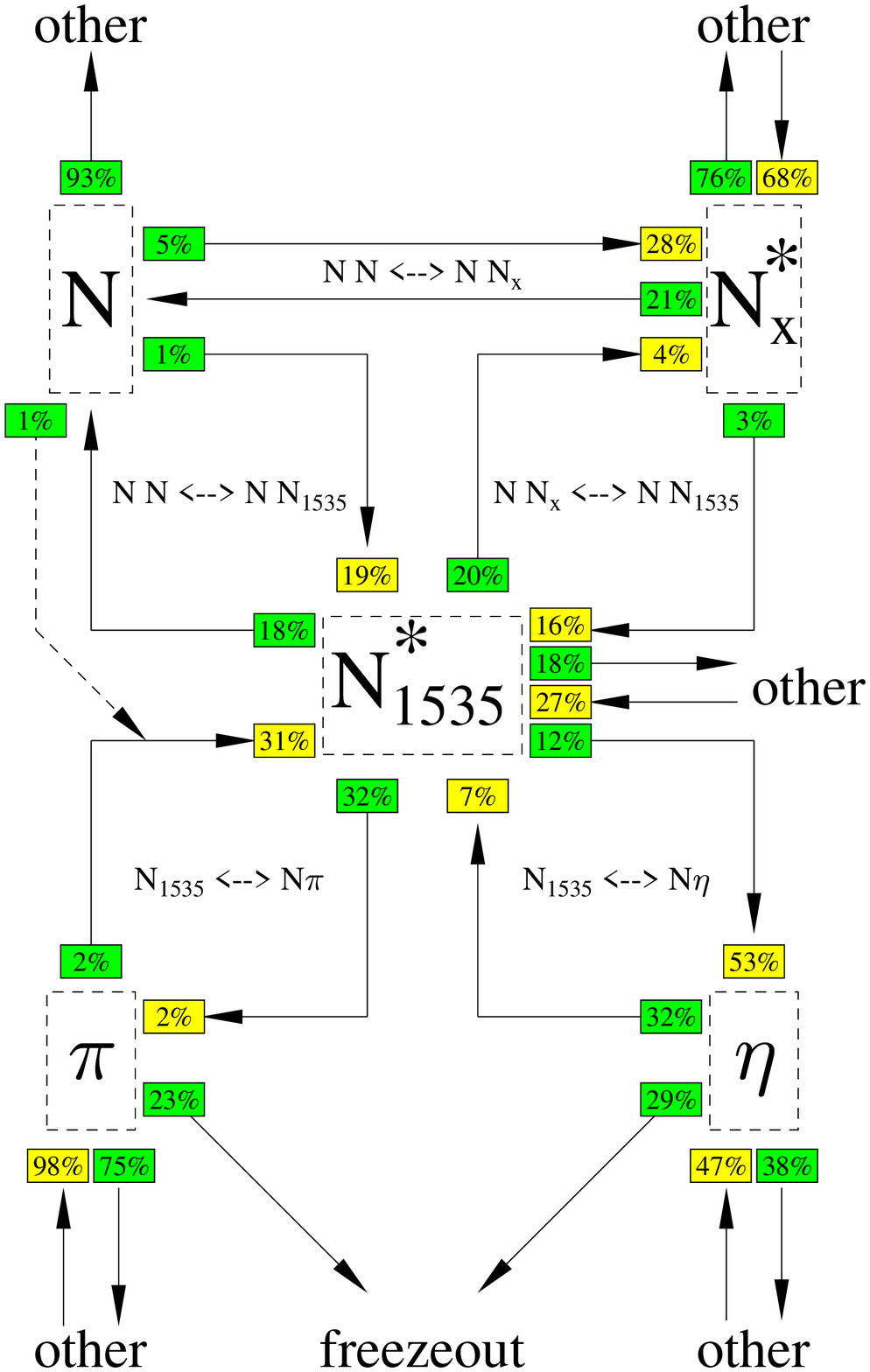,width=10cm}}
\caption{\label{eta_pump} N$^*_{1535}$-pump as obtained in UrQMD calculations 
for Ni + Ni collisions at 2 GeV/nucleon. 
The scheme describes (for b
$\le 5$ fm and time-averaged) all possible processes linked to the
creation of N$^*_{1535}$. The probabilities in the boxes always refer to the
vertices they are directly connected with.}
\end{figure}

Unfortunately, the probability for a nucleon to undergo inelastic
scattering and to form a $\Delta$ during the heavy ion collision 
at 1 GeV/nucleon is as low as
 10\%. The main process for sustaining
{\em $\Delta$-matter} is the $\Delta \to N \pi \to \Delta$ loop, 
which, however,
first has to be fueled by the $N N \to \Delta N$ process.
The average pion passes approximately three times through this loop (it has
been created by the decay of a {\em hard} $\Delta$). However, 
30\% pass more than 6 times through the loop.
For nucleons the probability of forming a {\em soft} $\Delta$ i.e. via
$\pi N \to \Delta$ 
is almost twice as high ({\em $\Delta$-matter pump}) 
than the probability of forming a {\em hard} $\Delta$
via $N N \to N \Delta$. This picture is however much more complicated when
going towards higher energies and higher resonances are considered. We show for
example in fig. \ref{eta_pump}, the N$^*_{1535}$-pump (time integrated) as
obtained in UrQMD calculations for Ni + Ni collisions at 2 GeV/nucleon. One
sees from the figure that only half of $\eta$-mesons produced at this energy
are coming from N$^*_{1535}$ decays. The other half comes mainly from 
meson-baryon collisions like N$\pi \longrightarrow \Delta \eta$ or from other 
resonances like the N$^*_{1710}$.

\begin{figure}
\centerline{\epsfig{figure=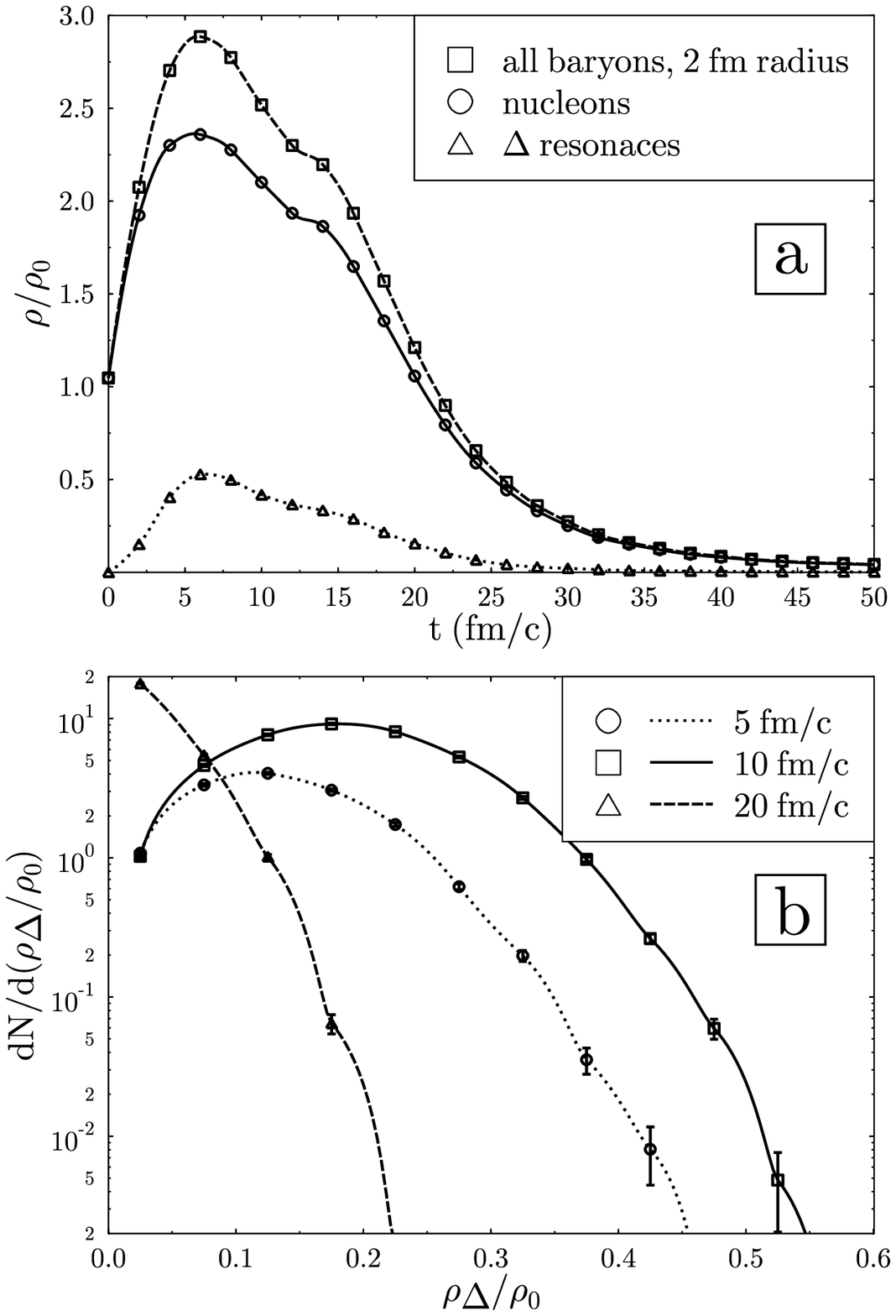,width=9cm}}
\caption{\label{tevol} IQMD time evolution of the
total baryon, nucleon and $\Delta$-resonance
density in units of $\rho/\rho_0$ (a) and $\Delta$ density distribution the
respective $\Delta$s experience for 5, 10 and 20 fm/c (b) for Au+Au collisions
at 1 GeV/nucleon.
The densities in the upper frame (a) are calculated in a
sphere of 2 fm radius around the
collision center. 
The densities in the lower frame (b) were calculated by summing over all
contributing Gaussians of all
$\Delta$'s in the system at the locations of the respective $\Delta$'s.
}
\end{figure}

Figure \ref{tevol}a shows the IQMD time evolution of 
the total baryon, nucleon and $\Delta$ densities 
in units of $\rho/\rho_0$ for Au+Au collisions at 1 GeV/nucleon. 
The densities are calculated in a sphere of 2 fm radius around the
collision center. 
Between 5 and 20 fm/c
more than 20 $\Delta$-resonances can be found in the whole system:
This time interval coincides with the hot and dense reaction phase.
At 10 fm/c up to 55 
resonances are present in the {\em total} reaction volume (keep in mind 
this is {\em not} in the 2 fm test sphere). 
A $\Delta$ multiplicity of $> 40$ can be sustained for an interval of 10 fm/c, 
6 times longer than the lifetime of a free $\Delta$-resonance. However, this 
is not pure {\em $\Delta$-matter}: 
in the small {\em test} volume shown in figure \ref{tevol}a the
resonance {\em density} is 0.5 $\rho_0$ and the
nucleon density is 2.2 $\rho_0$: the $\Delta$-contribution is 20\% in the
test volume which contains, as a matter of fact, only 2.5 resonances.
The total multiplicity of $\Delta$ resonances  is just about
10\% of the total nucleon multiplicity.

However,
it is obvious that the other $\Delta$s can be distributed all over the
reaction volume.
Figure \ref{tevol}b shows the $\Delta$ density 
distribution as experienced by the
$\Delta$'s in the system at 5, 10 and 20 fm/c.
The densities were calculated by summing over all contributing Gaussians of all
$\Delta$s in the system at the locations of the respective $\Delta$s. 
We would like to point out that the mean $\Delta$-density experienced by 
the $\Delta$s is about 0.25 $\rho_0$. Less than 1\% of the $\Delta$s 
experience $\Delta$ densities around 0.5 $\rho_0$.
$\Delta$s show {\em collective flow} in the reaction plane. Its measurable
signature (the pion $\vec{p_x}(y)$ distribution in central collisions) 
is discussed in section \ref{flowsection}.

\begin{figure}
\centerline{\epsfig{figure=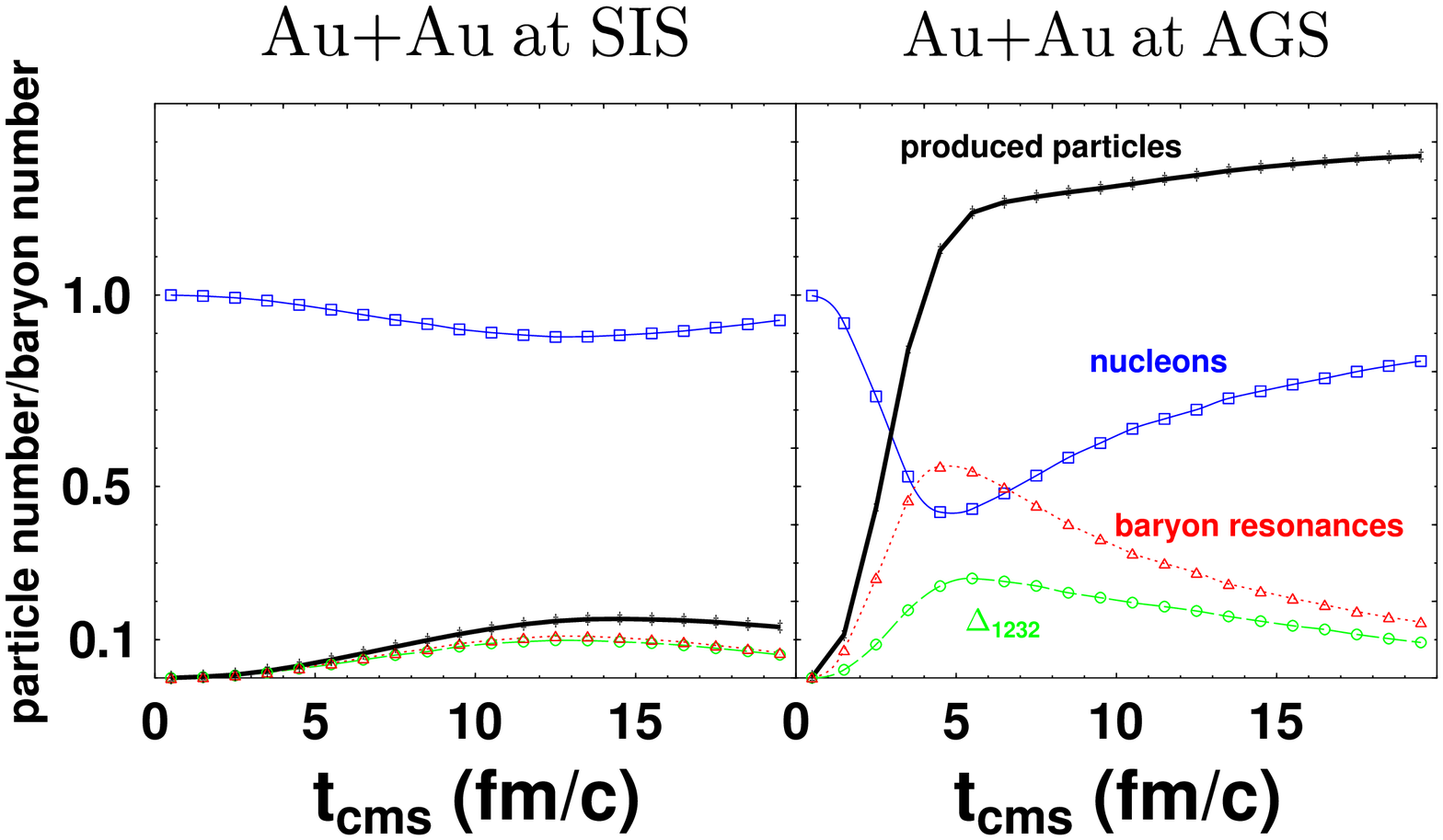,width=13cm}}
\caption{ \label{deltamatter}  Time evolution of particle multiplicities
(scaled with the number of incident nucleons) for central Au+Au collisions
at 1 GeV/nucleon (SIS) and at 10.6 GeV/nucleon (AGS). At SIS energies,
only about 10\% of the nucleons are excited to resonances whereas at
AGS energies the degree of excitation exceeds 50\%. For a time-span
of up to 10 fm/c the baryons are in a state of $\Delta-$matter.
The figure has been taken from \protect \cite{hofmann95a}.}
\end{figure}

\begin{figure}
\centerline{\epsfig{figure=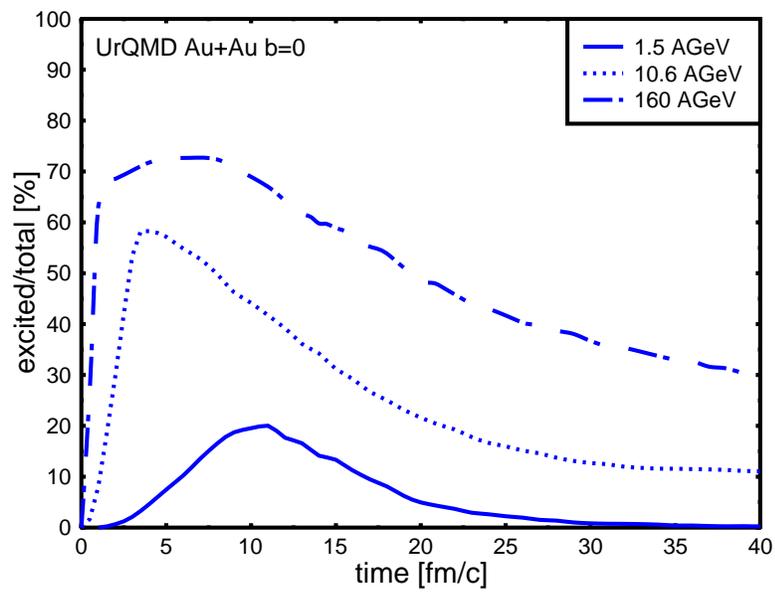,width=13cm}}
\caption{ \label{henning}  Fraction of excited resonances vs. time for central
Au+Au collisions at SIS (solid line), AGS (dotted line) and SPS (dashed-dotted
line) energies.}
\end{figure}

At AGS energies, however, RQMD calculations
predict an excited state of baryonic
matter, dominated by the $\Delta_{1232}$ resonance. 
Analysis show a long apparent lifetime ($>$ 10 fm/c) and a rather large volume
(several 100 fm$^3$) for this $\Delta$ matter state in central 
Au+Au collisions at the AGS \cite{hofmann95a} (see figure~\ref{deltamatter}).
At higher energies (SPS), UrQMD calculations shown in fig. \ref{henning},
indicate that in the early stages of the reaction, about 75\% of baryons are 
excited resonances (among which 55\% are nucleon and delta resonances and 20\%
are hyperons). This fraction is even higher at mid-rapidity (about 85\%). The
excited hyperons are created in the first 3 fm/c and remain almost constant for
the whole propagation. This fast strangeness equilibration is different from
the time scales for strangeness equilibration predicted by UrQMD box 
calculations at an energy density equivalent to that obtained in the central
region of a Au+Au collision at 160 GeV/nucleon \cite{brandstetter98a}. 
Because of the large fraction of hyperons during the whole propagation, 
there is
a finite probability for the formation of hyperon clusters (hyperon-matter).

\clearpage
		\subsection{Antimatter and strange matter}

\label{composite}

As mentioned above, the dominant reaction mechanism
in the early stage of a reaction is the excitation
of collision partners to resonances or strings \cite{sorge89a}. 
Then secondary interactions, i.~e.
the annihilation of produced mesons on
baryons, lead also to the formation
of $s$ channel resonances or strings,
which may explain the strangeness enrichment 
\cite{mattiello89a,sorge91a,sorge92a} and 
(for masses larger than 3$m_N$) allow for
$\overline{N}N$ creation \cite{jahns94a}. It is clear, however, that any
subsequent interaction of the newly produced particle can change the final
yields and spectra, which has to be taken into account.
E.g. the escape probability for $\bar p$'s
from the exploding nuclear matter enters into microscopic models usually 
via the free $N\overline{N}$ annihilation cross
section. 

The two counter--acting effects (production vs. absorption) may be 
measured by the directed ``antiflow'' of antimatter.
Varying the formation time
$\tau$, i.~e. applying $\tau=6$~fm/c instead of the default 
string fragmentation formation time 
($\tau_{\rm (anti-)baryons}\approx
1.5$~fm/c) reduces the absorption from $\approx 90$~\% to $\approx 50$~\% 
and the flow (to one half) for heavy systems. 

\begin{figure}
\begin{minipage}[t]{9cm}
\centerline{\epsfig{width=8cm,figure=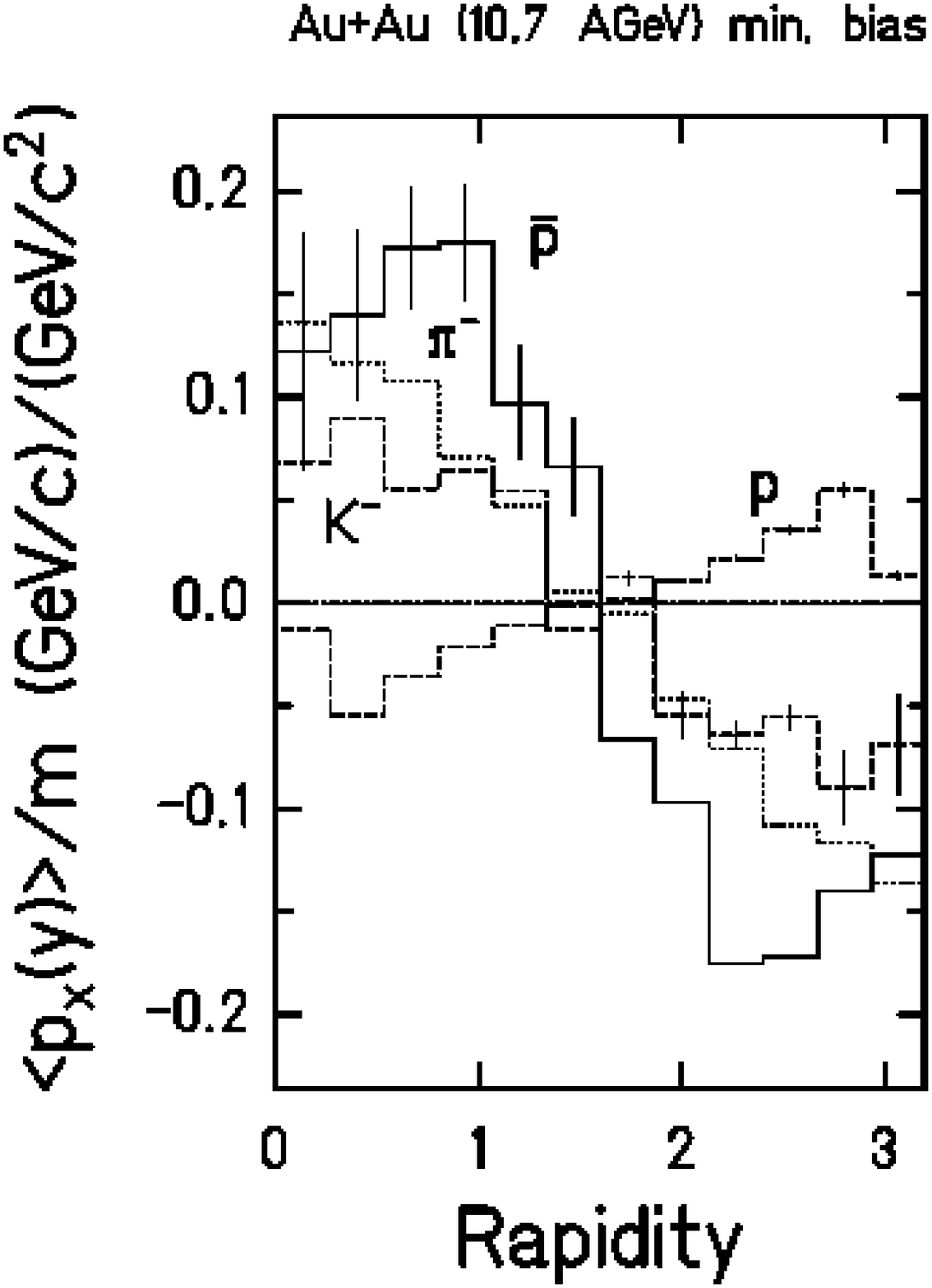}}
\caption{Mean directed transverse momentum  (in--plane) of 
various hadrons in Au+Au collisions at
10~GeV/u as a function of rapidity (RQMD calculation). 
Fig. taken from \protect \cite{jahns94a}.
\label{agsmpx}}
\end{minipage}
\hfill
\begin{minipage}[t]{9cm}
\centerline{\epsfig{width=9cm,figure=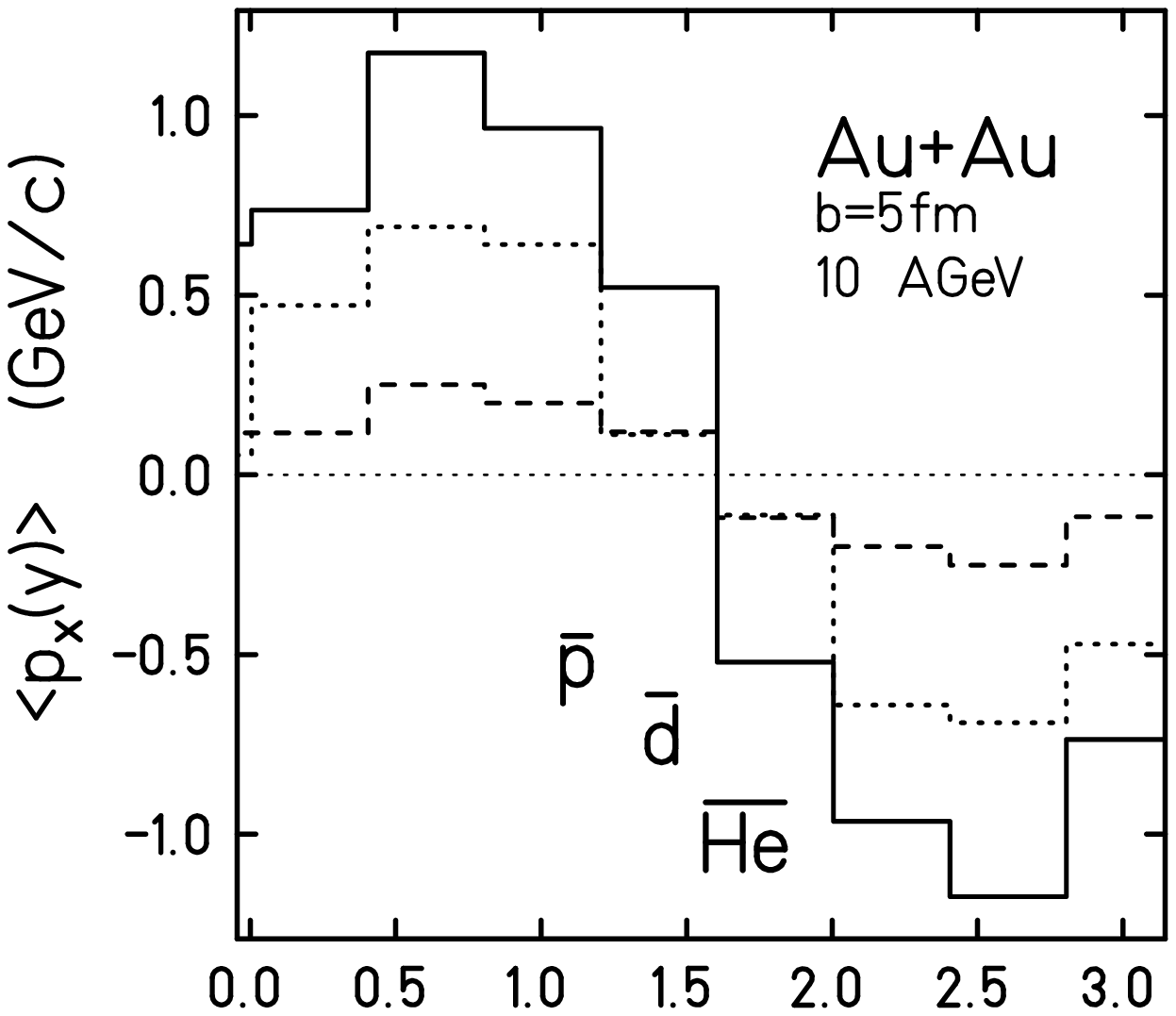}}
\caption{Mean directed transverse momentum  (in--plane) of anti--clusters in Au+Au collisions at
10~GeV/u as a function of rapidity (RQMD calculation). 
Fig. taken from \protect \cite{spieles95a}.
\label{agsf1.ps}}
\end{minipage}
\end{figure}

The observable asymmetry
for antimatter can be quantified by dividing
the yields of the upper and lower hemispheres  
for target and projectile rapidity separately. The ratio 
$ R=\frac{\displaystyle \left. \phi ^{\bar{\rm p}}<90^{\circ}\right|_{y<y_{\rm mid}}
+\left. \phi ^{\bar{\rm p}}>90^{\circ}\right|_{y>y_{\rm mid}} }
{\displaystyle \left. \phi ^{\bar{\rm p}}>90^{\circ}\right|_{y<y_{\rm mid}}
+\left. \phi ^{\bar{\rm p}}<90^{\circ}\right|_{y>y_{\rm mid}} }$
then gives the
probability for a {\em single} particle to follow the {\em collective} 
(directed) flow.
The following table shows the sensitivity of the
azimuthal asymmetry $R$ and the antiflow--parameter (the mean $p_x$,
integrated over the whole target or projectile rapidity)
on the impact parameter for
Au+Au reactions at 10.7~GeV/u within the RQMD model. The anti--flow of antiprotons appears to be
strongest for semi-central collisions of $b\approx5-7$~fm, while for protons
the maximum $p_x$ is at considerably smaller $b$--values. The latter is due
to the pressure (i.~e. the EoS), the former one due to absorption and
geometry. A similar absorption mechanism is at work for all hadrons with
valence anti-quarks, e.g. $\pi$'s and anti-kaons, resulting in mesonic
anti-flow as depicted in Fig. \ref{agsmpx}. 
\begin{center}
\begin{tabular}{|c|c|c|c|c|c|c|c|}
\hline
$b [{\rm fm}] $& 0 & 2.5 & 5.0 & 7.5 & 10.0 & 12.5 & 15.0 \\ \hline
$R$      & $\simeq$ 1:1 & 1:1.47 & 1:1.78& 1:1.70 & 1:1.54 & 1:1.28&1:1.17\\ \hline
 $<p_x^{\rm dir}> [{\rm MeV/c}]$ & $\simeq 0$ & 100  & 150& 159 & 119 & 53 &
30 \\ \hline
\end{tabular}
\end{center}

Now, let us turn to even more complex and sensitive observables, e.g.
{\it clusters} of antinucleons. 
The antideuteron
formation rate ($\overline{d}/\overline{p}^2$--ratio) in Si+Au
collisions at the AGS is sensitive to
the shape and size of the antinucleon source \cite{aoki92a,mrowczynski93a}. 
It is different from
the nucleon source.
Annihilation strongly distorts a homogeneous
coordinate space distribution of antibaryons. 
Asymmetric phase space distributions of anticlusters result for finite
impact parameter $b$, visible as directed transverse 
anti--flow (fig.~\ref{agsf1.ps}). Antimatter cluster formation is strongly suppressed.

In analogy to the deuteron case (see figure \ref{mattfig1}), 
the (anti-)deuteron formation is calculated by projecting the
(anti-)nucleon pair phase space distribution
 on the (anti-)deuteron wave function via
the Wigner--function method. The (anti-)deuteron
Wigner--density ($\rho^{^{\rm W}}_{\overline {d}}$) is again given by
the
Wigner--transformed Hulthen wave function. 
Thus the recently measured 
Si(14.6AGeV)Au antideuteron data at the AGS \cite{nagle94a} can be
reproduced!

\begin{figure}
\begin{minipage}[t]{9cm}
\centerline{\epsfig{width=9cm,figure=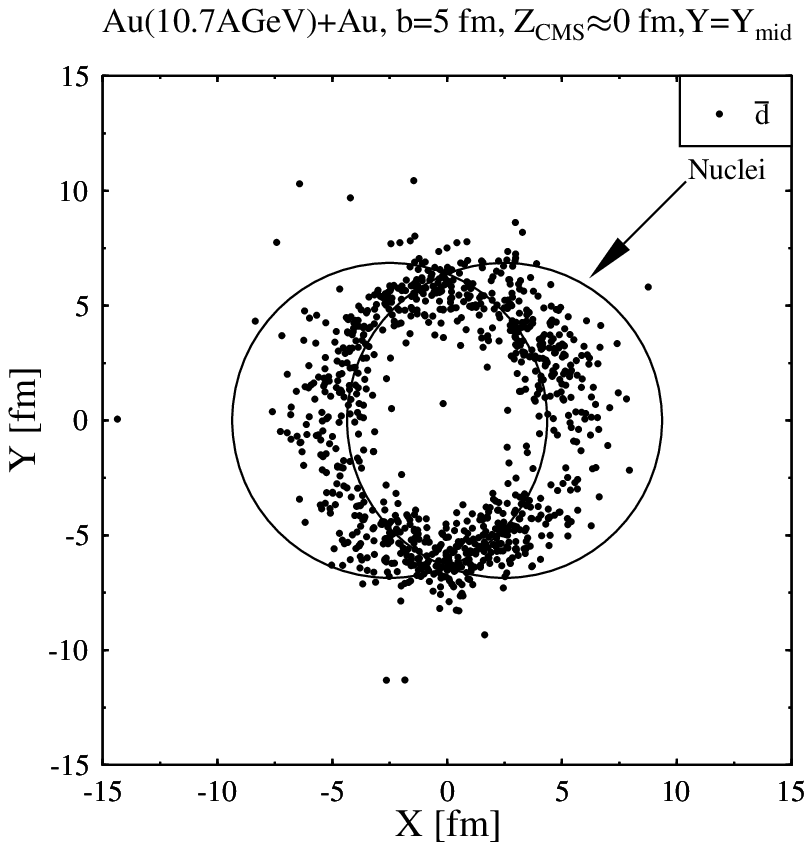}}
\caption{Coordinate space distribution of mid-rapidity anti-deuterons 
as obtained in RQMD calculations for the reaction Au(10.7~AGeV)+Au at 
freeze-out in a cut perpendicular
to the reaction plane. (Fig. taken from \protect \cite{bleicher95a})
\label{xx1}}
\end{minipage}
\hfill
\begin{minipage}[t]{9cm}
\centerline{\epsfig{figure=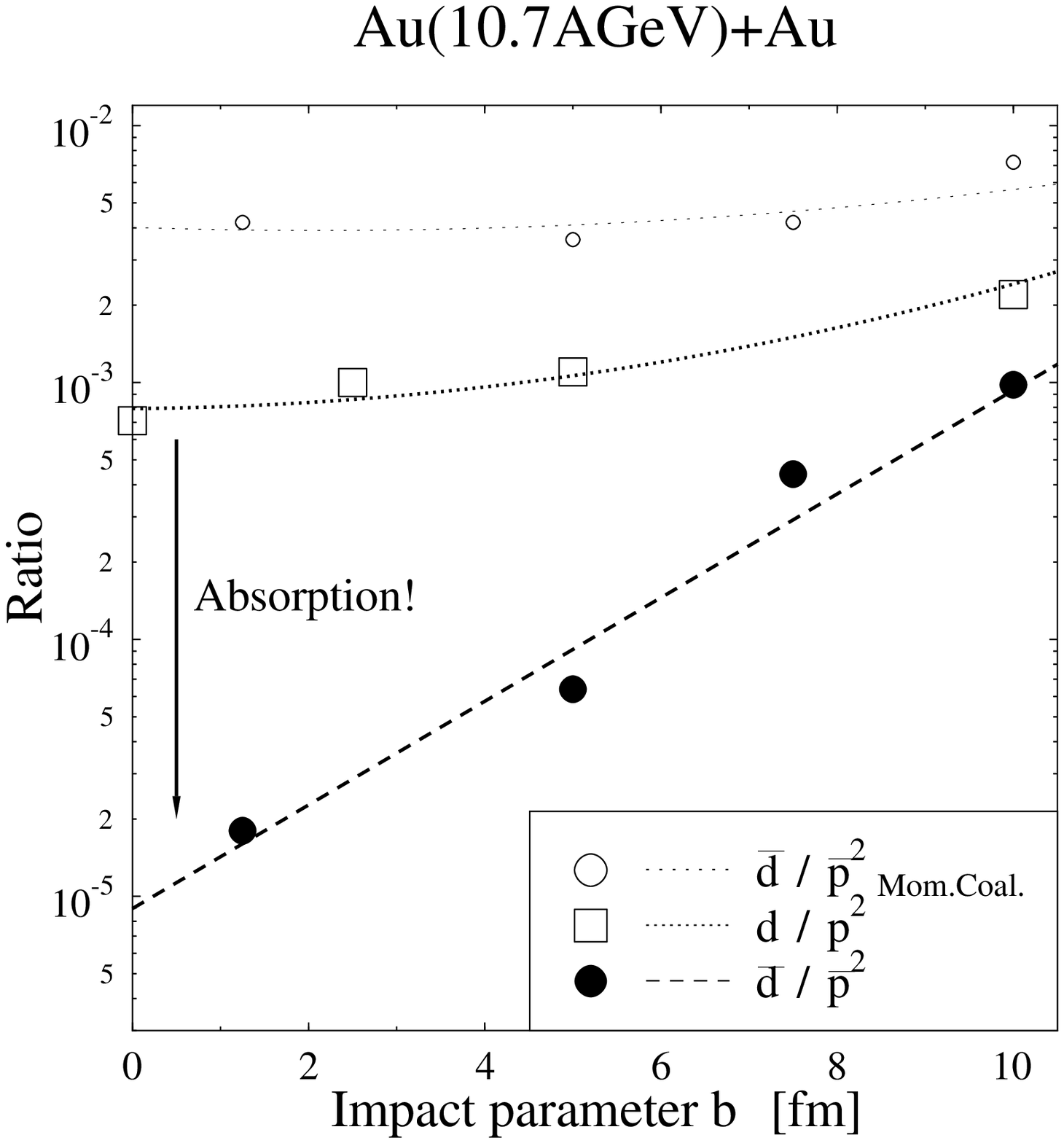,width=9cm}}
\caption{Impact parameter dependence of the anti-deuteron to the 
anti-proton ratio squared (full circles)
and of the deuteron to the 
proton ratio squared (open squares) both at mid-rapidity.
Open circles: the result of a simple momentum coalescence model.
(Fig. taken from \protect \cite{bleicher95a}) \label{stossp}}
\end{minipage}
\end{figure}

Fig.~\ref{xx1} shows
the calculated coordinate space distribution of mid-rapidity  antideuterons
in a cut perpendicular to the reaction plane for the 
Au(10.7 AGeV)+Au, $b=5$~fm reaction. 
Anti-matter is only emitted from the surface 
of the fireball -- i.e. the region of hot and dense matter -- in clear 
contrast to the source of baryons, which spreads over the whole reaction
volume. 
This can be understood as a consequence of antibaryon absorption, which
produces deep cuts in the momentum- and even more so in the coordinate 
space distribution
of the antibaryons at freeze-out \cite{jahns94a,mrowczynski93a}. 
$\overline{p}$'s  in--plane are shadowed more effectively
due to the presence of spectator matter.
This leads to an antimatter cluster
squeeze--out. 

Let us explore the reaction volume dependence of the
(anti-)deuteron formation by varying $b$
as shown in fig.~\ref{stossp}
 ($\overline{d}/\overline{p}^2$-ratio, full circles).
The deuteron formation rate ($d/p^2$-ratio, open squares) is
proportional to $V^{-1}$. Those rates measure the average phase--space distance of
$N$ and $\overline N$'s, respectively. 

In central collisions up to $95\%$ of the
initially produced antinucleons are reabsorbed. 

 The $\overline{d}$
formation rate is predicted to be roughly two orders of magnitude 
lower than the 
formation rate of deuterons.
This difference vanishes 
when going to high impact parameters or small systems (like Si+Al).
The assumption of the independent production
of both antinucleons becomes less justified in very small systems.  
Calculating the antideuteron 
formation rate within a
momentum coalescence model with a $\Delta p$--parameter of 120~MeV
($\overline{d}/\overline{p}^2$-ratio, open circles) one finds only small
sensitivity to the chosen impact parameter. This proves the pure 
coordinate space nature of the effect. 

\begin{figure}
\centerline{\epsfig{width=17cm,figure=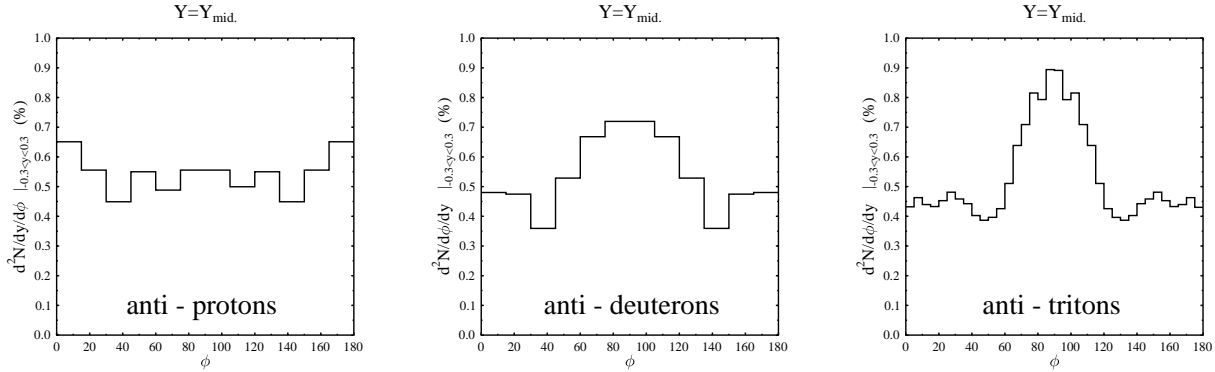}}
\caption{Correlations of antiprotons lead to a
squeeze-out for $\overline d$ and $\overline t$  
in peripheral Au$(10.7$AGeV$)$+Au reactions 
due to a higher anticluster
formation probability at the ends of the almond shaped
fireball (RQMD calculation). Fig. taken from \protect \cite{bleicher95a}.
\label{squeeze}}
\end{figure}

Fig.~\ref{squeeze} shows the azimuthal distribution of mid-rapidity 
$\bar p$, $\bar d$, $\bar t$  in peripheral Au+Au collisions. 
The momentum
distribution of $\overline{p}$ in the $p_x-p_y$-plane is nearly
isotropic. It reflects the geometry of the almond
shaped reaction zone. 
In contrast the $\overline{d}$ distribution is 
 shifted towards $\phi=90^{\circ}$ in line with the described
coordinate space distribution of the $\overline{d}$. 
This looks like a squeeze-out effect. It is even more pronounced for
$\overline{t}$. 

The predicted reaction volume dependence 
and the squeeze-out of anti--fragments reflect the
spatial distribution of antibaryons.
This suppression of antideuterons
may mask a possible antimatter cluster enhancement in a quark--gluon
plasma event \cite{heinz86a}.

An enhancement of the production of anti-nucleons due to 
the presence of relativistic mean fields was
predicted in~\cite{schaffner91a,greiner96b}. It has been calculated in the 
framework of
RBUU~\cite{teis94a,sibirtsev97a}, QMD~\cite{batko94a} and RVUU~\cite{liGQ94a} 
that enhanced
antiproton production due to reduced
effective masses may explain subthreshold antiproton 
measurements at SIS energies.
On the other hand, these transport model calculations show that 
most of the additional anti-nucleons get reabsorbed. The 
absorption rate is much higher than in a cascade
without mean fields~\cite{spieles93a}, because this additional production 
mechanism 
is most effective in regions of high baryon density and 
because of the additional attractive interaction between nucleons and 
anti-nucleons.

\begin{figure}
\begin{center}
\centerline{\psfig{figure=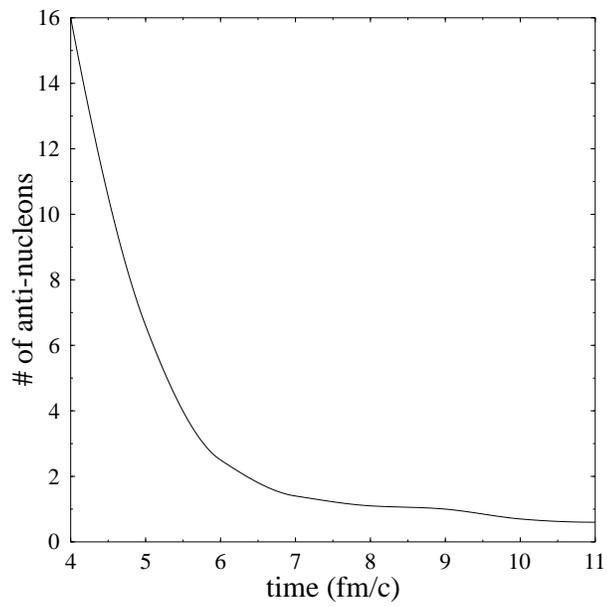,width=10cm}}
\vspace*{-1cm}
\caption{\label{anti}
Time evolution of 16 anti-nucleons ($8\cdot \bar p,\;8\cdot \bar n$)
in a Si-Al collision at $E_{lab}=13.7$ GeV/nucleon calculated in the
UrQMD-model (see text).}
\end{center}
\end{figure}

Enhanced production of anti-nucleons due to mean fields or a
quark-gluon plasma phase is conceivable at higher energies, i.e. at
AGS and SPS, too.
However, UrQMD calculations of a Si-Al collision
at $E_{lab}=13.7$ GeV/nucleon~\cite{gerland97a} without mean fields indicate 
that 
even at these energies, an enhanced production of anti-nucleons 
will be difficult to observe. Figure~\ref{anti} shows the time evolution of 
16 anti-nucleons 
($8\cdot \bar p,\;8\cdot \bar n$) that 
are added by hand after
4 fm/c in the central region of the collision.
One can see that nearly no anti-nucleon survives the 
further evolution of the system (less than 4\%). 
In Au+Au collisions, where higher baryon densities 
are predicted, all additional anti-nucleons are reabsorbed
after 1-2 fm/c.

We have learned that $\overline N$'s
suffer strong final--state interactions. These interactions have
in principle two components
which can be related to the $\overline N$ self--energy in matter:
 collisions
and annihilation on baryons\cite{gavin90a} (imaginary part,
semi-classically given by  2 Im $V=\sigma v \rho $) 
and a piece in the real part
($ {\rm Re}V=t_{\rm NN}\rho$ in the impulse approximation).
 In the semi-classical limit the real part of the self-energy
 can be approximated by
potential--type interaction\cite{buck79a,dover79a} or a mean field.

Here we will focus on the effect of the real part. The motivation
is that
the long--range force of baryons acting on a $\bar p$ is expected
 to be
stronger than for protons since the Lorentz--scalar and the
 Lorentz--vector
parts of a meson exchange potential now have the same sign.
The influence of baryonic mean--fields on baryons and mesons
is well established. Therefore there should also be some influence
 on $\bar p$'s.

\begin{figure}
\begin{minipage}[t]{9cm}
\centerline{\epsfig{figure=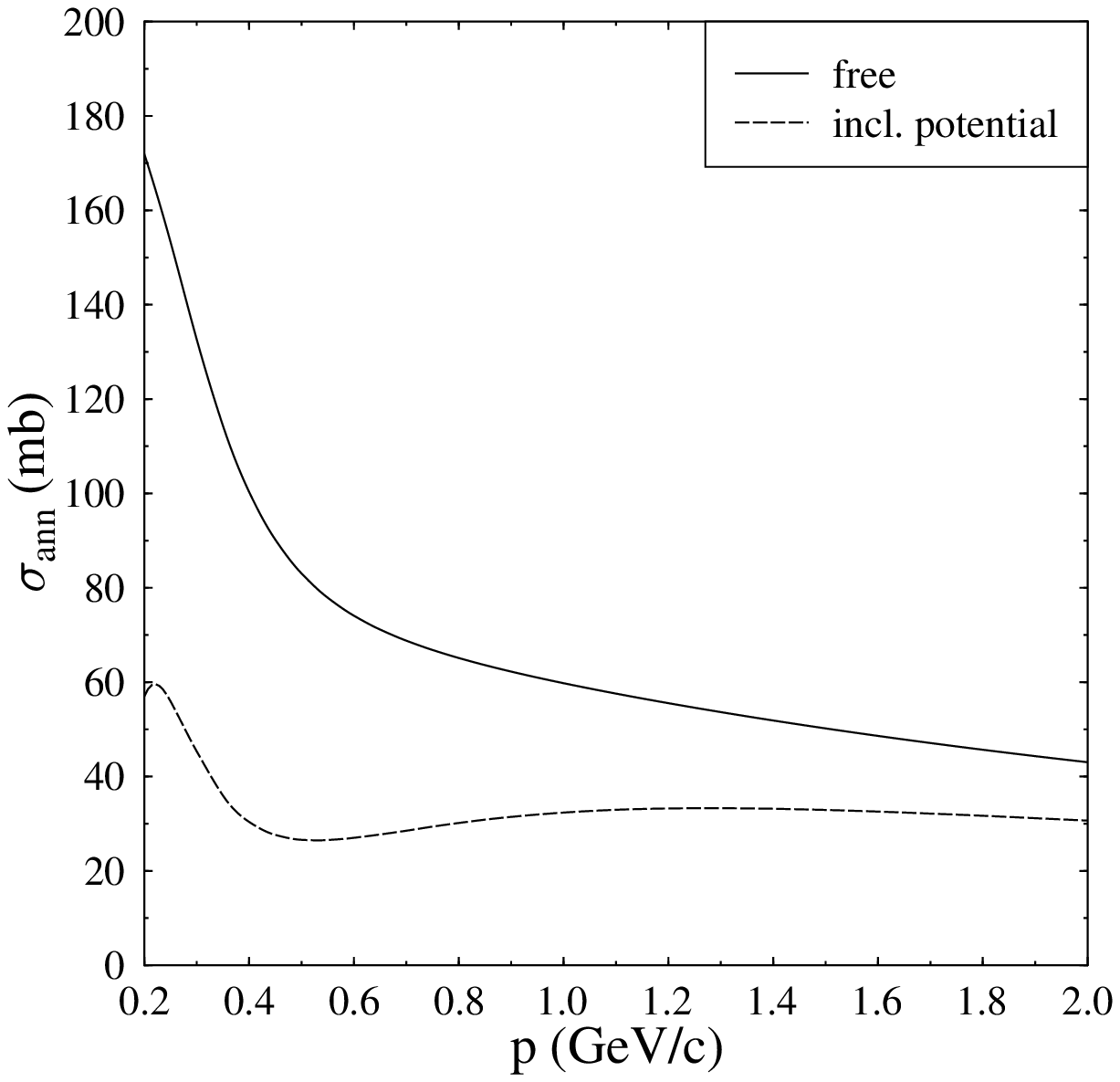,width=9cm}}
\caption{$\overline p p$ annihilation cross section as a function of the
lab-momentum.
 Parameterization of the free measured cross section (full line) and
the corrected value, if the described potential interaction is added (dashed line).
(Fig. taken from \protect\cite{spieles96c})\label{RANNTRUE}}
\end{minipage}
\hfill
\begin{minipage}[t]{9cm}
\centerline{\epsfig{width=9cm,figure=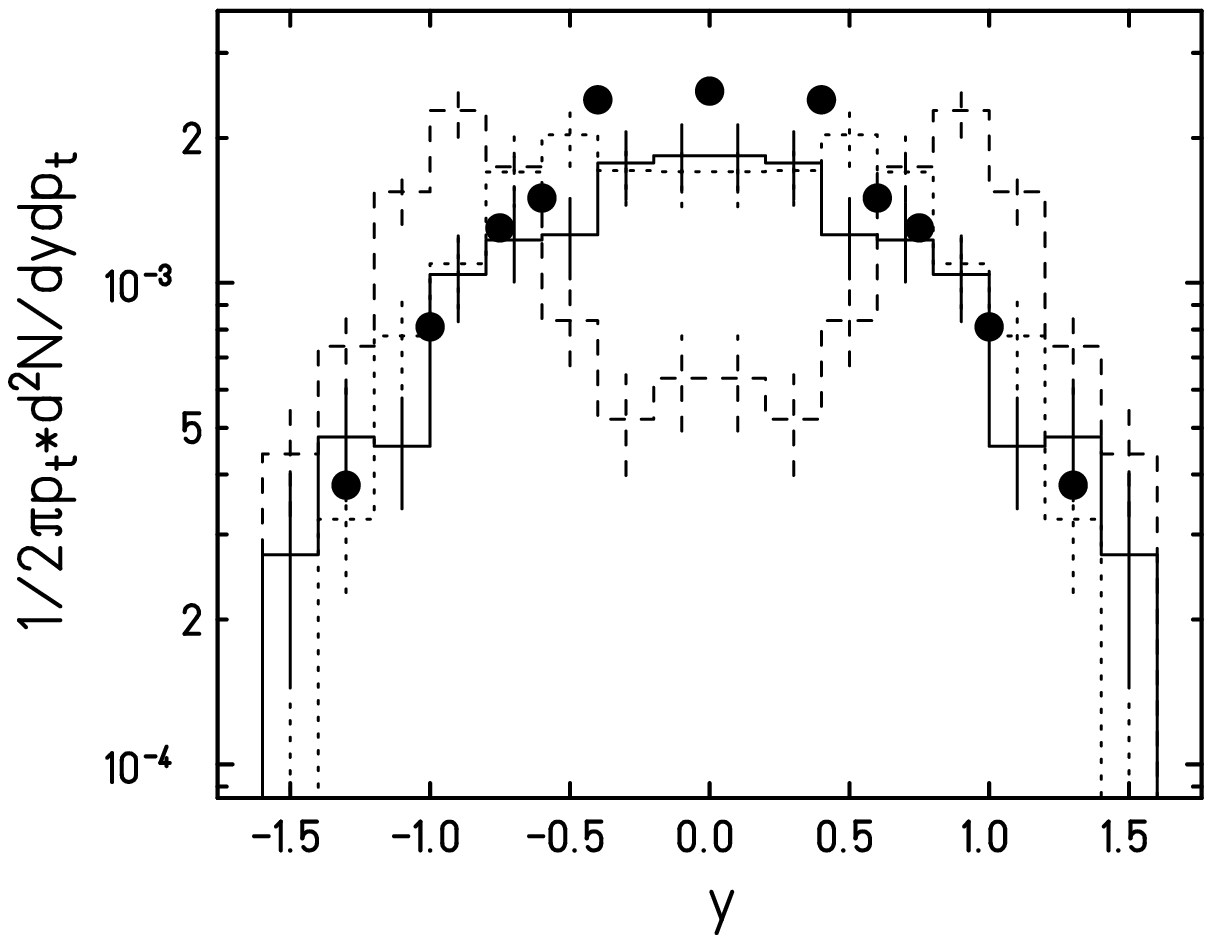}}
\caption{Invariant rapidity--distribution of the $\overline p$'s 
with $p_t<200$~MeV for Au+Au ($b<4$~fm) at 10.7~GeV/u.
Shown is the RQMD calculation (dashed line), with the
additional optical potential (full line) and the 50~\% reduced potential
(points). Preliminary data
(full circles) from \protect \cite{kumar94a}.
(Fig. taken from \protect \cite{spieles96a})\label{DNDY}}
\end{minipage}
\end{figure}

The strong {\em imaginary}
part of the optical $\bar NN$--potential has been taken into account by 
using the geometric annihilation
cross section. The {\em real} part may also be substantial \cite{kochV91a}.
However, it would be
fatal to ``improve'' the description
by just adding a mean field interaction,
since a part of the real potential has already been taken
into account by the (parameterized) free elastic and inelastic cross
sections. Fig.~\ref{RANNTRUE} shows this effect for our model potential. Due
to the strong attraction for the $\bar B B$ case, a reduced geometric
annihilation cross section suffices to account for the measured free
annihilation probability in binary $\bar p p$ reactions.
To avoid the inevitable double counting  in fully dynamical calculations
this must not be neglected.
In the case of a homogeneous nuclear medium the attractive potentials 
can partly cancel, leaving only the geometrical cross section effective for
$\bar N$ annihilation. The importance of the real potential for the
annihilation cross section results in an additional uncertainty of the $\bar
Y$ and $\bar \Delta$ annihilation probabilities, since the real part of the
self-energy is even less known than for $\bar p$'s. This should be taken into
account for the analyses of (preliminary) reports of surprisingly high $\bar
\Lambda/\bar p$ ratios at AGS and SPS experiments
\cite{lajoie96a,armstrong97a,roehrich97a}.

As a first step, we restrict our study on the residual potential interaction with its 
finite range to the dynamics {\it after} the hadronic freeze--out.
The success of Dirac equation optical model calculations for $pA$
scattering \cite{arnold81a} leads us to using 
 these relativistic
potentials with Yukawa functions as
interaction form factors--- applying G--parity transformation --- for the $\overline
p$ case:
The mass parameters are $\mu_V=3.952\;{\rm fm}^{-1}$ and
$\mu_S=2.787\;{\rm fm}^{-1}$, the coupling constants are 
$g_V=2674.5$~MeV fm and $g_S=2158.2$MeV fm.
In line  with \cite{arnold81a} Gaussians are used as baryon profiles with a
mean square radius of $0.8$~fm. 
The central part of
an effective Schr\"odinger equivalent potential (SEP) is constructed from the
above potentials:
\begin{equation}
U_{\rm SE}=\frac{1}{2E}(2EU_V+2mU_S-U_V^2+U_S^2) \quad,
\end{equation}
where $E$ is the total energy of the incident particle.

For small $\overline N N$ distances 
the real potential should not show an effect, since the huge
imaginary part absorbs the particular $\overline p$ anyhow. We have chosen 1.5~fm 
as cutoff distance  corresponding
 to an averaged $\overline p p$-annihilation cross section
of $\approx 70$~mb.

The Schr\"odinger equivalent potential with the above parameters results in
a mean $\overline p$-potential in nuclear matter of about -250~MeV
($p_{\overline p}=0$ GeV/c), increasing with energy towards -170~MeV
at $p_{\overline p}=1$~GeV/c. 
The actual (averaged) SE-potential of the $\overline p$'s 
at freeze--out is about -70~MeV. 

Due to the strong momentum dependence of the $\overline p p$-annihilation cross
section, $\overline p$'s with low transverse momentum are suppressed in RQMD
1.07.
Figure~\ref{DNDY} shows the invariant multiplicity 
 of $\overline p$'s with
$p_t<200$~MeV for central collisions of Au+Au at 10.7~AGeV. 
Calculations with and without potentials are
compared to preliminary data of the E878 collaboration\cite{kumar94a}. Besides
the proposed model interaction we calculated the effect for the same potentials,
arbitrarily reduced by 50~\%. Still the dip at mid-rapidity vanishes although
the change is less pronounced.

Note that the $p_t$-integrated spectrum is not considerably affected by the
potential interaction, but
the final phase space distribution of the $\overline p$'s at low--$p_t$
deviates substantially from
the standard RQMD calculation: A clearly non-thermal spectrum with a dip at
mid-rapidity for $p_t=0$ gets distorted by the potential interaction
during the last stage of the collision. Up to now, the possible
influence on the formation of anti--clusters, 
which have just proved to be a delicate probe of the final phase--space
distribution, is unclear.

\begin{figure}
\centerline{\epsfig{figure=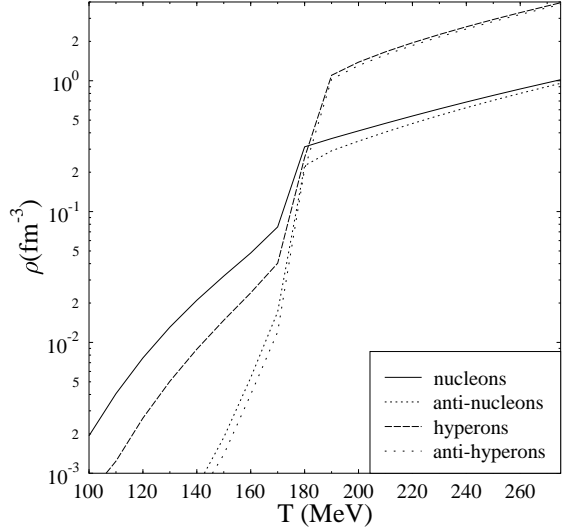,width=9.0cm}}
\caption{\label{allhyp} 
Densities of (anti--)nucleons and (anti--)hyperons as functions of
temperature for fixed $\mu_{\rm q}=100$~MeV and
strangeness fraction $f_{\rm s}=0$,
calculated with a relativistic meson--baryon field theory
(RMF) which implements hyperon--hyperon interactions
(figure taken from \cite{spieles96b}).
} 
\end{figure}

Mean-fields are not only expected to influence the phase space distributions of
hadrons, but also might have a strong impact on the production probabilities
(see above). Relativistic
meson-field models, which, at high temperature $T_{\rm C}\approx 180$~MeV
qualitatively simulate chiral behavior of
the nuclear matter, exhibit a transition into a
phase of massless baryons \cite{theis83a}.
Every \mbox{(anti-)}baryon species
(hyperons included \cite{schaffner92a}) shows approximately the number density 
of normal nuclear matter ($\rho_0\approx 0.16$~fm$^{-3}$)
near $T_{\rm C}$. 
Thus, the fraction of \mbox{(anti-)}strange hyperons
increases by 1-2 orders of magnitude at $T_{\rm C}$.
Several hundred \mbox{(anti-)}baryons, the majority being
\mbox{(anti-)}hyperons, may then fill the hot
 mid-rapidity  region. 
Fig.~\ref{allhyp} shows this transition for \mbox{(anti-)}nucleons and
(anti-)hyperons at $\mu_q$=100 MeV. 
Above $T_{\rm
C}$ all effective masses are small and the relative yields in the medium
are dictated
by the isospin degeneracy, thus favoring (anti-)hyperonic matter.
This scenario could enhance the production probability of (anti--)\-hyperons. 
Even the formation of meta-stable exotic multi-strange
objects (MEMO's \cite{schaffner92a}) is conceivable via this mechanism. 
However, the high temperature will suppress the
formation of clusters of mass $A$ by $e^{-A(m-\mu_{\rm B})/T}$.

It was speculated  that strange matter might
exist either as meta-stable exotic multi-strange 
objects (MEMO's \cite{schaffner92a}) or in form of {\em strangelets}. 
MEMO's are baryonic clusters which contain
hyperons, thus constituting lumps of hypermatter.
Strangelets, on the other hand, can be thought of as (meta-)stable
multi-strange quark clusters \cite{bodmer71a,chin79a,bjorken79a}.

The possible creation --- in heavy ion collisions ---
 of long-lived remnants of the quark-gluon-plasma, cooled and charged
up with strangeness by the emission of pions and kaons, 
was proposed in \cite{greinerC87a,liu84a,greinerC88a,greinerC91a}.
Strangelets can serve as signatures for the
creation of a quark gluon plasma.
The detection of strangelets would verify exciting
theoretical ideas with consequences for our knowledge of
the evolution of the early universe \cite{farhi84a,witten84a}, the dynamics of 
supernova explosions
and the underlying theory of strong interactions.
Currently, both at the BNL-AGS and at the CERN-SPS, experiments are carried 
out to search for MEMO's and strangelets, e.~g. by the E864, E878 and the NA52
collaborations \cite{akiba96a,armstrong97a}.

The production of strangelets and anti-strangelets might be possible
via the strangelet distillation mechanism \cite{greinerC88a,greinerC91a} 
if low values of the
bag-constant $B^{1/4} \leq 180 MeV$ are assumed. However hadronic ratios at 
SPS energies can be reproduced with the same hadronization model only under the
assumption of a bag constant of $B^{1/4} \geq 230 MeV$ \cite{spieles97b} (see
also \cite{barz90a,barz91a,braun-munzinger93b}). 
This would
exclude the formation of meta-stable strangelets within this model.

Be also reminded that the question, whether strangelets or MEMO's
can exist as bound states at all, is very speculative and
thus still a controversial point, on which
we did not focus here.
Special (meta-)\-stable candidates for experimental searches 
are the quark-alpha \cite{michel88a} with $A_B=6$ 
and the H-Dibaryon with $A_B=2$ \cite{jaffe77a}.

	\subsection{Particle freeze-out }
	\label{freezeout}

The previous sections have dealt with the description of the final
state of heavy ion reactions. Different analyses of data at AGS and
CERN/SPS energies have shown that the experimental results may be
described as a transversally and longitudinally expanding 
chemically and thermally equilibrated hadron-gas
\cite{letessier94a,braun-munzinger95a,braun-munzinger96a,sollfrank91a}.
The assumption of a chemically and thermally equilibrated source,
however, implies a uniform freeze-out for all particles; i.e. all
particle species have the same freeze-out times, radii and densities.
Such a scenario is easily tested in the framework of a microscopic
calculation. This section therefore deals with the investigation
of freeze out properties of different particle species in the 
framework of microscopic transport models and investigates model-independent
methods to gain access to freeze-out information
contained in the final state of the heavy ion reaction.

\begin{figure}
\centerline{\epsfig{figure=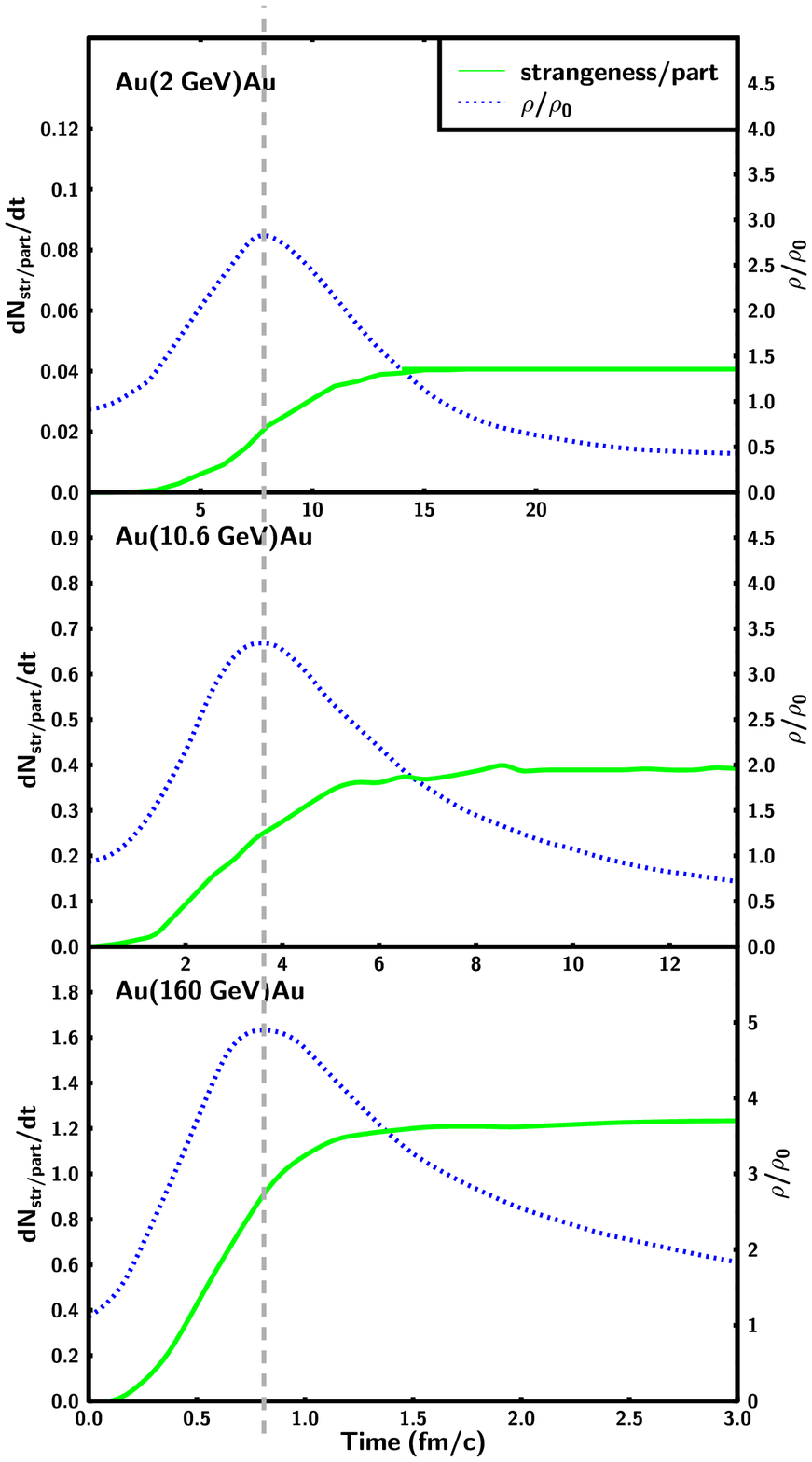,height=12cm}}
\caption{\label{allstr} 
Time evolution of the mean baryon 
density and total produced strangeness
per participant baryon (UrQMD calculation).
The mean baryon density is defined as the average over the baryon 
densities computed at the locations of all individual baryons. The 
total strangeness is the sum of all produced quarks and antiquarks.
The time axis has been rescaled in order to overlay the respective
times of highest baryon density (dashed grey line).}
\end{figure}

Figure~\ref{allstr} shows the time-evolution of the baryon density 
at the collision center and  of the total strangeness
per particle (i.e. the sum of $\bar s$ and $s$ quarks per particle) for
Au+Au reactions at 2, 10.6 and 160 GeV/nucleon. The time-axis has
been scaled in order to synchronize the time-evolution to 
the time of maximum  baryon-density. At 2 GeV/nucleon only about
50\% of the total strangeness has been produced. It saturates first
in the late stages of the reaction at $\approx 1.4 \rho/\rho_0$.
At AGS energies already 60\% of the total strangeness has been produced
at the point of maximum density -- saturation again takes places
at  $\approx 1.5 \rho/\rho_0$.
Between SIS- and AGS-energies no significant differences are visible
for the time-evolution of strangeness saturation. At SIS- and low AGS-energies
the dominant production mechanisms for strangeness are multi-step excitation
of heavy resonances and their subsequent decay into a hyperon and a
kaon. At AGS-energies string excitation may also contribute to 
strangeness production, but is not yet a dominant factor.
The situation changes for SPS-energies: Here already 75\% percent 
of the total strangeness is produced at the time of maximum
baryon density and saturation takes place at $\approx 3 \rho/\rho_0$.
The dominant production mechanism for strangeness at SPS energies
is string-excitation.

The time-evolution depicted in figure~\ref{allstr} characterizes
the {\em chemical} freeze-out of strangeness in heavy collision systems.
We shall now turn to the {\em kinetic} freeze-out, i.e. the time
of the last interaction of the particle (i.e. scattering or its production
via the decay of a resonance):

\begin{figure}[thb]
\begin{minipage}[t]{9cm}
\centerline{\epsfig{figure=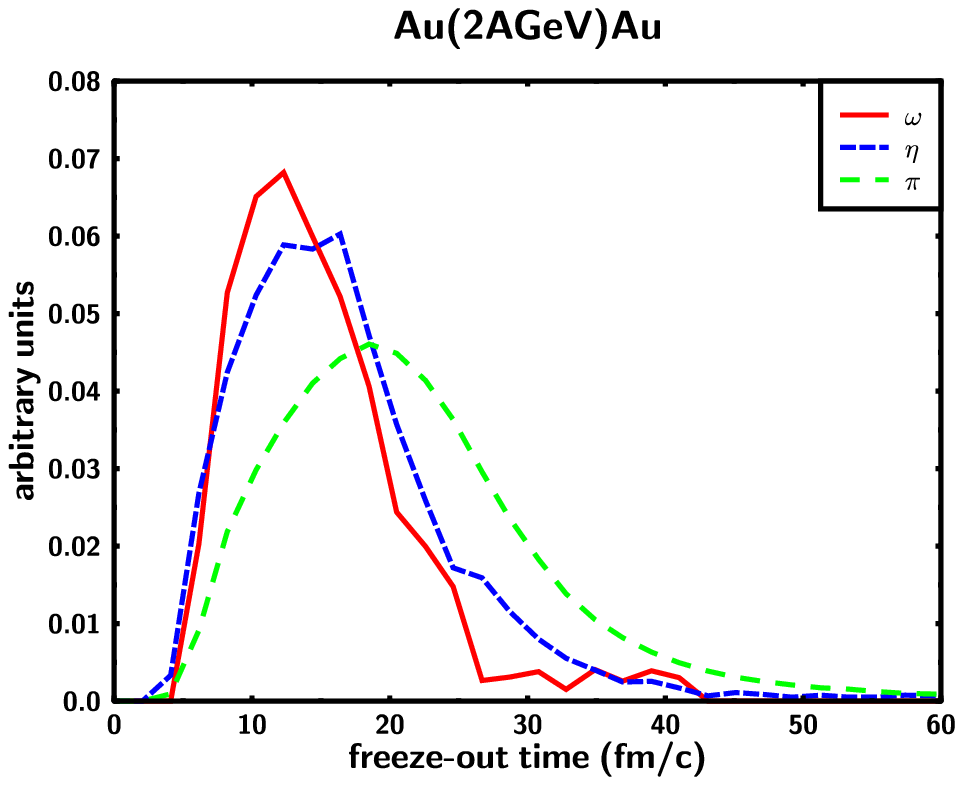,width=9cm}}
\caption{\label{fzmes} 
Normalized freeze-out time distributions for
$\pi$, $\eta$ and $\omega$ mesons in UrQMD central Au+Au reactions
at 2 GeV/nucleon. Heavy mesons freeze out earlier than light ones.
}
\end{minipage}
\hfill
\begin{minipage}[t]{9cm}
\centerline{\epsfig{figure=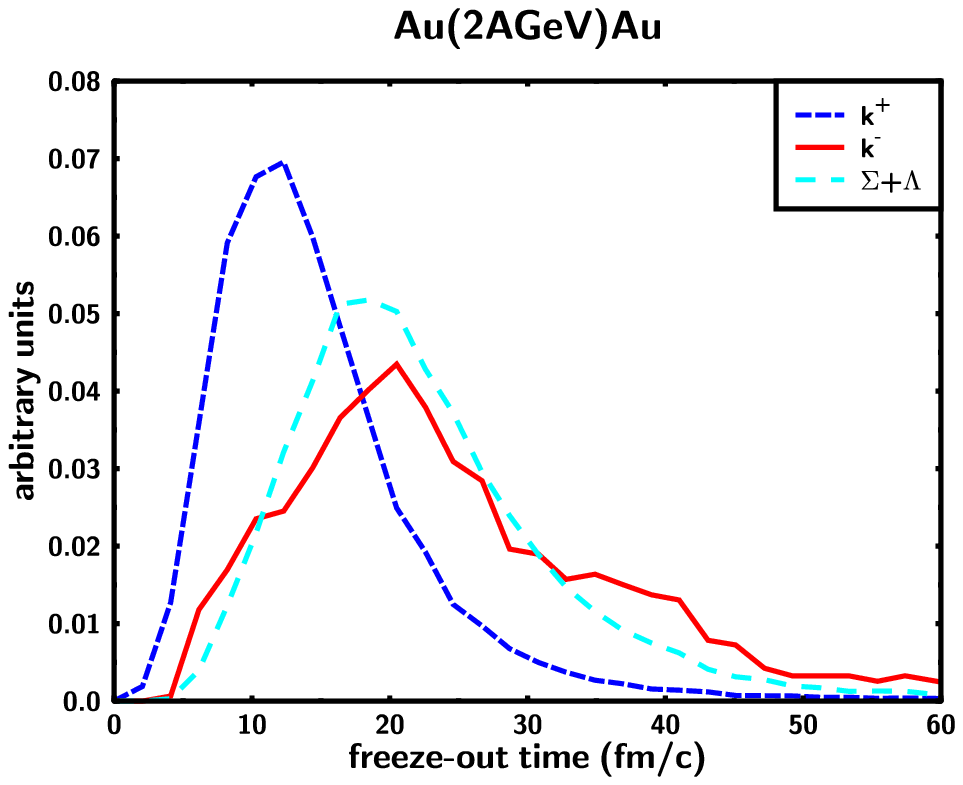,width=9cm}}
\caption{\label{fzksig} 
Normalized freeze-out time distributions for
$K^+$, $K^-$ and hyperons in UrQMD central Au+Au reactions
at 2 GeV/nucleon. Particles with $\bar s$-quarks freeze out
earlier than those with $s$-quarks.
}
\end{minipage}
\end{figure}
Freeze-out radii are closely correlated to freeze-out times. Although
the latter are definitely not measurable (for the radii, two-particle
correlation techniques may yield similar information) they are
the cleaner observable for a transport model analysis. Freeze-out
radii may be contaminated by low energy particles 
produced close to the geometric center of the reaction during the
late dilute expansion stage -- such particles would have small
freeze-out radii but large freeze-out times.

The mass dependence of freeze-out times is studied in
figure~\ref{fzmes}. It shows the freeze-out time distributions
for $\pi$, $\eta$ and $\omega$ mesons in central Au+Au reactions
at 2 GeV/nucleon. The heavier the meson, the earlier its freeze-out.
The reason for this correlation is that
heavy mesons are solely produced by the decay of massive baryon-resonances
which can only be excited in the early hot and dense reaction zone.
Of course freeze-out times do not only depend on the mass of the
produced particle, but also on its interaction cross section:
Figure~\ref{fzksig} shows freeze-out time distributions for $K^+$,
$K^-$ and hyperons. Despite their large mass difference,
the freeze-out times for hyperons and $K^-$ are
very similar, whereas the $K^+$ freeze out much earlier. This is due
to the different quark content of $K^+$ and $K^-$: Both, $K^-$ and
hyperons contain a $s$-quark and have relatively large hadronic
interaction cross sections. The $K^+$, however, contains a
$\bar s$-quark and has a far lower cross section.

So far, we have only studied the freeze-out behavior of mesons and
hyperons at relatively low bombarding energies around 2 GeV/nucleon.
In that domain, baryon-baryon and meson-baryon interactions are dominant
and available phase-space may strongly influence the freeze-out behavior
of the particles under investigation. Let us now turn to CERN/SPS
energies -- here meson production is so abundant that meson-meson
and meson-baryon interactions dominate over baryon-baryon interaction
and threshold or phase-space restrictions do not anymore apply.

\begin{figure}[thb]
\begin{minipage}[t]{9cm}
\centerline{\epsfig{figure=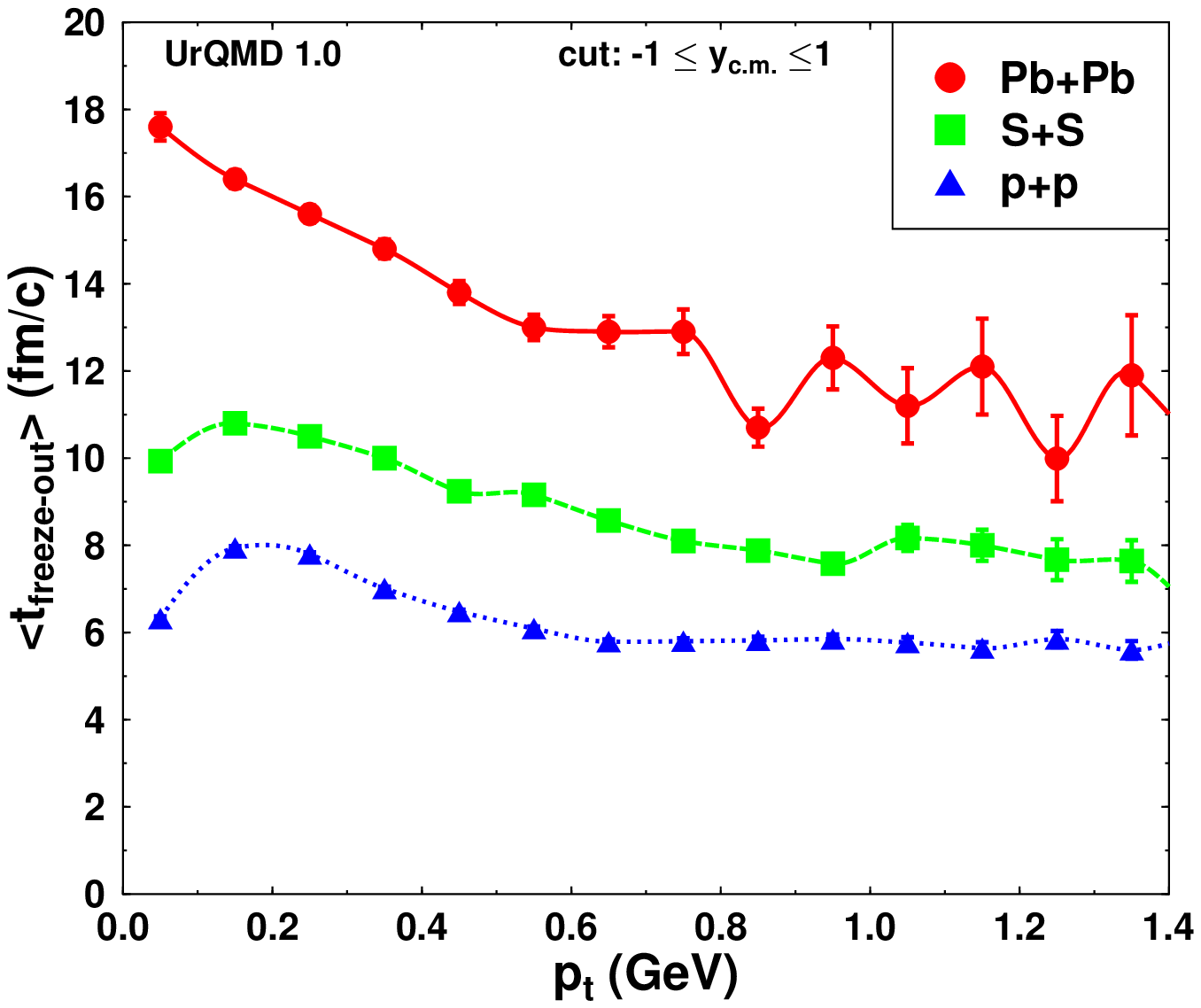,width=9cm}}
\caption{\label{tf-pt} 
Freeze-out time of pions as a function of transverse momentum for
p+p, S+S and Pb+Pb reactions at CERN/SPS energies. For heavy systems
early freeze-out is correlated to high $p_t$.}
\end{minipage}
\hfill
\begin{minipage}[t]{9cm}
\centerline{\epsfig{figure=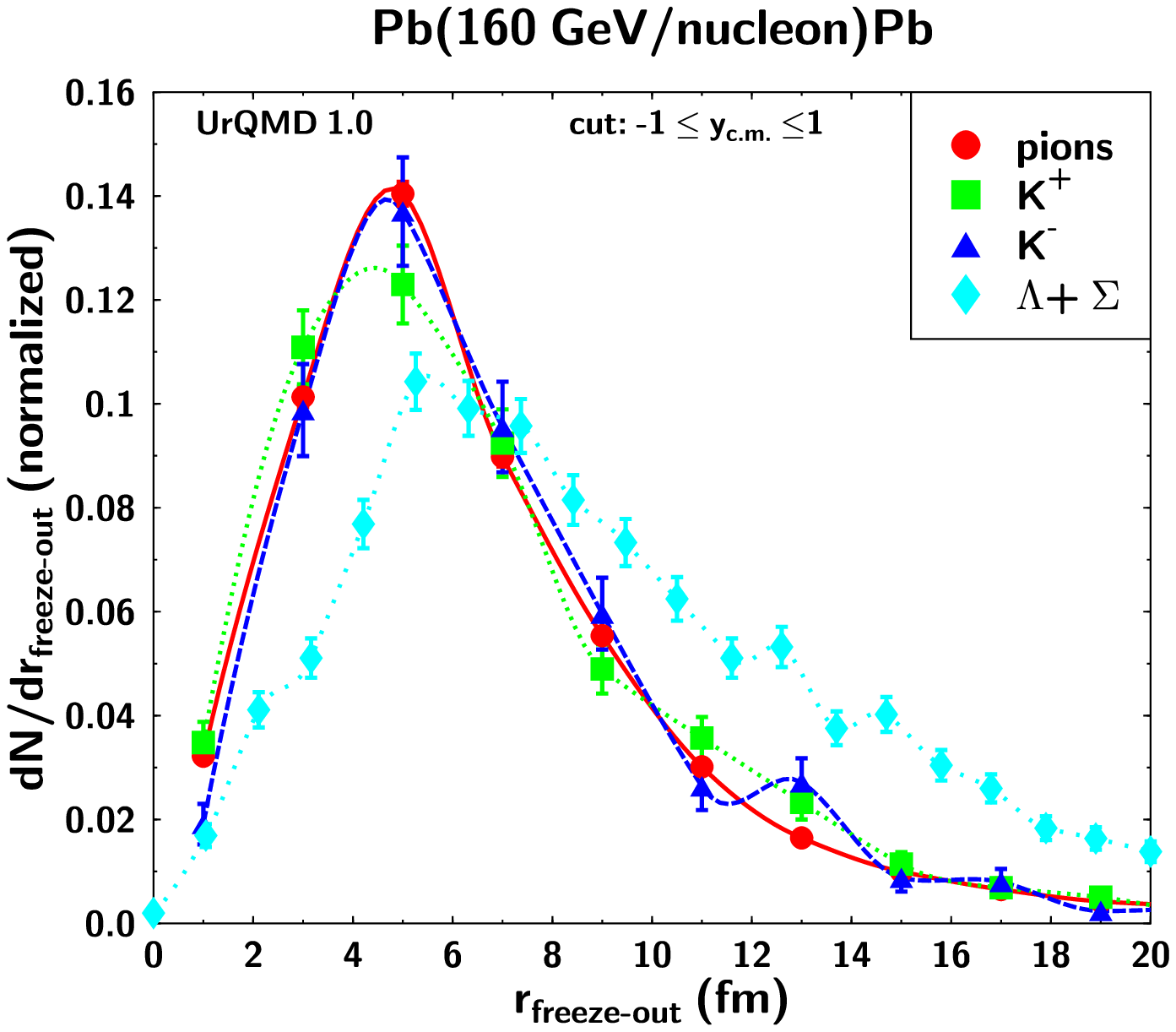,width=9cm}}
\caption{\label{dndrt_pbpb} 
Normalized transverse freeze-out radius distribution 
for pions, kaons, antikaons and hyperons.
Whereas all mesons show similar distributions, the hyperons freeze out 
at larger radii.
}
\end{minipage}
\end{figure}

Unfortunately, neither the freeze-out density, nor the 
freeze-out time is directly observable. However, figure~\ref{tf-pt}
shows that we can establish a correlation between high transverse
momenta and early freeze-out times, at least in heavy colliding systems.
In figure~\ref{tf-pt} the freeze-out time of pions is plotted versus
their transverse momenta for p+p, S+S and Pb+Pb reactions at SPS energies.
Naturally, the proton-proton system does not show any correlation, whereas
in the heavy Pb+Pb system a strong $p_t$-dependence of the freeze-out
time is visible.
Selecting particles with high transverse momenta thus yields a sample
of particles with predominantly early freeze-out times and high
freeze-out densities.

\begin{figure}[thb]
\begin{minipage}[t]{9cm}
\centerline{\epsfig{figure=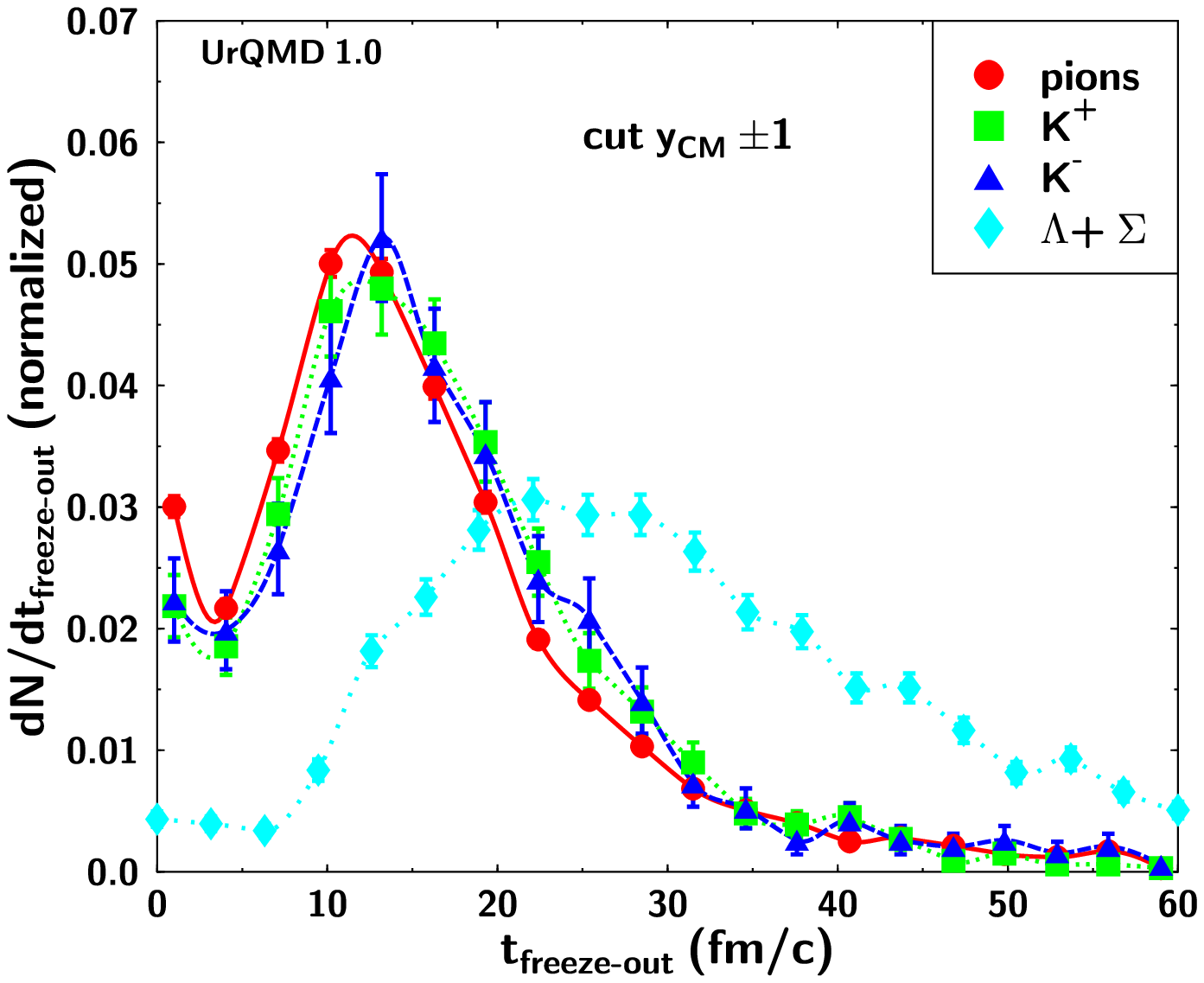,width=9cm}}
\caption{\label{dndtf_pbpb} 
Normalized freeze-out time distribution 
for pions, kaons, antikaons and hyperons.
As with the freeze-out radii, the times for the meson species are very
similar. The hyperons again show a different behavior.
}
\end{minipage}
\hfill
\begin{minipage}[t]{9cm}
\centerline{\psfig{figure=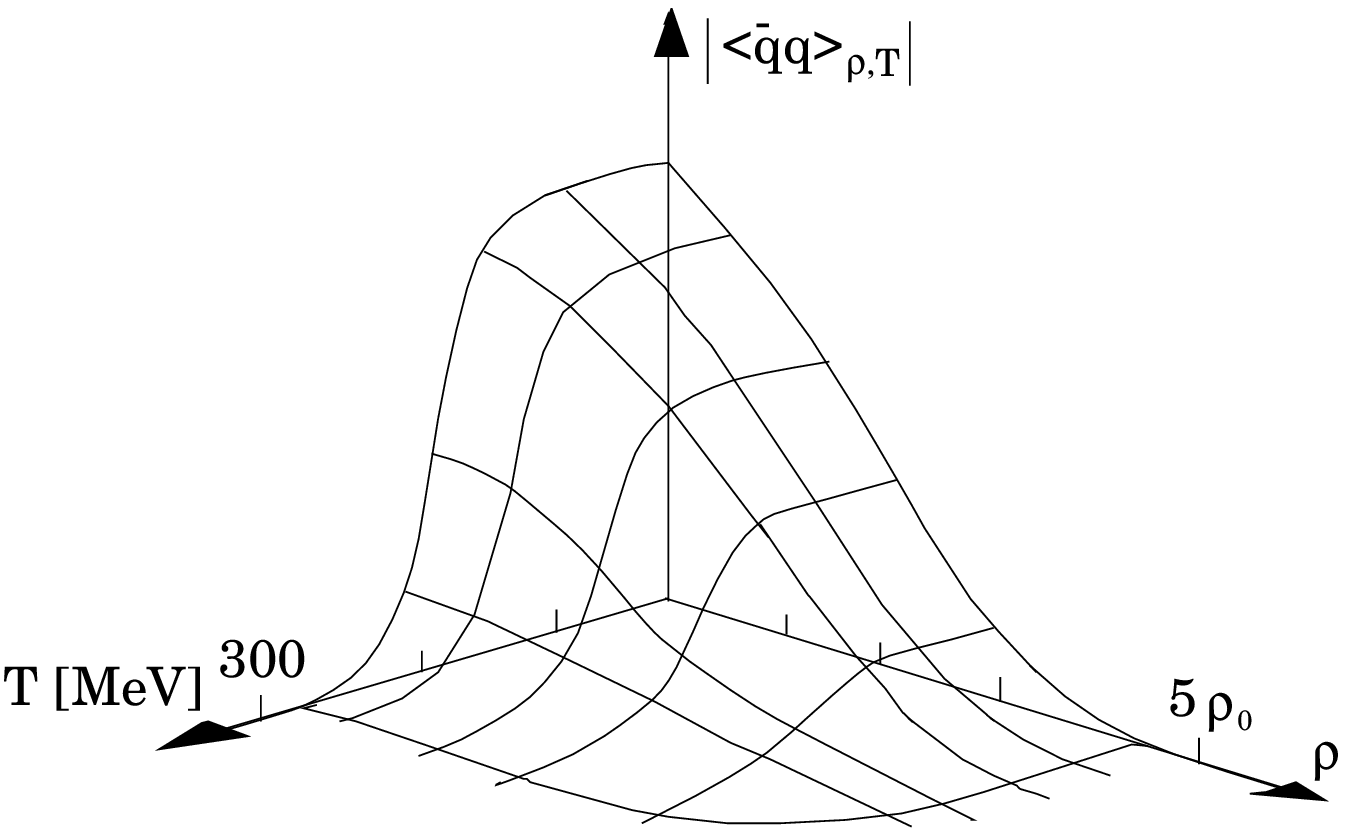,width=9cm}}
\caption{\label{weisebild} quark condensate $-\langle q \bar{q} \rangle$
as a function of temperature $T$ and baryon density $\rho/\rho_0$. The
figure has been adapted from
\protect \cite{weise93a}.}
\end{minipage}
\end{figure}
Do all hadron species exhibit a uniform freeze-out behavior -- or does
each species have its own complicated space-time dependent freeze-out
profile, as observed at SIS-energies?
Figures~\ref{dndrt_pbpb} and~\ref{dndtf_pbpb} show the freeze-out radius
and freeze-out time distributions for pions, kaons, antikaons and hyperons
at mid-rapidity in central Pb+Pb reactions at 160 GeV/nucleon.
The distributions have been normalized in order to compare the shapes
and not the absolute values.
In contrast to the situation at 2 GeV/nucleon, all meson species show
surprisingly the same freeze-out behavior -- the transverse freeze-out radius
and freeze-out time distributions all closely resemble each other.
Only the hyperons show an entirely different freeze-out behavior. 
Whereas the common freeze-out characteristics of the mesons seem to 
hint at a thermalization, the hyperons show that even at SPS energies
there exists no common global freeze-out for all hadron species.

Since the freeze-out distributions have a large width, the average
freeze-out radius clearly does not define a freeze-out volume and
therefore estimates of the reaction volume or energy density based
on average freeze-out radii have to be regarded with great scepticism.

In this section we have investigated freeze-out properties of hadrons
produced in relativistic heavy-ion reactions. At SIS energies each
particle species has its own complex freeze-out characteristics, governed
by interaction cross sections, available phase-space and resonance lifetimes.
At CERN/SPS energies, similar freeze-out distributions for various meson
species hint at a system closer to thermal equilibrium. A comparison
with the hyperon freeze-out distributions, however, indicates
that no global thermalization and freeze-out has been achieved. 
Therefore, thermal model fits to hadron ratios and spectra -- even though
they work pretty well -- may result in a misleading interpretation of
the final state of a relativistic heavy ion reaction.

	\subsection{Dilepton production}

One basic problem of heavy ion physics is
actually $how$
to investigate the hot and dense phase that is
presumably formed in a collision of two
relativistic nuclei. Since hadrons may
interact several times, the intermediate stages
are so to speak ''shadowed'' by the freeze-out
distributions. 
Dileptons are of great interest, because they do not
interact with the hadronic matter. Furthermore, they are
emitted by various mechanisms at all stages of a heavy ion
collision. Thus, the dilepton signal yields time
$integrated$ information on the reaction dynamics.

Dileptons were initially proposed as highly penetrating probes of the
QGP state \cite{feinberg76a,shuryak78a,domokos81a}. They
can be produced via a virtual photon in $q\bar{q}
\rightarrow \gamma^\ast g$ annihilations and in the QCD Compton process $g q
\rightarrow \gamma^\ast q$. If one assumes the QGP to be an
equilibrated gas, then dileptons (as well as real photons)
can probe its thermodynamic conditions \cite{kapusta91a}.
However, due to large background contributions of hadronic
radiation at low masses and the Drell-Yan process
\cite{drell70a} at high masses, dileptons stemming from the QGP 
will possibly  be overshined.

Recent interest has focussed on low mass dileptons. They
are emitted in hadronic decays and collisions. Especially
the Dalitz decays of light mesons and the direct decays of
vector mesons are supposed to contribute significantly
for invariant dilepton masses $M_{ll}<1GeV$. In this mass
region also various bremsstrahlung mechanisms might be important
\cite{mishustin97a,gale87c,haglin94a}
as well
as a number of direct production channels, e.g. $\pi\rho\to
ll\pi$ \cite{murray96a}. Because the vector mesons have rather small
life-times, they can supposedly resolve the rapid changes
during the hot and dense phases of the collision. Their
detection can proceed via low-mass dileptons.

In conjunction with the chiral symmetry restoration 
\cite{weise93a,birse94a,brown96a,weise96a}, the QCD condensates
(e.g. $\langle\bar q q\rangle$) should lower their values
at high temperatures and/or densities.
The dependence of $\langle q \bar{q} \rangle$ on the temperature
$T$  has been studied in the framework of lattice QCD \cite{bernard92a} and
chiral perturbation theory \cite{gerber89a}. Up to 0.7 - 0.8$T_C$,
$\langle q \bar{q} \rangle$ remains nearly constant and then 
its absolute value decreases rapidly (see figure \ref{weisebild}).
The behavior of the quark condensate at finite baryon densities 
is described in a model independent fashion
by the Hellman-Feynman-theorem \cite{drukarev90a}. 
A model calculation of the  dependence of $\langle q \bar{q} \rangle$ 
on both, the baryon density $\rho/\rho_0$ and temperature
$T$, can be seen in figure \ref{weisebild} -- the drop of 
$\langle q \bar{q} \rangle$ with $\rho$ and $T$ is quite analogous
to the temperature and density dependence of 
the nucleon effective mass in the $\sigma - \omega$ model as noted
in \cite{theis83a}.

The connection between the QCD condensates and 
phenomenological hadronic quantities (masses, width etc.)
can be drawn via the QCD sum rule technique
\cite{shifman79a,reinders85a,cohen95a}.
Under simplifying assumptions on the spectral function,
such calculations find lowering vector meson masses as
indicators of chiral symmetry restoration 
\cite{hatsuda92a,hatsuda93a}. More refined calculations,
which try to evaluate the medium contributions to
the self energy of the vector mesons, find
broadening spectral functions in matter
\cite{asakawa92a,herrmann93a,friman97a,rapp97a,klingl97a,peters97a}.

Various experiments have focussed on low mass lepton pairs:
the DLS spectrometer at the BEVALAC
\cite{porter97a,armstrong97b}, the CERES \cite{agakishiev95a,agakishiev97b}
and HELIOS \cite{mazzoni94a,masera95a} experiments
at the SPS at CERN. The 
dilepton spectrometers HADES at SIS (GSI) \cite{stroth95a} and PHENIX at
RHIC (BNL) \cite{gregory94a} are under construction.

\begin{figure}[thb]
\begin{minipage}[t]{7cm}
\centerline{\psfig{figure=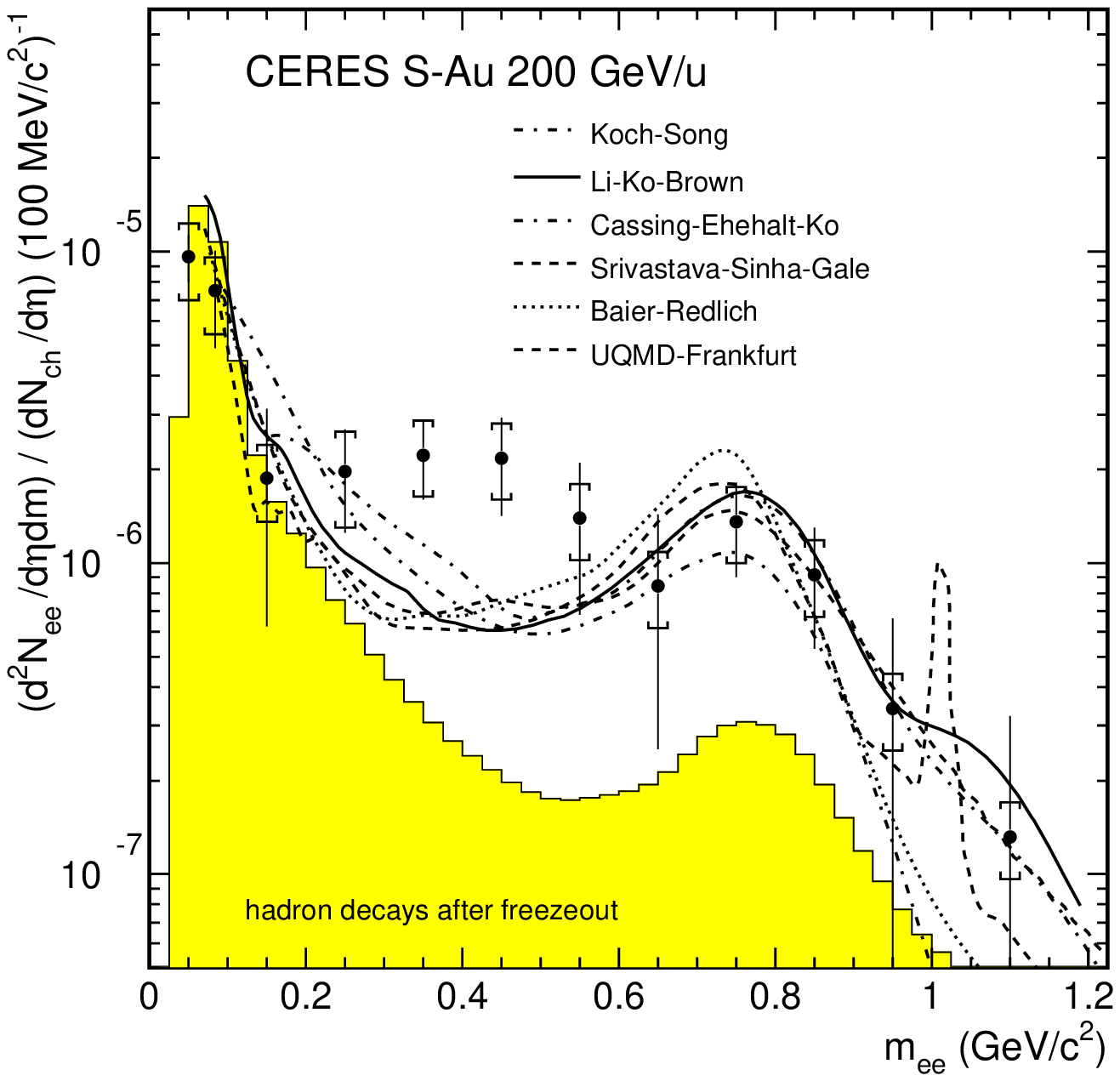,angle=270,width=7.5cm}}
\end{minipage}
\hfill
\begin{minipage}[t]{7cm}
\centerline{\psfig{figure=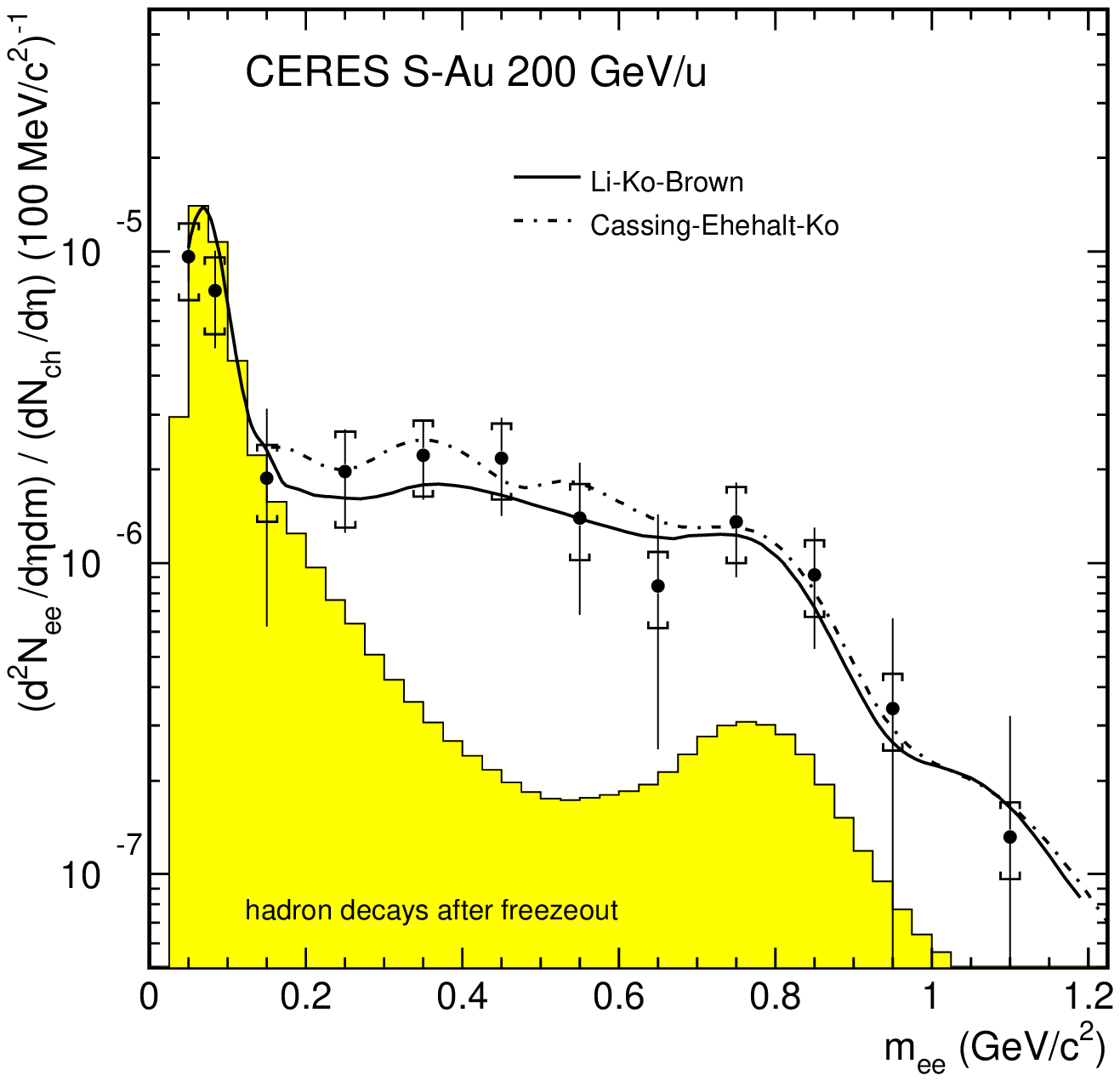,angle=270,width=7.5cm}}
\end{minipage}
\caption{\label{ceresfig1}
Inclusive $e^+e^-$ mass spectra in 200 GeV/nucleon S+Au collisions 
as measured by the CERES collaboration 
\protect \cite{agakishiev95a}. 
The figures have been taken from \protect \cite{drees96a}.
The shaded area depicts hadronic
contributions from resonance decays. The l.h.s. shows a comparison with
calculations based on a purely hadronic scenario 
\protect \cite{winckelmann96a,srivastava96a,cassing95a,ligq95a,koch96a} 
whereas the r.h.s shows calculations including either a QGP phase 
transition or medium dependent vector meson masses
\protect \cite{cassing95a,ligq95a}. }
\end{figure}

Experiments at CERN have reported an enhanced
production of dilepton pairs in nucleus-nucleus
collisions over a broad invariant mass region
around $M\sim 0.5~GeV$ \cite{drees96a}.
The enhancement is relative to the known sources as
measured in $pp$ or $pA$ collisions after scaling
to the nucleus-nucleus case (see l.h.s. of fig. \ref{ceresfig1}). 
The data have been however reproduced by
assuming density dependent masses of vector mesons 
as a consequence of partial chiral symmetry restoration 
\cite{cassing95a,cassing96a,ligq95a,ko96a,winckelmann96a} 
(r.h.s. of fig. \ref{ceresfig1}).

However, even bare hadronic transport model calculations, without
any mass shift included,  miss only the data in the 400 to 600 MeV bins 
(by 2 to 3 standard deviations) \cite{ko96a,winckelmann96a}. 
Since hadronic transport models did -- so far -- neglect some 
contributions, e.g. from bremsstrahlung, it has yet to be determined
whether partial
restoration of chiral symmetry is the only possible explanation of these
interesting new data. Calculations evaluating in-medium spectral functions,
due to the coupling of the $\rho$ with nucleon resonances and
particle-hole excitations, also achieve a satisfactory reproduction
of the CERES data \cite{rapp97a}, without requiring a dropping
$\rho$-mass. One has to conclude that, up to now,
comparisons of dilepton spectra with
hadronic models do not give evidence for (partial) restoration of chiral
symmetry \cite{wambach98a}.
Another, up to now neglected contribution to the low mass dilepton spectrum in
nucleus-nucleus collisions, are secondary Drell-Yan processes, which are
described in greater detail in the next sub-section.

\begin{figure}[hbt]
\vspace*{-0.5cm}
\centerline{\hbox{\psfig{figure=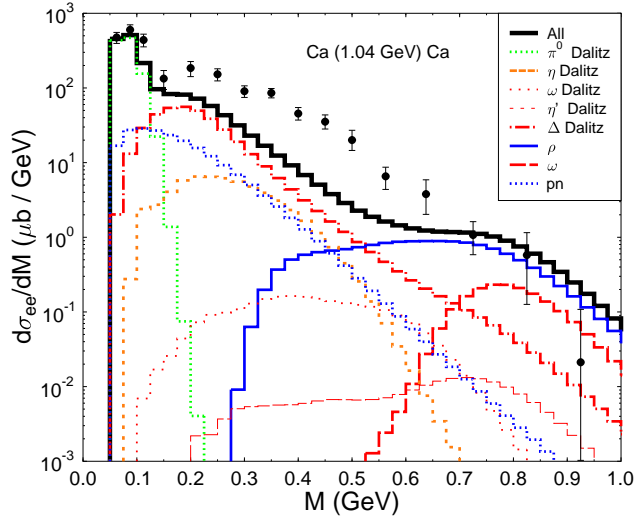,width=9cm}}}
\vspace*{-0.5cm}
\caption{UrQMD dilepton cocktail plot for Ca+Ca in
comparison to recent DLS data \cite{porter97a}.
The upper solid curve is the sum of all contributions.
The detector acceptance has been modeled with
the DLS Filter 4.1 and a mass resolution of 10\%
is adopted.}
\label{ernstfig}
\end{figure}

The production of low mass dileptons
is included in the UrQMD model via Dalitz decays of
$\pi^0,\eta,\omega,\eta'$ mesons and of the
$\Delta(1232)$ resonance, direct decays of
neutral vector
mesons and incoherent $pn$ bremsstrahlung \cite{ernst98b}.
This approach has been used to analyze
data of the DLS collaboration at the BEVALAC
accelerator \cite{porter97a,armstrong97b}. 
The recent AA data all show a large
enhancement as compared to model
calculations (see fig. \ref{ernstfig}).
The $pd$ spectra imply a high $pn\to\eta X$ cross
section, but TAPS data delimit this as an
explanation for the enhancement.
So far, these data resist any explanation by
medium-dependent spectral functions 
\cite{ernst98b,bratko97a}.
Note that the recent data strongly exceed earlier
published measurements of the same collaboration
\cite{roche89a}
which have been
reproduced by various transport models 
\cite{wolf90a,xiong90a,bratko96b,bratko97a} and are also favored by 
the UrQMD model. A satisfying explanation of the recent DLS data is still
missing.

\subsubsection{Drell-Yan pairs from secondary collisions}

\begin{figure}[htb]
\centerline{\psfig{figure=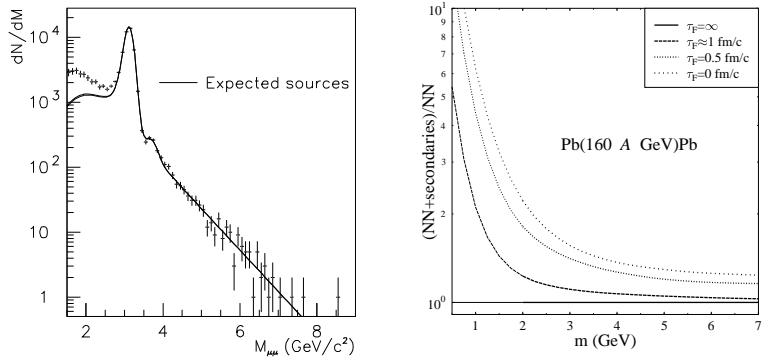,width=12cm}}
\caption{\label{scomp} 
Left: Comparison of measured dimuon yield with expected sources in Pb+Pb
collisions. Figure taken from \protect \cite{abreu96a}.
Right: Calculation of secondary Drell-Yan processes based on a UrQMD
simulation. Shown is the ratio between the total dilepton mass spectrum and the
  nucleon-nucleon scattering alone, at $y_{\rm cms}=0.5$,
  for Pb+Pb at 160~$A$~GeV
  assuming different hadronic formation times $\tau_F$
(Figure taken from \protect\cite{spieles97c}.}
\end{figure}

Reports on intermediate mass muon pairs 
\cite{lourenco94a,mazzoni94a,abreu96a}
 produced in heavy ion collisions have recently
attracted much attention.  
The measured  spectrum is enhanced relative to the
expected `cocktail' dilepton sources (Fig.~\ref{scomp}, left), one of them
being the Drell-Yan pair production \cite{drell70a}.

Rate estimates of this process (e.g. \cite{vogt92a}) are essentially based on
extrapolations of p+A reactions where a linear scaling with A is
observed for the Drell-Yan production cross section~\cite{alde90a}: 
\[\sigma_{\rm pA}= \sigma_{\rm 0}~\rm A \, . \]  This linear scaling can be
understood in the Glauber picture of the hadron-nucleus cross section,
constructed at high energies using the AGK cutting
rules~\cite{abramovskij73a}. 

But also the hot matter produced in ultrarelativistic
heavy ion collisions contributes to the total dilepton radiation 
\cite{kapusta92a,kaempfer92a,dumitru93b,winckelmann95a,geiger93b,gavin96aa}.
It is shown that emission from thermal sources such as a quark-gluon plasma or 
hadron gas, whether in or out of equilibrium, cannot be neglected
relative to nucleon-nucleon contributions to the Drell-Yan process for masses
below the J/$\psi$ peak --- at least at higher energies than those presently
available.  

Another source of dileptons which may already play a role at SPS energies
and could account for the observed intermediate mass dilepton enhancement
has been proposed in \cite{spieles97c}:
Drell-Yan production by interactions involving produced, or secondary,
hadrons.  
In order to investigate secondary dilepton production at SPS
energies UrQMD has been employed to
obtain a realistic collision spectrum of secondary hadrons.  The
differential Drell-Yan cross section is computed at leading order (LO)
using the standard equation~\cite{halzen78a}:
\begin{eqnarray}
  \frac{{\rm d}^2\sigma}{{\rm d}m^2{\rm d}y}({\rm AB}\rightarrow l\bar l X)=
  \frac{4\pi\alpha^2}{9m^2s}\sum_q e_q^2 \left[
  q^{\rm A}(x_{\rm A},m^2) {\overline q}^{\rm B}(x_{\rm B},m^2) +
  {\overline q}^{\rm A}(x_{\rm A},m^2) q^{\rm B}(x_{\rm B},m^2)\right] \; ,
\end{eqnarray}
where $q(x,m^2)$ and $\overline q(x,m^2)$ denote the quark and antiquark
densities (according to \cite{glueck92a,glueck95a}), $\sqrt{s}$ is the 
center-of-mass energy
of the colliding hadrons, $m$ is the invariant mass of the
lepton pair, $x_{\rm A}=\sqrt{\tau}~e^y$ and $x_{\rm B}=\sqrt{\tau}~e^{-y}$
with $\tau=m^2/s$, and $y$ is the dilepton rapidity in the \emph{cms}
frame.  
The lepton pair production cross section is
calculated for each hadron-hadron collision and weighted by the
inverse of the total hadron-hadron cross section. The distributions
from these elementary hh-collisions are then summed.
It is clear that in pion-nucleon
collisions, \emph{valence} quark-antiquark annihilation can play a significant
role in the Drell-Yan process. 
Pion-nucleon dilepton production cross
sections are consequently higher than
nucleon-nucleon cross sections, especially when $m/\sqrt{s} \stackrel{>}{\sim}
0.1$.
It is shown that these secondary collisions can serve as an important source
of $m\sim 2$~GeV dileptons due to the availability of valence antiquarks
in mesons and antibaryons (Fig.~\ref{scomp} right). 

The standard Drell-Yan process corresponds to the interaction of fully
formed hadrons.  However, it was shown \cite{kharzeev96e,satz96a,kharzeev97a}
that, during the early stages of the system evolution, partons
can scatter and annihilate before they have come on mass-shell.
To estimate the importance of these ``primordial'' or ``pre-resonance'' 
$q\bar q$ annihilations, the contribution of such
processes have been calculated assuming that the asymptotic parton distribution functions
are also valid for the primordial states.  
This is done very simply in the
UrQMD calculation by decreasing the formation time of
the produced hadrons within string excitations, $\tau_F$, 
from the ``default'' value of around 1~fm/$c$.

The importance of this primordial (pre-resonance) contribution to the dilepton 
mass spectra is shown in Fig.~\ref{scomp} (right).  
The secondary dilepton yield for $\tau_F=0$increases by a factor of $\sim
5$ at all masses compared to the calculations with the default
$\tau_F\approx 1$~fm/$c$.  
With a value of $\tau_F=0.5$~fm/$c$
the enhancement in the range $1.5<m<2.5$ GeV shows quite good
agreement with the mass data which are about a factor of two higher than
naively expected \cite{abreu96a,lourenco94a,mazzoni94a}.

	\subsection{Charmonium production and suppression}

\begin{figure}
\vspace*{-5cm}
\centerline{\psfig{figure=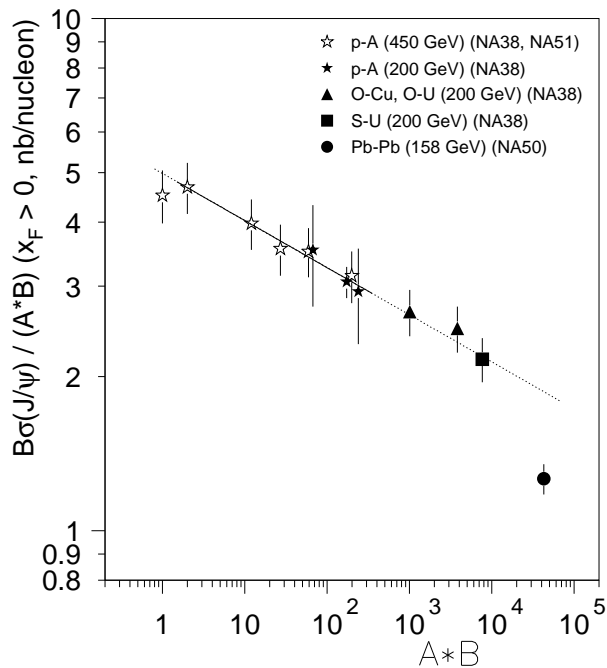,width=10cm}}
\caption{\label{lorencof} $J/\psi$ production cross sections versus $A\times
B$. Figure taken from \protect \cite{lourenco96a}. }  
\end{figure}

$J/\Psi$'s, $\chi$'s and $\Psi'$'s, consisting of a $c\bar{c}$ pair, are
formed in pp-collisions by the fusion of two gluons, or  to a larger
extend by the annihilation
of a quark-antiquark pair  \cite{baier83a}. 
In a QGP of sufficiently high density and temperature $c\bar{c}$ pairs
cannot bind together --
the presence of gluons and quarks in the plasma causes the color
charge of quarks to be screened. Alternatively, hard gluons may be able to
dissociate the $c\bar c$ pair (QCD photo-effect) in the QGP.
Suppression of quarkonia-mesons has been suggested as a 
possible signature for deconfinement \cite{matsui86a}.
Lattice SU(3) gauge theory \cite{degrand86a,kanaya86a} indicates that the 
necessary conditions for the screening are 
already fulfilled shortly above $T_C$ (the screening length $\lambda_D$ is
inversely proportional to the temperature).
Inclusion of dynamical fermions decreases the screening 
length \cite{karsch88a,blaschke91a}. 

However, the suppression of quarkonia meson production is no unambiguous 
signal for deconfinement. They are no weakly interacting probes which
can traverse hadronic matter unhindered. On the contrary, heavy quarkonia
mesons suffer from low production cross sections and rather large
hadronic dissociation cross sections. I.e. charmonium is already suppressed
in pA-collisions, where the conditions for a QGP are not fulfilled.
Figure~\ref{lorencof} depicts the suppression of J/$\Psi$'s in pA- and 
AB-collisions.
The line shows the suppression due to the nuclear constituents of the projectile
and target nuclei. One can see that this effect reproduces the data from pp- up
to SU-collisions, but to explain the new lead on lead data one needs additional
suppression.
Purely hadronic scenarios have been suggested
\cite{gavin90a,gavin88a,vogt88a,gavin96a,gavin96b}, 
which can explain $J/\Psi$ and $\Psi'$
suppression up to PbPb-collisions without the need for a deconfined phase.
The most successful 
hadronic scenarios for $J/\Psi$ and $\Psi'$ suppression explicitly take 
interactions
with  produced particles into account and are referred to as
{\em comover models} - the {\em comovers} being the produced mesons which 
are energetic enough to destroy a $J/\Psi$ in the case of a collision.  
For the $\Psi'$, all mesons can be {\em comovers}, but there is a threshold 
for the reaction $J/\Psi\, +\, X \rightarrow D\bar D$, because 
$2m_D-m_{J/\Psi}\approx 640$ MeV. So, "slow" pions (in the $J/\Psi$ rest frame) 
cannot destroy a $J/\Psi$. 

Hadronic transport
model calculations which incorporate the full collision dynamics and
go far beyond the commonly used simplified version of the Glauber theory  
yield conflicting results
\cite{cassing96b,gerland97a}. 
These transport model calculations
are very sensitive to certain input
parameters such as the formation time of the $J/\Psi$ and the comovers. The
HSD transport model \cite{cassing96b} can fully reproduce the NA50 lead
data while assuming a fixed formation time of 0.7--0.8 fm/c for both,
$J/\Psi$ and comovers. A second calculation with the HSD model
using a formation time of 0 fm/c for the $J/\Psi$ reproduces also the 
data~\cite{cassing97a}.
The UrQMD model, however, uses for the comovers 
a variable formation time  emerging from
the Lund string fragmentation formalism (here the formation time 
depends on the hadron mass) and zero formation time for the $J/\Psi$.
The assumption of zero formation time is valid if the $J/\Psi$ is
considered as a pre-resonance $c \bar{c}_8 - g$ state with a hadronic
dissociation cross section of 7 mb.
However, the UrQMD model does not reproduce the additional suppression 
of the Pb+Pb experiment \cite{gerland97a}.  
The question of formation time might be a central issue since in the
color octet model the dissociation cross section is actually higher during
the lifetime of the pre-resonance $c \bar{c}_8 - g$ state 
\cite{braaten96a,kharzeev96b} than after
hadronization. Furthermore the amount of comover-charmonium interaction
will crucially depend on the formation time of the comovers.

One has to bear in mind, however, that hadronic transport models
do not contain partonic degrees of freedom explicitly and are therefore 
incapable of correctly describing hard processes like charm production or
the Drell-Yan process.
From the measurements of Drell-Yan lepton pairs
we know that the cross section in pA-collisions is proportional to $A$,
as predicted by the Glauber model.
Hadronic models on the other hand take into account the energy loss of
incident nucleons due to the production of secondaries (which then could 
play an important role as comovers). If, however, the elementary hard
scattering cross section is strongly energy dependent -- as the charm
production at SPS energies and below -- the effect of nuclear stopping leads
to a considerable underestimation of the AB cross section.
So, at the moment
it is not clear, how it is possible to include hard 
processes consistently in a hadronic transport model.

Another point under discussion is whether charmonium states might be created
as a pre-resonance $c \bar{c}_8 - g$ state \cite{kharzeev97a}, which is not an 
eigenstate of the QCD Hamiltonian.
Thus, one would expect a time evolution of this state and its cross section
until it is a physical state, i.e a $J/\Psi$. The pre-resonance should evolve   
during the formation time which would lead to different effective absorption
cross sections in small and heavy nuclei.

To summarize this section, we can say that the suppression of charmonium
states in heavy ion collisions and even in proton-nucleus is not yet 
understood.
Although, the assumption of a pre-resonance state with a constant cross section
seems to reproduce the data, it contradicts principle ideas of quantum
mechanics.

	\section{Collective flow}
	\label{flow}

The excitation function of transverse collective flow is 
the earliest predicted signature for probing the formation of 
compressed nuclear matter
\cite{scheid68a,scheid74a}.
It has been shown that the excitation function of flow is sensitive
to the EoS and can be used to search for abnormal matter states 
and phase transitions \cite{hofmann76a}.

In the fluid dynamical approach the transverse collective flow is directly
linked to the pressure of the matter  in the reaction
zone. With $P(\rho, S)$ being the pressure (depending on the density $\rho$
and the entropy $S$) the generated collective transverse momentum can be written
as an integral of the pressure over surface and time \cite{stoecker81a}:
\begin{equation}
\label{perpeqn}
P_{\perp} \,=\, \int_t \int_A P(\rho,S) \, {\rm d}A \, {\rm d}t \, ,
\end{equation}
where d$A$ represents the surface element between the participant and
spectator matters and the total pressure is the sum of the
potential pressure and the kinetic pressure:
The transverse collective flow depends directly on the equation of state,
$P(\rho,S)$.

Collective flow has been  predicted by nuclear fluid dynamics (NFD) 
\cite{scheid68a,scheid74a,stoecker80a,stoecker82a,buchwald84a,amelin92a}. 
It is well established 
experimentally at
the BEVALAC \cite{gustafsson84a,doss86a,gutbrod89} 
for charged particles by the Plastic-Ball
and streamer chamber  collaborations and at GSI by the FOPI collaboration
\cite{herrmann96a}. Microscopic models such as VUU
(Vlasov-Uehling-Uhlenbeck) and QMD (Quantum Molecular Dynamics) have
predicted smaller flow than NFD. These models, however, show good 
agreement with the experimental findings 
\cite{molitoris84a,hartnack89a,schmidt93a,hartnack92a,molitoris85a}.
One has to distinguish between different signatures of collective flow:
The undirected {\em radial} flow which is best observed in ultracentral
collisions and
the {\em bounce--off} \cite{stoecker80a} of compressed matter 
{\em in the reaction plane} as well as the 
{\em squeeze--out} \cite{stoecker82a} of the participant matter 
{\em out of the reaction plane} in semi-central to semi-peripheral collisions.

The most strongly stopped, compressed matter
around mid-rapidity is seen directly in the {\em squeeze--out} 
\cite{hartnack90a}.
A strong dependence of these collective effects  
on the nuclear equation of state
is predicted \cite{hartnack92a}. For higher beam energies, however,
projectile
and target spectator decouple quickly from the reaction zone, giving
way to a preferential emission of matter in the reaction plane, 
even at mid-rapidity \cite{ollitrault93a}.

\begin{figure}[thb]
\begin{minipage}[t]{9cm}
\centerline{\psfig{figure=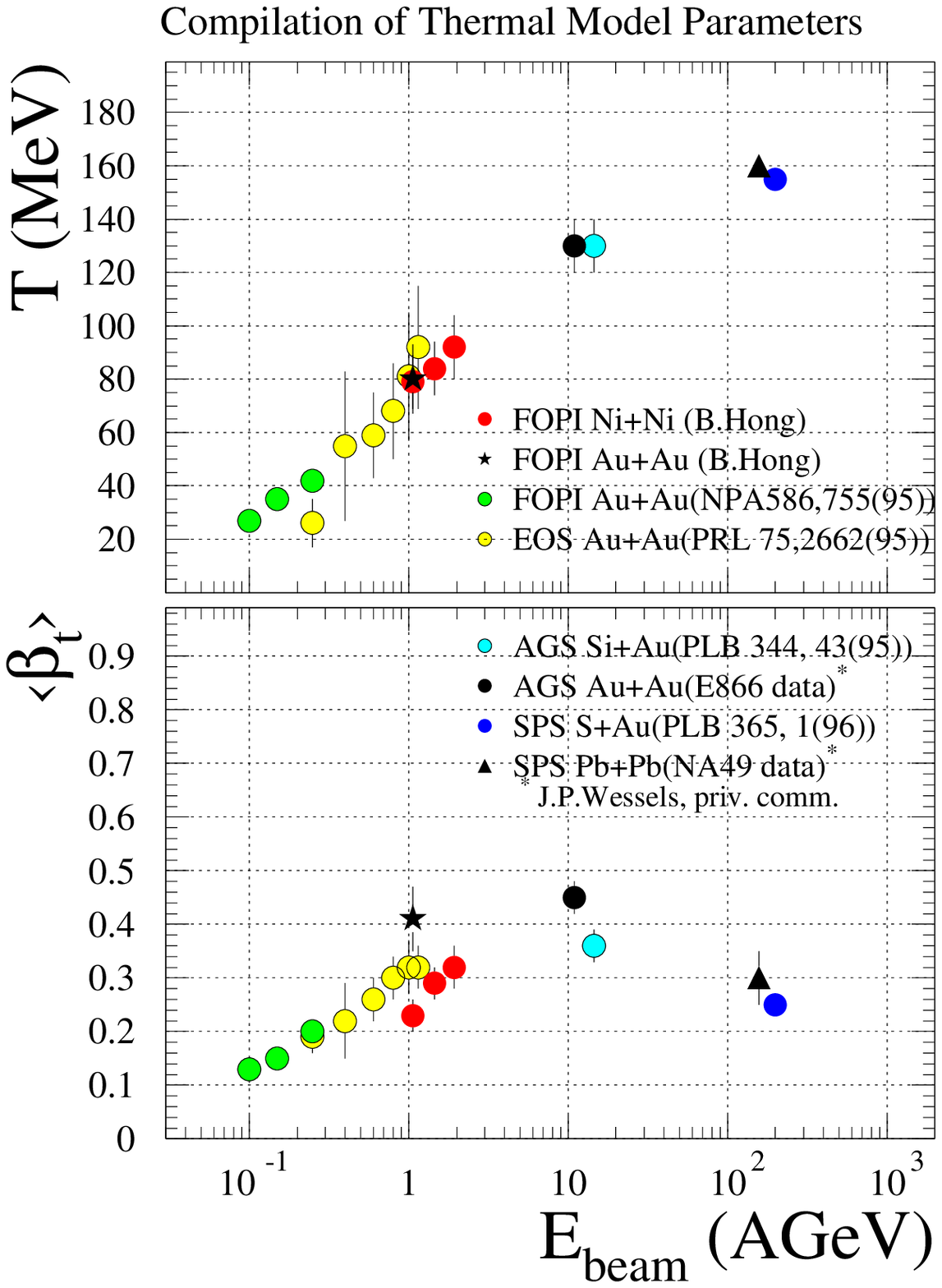,width=12cm}}
\end{minipage}
\hfill
\begin{minipage}[t]{9cm}
\centerline{\psfig{figure=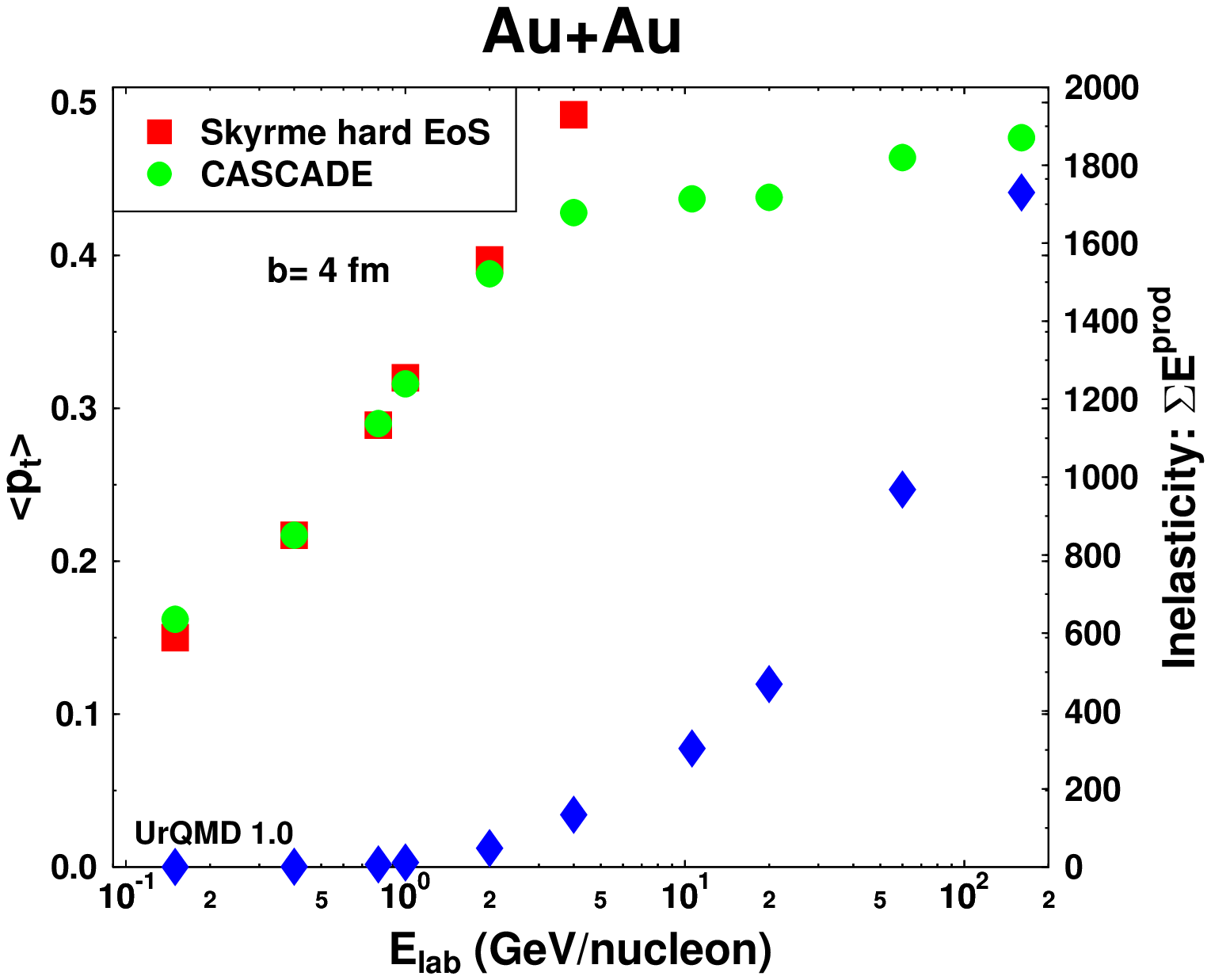,width=9cm}}
\end{minipage}
\caption{\label{spekfig3}Left: Excitation function of temperature T and
average transverse expansion velocity $\beta_t$. The figure
has been taken from \protect \cite{herrmann96a}. The data are from
\cite{hong98a,poggi95a,lisa95a,braun-munzinger95a,braun-munzinger96a}.
Right: \label{ptinel} Excitation function of the mean transverse momentum
(squares and circles, left ordinate)
in comparison with the total energy of produced particles (diamonds, right
ordinate). Shown are the UrQMD
results for Au+Au collisions with impact parameter $b=4$~fm.}
\end{figure}

Apart from the above discussed {\em directed} flow, 
the so-called ``radial'', i.e.
undirected, flow component can be used for simplicity (spherical symmetry)
\cite{siemens79a,stoecker81b}. 
It changes drastically
the  interpretation of particle spectra used for temperature extraction
which may drop by as much as a factor of 2 
(see the l.h.s. of figure~\ref{spekfig3}).
The mean (undirected) transverse velocity ($\langle \beta_t \rangle$) 
or momentum ($\langle p_t \rangle$) can be used as a measure for this
undirected flow in central collisions. The r.h.s. of figure~\ref{spekfig3}
shows a UrQMD excitation function for $\langle p_t \rangle$. Up to
an incident beam energy of 2~GeV/nucleon, $\langle p_t \rangle$ rises
steadily and then almost saturates. The inclusion of potentials does 
not play a major role for this observable. The saturation at 2~GeV/nucleon
(which is also observed experimentally, see the l.h.s. of the figure)
can be explained in the following scenario: Up to 2~GeV/nucleon the
longitudinal momentum of the incoming nucleons is predominantly transferred
into transverse degrees of freedom; particle production only accounts for
a minor fraction of the total energy of the system at the end of the reaction.
Above 2~GeV/nucleon, however, particle production gets more important and
finally most of the energy of the final state is stored in produced particles,
leading to a saturation in the amount of energy carried by the 
``primordial'' constituents of the collision system, the nucleons.
The increase of energy stored in produced particles can be seen
on the r.h.s. of figure~\ref{spekfig3} - here the excitation function for 
total energy of all produced particles (diamonds) is plotted into the
same frame as the $\langle p_t \rangle$ excitation function.

\begin{figure}[b]
\vspace*{\fill}
\centerline{\psfig{figure=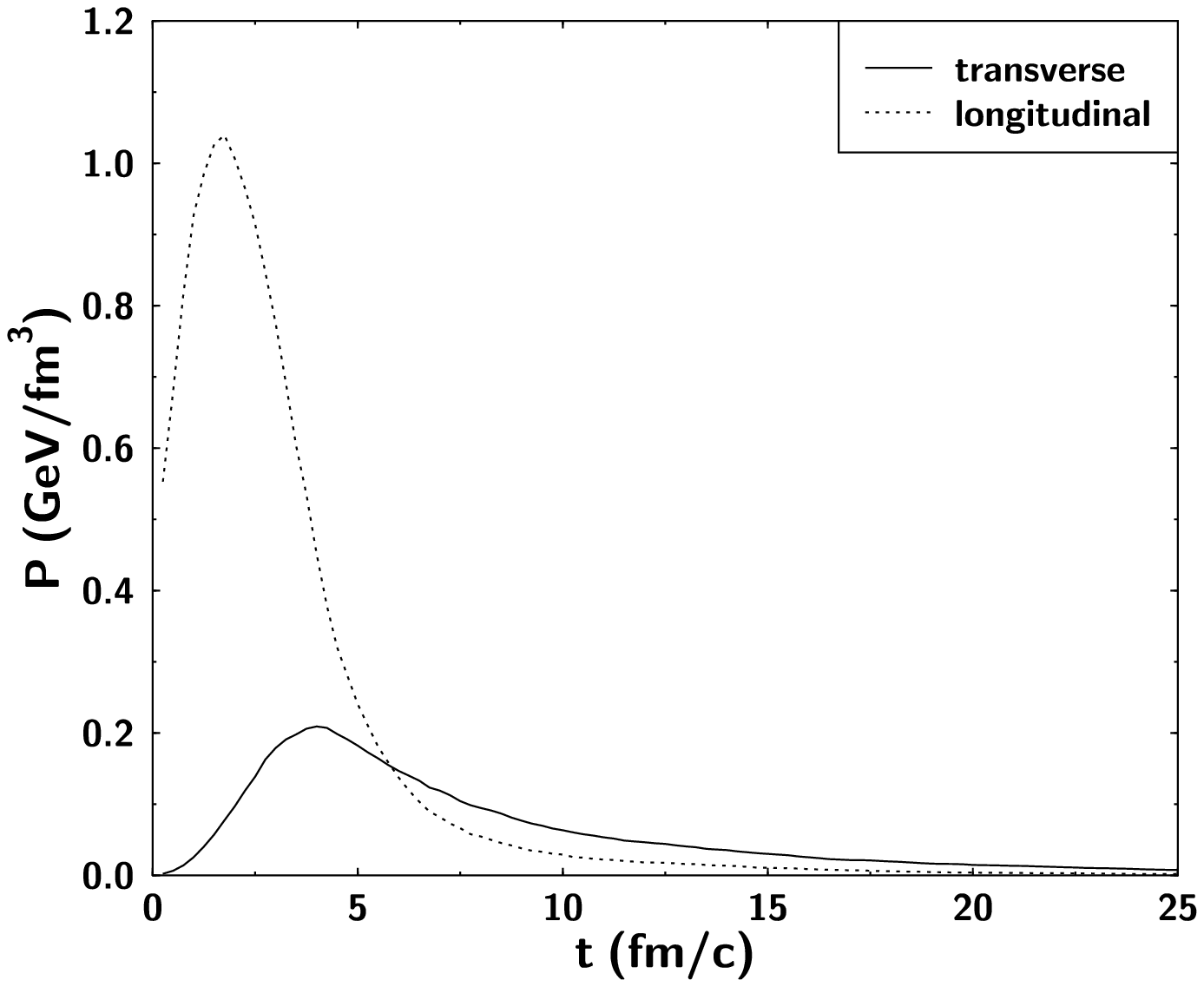,width=12cm}}
\caption{Time evolution of the longitudinal and the transverse pressure in
the central region of a Au(11GeV)Au collision (b=0) calculated with UrQMD.
\label{press}}
\vspace*{\fill}
\end{figure}

The origin of nuclear flow in ultrarelativistic heavy
ion collisions can be seen in
Fig.~\ref{press}, where the time evolution of the longitudinal and the
transverse pressure in the central region of Au(11GeV)Au (b=0), 
simulated with UrQMD, is depicted. A cylindrical volume of $r=6$~fm and
$\Delta z=2$~fm in the center of the reaction is considered. The pressure is
calculated according to the virial theorem:
\[
V P^i=\frac{1}{\rm N_{events}}\sum_{\rm n=1}^{\rm N_{events}} 
\sum_{\rm k=1}^{\rm N_{particles}} p^i_k p^i_k/p^0_k \quad ,
\]
where $i=x,y,z$ denotes the three spatial directions. 
The contribution of this kinetic part and a part from quasi-potentials have 
been studied separately in the framework of RQMD \cite{sorge97a}. 
It takes about 5~fm/c until the longitudinal pressure reaches its maximum
value. The longitudinal pressure has decreased considerable at that time.
Mainly not due to scattering and momentum transfer but due to the particles
that simply have passed the central region and are therefore not considered
any more. If {\em all} particles are taken into account, the ratio of 
transverse to longitudinal pressure does not exceed 0.5. This is in line
with the observation of
larger longitudinal than transverse flow after freeze-out.

		\subsection{Bounce--off: collective flow in the reaction plane}
		\label{flowsection}

Due to its direct dependence on the EoS, $P(\rho,T)$, flow excitation
functions can provide unique information about phase transitions:
The formation of abnormal nuclear matter, e.g., yields a reduction 
of the collective flow \cite{hofmann76a}.
A directed flow excitation function as
signature of the phase transition into the QGP has been proposed 
by several authors \cite{stoecker86a,amelin92a}.
A microscopic analysis showed that the existence of
a first order phase transition 
can show up as a reduction in the directed transverse
flow \cite{hartnack90a}.

\begin{figure}[thb]
\centerline{\epsfig{figure=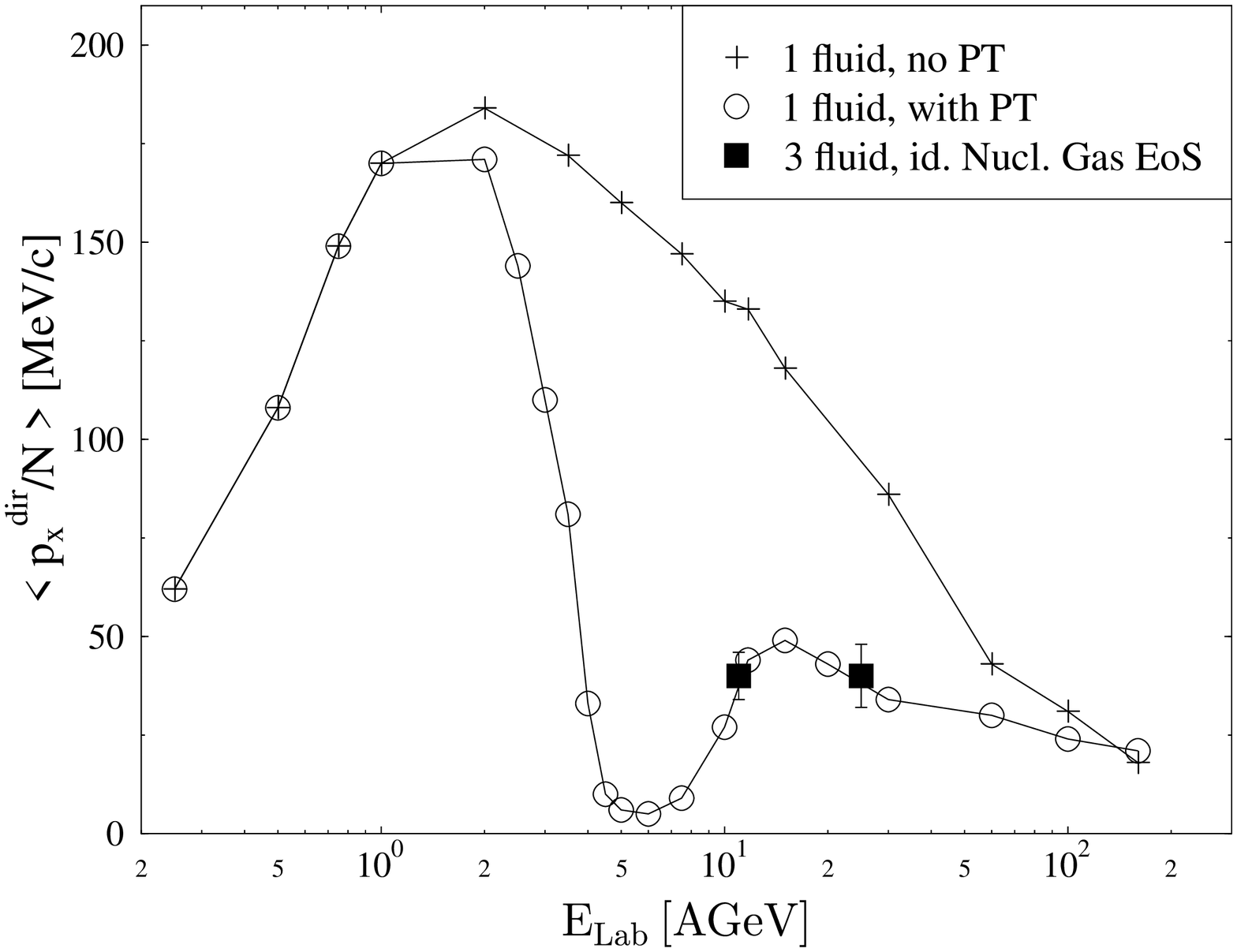,width=12cm}}
\caption{ \label{flowfig1} Excitation 
function of directed transverse flow, calculated in the
framework of  nuclear hydrodynamics 
\protect \cite{rischke95b,rischke96a}, 
with and
without deconfinement phase transition. In the case of a phase 
transition a minimum in the excitation function is clearly visible. Solid
squares correspond to a 3-fluid model calculation with ideal nucleon gas EoS.}
\end{figure}

For first order phase transitions, the pressure remains constant
in the region of the phase coexistence. This results in a vanishing 
velocity of sound $c_s= \sqrt{\partial p/\partial \varepsilon}$.
The expansion of the system is driven by the pressure gradients, therefore
expansion depends crucially on $c_s^2$. Matter in the  mixed phase
expands less rapidly than a hadron gas at the same energy density
and entropy.
In case of rapid changes in the EoS without phase transition, the
pressure gradients are finite, but still smaller than for an ideal gas
EoS, and therefore the system expands more slowly \cite{kapusta85,gersdorf87}.

This reduction of $c_s^2$ in the transition region is commonly referred to
as {\em softening} of the EoS. The respective region of energy densities
has been called the 
{\em soft region} \cite{hung95a,rischke95a,rischke96a,rischke96b}.
Here the
flow will temporarily slow down (or possibly even stall).
Consequently a {\em time delay} is expected in the expansion
of the system.
This prevents the deflection of spectator matter
(the {\em bounce--off}) and, therefore, causes 
a reduction of the directed transverse 
flow \cite{bravina94a,rischke95b}.
The softening of the EoS should be observable in the excitation function of the
transverse directed flow of baryons (see figure \ref{flowfig1}).

The overall decrease of $p_x$ seen in Fig. \ref{flowfig1} 
for $E_{lab} > 2$ AGeV both for
the hadronic and the QGP equation of
state demonstrates that faster spectators are less easily
deflected (because $A$ and $t$ in equation~\ref{perpeqn} are decreasing
with $E_{lab}$) by the hot, expanding participant matter. 
For the QGP equation of state, however, these one-fluid calculations
show a {\em local minimum} in the excitation function, at about
6~GeV/nucleon. 
This can be related to  the
QGP phase transition, i.e. to the existence of the 
{\em soft region} in the EoS.

However, one-fluid hydrodynamic calculations assume instantaneous
thermalization. This becomes unrealistic for increasing beam energies since
due to the average rapidity loss of only one unit per proton-proton 
collision, nucleons require several collisions for thermalization.
A more realistic three-fluid calculation without a first order phase-transition,
in which only local thermal equilibrium within each fluid
is assumed, yields similar flow values as the one fluid model with a phase
transition (solid squares in fig. \ref{flowfig1}) \cite{brachmann97a,dumitru97a}. The position of the minimum (the magnitude of
the overall effect) therefore strongly depends on the degree of stopping
(i.e. which type of fluid-dynamical model is employed)
and on the details of the chosen EoS and phase transition parameters.

\begin{figure}[hbt]
\begin{minipage}[t]{9cm}
\centerline{\psfig{figure=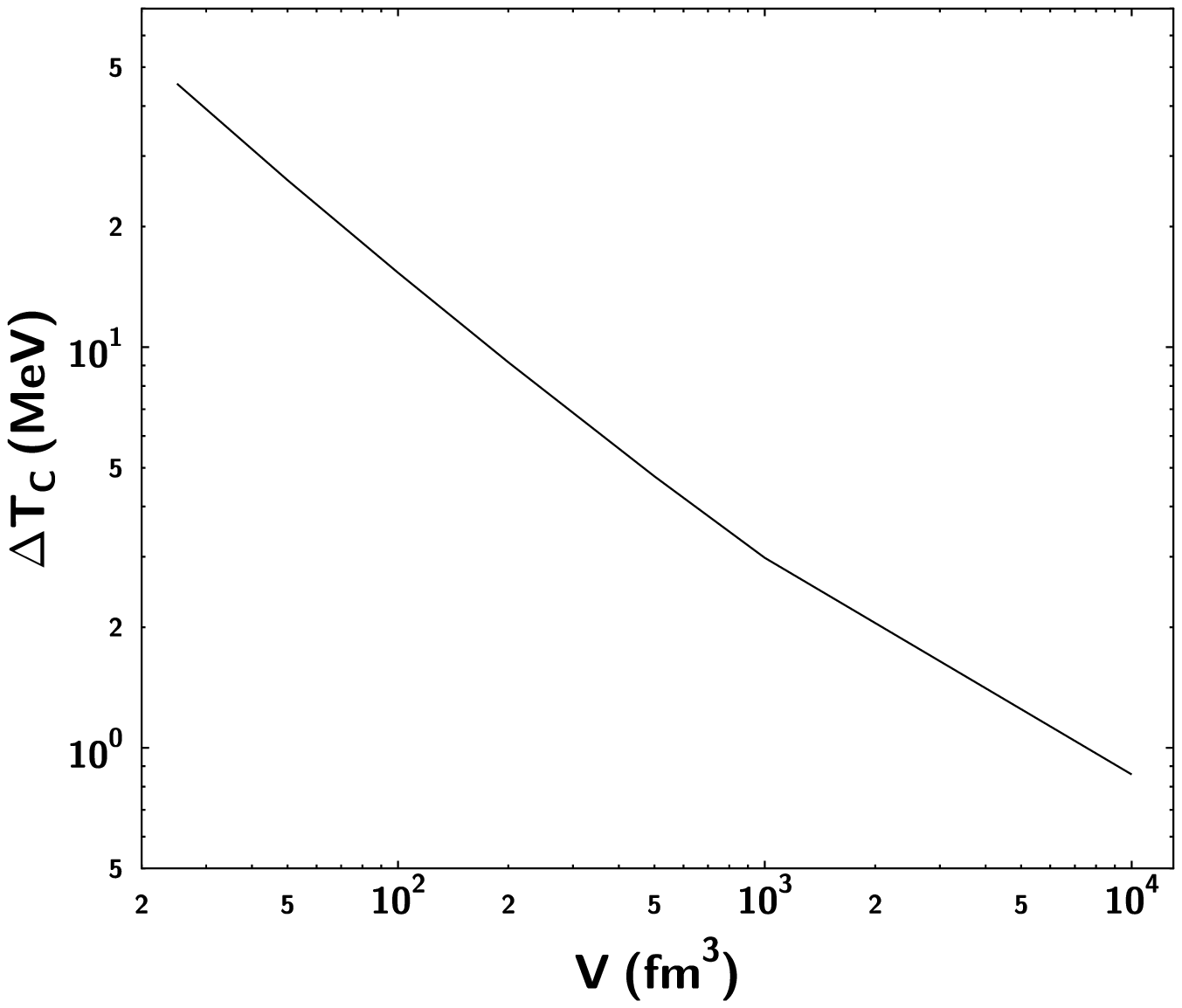,width=8cm}}
\end{minipage}
\hfill
\begin{minipage}[t]{9cm}
\centerline{\psfig{figure=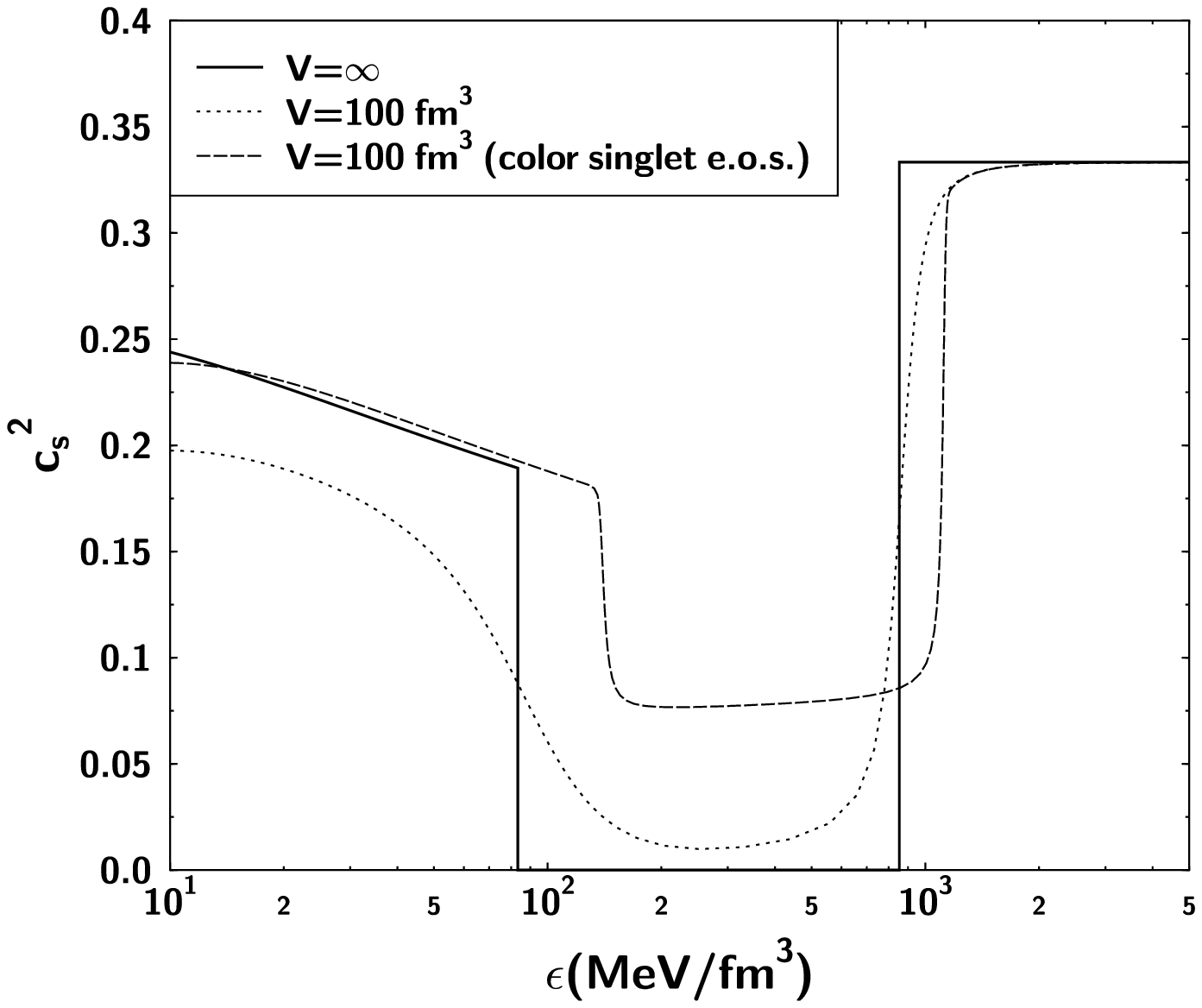,width=8cm}}
\end{minipage}
\caption{\label{finite_size}
Left: Shift of the critical temperature $\Delta T_{\rm C}$ vs.
the systems size $V$. The bag constant is
$B^{1/4}=200$~MeV. The color singlet constraint is taken into account for
the QGP equation of state.
Right: Speed of sound (squared) $c_s^2$ as a function of energy density
$\epsilon$ for three different
cases:\protect\newline 
1) infinite volume of the system (full line). \protect\newline
2) $V=100\,\rm fm^3$ using the infinite matter EoS (dotted). \protect\newline
3) $V=100\,\rm fm^3$ using the EoS with color singlet constraint
(dashed).\protect\newline Figures taken from \protect\cite{spieles97b}
}
\end{figure}
Moreover, for finite volumes, $V<100\; \rm fm^3$, corresponding to
expected plasma volumes in heavy ion collisions, there is a considerable
rounding in the variables $\epsilon/T^4$ and
$s/T^3$ around $T_C$, even if there is a first-order phase transition for
the infinite volume limit \cite{spieles97b}. 
This is inferred, under simple
assumptions, from basic thermodynamic considerations:
Fluctuations of the two phases in a finite
system lead to a smooth transition between the low temperature regime ---
where
the hadronic phase dominates the system --- and the high temperature regime
--- where the pure quark phase is most probable. 
Such a behavior has severe implications on the proposed signal 
as was shown in 
\cite{rischke96b}: a smooth crossover transition within an
assumed interval of $\Delta T=0.1 T_C$
results in drastically reduced time delays as compared to a sharp transition.

Another effect is present if explicit finite size effects and
the necessary requirement of exact color-singletness within the quark phase
\cite{elze83a,elze84a} is included \cite{spieles97b}.
The model exhibits a barrier in the free energy between the two phases 
near the phase transition. 
This leads to a shift of the critical
temperature to higher temperatures for finite volumes (see Fig. 
\ref{finite_size}).
The speed of sound is considerably increased
in the mixed phase.
The significance of the time delay signal for
the experimental detection of a QGP phase in heavy ion collisions, in turn,
becomes questionable.

\begin{figure}[htb]
\begin{minipage}[t]{9cm}
\centerline{\epsfig{figure=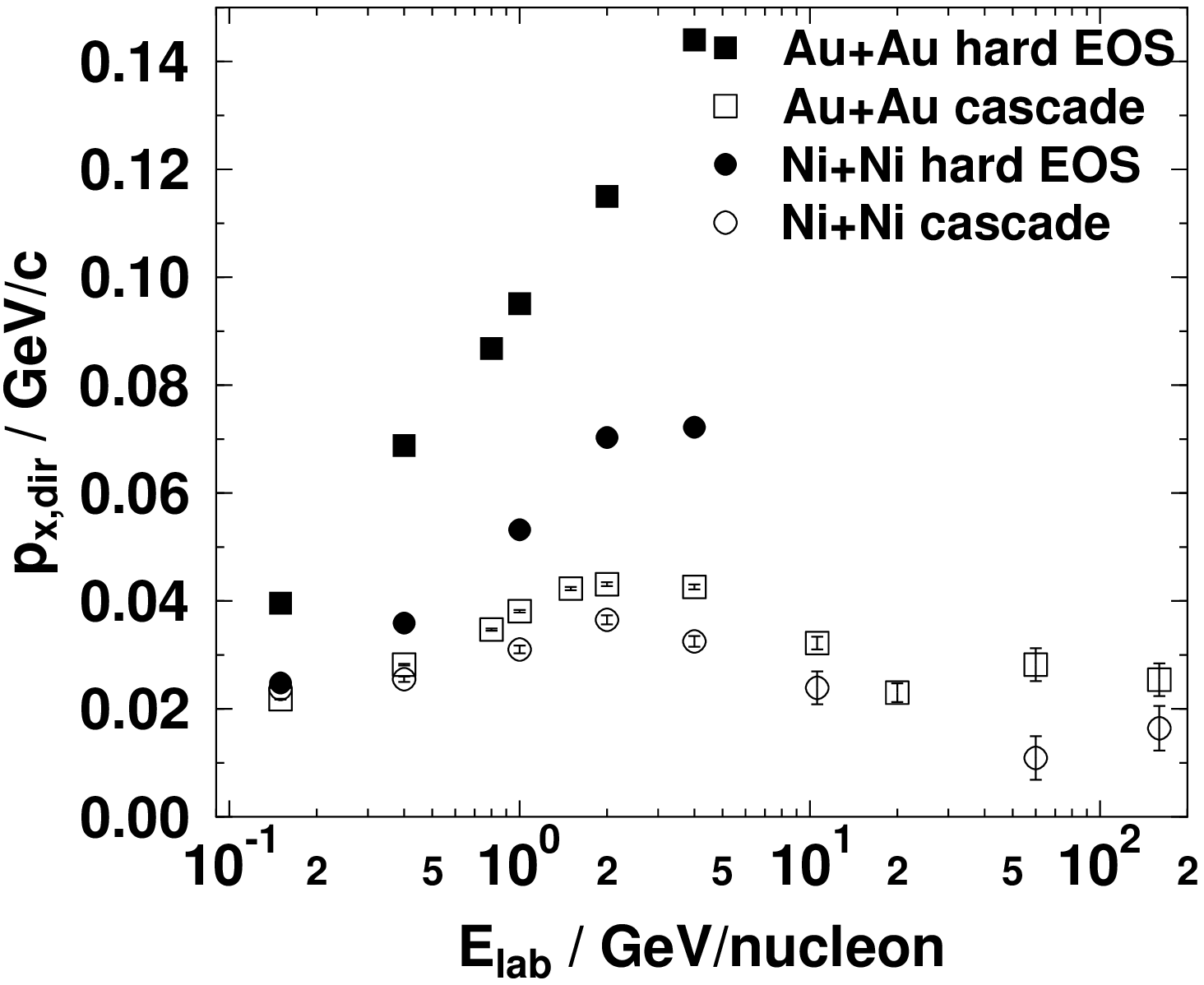,width=9cm}}
\caption{\label{fig:trans1} Excitation function of the total 
directed transverse momentum transfer
$p_x^{dir}$ for the Au+Au and Ni+Ni systems. UrQMD calculations 
including a hard equation of
state (full symbols) are compared to UrQMD cascade calculations
(open symbols).  }
\end{minipage}
\hfill
\begin{minipage}[t]{9cm}
\centerline{\epsfig{figure=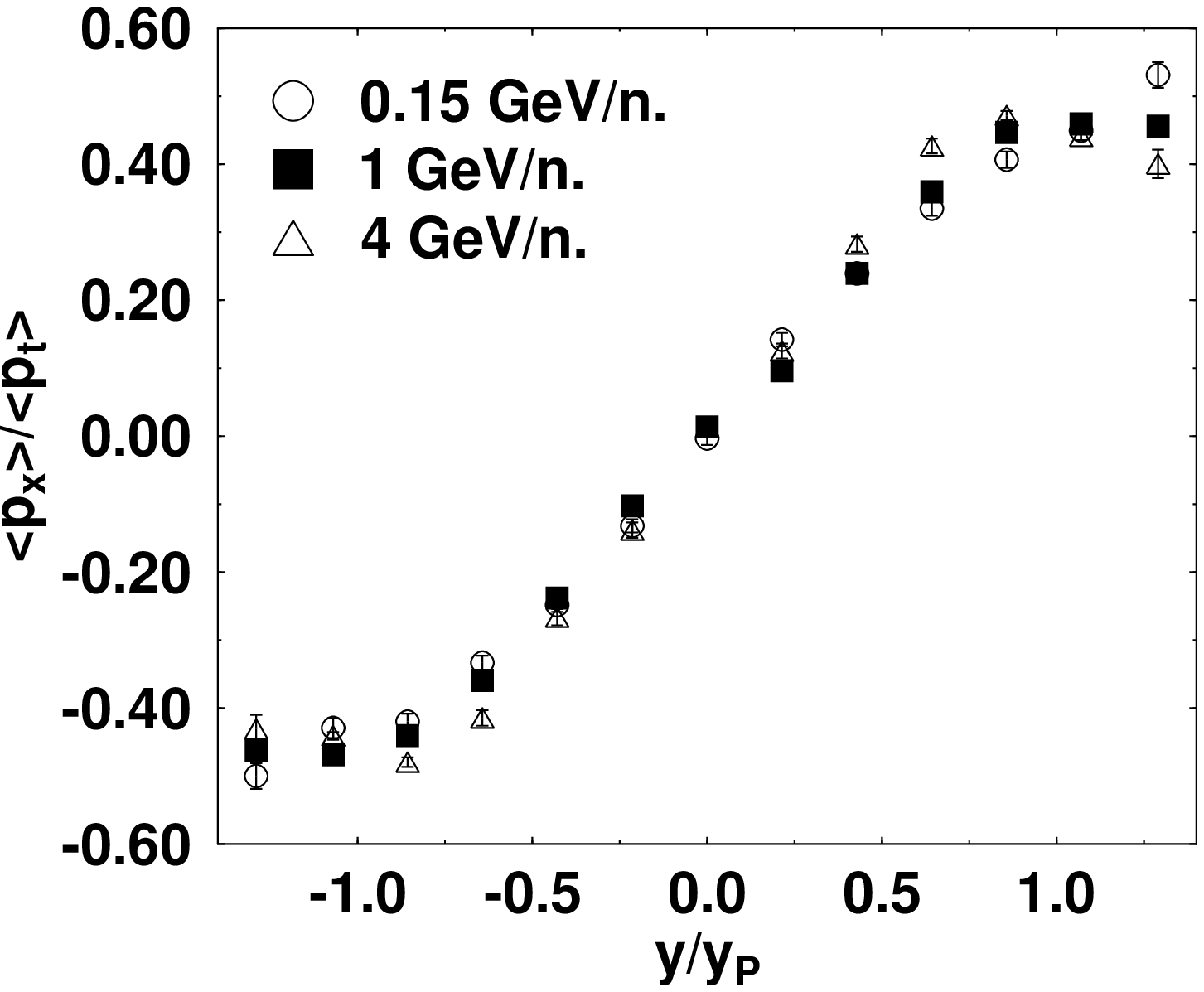,width=9cm}}
\caption{\label{fig:trans2} Mean directed transverse momentum as a 
function of the scaled rapidity for Au+Au at b=4 fm (UrQMD calculation).
If the transverse flow is scaled with the mean transverse momentum,
the sidewards does not depend on the bombarding energy. }
\end{minipage}
\end{figure}

A second order phase transition may not exhibit
this minimum in the flow excitation function:
The existence of a minimum in $p_{x,dir}(E_{lab})$ is rather a
{\em qualitative} signal for a strong first order transition.
If such a drop of $p_{x,dir}(E_{lab})$ is observed, it remains
to be seen which phase transition caused this behavior: a 
hadron--quark-gluon phase transition or, e.g., a nuclear matter to resonance matter
transition
in confined hadronic matter \cite{waldhauser87a,boguta81a}.

What does the purely hadronic UrQMD model predict with respect
to directed transverse flow?
Figure~\ref{fig:trans1} shows the averaged in plane transverse momenta
for Ni+Ni and Au+Au in the 0.1 -- 4 GeV/nucleon region which is accessible
through experiments at SIS and AGS.
Calculations employing a hard equation of state
(full symbols) are compared to cascade simulations (open symbols). In the
latter case only a slight mass dependence is observed. For the calculation
with potentials the integrated directed transverse momentum push per baryon
is more than twice as high for the heavier system which corroborates the 
importance of a non-trivial equation of state of hadronic matter.

For beam energies below 5 GeV/nucleon the amount of directed 
transverse momentum 
scales  in the very same way as the total transverse momentum in the 
course of the reaction. Therefore the $\langle p_x \rangle$ versus rapidity
divided by the $\langle p_t \rangle$ is identical for all beam energies in the
range of scaling as can be seen in figure~\ref{fig:trans2}.

		\subsection{Squeeze--out: flow perpendicular to the reaction
			plane}

The most strongly stopped, compressed matter
at mid-rapidity is responsible for the {\em squeeze--out} \cite{hartnack90a}.
A strong dependence on the nuclear equation of state for this
collective effect is seen \cite{hartnack92a,hartnack94a}.

\begin{figure}
\begin{minipage}[t]{9cm}
\centerline{\epsfig{figure=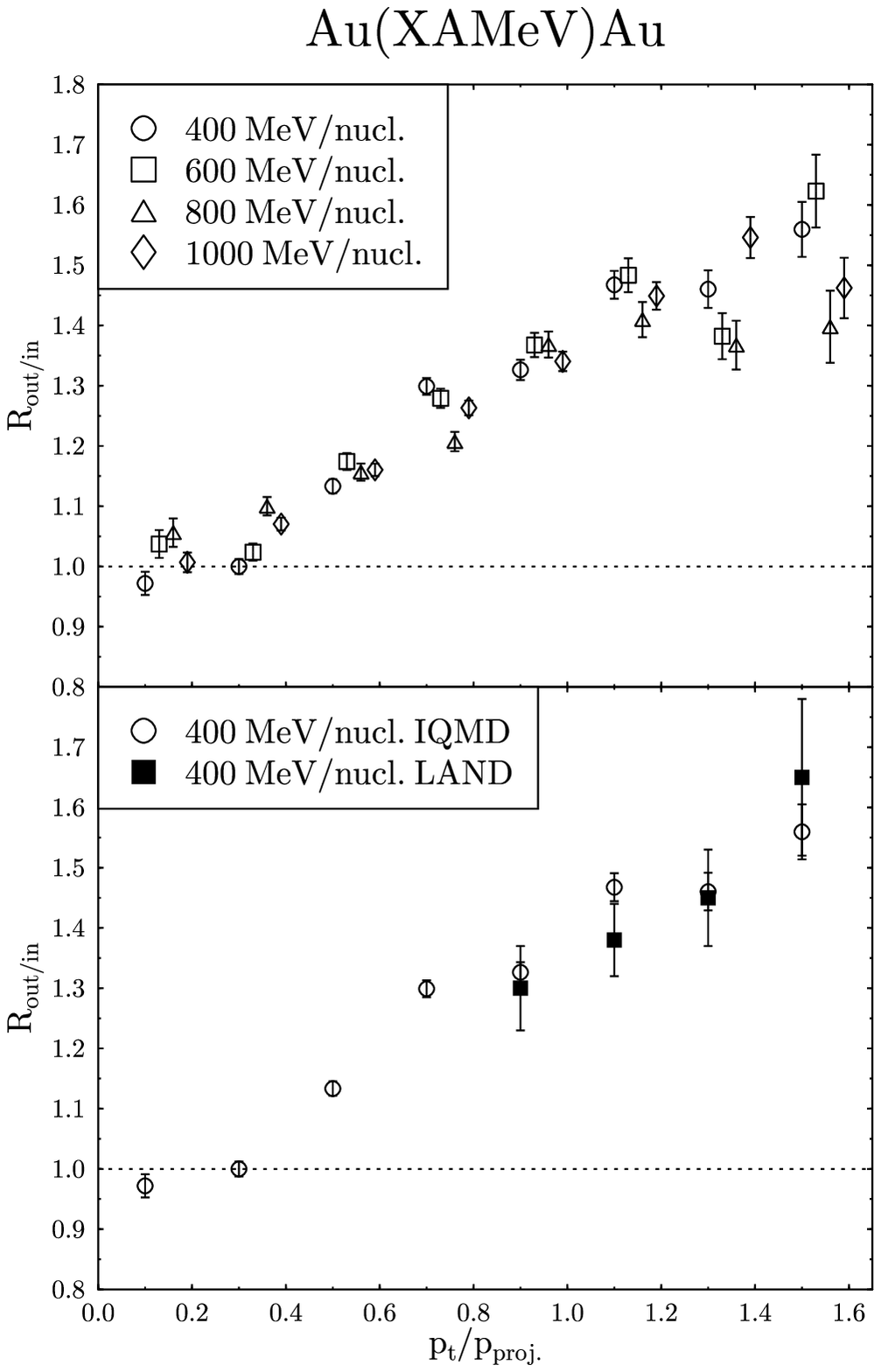,width=9.0cm}}
\caption{\label{nsqueeze2} {\em Squeeze--out ratio} 
$R_{out/in}$ versus scaled transverse momentum
$p_t/p_{proj}$ for neutrons in Au+Au collisions at 400, 600, 800 and 1000 
MeV/nucleon calculated with a hard equation of state without momentum 
dependent interaction (top) and a comparison between the IQMD calculation and
data by the LAND collaboration for 400 MeV/nucleon (bottom).
By scaling $p_t$ with
$p_{proj.}$ the ratio becomes independent of the incident beam energy (top).
The binning is identical for all 4 energies, the symbols were shifted to
increase the readability of the figure. The calculation shows good agreement
with the data (bottom).}  
\end{minipage}
\hfill
\begin{minipage}[t]{9cm}
\centerline{\epsfig{figure=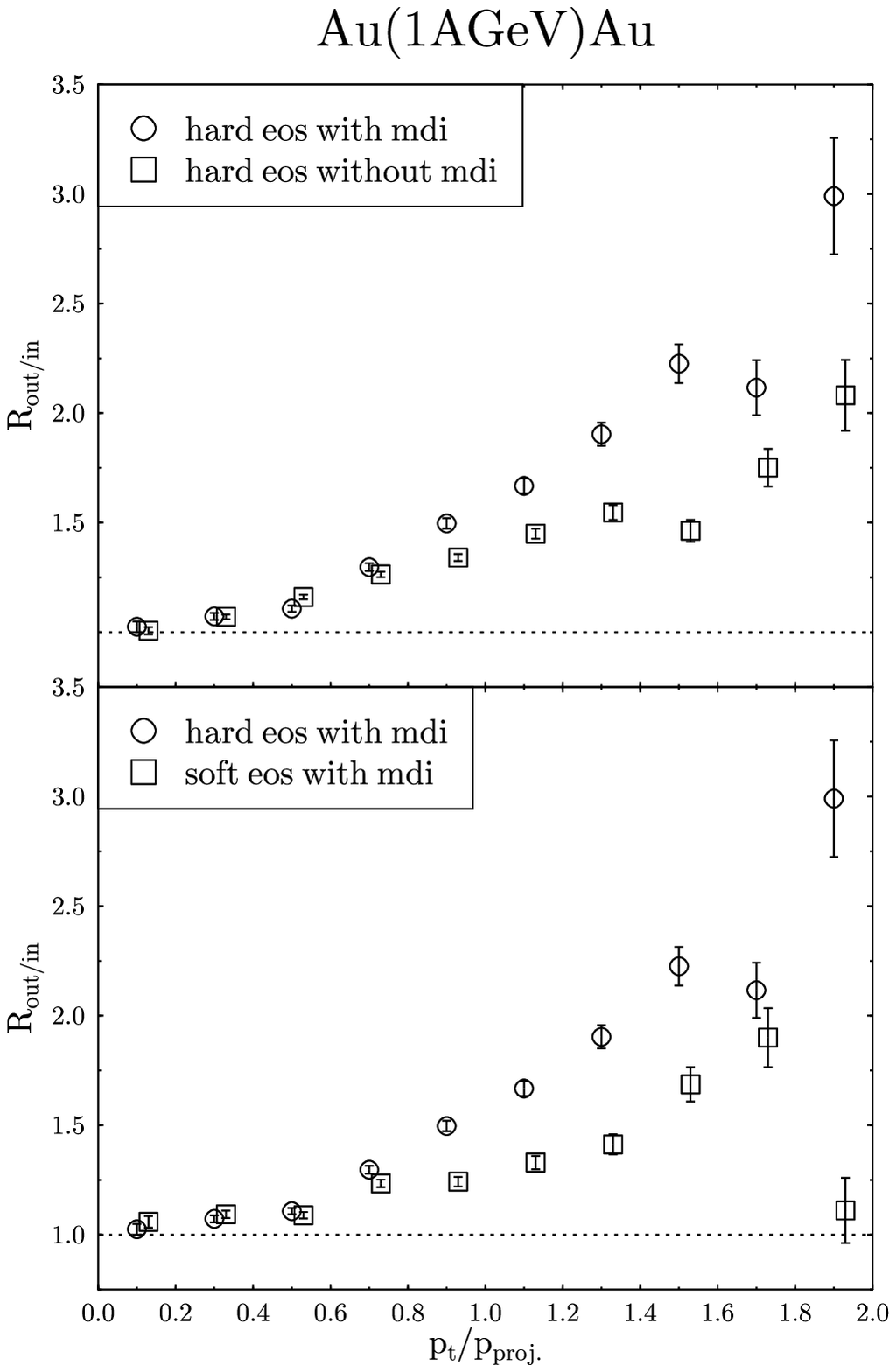,width=9.0cm}}
\caption{ \label{nsqueeze3} {\em Squeeze--out ratio} 
$R_{out/in}$ versus scaled transverse momentum
$p_t/p_{proj}$ for neutrons in Au+Au collisions at 1 GeV/nucl.. The upper frame
shows a comparison between IQMD calculations with a hard equation of state with and
without momentum dependent interaction (mdi). For large transverse momenta
$p_t/p_{proj} \ge 1$ the calculation with mdi exhibits a 50\% larger
{\em squeeze--out ratio}. The lower frame shows a comparison between calculations
with hard equation of state with mdi and soft equation of state with mdi:
Again, for large transverse momenta $p_t/p_{proj} \ge 1$ the hard equation
of state shows a 50\% higher {\em squeeze--out ratio} than the soft 
equation of state.}
\end{minipage}
\end{figure}

Let us now show the dependence of the observed {\em squeeze--out } on 
increasing beam energy, transverse momentum and impact parameter.
We define a {\em squeeze--out ratio} \cite{hartnack90a}
\begin{displaymath}
R_{out/in} \,=\, \left.\frac{ \frac{dN}{d\varphi}(\varphi=90^{\circ}) +
        \frac{dN}{d\varphi}(\varphi=270^{\circ})}
                            { \frac{dN}{d\varphi}(\varphi=0^{\circ}) +
\frac{dN}{d\varphi}(\varphi=180^{\circ})} \right|_{y=y_{CM}} \, .
\end{displaymath}
For values $R_{out/in} > 1$  neutrons are emitted preferentially perpendicular 
to the reaction plane.

\begin{figure}
\centerline{\epsfig{figure=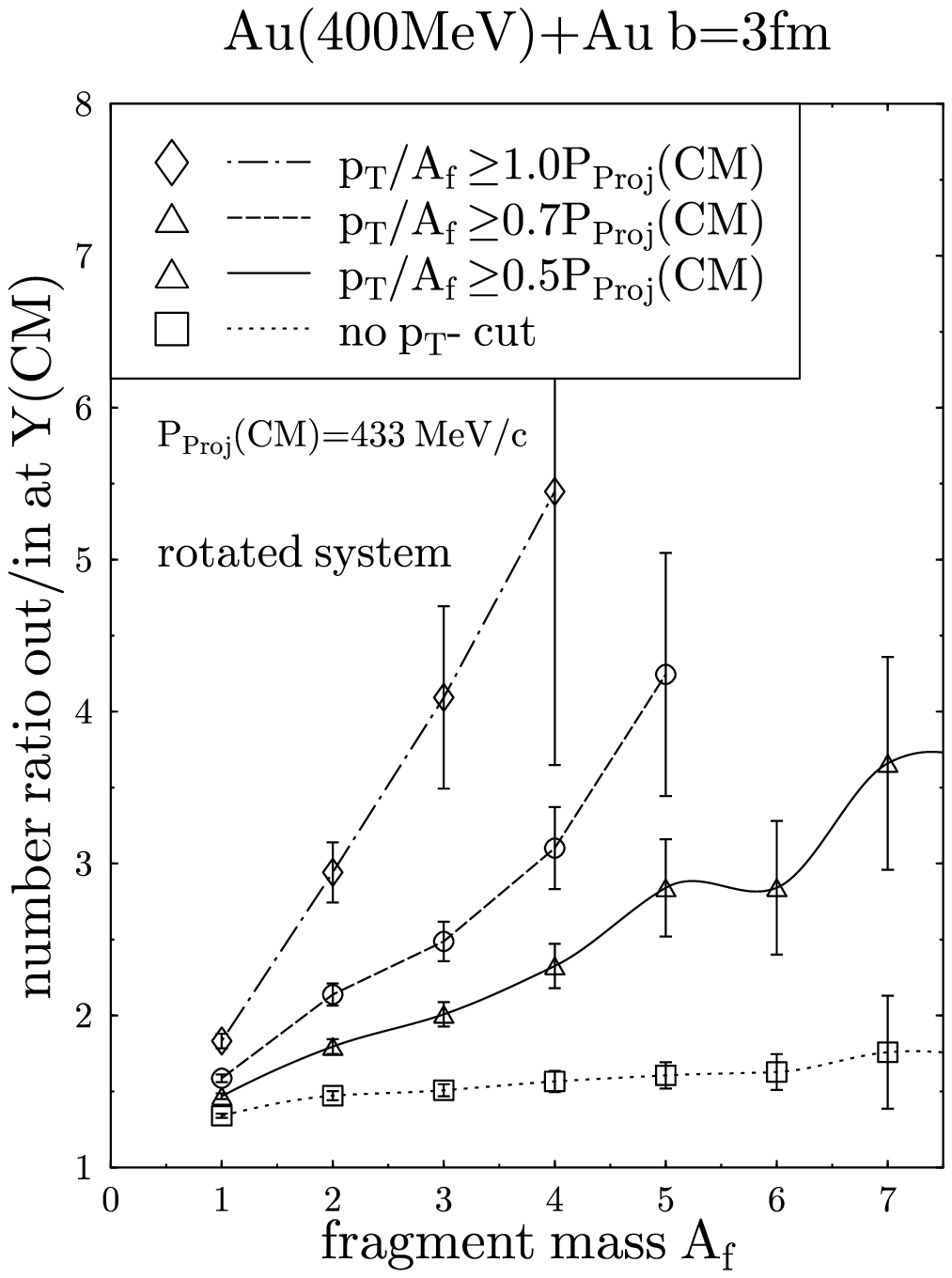,width=9.0cm}}
\caption{\label{fragsqueeze1} {\em Squeeze--out ratio} 
$R_{out/in}$ versus fragment mass $A_f$ for Au+Au collisions
at 400 MeV/nucleon (IQMD calculation). The system has been rotated by $\Theta_{flow}$ into
the principal axis system, which enhances $R_{out/in}$. 
The rise of $R_{out/in}$ with increasing
fragment mass is clearly visible, especially for high transverse
momenta $p_t/A_f$. The figure has been taken from
\protect \cite{hartnack94a}. }  
\end{figure}

The top frame of figure \ref{nsqueeze2} 
shows $R_{out/in}$ for Au+Au collisions versus  scaled transverse momentum
$p_t/p_{proj}$ for 
beam energies of 400, 600, 800 and 1000 MeV/nucleon
with cuts on  rapidity ( $-0.15 \le y_{CM} \le 0.15$) and
impact parameter ($ 3 \le b\; [fm] \le 9$).
The ratio increases monotonously with the transverse momentum and
is independent of the beam energy if the transverse
momenta are scaled with the projectile momenta.  
Therefore, higher transverse momenta must be
probed with increasing beam energy if a clean {\em squeeze--out} signal
is to be isolated.
Such a scaling behavior has already   been extracted for the
{\em in--plane bounce--off} in the hydrodynamic model  \cite{bonasera88a}.
It has been experimentally  observed in the case of the
neutron {\em squeeze--out} by the LAND collaboration \cite{lambrecht94a}. 
The respective comparison between the IQMD model and LAND data
(mid-rapidity, ERAT2 centrality criterion) is shown
for 400 MeV/nucleon in the lower frame of figure \ref{nsqueeze2}. 
One should note, however,
that in the IQMD calculation the reaction plane is always known, whereas
the experimental determination of the reaction plane yields fluctuations 
which might underestimate the measured ratio by 15 to 30\%. 
The $p_t$ dependence of the squeeze-out ratio is even enhanced for heavier
clusters, as can be seen in figure~\ref{fragsqueeze1}.

The dependence of the neutron {\em squeeze--out} on the momentum dependent 
interaction and the equation of state  is shown in 
figure \ref{nsqueeze3}  (for Au+Au collisions at 1 GeV/nucleon): 
The upper frame shows the hard equation of state with and without momentum
dependent interaction (mdi). With mdi, $R_{out/in}$ increases
by 50\% for large transverse momenta! A difference of 50\% can also be
observed by comparing the hard and soft equations of state (both including mdi)
in the lower frame of figure \ref{nsqueeze3}:
At high transverse momenta the hard equation of state with mdi exhibits a
50\% higher {\em squeeze--out ratio} than the soft equation of state with mdi.
The hard equation of state without mdi shows a $R_{out/in}$ vs. $p_t$
dependence similar to the soft equation of state with mdi. 
It is important to note, however, that these differences are only seen
for high transverse momenta $p_t/p_{proj} \ge 1$. 
Analyzing $p_t$-integrated data will therefore
severely diminish the sensitivity towards the nuclear equation of state
and the momentum dependent interaction.  

\begin{figure}
\centerline{\epsfig{figure=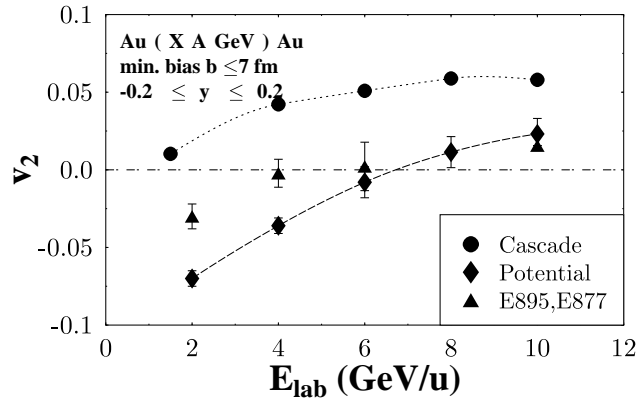,width=9.0cm}}
\caption{\label{svenfig} UrQMD calculation of the elliptical flow parameter 
$v_2$ for Au+Au collisions with (diamond) and without (circle) potentials. Also
shown (triangle) are the preliminary data of the E877 and E895 collaborations 
\cite{qm97}.}  
\end{figure}

In fig. \ref{svenfig},
the excitation function of the coefficient $v_2$ is shown representing the 
so-called elliptical flow. $v_2$ is 
extracted from a fit to the azimuthal distribution of nucleons according to 
$dN/d\phi=v_0[1+v_1\cos(\phi)+v_2\cos(2\phi)]$. Positive values of 
$v_2$ correspond to a preferential in-plane enhancement of the emitted 
particles while negative values describe 
preferred emission perpendicular to the reaction plane. 
 
Clearly, large differences are seen when comparing calculations with and 
without potentials. On the one hand, cascade calculations only show in-plane 
enhancement of the 
emitted particles. On the other hand, calculations including nucleonic 
potentials show a transition from in-plane to out-of-plane 
emission with increasing 
bombarding energy which is seen by current experiments of the EOS collaboration 
\cite{qm97}. This big sensitivity to the model ingredients gives a handle to 
pin them down further.

\chapter{Summary and conclusions}
\label{conclusion}

This review has by no means  covered the entire field of 
microscopic transport model applications to (ultra-)relativistic
heavy ion collisions. To cover the entire field in one review is
almost impossible due to its fast development pace and the width
of applications which have been added in the last couple of years.
Instead, we have selected a sample of topics which allow to elucidate
different important features of microscopic non-equilibrium 
transport theory.
Of course a ``good'' transport model must be able to come close to known
correct data. It should 
predict the outcome of future experiments. However, it's
usefulness must stretch far beyond that of a mere event-generator:
it allows for the analysis of the underlying physics  (the ``input'' of
the transport model) and shows how the various components contribute 
to the final result of the calculation.

In chapter~\ref{manybody}
the concepts of microscopic transport theory have been introduced  and
the features and shortcomings of the most commonly used
ansatzes are discussed. For pedagogical reasons transport theory
is first discussed extensively for the non-relativistic case and
the relativistic extension is introduced in a subsequent section.
An introduction to the Quantum Molecular Dynamics model concludes
this chapter.

The UrQMD transport
model has been described in great detail in chapter~\ref{urqmdkap}.
Based on the same principles as QMD and RQMD it incorporates a
vastly extended collision term with full baryon-antibaryon symmetry,
55 baryon and 32 meson species. Isospin is explicitly treated for
all hadrons. The range of applicability stretches from $E_{lab}< 100$ MeV/nucleon
up to $E_{lab}> 200$ GeV/nucleon, allowing for a consistent
calculation of excitation functions from the intermediate energy domain up
to ultrarelativistic energies. 
The UrQMD model as described in chapter~\ref{urqmdkap} is no static 
structure. Its most important purpose is to serve as a framework
into which new transport theoretical concepts and physics ideas can
be incorporated, so that their effects on (ultra-)relativistic heavy ion
reactions may be explored.

Among the tasks to be tackled in the near future are 
the implementation of a relativistic generalization of n-body forces,
an improved treatment of the imaginary part of the relativistic hadron
self-energy
(off-shell propagation), medium-dependent cross sections, an
improved treatment of resonance lifetimes and time-delays as well
as the inclusion of partonic degrees of freedom. The latter is 
important for the extension and continued application of 
transport models at collider energies (RHIC and LHC). The intriguing role
of color coherence phenomena (transparency and opacity) for fluctuations,
stopping and charmonium production can be studied best at these energies.
However,
even at these high energies, hadronic degrees of freedom must not
be neglected since hadronic interactions in the late reaction
phase may considerably change the hadrochemical cocktail and phase space
distribution of the system (e.g. via feeding).

Applications of transport models to (ultra-)relativistic heavy ion 
collisions have then been reviewed in chapter~\ref{hotdense}.
The main topics under discussion have been stopping, particle production
and collective flow. The main emphasis in the stopping section has
been on the scaling behavior of the rapidity distribution with 
respect to the incident beam energy and collision system mass as well
as the dependence of the stopping behavior on the (di-)quark hadron
cross section (i.e. the formation time dependence). 

The section on particle production has carried the largest weight in
this review:  subthreshold production of antiprotons has been used
as an example of subthreshold  particle production in heavy ion collisions
-- the ideas and concepts being very similar for other particle species
such as etas and kaons. Multi-step processes and the excitation of heavy
resonances dominate this domain of particle production. Therefore
subthreshold particle production is a sensitive probe for the investigation
of in-medium cross sections and the nuclear equation of state. 
Single particle spectra have been discussed for
protons, deuterons, tritons and pions. The proton and cluster spectra
have been used to point out problems with the source temperature extraction
from single particle spectra. Pion energy spectra show how different regions of the spectra may be 
sensitive to different sources and reaction stages.

Particle ratios have been used to discuss problems of temperature and chemical
potential extraction via chemical equilibrium assumptions. 
The rapidity dependence of certain particle ratios may allow
for the in-depth investigation of the baryon stopping mechanism
in heavy ion collisions. The UrQMD exhibits
the properties of a free hadron gas, however with a limiting temperature
of approximately 140 MeV. The central cell in Au+Au reactions
at AGS seems to be close to local thermal and chemical equilibrium.  
  
The formation of resonance matter has been discussed in
great detail. The fraction of excited matter in the collision
system rises continuously from 20\% at SIS up to 70\% at CERN SPS energies.
The properties and interactions
of highly excited resonance matter is an important topic for further research.
Creation of strange matter is discussed,
although explicit {\em strangelet}-formation has yet to be included into
microscopic transport model calculations.
Hyperclusters and (anti-)deuterons have served as examples for
the formation of composite particles. The main emphasis of the 
discussion has been on the phase space distributions of anti-deuterons
and the sensitivity of the respective (centrality dependent) yields
on parameters such as the formation time and the treatment of the 
baryon-antibaryon annihilation cross section.

Freeze-out distributions of mesons and hyperons at SIS and SPS energies 
show that global equilibration is not achieved
in heavy ion reactions. The distributions indicate 
a complex non-equilibrium time evolution of the hadronic system. 
A thermal model although quite successful in fitting the data, 
may not be yielding the correct physics interpretation.

Electromagnetic probes allow for an unhindered view into the hot and dense
reaction zone. This review focused on dilepton production
(both at BEVALAC and SPS energies)
and the creation of secondary Drell-Yan pairs at CERN/SPS energies, two
of the most current topics. The latter offer a novel 
explanation for the observed dimuon excess in the intermediate mass
range of $1.5 \le m \le 2.5$ GeV of the dimuon-spectrum.
A brief discussion of charmonium production and suppression concludes
the section on particle production.

The final section of chapter~\ref{hotdense} deals with collective flow. 
The excitation function of the directed
flow in the reaction plane can provide a signature for the hadron gas
to quark gluon plasma phase transition. However, finite size effects
may severely diminish the signal. Transport model calculations
(which in most cases do not contain a phase transition) serve here again 
as an important smooth baseline to check the occurrence of possible 
QGP signatures, namely as irregularities in the excitation function, taken in
small steps, $\Delta E \approx$ 5 GeV or so.
The sensitivity of flow perpendicular to the reaction plane
to the nuclear equation of state is analyzed. Especially the squeeze-out
excitation function seems to offer new possibilities for the investigation
of the density dependence of the nuclear equation of state.

Microscopic transport models like UrQMD are a unique tool to further develop
the general understanding of the dynamics of heavy
ion collisions and interactions over a vast energy range
from the Coulomb barrier (several MeV per nucleon) to the
highest energies currently available or planned for the future.
Their development is by far not complete. Nevertheless they allow
for important insights into the field of hot, dense hadronic matter created in 
(ultra-)relativistic heavy ion reactions.

\bibliography{main}

\end{document}